%% file: Main.tex
\documentclass{BaylorThesis} % based on REPORT class, using these options: 12pt,letterpaper,oneside,onecolumn
\usepackage{lipsum}

\usepackage{amsfonts, amssymb, amsmath, latexsym, mathrsfs}
\usepackage{amsthm}
\allowdisplaybreaks[1]

\usepackage{url}

\usepackage[T1]{fontenc}
\usepackage[utf8]{inputenc}

\usepackage[all]{nowidow}

\usepackage{ragged2e}
\usepackage[numbers]{natbib} 
\usepackage{float}

\usepackage{graphicx} %Graphics/figures

\usepackage{rotating} 

\usepackage{booktabs,threeparttable} 

\usepackage{tikz}
\usetikzlibrary{calc,arrows,positioning,shapes,patterns}

\usepackage{longtable} 
\usepackage{array}
\newcolumntype{C}[1]{>{\centering\let\newline\\\arraybackslash\hspace{0pt}}p{#1}}
\newcolumntype{L}[1]{>{\raggedright\let\newline\\\arraybackslash\hspace{0pt}}p{#1}}

\usepackage{listings}
\sectionnumbering{\secnumtrue} %section numbering
\usepackage{subfiles}
\usepackage{fix-cm}
\usepackage{enumitem}
\usepackage{multicol}
\usepackage{relsize}
\usepackage{xcolor}
\usepackage{lastpage}
\usepackage{blindtext}
\usepackage{subcaption}
\usepackage{wrapfig}
\usepackage{indentfirst}
\usepackage{soul}
\usepackage{empheq}
\usepackage{tensor}
\usepackage[section]{placeins}
\setlist{parsep=0pt,listparindent=\parindent}
\setlist[itemize]{noitemsep, topsep=0pt}
\setlist[enumerate]{noitemsep, topsep=0pt}
\setlist{parsep=0pt,listparindent=\parindent}

\stepcounter{footnote}
\catcode`\^=13\def^#1{\sp{#1}{}}
\catcode`\_=13\def_#1{\sb{#1}{}}
\graphicspath{{Images/}{../Images/}}

\title{Curvature Invariants for Wormholes and Warped Spacetimes}
\author{B.~Mattingly}
\degrees{B.A., M.S.} % degrees held before this Ph.D.

\mentor{G. Cleaver, Ph.D.}
\reader{G. Benesh, Ph.D.}% your committee members and degree in alphabetical order
\readerThree{J. Dittmann, Ph.D.}
\readerFour{Q. Sheng, Ph.D.} 
\readerFive{A. Wang, Ph.D.}

\confDate{May 2020} % month and year of actual graduation

\makeCopyrightPage

\graduateDean{J. Larry Lyon, Ph.D.}
\deptChair{D. Russell, Ph.D.}

\abstract{
    The Carminati and McLenaghan (CM) curvature invariants are powerful tools for probing spacetimes. Henry et al. formulated a method of plotting the CM curvature invariants to study black holes. The CM curvature invariants are scalar functions of the underlying spacetime. Consequently, they are independent of the chosen coordinates and characterize the spacetime. For Class B$_1$ spacetimes, there are four independent CM curvature invariants: $R$, $r_1$, $r_2$, and $w_2$. Lorentzian traversable wormholes and warp drives are two theoretical solutions to Einstein’s field equations, which allow faster-than-light (FTL) transport. The CM curvature invariants are plotted and analyzed for these specific FTL spacetimes: (i) the Thin-Shell Flat-Face wormhole, (ii) the Morris-Thorne wormhole, (iii) the Thin-Shell Schwarzschild wormhole, (iv) the exponential metric, (v) the Alcubierre metric at constant velocity, (vi) the Nat\'ario metric at constant velocity, and (vii) the Nat\'ario metric at an accelerating velocity. Plots of the wormhole CM invariants confirm their traversability and show how to distinguish the wormholes. The warp drive CM invariants reveal key features such as a flat harbor in the center of each warp bubble, a dynamic wake for each warp bubble, and rich internal structure(s) of each warp bubble.  \\
}

\abbreviation{ %make rows that continue definition single spaced
\singlespacing %make rows between definitions double spaced
\setlength{\extrarowheight}{1\baselineskip}
\begin{tabular}{L{.2\textwidth}L{.75\textwidth}}
CM & 	Carminati and McLenaghan\\
CTC &   Closed-timelike-curves\\
FTL &   Faster-than-Light\\
GR & 	General Relativity\\
MT &    Morris-Thorne\\
NEC &   Null-energy-condition(s)\\
PhD & 	Doctor of Philosophy\\
QM &    Quantum Mechanics\\
SP &    Scalar-Polynomial\\
SR &    Special Relativity\\
TS &    Thin-Shell
\end{tabular}
}

\acknowledgements{
Thank you, Dr. Davis for inspiring this research and helping me understand the basics of wormholes and warp drives.

Thank you, Dr. McNutt for helping me to understand SP invariants.

Thank you, Dr. Cleaver for this amazing opportunity to study the boundaries of Physics.

Thank you, Abinash, Cooper, Mat, Will, and Caleb for helping me at each step to complete this research.

Thank you, A.~Alexander~Beaujean, Ph.D. (Associate Professor of Psychology \& Neuroscience) for providing the \LaTeX\ template for this dissertation.}

\dedication{To my parents, Ed and Jill. 
You two supported me every step along this journey.
During the highs, y'all were there to celebrate.
During the lows, y'all were there to encourage.
I love both of y'all so much.
Godspeed.}

\theoremstyle{definition}
\newtheorem{definition}{Definition}[section]

\begin{document}
\pagenumbering{arabic}

%%%%% INTRODUCTION %%%%%%%
\input{ch1.tex}
% 
% %%%%%% Chapter 2 %%%%%%%%
\input{Ch2.tex}
% 
% %%%%%% Chapter 3 %%%%%%%%
\input{Ch3.tex}
% 
% %%%%%% Chapter 4 %%%%%%%%
\input{Ch4.tex}
% 
% %%%%%%% Chapter 5 %%%%%%%%
\input{Ch5.tex}

% %%%%%%% Chapter 5 %%%%%%%%
\input{Ch6.tex}

%%%%%%%%%%%%%%%%%%%%%%%%%%%%%%%%%%%%%%%%%%%%%%%%%%%%%%%%%%%%%%
% APPENDIX
%%%%%%%%%%%%%%%%%%%%%%%%%%%%%%%%%%%%%%%%%%%%%%%%%%%%%%%%%%%%%%

\clearpage
\vspace*{3.25in}
 \begin{center}
      APPENDICES
 \end{center}
 \pagestyle{plain}
  \pagebreak

\newpage
\appendix
\renewcommand\thesection{\Alph{chapter}.\arabic{section}}
\input{Appendix.tex}

%%%%%%%%%%%%%%%%%%%%%%%%%%%%%%%%%%%%%%%%%%%%%%%%%%%%%%%%%%%%%%
% REFERENCES
%%%%%%%%%%%%%%%%%%%%%%%%%%%%%%%%%%%%%%%%%%%%%%%%%%%%%%%%%%%%%%

\newpage
\raggedright

%\bibliographystyle{BibTeX} % You may have to select another style. Remember: LaTeX, BibTeX, LaTeX, LaTex to get the citations to appear
%\raggedright
%\urlstyle{same}
%\bibliography{references.bib}
\input{References.tex}

\end{document}

%% file: ch1.tex
\chapter{Introduction}
\section{Historical Foundations}
\label{chp1:HF}
Faster-Than-Light (FTL) travel fascinates and delights the imaginations of scientists and science enthusiasts alike.
FTL promises a voyage into the final frontier.
It tempts us with possibilities and answers beyond what we can either find or conceive here on Earth.
Unfortunately, physical reality constrains what is possible.
Einstein and his theories of relativity teach us that the speed of light, $c=299,792,458 \frac{\mathrm{m}}{\mathrm{s}}$, is the speed limit for travel inside of the universe.
Matter and energy absolutely cannot move through spacetime faster than $c$.
However, Einstein did leave us a loophole.
Spacetime itself has no such limitation.

General Relativity (GR) successfully describes classical physics on the largest of scales.
It precisely lays the theoretical foundation for how the stars dance through the galaxies, conducted by gravity.
Its every major prediction has been rigorously tested.
Most recently, its prediction of gravitational waves was confirmed by LIGO.
But, there exist solutions in Einstein's equations
\begin{equation}
    G^{\mu\nu}=\frac{8\pi G}{c^4}T^{\mu\nu} \tag{1.1} \label{eq:1.1}
\end{equation}
that allow extraordinary consequences.
By considering spacetimes with interesting properties, Eq.~\eqref{eq:1.1} may be solved in reverse for its possible matter source.
In this manner, ``exotic matter'' sources allow FTL spacetimes in GR without requiring new physics such as the unification of GR and Quantum Mechanics (QM).

It is emphasized at this point that these FTL solutions are mathematically reasonable, but physically extreme.
The ``exotic matter'' at the source of each solution violates every known null-energy-condition (NEC).\footnote{A NEC asserts that for any null vector, $k^\mu$, the stress energy tensor, $T^{\mu\nu}$ in Eq.~\eqref{eq:1.1} must satisfy $T_{\mu \nu} k^\mu k^\nu \geq 0$ \cite{Lobo:2017}.
Matter that violates the NEC is called ``exotic matter.''}
Moreover, the solutions have physical implications like closed-timelike-curves (CTC), which may violate causality.
Instead of dismissing FTL solutions as physically unrealistic and impossible, their best use is in exploring the theoretical limits of GR in the same way that the black-body radiation experiments and the photoelectric effect probed the limits of classical physics.

In the last one hundred years, physicists identified two main possibilities for FTL travel.
The first is the Einstein-Rosen Bridge, colloquially known as a wormhole.
The second is the warp drive developed by Alcubierre, Van Den Broeck, Krasnikov, Nat\'ario and others.
To begin this dissertation, two gedanken experiments will be presented for each FTL possibility.
Each one will be constructed in GR alone to demonstrate that these possible means of FTL require no exotic physics.

\subsection{Einstein-Rosen Bridge}
\label{chp1.1:EPR}
The first hint of FTL travel dates to the early development of GR \cite{Flamm:1916}.
It was the seminal paper by Einstein and Rosen which introduced the first series of mathematical calculations for a wormhole \cite{Einstein:1935}.
In it, they showed a "bridge" which connected distant regions of asymptotically flat spacetime.
While traveling through the bridge is fatal, it is a good test example of FTL.

The Einstein-Rosen Bridge is a submanifold of the Kruskal black hole in GR \cite{Visser:1995,DInverno:1992}.
The Kruskal black hole is a maximal and unique solution to the Schwarzschild solution. 
Its line element is
\begin{equation}
    ds^2=\frac{16m^2}{r} e^{-\frac{r}{2m}}d{t'}^2-\frac{16m^2}{r} e^{-\frac{r}{2m}}d{x'}^2-r^2(d\theta^2+\sin^2\theta d\phi^2), \tag{1.2} \label{eq:1.2}
\end{equation}
with $r$ being a function determined by ${t'}^2-{x'}^2=-(r-2m) e^{-\frac{r}{2m}}$, $t'$ and $x'$ being related to the advanced null coordinate, $v'$, and the retarded null coordinate, $w'$, by $t'=\frac{1}{2}(v'+w')$ and $x'=\frac{1}{2}(v'-w')$.
Making the free choice, $t'=0$, and taking the cross section, $\theta=\frac{1}{2}\pi$, the line element then becomes 
\begin{equation}
    ds^2=-F^2d{x'}^2+r^2d\phi^2. \tag{1.3} \label{eq:1.3}
\end{equation}

\begin{figure}[ht]
\centering
	\includegraphics[scale=0.5]{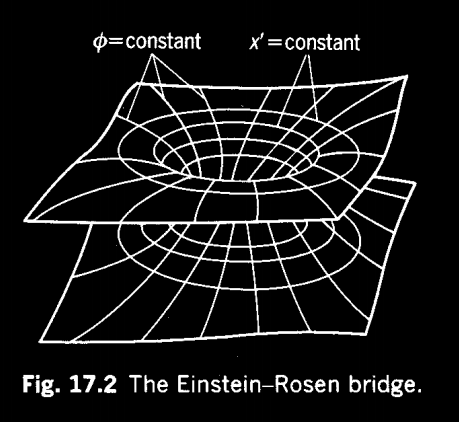} %this figure could be made bigger
	\caption{Figure 17.2 from \cite{DInverno:1992}. It depicts the 2D analog of the time evolution of Eq.~\ref{eq:1.2} rotated about the central vertical axis.} \label{fig:1.1}
\end{figure}

\begin{figure}[ht]
\centering
	\includegraphics[scale=0.70]{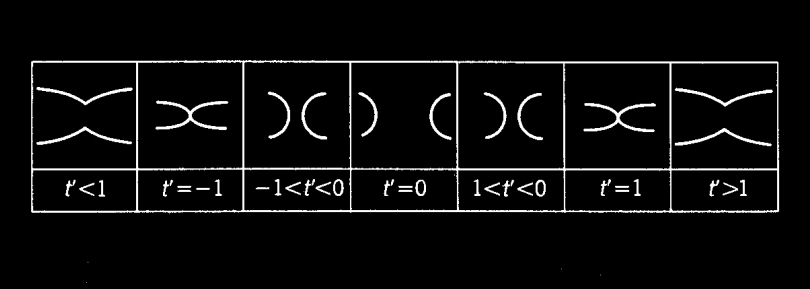}
	\caption{Figure 17.3 from \cite{DInverno:1992}. It depicts the full time evolution of Eq.~\ref{eq:1.2}.} \label{fig:1.2}
\end{figure}

From Fig.~\ref{fig:1.1}, the chosen cross section of the Kruskal manifold can be described as a "bridge" in Einstein-Rosen's words.
In modern terms, it is a wormhole connecting different parts of an asymptotically flat universe.
In the time period of $t'<1$ in Fig.~\ref{fig:1.2}, two distinct, asymptotically flat Schwarzschild manifolds exist before connecting the wormhole. 
At $t'=-1$, the previously separate manifolds pinch and connect. 
During $-1<t'<1$, the manifolds are connected and particles may pass uninhibited from one manifold to the other.
At $t'=1$, the manifolds disconnect.
For $t'>1$, the manifolds remain separate, and the particles that transferred during the pinch will remain in their new manifold.
To a distant observer, the particles will appear to have traveled through vast distances of spacetime, though from the particles' perspective they traveled the short distance of the pinch.

The EPR bridge naturally leads to many of the terms used to describe a wormhole.
The narrowest part of the geometry that allows passage is often called the throat of the wormhole and can be seen in the $t'=0$ slice of Fig.~\ref{fig:1.2}.
The region nearby is called the bridge or wormhole, where travel from one manifold to the other is possible.
The two asymptotically flat regions of space connected by the bridge are labeled patches.

Due to its construction, the EPR bridge is equivalent to the maximally extended Schwarzschild solution.
It is merely a coordinate artifact due to a special choice of coordinate patches.
An observation of one would appear as a black hole through our telescopes.
While this is a nice toy model of how a wormhole could exist, it is a highly simplified form of the Schwarzschild equation and requires an extreme amount of exotic matter to maintain in reality.
The discovery and identification of a real wormhole in our universe remains as an open question in physics. 

	~

\FloatBarrier

\subsection{Tipping Light Cones}
\label{chp1:TLC}
\begin{figure}[ht]
\centering
	\includegraphics[scale=1.0]{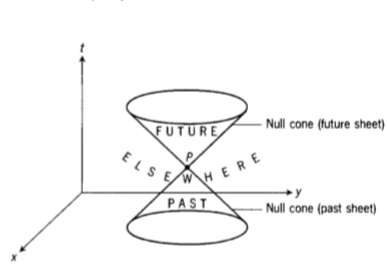}
	\caption{Figure 8.3 from \cite{DInverno:1992}. It shows the null cone or light cone of events relative to P.} \label{fig:1.4}
\end{figure}

%In this section, I will present Dr. Davis notion of Tipping Light Cones.
%The superluminal censorhip theorem provides a basis for the amount light cones may be tipped over.
In special relativity (SR), the light cone of an event, P, consists of all null geodesics\footnote{A geodesic is a curve that is straight and uniformly parameterized as measured by each local Lorentz frame along its length \cite{Misner:1974}.
It is a generalization of Euclid's statement that ``the shortest distance between two points is a straight line'' in flat space to curved space.} passing through the point P \cite{DInverno:1992}.
An example of a light cone is presented in Fig.~\ref{fig:1.4}.
The light cone divides spacetime into three distinct regions.
The Future can be connected to P by future directed timelike or null geodesics.
Points in the future will be mapped to the future possibilities of P by an orthochronous Lorentz transformation, hence the name.
The Past can be connected to P by past-directed timelike or null geodesics.
Points in the past will be mapped to the past history of P by an orthochronous Lorentz transformation, again hence the name.
Geodesics that pass through P from the past into the future, whether null or timelike, are consistent with causality and SR restricts itself to only considering these geodesics physical.
All other geodesics are spacelike and lie in the volume outside the light cone labeled Elsewhere.
They do not adhere with causality.
Assume there are two events, $P$ and $Q$, connected by a spacelike geodesic in Elsewhere and ordered such that $P$ happens before $Q$ in time.
Then, an orthochronous Poincar\'e transformation is not guaranteed to preserve the ordering of $P$ and $Q$ for all observers.
In this sense, spacelike geodesics violate causality.
As light travels along null geodesics, the slope of each null cone is $c$. 
For a warp drive to exist, it must travel along a non-physical spacelike geodesic, and that ends the possibility of warp drives in just SR.

~
\begin{figure}[ht]
\centering
	\includegraphics[scale=0.6]{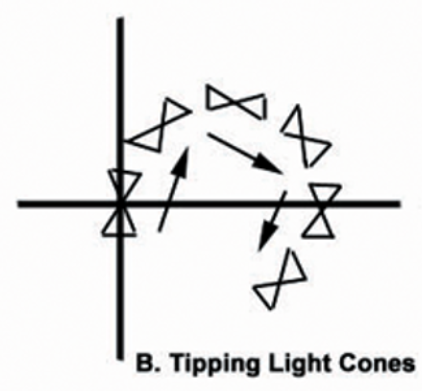}
	\caption{An example of a tipping light cone by a warped underlying spacetime.} \label{fig:1.5}
\end{figure}
~
\begin{figure}[ht]
\centering
	\includegraphics[scale=0.6]{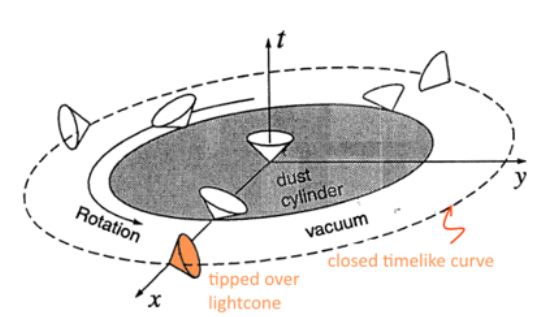}
	\caption{The light cone of an object rotating around the axis of a van Stockum spacetime will tip over.} \label{fig:1.6}
\end{figure}

\newpage

But GR allows another possibility: the light cones themselves may be tipped over in significantly warped spacetimes as in Fig.~\ref{fig:1.5}.
A sequence of such tipping may wind a timelike geodesic without even leaving its light cone.
An example is the van Stockum spacetime consisting of a cloud of dust rotating cylindrically about an axis though it is a general feature of any cylindric spacetime \cite{vanStockum:1937,Lobo:2010}.
As an object rotates about the azimuthal coordinate $\phi$, its light cone will begin to tip as depicted in Fig.~\ref{fig:1.6}.
For a large enough angular velocity, the light cone will inevitably become a CTC and contain a causality violating region outside the dust cylinder.

Now, the van Stockum solution is physically unreasonable.
Its toy construction of an infinitely long cylinder of dust is present nowhere in our physical universe.
Moreover, it requires non-asymptotically flat spacetime, which matches not a single realistic model of our universe.
Instead, it illustrates how light cones may be manipulated by strong gravitational fields.
The theoretical limit on the rate at which gravity may expand and contract spacetime is unknown and potentially greater than $c$.
The possibility of warping spacetime at a rate greater than $c$ sparks the potential for a warp drive.

It is an open question in physics how deeply one can tip a light cone.
The Superluminal-Censorship Theorem predicts that ``if an asymptotically flat spacetime, whose domain of outer communication is globally hyperbolic, possesses a FTL null curve $\gamma$ from past null infinity to future null infinity, then there exists a fastest null geodesic $\gamma_{max}$ from past null infinity to future  null infinity that is in the same  homotopy class and does not have any conjugate points'' \cite{Visser:1998}. 
The Averaged NEC will be violated by $\gamma_{max}$.
Now, both quantum and classical systems violate the ANEC, so the Superluminal-Censorship Theorem does not immediately prohibit FTL \cite{Barcelo:2002}.
It remains to discover whether $\gamma_{max}$ is above or below $c$.

\section{Wormholes}
\label{chp1:Worm}
    %Where the current understanding on wormholes is...
    A physically more reasonable alternative to the EPR bridge presented in Section~\ref{chp1.1:EPR} is that of the Lorentzian traversable wormholes.
    They were first described by Kip Thorne and his collaborators who used Einstein's general relativistic field equations to explore the possibility of FTL spaceflight without violating SR \cite{Visser:1995,Morris:1988A,Morris:1988B}. 
    Earlier studies demonstrated the possibility of traversable wormholes in GR \cite{4,5}.
    A precise definition of a Lorentzian traversable wormhole is a topological opening in spacetime, which manifests traversable intra-universe and/or inter-universe connections, as well as possible different chronological connections between distant spacetime points.
    In simpler terms, it is a shortcut in spacetime created by an extreme warping and/or folding of spacetime due to a powerful gravitational field.
    The condition for a Lorentzian wormhole to be traversable is that it is free of both event horizons and singularities  \cite{Visser:1995}.
    Such a wormhole is fully traversable in both directions, geodesically complete, and possesses no crushing gravitational tidal forces. 
    Consequently, Lorentzian traversable wormholes are unlike the non-traversable Schwarzschild wormhole, or Einstein-Rosen bridge considered in Section \ref{chp1.1:EPR}. 
    Exotic matter, which violates the point-wise and averaged energy conditions, is required to open and stabilize a Lorentzian traversable wormhole. 
    A comprehensive technical overview of this subject is found in \cite{Visser:1995}.
    
    %Paragraph going over the main papers on Wormholes
    The seminal paper on the EPR Bridge prompted the discovery of the existence of many  wormhole solutions in GR.
    Wheeler introduced the name "wormhole" in his discussion of spacetime foam and the Schwarzschild solution \cite{Wheeler:1955,Wheeler:1957}.
    Morris and Thorne popularized the concept of a wormhole with their exploration of the Morris-Thorne(MT) wormhole \cite{Morris:1988A,Morris:1988B}.
    They looked at Kerr's solution for a rotating black hole and saw that impassable, intra-universal passageways must exist in it.
    Continuing, they predicted the basic requirements for a traversable wormhole and then penned their famous metric presented later as Eq.~\eqref{eq:MT}.
    Since then, many more types of wormholes have been presented such as rotating traversable wormholes and wormholes whose throats change dynamically in time \cite{Lobo:2017,Visser:1995,Teo:1998}.
    Furthermore, well known line elements have been demonstrated to contain wormholes \cite{Boonserm}.
    Wormhole physics is a rich field that probes the limits of GR.
    
\begin{figure}[ht]
    \centering
    	\includegraphics[scale=0.55]{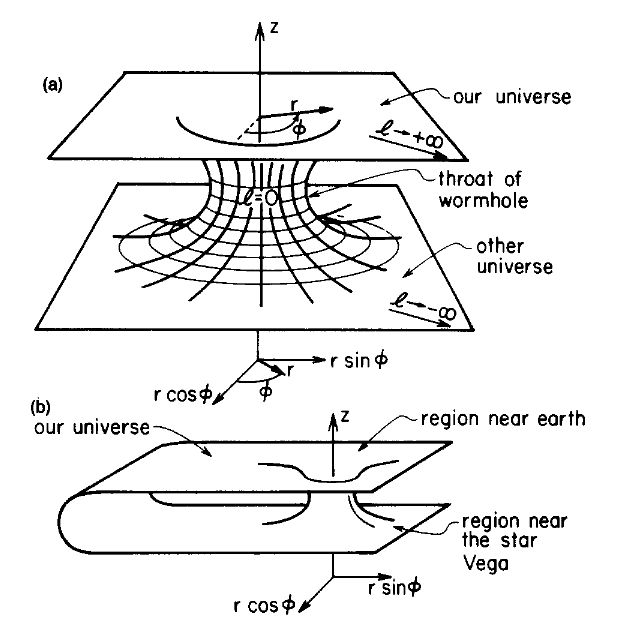}
    	\caption{Fig.~1 from \cite{Morris:1988A}. (a) is an embedding diagram for a wormhole that connects two different universes. (b) is an embedding diagram for a wormhole that connects two distant regions of our own universe. Each diagram depicts the geometry of an equatorial ($\theta=\frac{\pi}{2}$) slice through  space at a specific moment in time ($t$ = constant).} \label{fig:1.7}
    \end{figure}

    Previous studies of Lorentzian traversable wormholes rely on either calculating the elements of the Riemann curvature tensor, $R^i_{jkl}$, to ``observe'' the effects of the wormhole’s spacetime curvature on photons and matter moving through it or by embedding diagrams. 
    However, the $R^i_{jkl}$ cannot be calculated in an invariant manner because it is a function of the chosen coordinates \cite{Penrose:1986}. 
    Analysis of $R^i_{jkl}$ can be misleading because a different choice of coordinate basis will result in different tensor components.
    These coordinate mapping distortions arise purely as an artifact of the coordinate choice.
    Embedding diagrams offer a narrow view of the spacetime manifold.
    In Fig.~\ref{fig:1.7}, the embedding diagrams depict the wormhole geometry along an equatorial ($\theta=\frac{\pi}{2}$) slice through space at a specific moment in time \cite{Morris:1988A}. 
    Embedding diagrams offer only a limited view of the physics involved in the wormhole.
    Curvature invariants allow a manifestly coordinate invariant characterization of certain geometrical properties of spacetime \cite{Stephani:2003,Zakhary:1997}.
    Using curvature invariants to probe black holes is a fruitful field of interest and research \cite{Overduin:2020,Baker:2000,Cherubini:2003,MacCallum:2006,Coley:2009,Abdelqader:2014,MacCallum:2015,Page:2015,Coley:2017}.
    In a similar fashion, the research presented herein seeks the best way to illustrate wormhole spacetimes by plotting their independent curvature invariants \cite{Mattingly:2020}.\\
    
\section{Warp Drives}
\label{chp1:Warp}
    A warp drive is a solution to the Einstein field equations that allows a spaceship to make a trip to a distant star in an arbitrarily short proper time \cite{Alcubierre:1994}. 
    The first propulsion mechanism for a warp drive was a dipole configuration connecting a local contraction of spacetime in front of the spaceship with a local expansion of spacetime behind the ship. 
    While locally the spaceship remains within its own light cone and never exceeds $c$, globally the relative velocity\footnote{defined as proper spatial distance divided by proper time} can be much greater than $c$. 
    For a distant observer, the effect of the expansion and contraction would cancel each other out and spacetime would be asymptotically flat.
    While an interesting theoretical concept, the original formulation was shown to violate every NEC, produce tachyonic matter, and require an amount of energy greater than the mass of our observable universe.
    Only naively could such an original idea be considered possible.

    Since then, there has been considerable research into realistic FTL warp drives.
    In \cite{Krasnikov:1995ad}, Krasnikov considered a non-tachyonic FTL warp bubble and showed it to be possible mathematically. 
    Van Den Broeck in \cite{VanDenBroeck:1999sn} modified Alcubierre's warp drive to have a microscopic surface area and a macroscopic volume inside.
    He demonstrated a modification that allowed a warp bubble to form with energy requirements of only a few solar masses.
    His geometry also has more lenient violations of the NEC.
    Later, Nat\'ario improved upon Alcubierre's work by presenting a warp drive metric such that zero spacetime expansion occurs \cite{Natario:2001}. 
    Instead of riding a contraction and expansion of spacetime, the warp drive may be observed to be ``sliding" through the exterior spacetime at a constant global velocity. 
    Finally, Loup expanded Nat\'ario's work to encompass an accelerating global velocity by modifying the distance between hypersurfaces.
    All of this research hints at the possibility in GR of a realistic warp drive \cite{Loup:2017}.
    
    FTL travel obeys eight general requirements \cite{Davis}. 
    First, the rocket equation does not apply. 
    Second, the travel time through the FTL space warp should take less than one year as seen both by the passengers in the warp and by stationary observers outside the warp. 
    Third, the proper time as measured by the passengers will not be dilated by any relativistic effects. 
    Fourth, any tidal-gravity accelerations acting on any passengers should be less than the acceleration of gravity near the Earth's surface, $g=9.81 \frac{m}{s^2}$. 
    Fifth, the local speed of any passengers should be less than $c$.
    Sixth, the matter of the passengers must not couple with any material used to generate the FTL space warp. 
    Seventh, the FTL space warp should not generate an event horizon.
    Finally, the passengers should not encounter a singularity inside or outside of the FTL warp.

    While the mathematics of a warp drive are well developed, mapping the spacetime around the warp drive remains underdeveloped.
    Considering that a ship inside of a warp bubble is causally disconnected from the exterior \cite{Lobo:2017,Natario:2001}, computer simulations of the surrounding spacetime are critical for the ship to map its journey and steer the warp bubble.
    In \cite{Alcubierre:1994}, Alcubierre uses the York time\footnote{The York time is defined as $\Theta=\frac{v_s}{c}\frac{x-x_s}{r_s}\frac{df}{dr_s}$} to map the volume expansion of a warp drive. 
    He plotted the York time to show how spacetime warped behind and in front of the spaceship. 
    While the York time is appropriate when the 3-geometry of the hypersurfaces is flat, it will not contain all information about the curvature of spacetime in non-flat 3-geometries such as the accelerating Nat\'atio warp drive spacetime.
    As an alternative to the York time, curvature scalars of the line element may be considered.

\section{Purpose of Research}
\label{chpt:PoR}
%Why did we do this research?
%What question does this research add?
%What knowledge does this research add to the field at large?
    In this dissertation, the CM curvature invariants for four different wormhole and three warped spacetimes will be derived, plotted and analyzed.
    The curvature scalars (or invariants) show the magnitude by which spacetime differs from being flat.
    They can be used to map the geometric structure of the spacetime independent of coordinate basis.
    They have values independent of the choice of coordinates and provide a manifestly coordinate-invariant characterization of spacetime \cite{Misner:1974,Stephani:2003,Zakhary:1997}.
    They are especially helpful in identifying intrinsic curvature singularities of the spacetime, the Petrov and Segre type of the eigenvalue problem for the Ricci tensor and Weyl tensor, and can answer the equivalence problem between spacetimes.
    By computing all independent curvature scalars, the geometry of any spacetime may be analyzed and plotted.
    
    % In this paragraph, I will go over what my research adds to the sum knowledge
    The focus of previous studies of wormholes and warp drives is their stress-energy tensor.
    The stress-energy tensor answers questions such as how realistic a warped spacetime is and what amount of exotic matter is required to generate and control one.
    Little attention has been given to what each spacetime looks like, how the spacetime curves, or what singularities are present in them.
    Computing and plotting the invariants will provide a window to answer these questions.
    In this dissertation, the independent set of CM curvature invariants for four wormhole spacetimes and four warp drive spacetimes are calculated and plotted.
    The identity of any intrinsic curvature singularities present will be found.
    An analysis of each plot will be provided to help describe the surrounding spacetime.

    The order of the research in this dissertation is as follows.
    Chapter~\ref{chapter2} details how to calculate the independent CM curvature invariants.
    It includes a basic review of GR, an introduction to the set of CM curvature invariants, and how to calculate the four independent invariants $R$, $r_1$, $r_2$, and $w_2$ for Class $B1$ spacetimes.
    Chapter~\ref{Chapter3} investigates the wormhole spacetimes.
    The non-zero curvature invariants for four wormhole spacetimes are presented: the Thin-Shell(TS) Flat-Face wormhole, the MT wormhole, the TS Schwarzschild wormhole, and the exponential metric.
    Chapter~\ref{Chapter4} explores warp drive spacetimes at a constant velocity.
    It starts with a description of a generic warp drive spacetime in $3+1$ ADM.
    Then, the Alcubierre and Nat\'ario warp drives are demonstrated  and the spacetimes variables, $v_s$, $\sigma$ and $\rho$, are varied to reveal their effect on the curvature invariants.
    Chapter~\ref{Chapter5} expands on Chapter \ref{Chapter4} by looking at the effects of a constant acceleration for the Nat\'ario spacetime.
    The warped drive spacetimes are expanded from a constant velocity to a changing velocity.
    Then, it investigates the Nat\'ario warped drive spacetime with a constant acceleration and the effect of the different variables $t$, $a$, $\rho$ and $\sigma$ on the curvature invariants.
    The conclusion in Chapter~\ref{Chapter6} looks at the different possible directions the research contained within can be expanded and what research is already being undertaken by the EUCOS group.

%% file: ch2.tex
\chapter{Curvature Invariants} 
\label{chapter2}
%This chapter details hot to calculate the curvature invariants needed for the dissertation.
GR provides the framework to study FTL spacetimes.
While it is impossible for an individual particle to exceed a velocity of $c$, the underlying spacetime has no such restriction.
Consequently, the study of FTL line elements examines spacetime manifolds inside the prescription of GR.

To analyze the underlying spacetime manifold, GR uses the language of tensors.
In this chapter, the basic principles of GR such as line elements and the basic tensors used in the dissertation will be presented.
Next, the set of CM curvature invariants will be demonstrated.
Finally, the minimal set needed for FTL line elements will be shown. \\

\section{Spacetime Manifolds}
\label{chp2:GR}
GR is a purely classical theory \cite{Visser:1995,DInverno:1992,Misner:1974,Penrose:1986,Stephani:2003,Groen:2007,Carroll:2014}.
It recognizes that the Newtonian concepts of space and time are a single object called spacetime.
Spacetime is a differentiable manifold endowed with a metric, $g_{\mu \nu}$,\footnote{Indices are purely a choice of convention.
In this dissertation, Greek indices range from 0 to 3, including both the single time component and the three space components, of the manifold.
Lowercase Latin indices that range from 1 to 3 will denote the space components of the geometry inside the manifold.
Uppercase Latin indices denote spinor components.} which is a symmetric covariant tensor of rank 2.
Formally,
\theoremstyle{definition}
\begin{definition}{Spacetime}
is a four-dimensional manifold equipped with a Lorentz\-ian (pseudo-Riemannian) metric.
The metric should have Lorentzian (pseudo-Riemann\-ian) signature ($-$,+,+,+). The signature is represented as $(3+1)$ dimensions with 3 spacelike dimensions and 1 timelike dimension.
The manifold is Hausdorff and paracompact.
\end{definition} 
Locally, a spacetime manifold resembles a small, Lorentzian patch around each point in it.
A single point in spacetime represents the physical event that occurs.
A basis set of coordinates $(e^\mu=x^1,x^2, ..., x^n)$ can be labeled at each point in the manifold.
The coordinates are not needed in an absolute sense, but they ease the questions of ordering, comparing, and/or relating the events between separate points.
The coordinates take real values. 
Their possible range is $(-\infty, \infty)$ though subsets will be often useful for coordinate patches.
In some cases, a unique set of coordinates can be defined at each point in the manifold.
In others, the spacetime manifold necessitates multiple coordinate patches with each only covering a portion.
A coordinate transformation may be defined for these cases in an overlapping region between the coordinate patches.

\begin{figure}[ht]
\centering
	\includegraphics[scale=1.00]{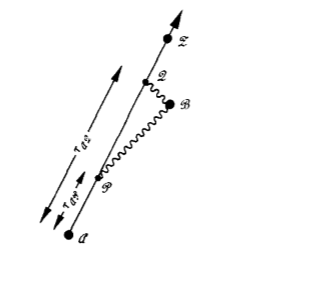}
	\caption{Figure 1.3.1 from \cite{Misner:1974}.} \label{fig:2.1.1}
\end{figure}

%Paragraph discussing a Lorentzian spacetime. For reference consider Misner, Thorne, and Wheeler page 21.
The geometry of spacetime is Lorentzian.
The intervals between any event and any other event satisfies the theorems of Lorentz-Minkowski geometry.
Consider the worldline of $\mathscr{A}\mathscr{L}$ and an event, $\mathscr{B}$, not on it as in Fig.~\ref{fig:2.1.1}.
For $\mathscr{B}$ to interact with $\mathscr{A}\mathscr{L}$, light rays need to be transmitted and/or reflected.
Assume one light ray transmitted from $\mathscr{A}\mathscr{L}$ interacts with $\mathscr{B}$ at the event $\mathscr{P}$ and a second light ray transmitted from $\mathscr{B}$ interacts with $\mathscr{A}\mathscr{L}$ at the event $\mathscr{L}$.
Then, the proper time, $\tau_{\mathscr{A}\mathscr{B}}$, between $\mathscr{A}$ and  $\mathscr{B}$ is the timelike separation given the difference between the two light rays.
The the proper distance $s_{\mathscr{A}\mathscr{B}}$ is the spacelike separation related to the proper time by $s_{\mathscr{A}\mathscr{B}}^2\equiv-\tau_{\mathscr{A}\mathscr{B}}^2=-\tau_{\mathscr{A}\mathscr{B}}\tau_{\mathscr{A}\mathscr{B}}$. 
Then, the coordinate system is locally Lorentzian, and the geometry of the spacetime is also locally Lorentzian.

\begin{figure}[ht]
\centering
	\includegraphics[scale=0.95]{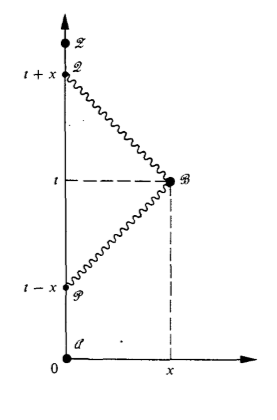}
	\caption{Figure 1.3.2 from \cite{Misner:1974}.} \label{fig:2.1.2}
\end{figure}

\begin{figure}[ht]
\centering
	\includegraphics[scale=1.0]{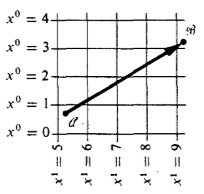}
	\caption{Figure 1.3.3 from \cite{Misner:1974}.} \label{fig:2.1.3}
\end{figure}

As an example, consider the world line $\mathscr{A}\mathscr{L}$ to be propagating along the $x=0$ hypersurface and the event $\mathscr{B}$ is separated from the world line by a spatial distance of $x$ as in Fig.~\ref{fig:2.1.2}.
Then, the proper time is $\tau_{\mathscr{A}\mathscr{B}}^2=t^2-x^2=(t+x)(t-x)=\tau_{\mathscr{A}\mathscr{P}}\tau_{\mathscr{A}\mathscr{L}}$.
As a second example, choose a coordinate system given by the parametrization $x^\mu=x^\mu(\mathscr{A})$ for any local event $\mathscr{A}$ as in Fig.~\ref{fig:2.1.3}.
The proper time between two locally, neighboring event pairs $\mathscr{A}$ and $\mathscr{B}$ is $s_{\mathscr{A}\mathscr{B}}^2\equiv-\tau_{\mathscr{A}\mathscr{B}}^2= -[x^0(\mathscr{B})-x^0(\mathscr{A})]^2+[x^1(\mathscr{B})-x^1(\mathscr{A})]^2+[x^2(\mathscr{B})-x^2(\mathscr{A})]^2+[x^3(\mathscr{B})-x^3(\mathscr{A})]^2$.
The examples above describe in GR the generalization of Euclid's famous statement, ``The shortest distance between two points is a straight line.''
~
~

The gravitational field of an object is the bending of spacetime near that object.
The spacetime metric describes the gravitational potential in any given region.
In components, the spacetime line element is
\begin{equation}
    ds^2=g_{\mu\nu}dx^{\mu}dx^{\nu}. \label{eq:2.1} \tag{2.1}
\end{equation}
The line element represents the squared length of the infinitesimal displacement between two neighboring points $x^\mu$ and $x^\mu+dx^\mu$.
The metric, $g_{\mu\nu}$, is a set of ten functions of position that gives the interval between any event and any nearby event as described above.
It defines the distances and lengths of vectors on the Riemannian manifold.
For a contravariant vector,\footnote{
A contravariant vector is a set of quantities at a point P that transforms under a change of coordinates as ${X'}^\mu=\frac{\partial {x'}^{\mu}}{\partial x^\nu} X^\nu$ evaluated at P \cite{DInverno:1992}.
It is written as ${X'}^\mu$ in the ${x'}^\mu$ coordinate system.
It is also called a contravariant tensor of rank (order) 1.} $X^\mu$, the length or norm is defined to be the scalar product
\begin{equation*}
    X^2=g_{\mu\nu}X^\mu X^\nu. \label{eq:2.2} \tag{2.2}
\end{equation*}
The metric components depend on the basis vectors, called a tetrad, of the manifold as well as several conditions.
The metric is symmetric $g_{\mu\nu}=g_{\nu\mu}$.
In the mixed representation $g_\mu^\nu$, it is identical to the Kronecker delta, $\delta_\mu^\nu$.
The inverse metric is defined as
\begin{align*}
    g^{\mu\lambda}g_{\lambda\nu} &=g^\mu_\nu=\delta^\mu_\nu, \label{eq:2.3} \tag{2.3} \\
    ||g^{\mu\nu}|| &= ||g_{\mu\nu}||^{-1}. \label{eq:2.4} \tag{2.4}
\end{align*}
For an arbitrary tensor, $T^{\mu \nu ...}_{\lambda \xi ...}$, the metric and inverse metric can raise and lower the tensor elements. Using the Einstein summation,
\begin{align*}
    g^{\rho \lambda} T^{\mu \nu ...}_{\lambda \xi ...} &=T^{ \rho \mu \nu ...}_{\xi ...}, \label{eq:2.5} \tag{2.5} \\
    g_{\rho \mu} T^{\mu \nu ...}_{\lambda \xi ...} &=T_\rho^{\nu ...}_{\lambda \xi ...}. \label{eq:2.6} \tag{2.6}
\end{align*}
For a Lorentzian spacetime, the line element can be decomposed into a vector field that forms its tetrad or basis.
An orthonormal tetrad consists of three spacelike vectors $V_i$ and one timelike vector $V_t$ that satisfy the relationships
\begin{align*}
    {V_\mu} &=\{V_t,V_i\}=\{t,x,y,z\}, \tag{2.7.a} \label{eq:2.7.a} \\
    g_{\mu\nu} &= x_\mu x_\nu + y_\mu y_\nu + z_\mu z_\nu - t_\mu t_\nu, \tag{2.7.b} \label{eq:2.7.b}\\
    V_i \cdot v_j &= \delta_{i,j}, \tag{2.7.c} \label{eq:2.7.c}\\
    t \cdot t &= -1, \tag{2.7.d} \label{eq:2.7.d}\\
    V_i \cdot t &=0. \label{eq:2.7.e} \tag{2.7.e}
\end{align*}
In some cases, an orthonormal tetrad may be read directly from the non-zero metric components.
Complex null tetrads form a second type of tetrad.
They consist of two real null vectors $\textbf{k}$ and $\textbf{l}$ and two complex conjugate null vectors $\textbf{m}$ and $\bar{\textbf{m}}$.
The scalar products of the tetrad vanish apart from $k^\mu l_\mu=-1$ and $m^\mu \bar{m}_\mu = 1$. 
The easiest way to find a null tetrad for a spacetime is by first finding an orthonormal tetrad and using the relationships
\begin{align*}
    l_\mu &= \frac{1}{2}(V_0+V_1), \label{eq:2.8.a} \tag{2.8.a}\\
    k_\mu &= \frac{1}{2}(V_0-V_1), \label{eq:2.8.b} \tag{2.8.b}\\
    m_\mu &= \frac{1}{2}(V_2+ i V_3), \label{eq:2.8.c} \tag{2.8.c}\\
    \bar{m}_\mu &= \frac{1}{2}(V_2- i V_3). \label{eq:2.8.d} \tag{2.8.d}
\end{align*}
Computing the null tetrad for a spacetime greatly increases the computational speed for the invariants found in this chapter.

Given a spacetime line element, the covariant derivative may be derived to compare the spacetime between any two points in it.
The affine connection, $\Gamma^i_{jk}$, parallel transports neighboring events to allow them to be compared. It is defined as
\begin{equation*}
    \Gamma^\mu_{\alpha \beta} = \frac{1}{2} g^{\mu\lambda} \left(\partial_\alpha g_{\lambda\beta}+\partial_\beta g_{\lambda\alpha}-\partial_\lambda g_{\alpha\beta}\right). \label{eq:2.9} \tag{2.9}\\
\end{equation*}
The connection links the acceleration a free-falling particle experiences with the surrounding gravitational field.
The particle will follow a geodesic. A geodesic computes the shortest distance between two points in curved space and is an extension of a straight line in Euclidean space.
The geodesic equation for an affine parameter is
\begin{equation*}
    \frac{d^2x^\mu}{ds^2}+\Gamma^\mu_{\alpha \beta}\frac{dx^\alpha}{ds}\frac{dx^\beta}{ds}=0. \label{eq:2.10} \tag{2.10}\\
\end{equation*}
The four-vectors $V^\mu=\frac{dx^\mu}{ds}$ may be defined to be timelike if $g_{\mu\nu}<0$, null if $g_{\mu\nu}=0$, and spacelike if $g_{\mu\nu}>0$.
For torsion-free connections, the Riemann tensor $R^\mu_{\alpha\beta\gamma}$ is related to the commutator of covariant differentiation.
It is defined to be 
\begin{equation*}
    R^\mu_{\alpha\beta\gamma} = \partial_\beta\Gamma^\mu_{\alpha\gamma}-\partial_\gamma\Gamma^\mu_{\alpha\beta}+\Gamma^\mu_{\nu\beta}\Gamma^\nu_{\alpha\gamma}-\Gamma^\mu_{\nu\gamma}\Gamma^\nu_{\alpha\beta}. \label{eq:2.11} \tag{2.11}
\end{equation*}
The Riemann tensor governs the difference in accelerations that two different, free-falling particles experience.
A necessary and sufficient condition for a metric to be flat is for its Riemann tensor to vanish everywhere.
Conversely, the curvature tensor is then the Riemann tensor.
The Ricci tensor $R_{\alpha \beta}$, the Ricci scalar $R$, and the trace-free Ricci tensor, $S_{\alpha \beta}$ can be obtained from Eq.~\eqref{eq:2.9} and the contractions in Eqs. \eqref{eq:2.5} and \eqref{eq:2.6}.
They are defined as the following
\begin{align*}
	R_{\alpha\beta} &= \partial_\gamma\Gamma^\gamma_{\alpha\beta}-\partial_\beta\Gamma^\gamma_{\alpha \gamma}+\Gamma^\delta_{\alpha\beta}\Gamma^\gamma_{\delta \gamma}-\Gamma^\delta_{\alpha \gamma}\Gamma^\gamma_{\delta \beta},\label{eq:2.12} \tag{2.12}\\
	R &= g^{\alpha \beta}R_{\alpha \beta}, \label{eq:2.13} \tag{2.13}\\
	S_{\alpha \beta} &= R_{\alpha \beta}-\frac{R}{4}g_{\alpha \beta}.\label{eq:2.14} \tag{2.14}
\end{align*}
The Ricci scalar or curvature scalar is the first example of an invariant.
In dimensions $n=3+1$, there are 20 independent components of the Riemann tensor: $10$ of the independent components of the Ricci tensor and $10$ of the independent components of the Weyl tensor. 
The Weyl tensor or conformal tensor $C_{\alpha\beta\gamma\delta}$ is 
\begin{multline*}
	C_{\alpha\beta\gamma\delta} = \\ R_{\alpha\beta\gamma\delta}+\frac{1}{2}\left(g_{\alpha\delta}R_{\beta\gamma}+g_{\beta\gamma}R_{\alpha\delta}-g_{\alpha\gamma}R_{\beta\delta}-g_{\beta\delta}R_{\alpha\gamma}\right) +\frac{1}{6}\left(g_{\alpha\gamma}g_{\beta\delta}-g_{\alpha\delta}g_{\beta\gamma}\right)R.\label{eq:2.15} \tag{2.15}
\end{multline*}
A necessary and sufficient condition for a metric to be conformally flat, $g_{\mu\nu}=\Omega^2\eta_{\mu\nu}$,\footnote{$\eta_{\mu\nu}$ is the Minkowski metric $diag(-1,1,1,1)$.}is that its Weyl tensor vanishes everywhere.
There are many additional tensors that may be derived from a given metric such as the Einstein $G_{\mu\nu}$ or the other tensors of the irreducible representation of the full Lorentz group $E_{\alpha\beta\gamma\delta}$ and $G_{\alpha\beta\gamma\delta}$. This includes the set of tensors needed for the following research.\\

\section{Curvature Invariants}
\label{chp2:Inv}
    %How to use the tensors in \ref{chp2:GR} to derive the Curvature Invariants. 
    The Riemann curvature invariants provide a manifestly coordinate invariant characterization of spacetime \cite{Misner:1974,Stephani:2003,Zakhary:1997,7}.
    A curvature invariant has a value independent of the choice of the coordinates.
    Riemann curvature invariants are scalar products of the Riemann Eq.~\eqref{eq:2.12}, Ricci Eq.~\eqref{eq:2.12}, the Weyl tensors Eq.~\eqref{eq:2.15} and their traces, covariant derivatives, and/or duals. 
    Invariants measure the curvature by which a spacetime geometry differs from being flat. 
    The prime example of the invariants are the scalar polynomial (SP) invariants such as the Kretschmann invariant, $R^{ijkl}R_{ijkl}$.
    Other types exist, such as the Cartan invariants which gives a unique coordinate-independent characterization, but this research focuses on the SP invariants.
    The SP prefix should be assumed throughout this dissertation.
    
    The complete set of invariants are important in the study of GR. 
    Invariants are critical for studying curvature singularities, the Petrov type of the Weyl tensor, the Segre type of the trace-free Ricci tensor, and the equivalence problem.
    If a singularity occurs in the curvature invariants, then the curvature singularity must be fundamental to the spacetime, instead of an artifact of the choice of coordinates.
    Studies of the Petrov type and the Segre type categorize the solutions of the eigenvalue problem of the Riemann tensor.
    Investigating the curvature invariants reveals the eigenvalue structure of the spacetime and relates it using the NP components. 
    The equivalence problem asks whether two different metrics describe identical spacetimes.
    Finally, a scalar invariant is the seed of GR.
    Motivated by Einstein, Hilbert saw that an action principle based on the scalar invariant $R$ gives the geometrodynamic law.
    Choosing a Lagrangian $L_{geom}=(\frac{1}{16})R$ gives the law with a very simple correspondence to Newtonian's Theory of Gravity.
    Any scalar invariant may be chosen in place of $R$, but the simple correspondence to Newton's theory will be lost as the action will no longer be second order in the derivatives from the metric components.
    In these manners, the curvature invariants are of fundamental importance in the study of GR.
    
    The number of free parameters in the Riemann tensor determines the number of curvature invariants.
    The Riemann tensor has 20 independent components after including its symmetries.
    The first and most famous component is the curvature scalar Eq.~\eqref{eq:2.13}.
    Nine independent components appear in the trace-free Ricci tensor Eq.~\eqref{eq:2.14}.
    Ten independent components appear in the Weyl tensor Eq.~\eqref{eq:2.15}.
    The Lorentz transformation represents 6 additional degrees of freedom and entangles these components.
    Its six parameters further reduce the number of free parameters and curvature invariants to $20-6=14$.
    In an arbitrary spacetime, the fourteen parameters determine the coordinate independent, local features of the curvature.
    A specific choice of line element can further reduce the number since the number of independent variables\footnote{The independent variables are the four coordinates plus any functions of integration that have physical meaning.} might be further reduced. \\
  
  \subsection{Scalar Polynomial Invariants}  
    In general, there are three classes of Riemann SP invariants \cite{Zakhary:1997}.
    The first is the set of four Weyl invariants.
    The two complex invariants are 
    \begin{align*}
        I &=\frac{1}{6}\mathfrak{I}(\Psi,\Psi) = \frac{1}{6} \ \Psi_{ABCD} \ \Psi^{ABCD}, \label{eq:2.16} \tag{2.16} \\
        J &=\frac{1}{6}\mathfrak{I}(\Psi,Q) = \Psi_{ABCD} \ \Psi^{CD}_{EF} \ \Psi^{EFAB}. \label{eq:2.17} \tag{2.17}
 \end{align*}
    where $\mathfrak{I}$ is the invariant product and $\Psi$, $Q$ are $\Psi$-like spinors derived from the Weyl tensor in Eq.~\eqref{eq:2.15}.
    The Weyl invariants come from the real and imaginary components of these two functions. They can be expressed as the following relationships,
    \begin{align*}
        I_1 &= \mathrm{Re}(I), \label{eq:2.18} \tag{2.18} \\
        I_2 &= \mathrm{Im}(I), \label{eq:2.19} \tag{2.19} \\
        I_3 &= \mathrm{Re}(J), \label{eq:2.20} \tag{2.20} \\
        I_4 &= \mathrm{Im}(J). \label{eq:2.21} \tag{2.21}
    \end{align*}
    Any other choice of Weyl invariants will be related to this choice.
    The second set is the set of four Ricci invariants.
    They are the following
    \begin{align*}
        I_5 &= R, \label{eq:2.22} \tag{2.22} \\
        I_6 &= \frac{1}{3} \mathfrak{I}(\Phi,\Phi), \label{eq:2.23} \tag{2.23} \\
        I_7 &= \frac{1}{6} \mathfrak{I}(\Phi,E), \label{eq:2.24} \tag{2.24} \\
        I_8 &= \frac{1}{12} \mathfrak{I}(E,E). \label{eq:2.25} \tag{2.25}
    \end{align*}
    where $\Phi$ is a $\Psi$-like spinor and $E$ is a $\Phi$-like spinor. 
    Any choice of Ricci invariants will be related to this choice.
    The final type is the Mixed invariants.
    The Mixed invariants are the hardest to construct. Most of the differences between the sets of invariants occur in the Mixed type. The complete set is 
    \begin{align*}
        I_9 &= \mathrm{Re}(K), \label{eq:2.26} \tag{2.26} \\
        I_{10} &= \mathrm{Im}(K), \label{eq:2.27} \tag{2.27} \\
        I_{11} &= \mathrm{Re}(L), \label{eq:2.28} \tag{2.28} \\
        I_{12} &= \mathrm{Im}(L), \label{eq:2.29} \tag{2.29} \\
        I_{13} &= \mathrm{Re}(M), \label{eq:2.30} \tag{2.30} \\
        I_{14} &= \mathrm{Im}(M), \label{eq:2.31} \tag{2.31} \\
        I_{15} &= M_1= \mathfrak{I}(C',C), \label{eq:2.32} \tag{2.32} \\
        I_{16} &= \mathrm{Re}(M_2), \label{eq:2.33} \tag{2.33} \\
        I_{17} &= \mathrm{Im}(M_2). \label{eq:2.34} \tag{2.34}
    \end{align*}
    where $K=\mathfrak{I}(\Phi,C)$, 
    $L=\mathfrak{I}(Q,\xi)$, 
    $M=\frac{1}{4} \mathfrak{I}(\Psi,\tilde{\xi})$, 
    $M_1$, and 
    $M_2=\mathfrak{I}(C',\tilde{C})$ 
    are spinors that mix $\Psi$-like and $\Phi$-like spinors according to the choice of line element.
    These are the general sets of CM invariants.
    Specific choices of line elements and the symmetries inherent to the spacetime will reduce the number of invariants.
    
    The list of invariants presented contains 17 elements, not 14, as certain non-degenerate cases are taken into account. 
    It is stressed that this set of invariants is a \textsc{complete} set as opposed to an \textsc{independent} set.
    A \textsc{complete} set of invariants includes the number of invariants that meet the requirements of the 90 different Petrov and Segre types.
    In contrast, an \textsc{independent} set of invariants contains only invariants that independent of each other.
    The CM invariants in Eq.~\eqref{eq:2.26} to Eq.~\eqref{eq:2.34} will not be linearly independent for a specific choice of a spacetime. 
    Algebraic and polynomial relationships will reduce the number of invariants in a \textsc{complete} set down to the six required by Lorentz invariance.
    These relationships are the syzygies of the invariant set.
    \theoremstyle{definition}
    \begin{definition}{Syzygy}
    is a  polynomial relationship between functions in a complete set. The \textsc{independent} invariants, I, satisfy 
    \begin{equation}
        c_0+c_1 I +c_2 I^2 + c_3 I^3 + ... + c_n I^n = 0. \label{eq:2.35} \tag{2.35}
    \end{equation}
    where the $c_i$'s are polynomials of the other non-independent invariants in the set.
    \end{definition} 
    Identifying the non-zero spinor components allows the syzygy relationships to be derived.
    Solving the syzygies between each spacetime's invariants will greatly reduce the number of independent invariants. \\
 
\subsection{Invariants of Sphere}
\label{chp2:Sphere}  
    A simple example can help illustrate the use of invariants.
    Consider a 2-sphere as given in \cite{MacCallum:2015}.
    The metric of a 2-sphere is 
    \begin{equation}
        g_{ij}=
        \begin{pmatrix}
        \ a^2&\ \ 0\\
        \ 0&\ \ a^2 \sin^2{\theta} 
        \end{pmatrix}.\tag{A.1} \label{A.1}
    \end{equation}
    where $a$ is the radius of the sphere. 
    There are two nonzero components of the Riemann tensor, $R^1_{221}=\sin^2{\theta}$ and $R^1_{212}=\sin^2{\theta}$, computed from Eq.~\eqref{eq:2.11}.
    The non-zero components fully determine the curvature of the sphere. 
    However, normally we think of the curvature in terms of the Gaussian curvature computed from Eq.~\eqref{eq:2.13}. 
    For the sphere, the Ricci scalar, $R=\frac{1}{a^2}$, is related to equation of the circle bounding the equator. 
    Alternatively, any other invariant for other characteristics of the curvature can be computed. 
    For example, the Kretschmann invariant is $R^{ijkl}R_{ijkl}=\frac{2}{a^4}$ and is connected with the surface area of the sphere. 
    Here, the curvature invariants measure the curvature of the manifold, and not the object's path through the manifold. 
    Unfortunately, the invariants in \ref{Chapter3}, \ref{Chapter4}, and \ref{Chapter5} are not as simple as the invariants of the sphere. 
    To gain a physical insight into the nature of the invariants, they will be plotted. \\

\subsection{CM Invariants}
\label{chp2:CM}
%Derive the CM Invariants.
%Be sure to cite the CM's paper.
%Display the full set in general form from CM's paper.
Carminati and McLenaghan (CM) proposed a set of invariants that is a complete and minimal set for the Einstein-Maxwell and perfect-fluid spacetimes \cite{CM}.
The CM invariants have highly desirable properties such as linear independence, the lowest possible degree, and containing a minimal set for both any Petrov type and specific choice of the Ricci tensor.
The complete set of invariants is the Ricci scalar from Eq.~\eqref{eq:2.13} and
\begin{align*}
    r_1 &:= \Phi_{AB\dot{A}\dot{B}} \ \Phi^{AB\dot{A}\dot{B}} = \frac{1}{4} S_\alpha^\beta S_\beta^\alpha, \label{eq:2.36} \tag{2.36} \\
    r_2 &:= \Phi_{AB\dot{A}\dot{B}} \  \Phi^B_C^{\dot{B}}_{\dot{C}} \ \Phi^{CA\dot{C}\dot{A}} = -\frac{1}{8} S_\alpha^\beta S_\beta^\gamma S_\gamma^\alpha, \label{eq:2.37} \tag{2.37} \\
    r_3 &:= \Phi_{AB\dot{A}\dot{B}} \  \Phi^B_C^{\dot{B}}_{\dot{C}} \ \Phi^C_D^{\dot{C}}_{\dot{D}} \ \Phi^{DA\dot{D}\dot{A}} = \frac{1}{16} S_\alpha^\beta S_\beta^\gamma S_\gamma^\delta S_\delta^\alpha, \label{eq:2.38} \tag{2.38} \\
    w_1 &:= \Psi_{ABCD} \ \Psi^{ABCD} = \frac{1}{4} \bar{C}_{\alpha \beta \gamma \delta} \bar{C}^{\alpha \beta \gamma \delta}, \label{eq:2.39} \tag{2.39} \\
    w_2 &:= \Psi_{ABCD} \ \Psi^{CD}_{EF} \ \Psi^{EFAB} = - \frac{1}{8} \bar{C}_{\alpha \beta \gamma \delta} \bar{C}^{\gamma \delta}_{\epsilon \zeta}\bar{C}^{\epsilon \zeta \alpha \beta}, \label{eq:2.40} \tag{2.40} \\
    m_1 &:= \Psi_{ABCD} \ \Phi^{CD}_{\dot{C}\dot{D}} \ \Phi^{AB\dot{C}\dot{D}} = \frac{1}{4} \bar{C}_{\alpha \gamma \delta \beta} S^{\gamma \delta} S^{\alpha \beta}, \label{eq:2.41} \tag{2.41} \\
    m_2 &:= \Psi_{ABCD} \ \Phi^{CD}_{\dot{C}\dot{D}} \ \Psi^{AB}_{EF} \ \Phi^{EF\dot{C}\dot{D}} = \frac{1}{4} \bar{C}_{\alpha \gamma \delta \beta} S^{\gamma \delta} \bar{C}^\alpha_{\epsilon \zeta}^\beta S^{\epsilon \zeta}, \label{eq:2.42} \tag{2.42} \\
    m_3 &:= \Psi^{AB}_{CD} \ \Phi^{CD}_{\dot{A}\dot{B}} \ \Bar{\Psi}^{\dot{A}\dot{B}}_{\dot{C}\dot{D}} \ \Phi_{AB}^{\dot{C}\dot{D}} = \frac{1}{4} \bar{C}_{\alpha \gamma \delta \beta} S^{\gamma \delta} C\textsuperscript{\textdagger}^{\alpha}_{\gamma \zeta}^\beta S^{\gamma \zeta}, \label{eq:2.43} \tag{2.43} \\
    m_4 &:= \Psi_{A}^{B}_{DE} \ \Phi^{DE}_{\dot{A}}^{\dot{B}} \ \Bar{\Psi}_{\dot{B}}^{\dot{C}}_{\dot{D}\dot{E}} \ \Phi_B^{C\dot{D}\dot{E}} \ \Phi_C^A_{\dot{C}}^{\dot{A}} = - \frac{1}{8} \bar{C}_{\alpha \gamma \delta \beta} S^{\gamma \delta} C\textsuperscript{\textdagger}^\beta_{\epsilon \zeta \eta} S^{\epsilon \zeta} S^{\alpha \eta}, \label{eq:2.44} \tag{2.44} \\
    m_5 &:= \Psi^{AB}_{CD} \ \Psi^{CD}_{EF} \ \Phi^{EF}_{\dot{E}\dot{F}} \ \bar{\Psi}^{\dot{E}\dot{F}}_{\dot{C}\dot{D}} \ \Phi_{AB}^{\dot{C}\dot{D}} = \frac{1}{4} \bar{C}_{\alpha \eta \theta \beta} \bar{C}^\alpha_{\gamma \delta}^\beta S^{\gamma \delta} C\textsuperscript{\textdagger}^\eta_{\epsilon \zeta}^\theta S^{\epsilon \zeta}. \label{eq:2.45} \tag{2.45}
\end{align*}
This set can be related to Eq.~\eqref{eq:2.26} through Eq.~\eqref{eq:2.34} by noting that:
\begin{align*}
    I&=\frac{1}{6} w_1, \label{eq:2.46} \tag{2.46} \\
    J&=\frac{1}{6} w_2, \label{eq:2.47} \tag{2.47} \\
    I_6&=\frac{1}{3} r_1, \label{eq:2.48} \tag{2.48} \\
    I_7&=\frac{1}{3} r_2, \label{eq:2.49} \tag{2.49} \\
    I_8&=\frac{1}{6} \Phi_{AB\dot{C}\dot{D}} \  \Phi^B_E^{\dot{C}}_{\dot{F}} \ ( \Phi^{AG\dot{H}\dot{D}} \ \Phi^{E}_{G}^{\dot{F}}_{\dot{H}}+\Phi^{AG\bar{H}\bar{F}} \ \Phi^{E}_{G}^{\dot{D}}_{\dot{H}}), \label{eq:2.50} \tag{2.50} \\
    K&= m_1, \label{eq:2.51} \tag{2.51} \\
    L&= \Psi_{(AB}^{EF} \ \Psi_{CD)EF} \ \Phi^{AB}_{\dot{G}\dot{H}} \ \Phi^{CD\dot{G}\dot{H}}, \label{eq:2.52} \tag{2.52} \\
    M&= \frac{1}{2} \Psi_{ABCD} \ \Phi^A_{E\bar{F}\bar{G}} \ \Phi^{BE\bar{F}}_{\bar{H}}(\Phi^C_E^{\bar{G}}_{\bar{F}} \ \Phi^{DE\bar{F}\bar{H}} + \Phi^C_E^{\bar{H}}_{\bar{F}} \ \Phi^{DE\bar{F}\bar{G}}), \label{eq:2.53} \tag{2.53} \\
    M_1&= m_3, \label{eq:2.54} \tag{2.54} \\
    M_2&= \Psi^{(AB}_{CD} \ \Psi^{EF)CD} \  \bar{\Psi}_{\dot{G}\dot{H}\dot{I}\dot{J}} \ \Phi_{AB}^{\dot{I}\dot{J}} \ \Phi_{EF}^{\dot{G}\dot{H}}. \label{eq:2.55} \tag{2.55}
\end{align*}

Several invariants do not have an explicit relationship, but $I_8$ is related to $R_3$, $L$ is the symmetrization of $m_2$, and $M_2$ is the symmetrization of $m_5$.
The invariant $M$ is an additional one designed to augment and complete the CM invariants for general spacetimes \cite{Zakhary:1997}.
It is emphasized that the specific set attributed to CM is sufficient to compute the invariants for the spacetimes considered in Chapters \ref{Chapter3}, \ref{Chapter4}, and \ref{Chapter5}.
The full set of CM invariants have been computed in the NP formalism using the computer program MAPLE.
The set is long and complicated, so only the non-zero ones in Class B spacetimes will be presented in the next subsection. \\

\subsection{Syzygies of CM Invariants} 
\label{chp2:Syzygies}
%Citing Santossuosso's paper, display the syzygies of the CM invariants and the four invariants that minimize them: $R$, $r_1$, $r_2$, $w_2$. Make sure to discuss what a syzygy is and what a Class B$_1$ spacetime is.
Each CM of the invariants Eq.~\eqref{eq:2.36} through Eq.~\eqref{eq:2.45} may be calculated for any possible spacetime.
But the set will be degenerate due to sets of internal relationships.
To reduce the complete set of invariants to the subset of independent ones, the syzygies of the CM invariants must be considered.
For Class B warped product spacetimes, the syzygies and independent set are known and will be presented in this subsection from \cite{Santosuosso}.

\theoremstyle{definition}
\begin{definition}{Class B Spacetime}
is the product of two 2-D spaces, one Lorentzian and one Riemannian, subject to a separability condition on the function which couples the spaces.
The metric is of the form
\begin{equation}
    ds^2=ds_{\Sigma_1}^2(u,v)+C(x^\gamma)^2ds_{\Sigma_2}(\theta,\phi), \label{eq:2.56} \tag{2.56}
\end{equation}
where $\Sigma_1$ is the Lorentzian manifold and $\Sigma_2$ is the Riemannian.
The restriction $C(x^\gamma)^2$ is of the form $C(x^\gamma)^2=r(u,v)^2w(\theta,\phi)^2$.
\end{definition}
Class B spacetimes are a specific case of the Petrov Class D metric.\footnote{Petrov Class D spacetimes have an eigenvalue equation of the Weyl tensor of the form 
$(\textbf{C}+\frac{1}{2}\lambda \textbf{I})(\textbf{C}-\frac{1}{2} \lambda \textbf{I})=0$. 
The eigenvalues are simple divisors and satisfy $\lambda_1=\lambda_2 \neq \lambda_3$.}
Class B spacetimes are further divided into two main categories.
Class B$_1$ spacetimes have $sig(\Sigma_1)=0$ and $sig(\Sigma_2)=2\epsilon (\epsilon=\pm1)$.
A sufficiently generic metric of Class B$_1$ is
\begin{equation}
    ds^2=-2f(u,v)du dv +r(u,v)^2 g(\theta,\phi)^2 (d\theta^2+d\phi^2). \label{eq:2.57} \tag{2.57}
\end{equation}
Class B$_2$ spacetimes have $sig(\Sigma_1)=2\epsilon$ and $sig(\Sigma_2)=0$.
A sufficiently generic metric of Class B$_2$ is
\begin{equation}
    ds^2=f(u,v)^2(du^2 + dv^2) -2r(u,v)^2 g(\theta,\phi) d\theta d\phi. \label{eq:2.58} \tag{2.58}
\end{equation}
Class B$_1$ warped product spacetimes include all spherical, planar, hyperbolic, while Class B$_2$ spacetimes include the non-null EM, $\Lambda$-term, or vacuum spacetimes.
All spacetimes considered in this dissertation are Class B$_1$.

Next, the number of independent invariants will be considered to restrict the number of needed syzygies.
The number of independent degrees of freedom in a spacetime leads to the required number of invariants.
General Petrov type D spacetimes allow a choice of our tetrad to be along the principal null directions of the Weyl tensor.
With this choice, the remaining degrees of freedom are the Weyl and Ricci freedoms, the Ricci scalar, and the dimension of the invariance group.
Consequently, the number of independent invariants is four for Class B$_1$ spacetimes.

There are twelve separate invariants in the CM set between Eqs.~\eqref{eq:2.13}, \eqref{eq:2.36} through \eqref{eq:2.45}, and \eqref{eq:2.53}.
After considering the degrees of freedom, the CM set should have eight syzygies leaving only four real, independent invariants.
The eight syzygies are
\begin{align*}
    0 &=6w_2^2-w_1^3, \label{eq:2.59} \tag{2.59} \\
    0 &=(3m_2-w_1 r_1)w_1-3m_1 w_2, \label{eq:2.60} \tag{2.60} \\
    0 &=(3m_5-w_1 \bar{m}_1)w_1-3m_3 w_2, \label{eq:2.61} \tag{2.61} \\
    0 &=6m_4+w_1 r_2, \label{eq:2.62} \tag{2.62} \\
    0 &=m_3-m_2, \label{eq:2.63} \tag{2.63} \\
    0 &=(-12r_3+7r_1^2)w_1 m_1-(12r_2^2-36r_1 r_3+17r_1^3)w_2, \label{eq:2.64} \tag{2.64} \\
    0 &=2(3m_6-m_1r_1)w_2+m_1^2w_1, \label{eq:2.65} \tag{2.65} \\
    0 &=(-12r_3+7r_1^2)^3-(12r_2^2-36r_1r_3+17r_1^3)^2. \label{eq:2.66} \tag{2.66}
\end{align*}
The syzygies Eqs.~\eqref{eq:2.60} through \eqref{eq:2.66} are well known for Class D spacetimes.
The remaining syzygy Eq.~\eqref{eq:2.59} is particular to Class B warped product spacetimes.
The syzygies allow the invariants Eqs.~\eqref{eq:2.38}, \eqref{eq:2.39}, and \eqref{eq:2.41} through \eqref{eq:2.45} to be expressed in terms of the remaining ones.
Syzygy Eq.~\eqref{eq:2.59} allows $w_2$ to be expressed in terms of $w_1$ or vice versa.
Since the sign of $w_2$ changes based on the signature of the metric, it is chosen as part of the independent set.
Consequently, the set of independent invariants for Class B warped product spacetimes is the four  
\begin{equation}
    (R,r_1,r_2,w_2). \label{eq:2.67} \tag{2.67}
\end{equation}
For reference in Chapters \ref{Chapter3}, \ref{Chapter4}, and \ref{Chapter5}, the three invariants $r_1$, $r_2$, and $w_2$ written in terms of the trace-free Ricci tensor, Weyl tensor, and in terms of the NP coordinates are
\begin{align*}
    \begin{split}
	r_1& = \frac{1}{4} S_\alpha^\beta S_\beta^\alpha \\ & = 2\Phi_{20}\Phi_{02}+2\Phi_{22}\Phi_{00}-4\Phi_{12}\Phi_{10}-4\Phi_{21}\Phi_{01}+4\Phi_{11}^2, \end{split} \label{eq:2.68} \tag{2.68}
	\\
	\begin{split}
	r_2& = -\frac{1}{8} S_\alpha^\beta S_\gamma^\alpha S_\beta^\gamma \\ & = 6\Phi_{02}\Phi_{21}\Phi_{10}-6\Phi_{11}\Phi_{02}\Phi_{20}+6\Phi_{01}\Phi_{12}\Phi_{20}-6\Phi_{12}\Phi_{00}\Phi_{21} -6\Phi_{22}\Phi_{01}\Phi_{10}+6\Phi_{22}\Phi_{11}\Phi_{00}, \end{split} \label{eq:2.69} \tag{2.69}
	\\
	w_2& = -\frac{1}{8} \bar{C}_{\alpha \beta \gamma \delta} \bar{C}^{\alpha \beta \epsilon \zeta} \bar{C}^{\gamma \delta}{}_{\epsilon \zeta} = 6\Psi_4\Psi_0\Psi_2-6\Psi_2^3-6\Psi_1^2\Psi_4-6\Psi_3^2\Psi_0+12\Psi_2\Psi_1\Psi_3, \label{eq:2.70} \tag{2.70}
\end{align*}
where the $\Psi$ and $\Phi$ are the following abbreviations for the tetrad components of the trace-free Ricci tensor Eq.~\eqref{eq:2.14} and the Weyl tensor Eq.~\eqref{eq:2.15}: 
\begin{align*}
    \Phi_{00}&\equiv\frac{1}{2}S_{\alpha \beta}k^\alpha k^\beta=\frac{1}{2}R_{44}=\bar{\Phi}_{00}, \tag{2.71.a} \label{eq:2.71.a} \\
    \Phi_{01}&\equiv\frac{1}{2}S_{\alpha \beta}k^\alpha m^\beta=\frac{1}{2}R_{41}=\bar{\Phi}_{10}, \tag{2.71.b} \label{eq:2.71.b} \\
    \Phi_{02}&\equiv\frac{1}{2}S_{\alpha \beta}m^\alpha m^\beta=\frac{1}{2}R_{11}=\bar{\Phi}_{20}, \tag{2.71.c} \label{eq:2.71.c} \\
    \Phi_{11}&\equiv\frac{1}{2}S_{\alpha \beta}(k^\alpha l^\beta+m^\alpha \bar{m}^\beta)=\frac{1}{4}(R_{43}+R_{12}) =\bar{\Phi}_{11}, \tag{2.71.d} \label{eq:2.71.d} \\
    \Phi_{12}&\equiv\frac{1}{2}S_{\alpha \beta}l^\alpha m^\beta =\frac{1}{2}R_{31}=\bar{\Phi}_{21}, \tag{2.71.e} \label{eq:2.71.e} \\
    \Phi_{22}&\equiv\frac{1}{2}S_{\alpha \beta}l^\alpha l^\beta=\frac{1}{2}R_{33}=\bar{\Phi}_{22}, \tag{2.71.f} \label{eq:2.71.f} \\
    \Psi_{0}&\equiv C_{\alpha \beta \gamma \delta}k^\alpha m^\beta k^\gamma m^\delta, \tag{2.71.g} \label{eq:2.71.g} \\
    \Psi_{1}&\equiv C_{\alpha \beta \gamma \delta}k^\alpha l^\beta k^\gamma m^\delta, \tag{2.71.h} \label{eq:2.71.h} \\
    \Psi_{2}&\equiv -C_{\alpha \beta \gamma \delta}k^\alpha m^\beta l^\gamma \bar{m}^\delta, \tag{2.71.i} \label{eq:2.71.i} \\
    \Psi_{3}&\equiv C_{\alpha \beta \gamma \delta}l^\alpha k^\beta l^\gamma \bar{m}^\delta, \tag{2.71.j} \label{eq:2.71.j} \\
    \Psi_{4}&\equiv C_{\alpha \beta \gamma \delta}l^\alpha \bar{m}^\beta l^\gamma \bar{m}^\delta. \tag{2.71.k} \label{eq:2.71.k}
\end{align*}

As some final comments about this set, it is neither complete nor unique.
A complete choice of invariants would cover every possible Petrov and Segre type.
Each CM invariant not contained in Eq.~\eqref{eq:2.67} may be written as a rational integer multiple of the independent invariants.
The best example of this statement is how $w_1 = 6^{\frac{1}{3}} w_2^{\frac{2}{3}}$ in Eq.~\eqref{eq:2.63}.
A choice was made to include $(R,r_1,r_2)$ in the set of independent invariants because of their relative simplicity compared to the other CM invariants.
An alternative set, $(R,m_2,m_4)$, would have also satisfied the syzygies, but $m_2$ and $m_4$ contain more summations of $S_\alpha^\beta$ and $\bar{C}_{\alpha \beta \gamma \delta}$ than $r_1$ and $r_2$. The chosen three are the most direct to calculate from a given metric Eq.~\eqref{eq:2.1}.
Next, the remaining CM invariants can be found from the set Eq.~\eqref{eq:2.67} by solving for them using the syzygies.

Finally, the CM invariants may be found either by use of the standard trace-free Ricci tensor Eq.~\eqref{eq:2.14} and the Weyl tensor Eq.~\eqref{eq:2.15} or by computing the NP indices Eq.~\eqref{eq:2.71.a} through Eq.~\eqref{eq:2.71.k}.
Two formulas for calculating the invariants have been presented in this section because each method has its own advantages.
The first set of formulas computes the CM invariants from the Trace-free Ricci tensor and the Weyl tensor.
It is the more straightforward procedure as it requires only a single input, the metric, Eq.~\eqref{eq:2.1}, of the spacetime manifold. 
The wormhole metrics presented in Chapter \ref{Chapter3} were computed using the first set.
The second set of formulas computes the NP indices from specific functions in that they are individual elements of the Ricci tensor and Weyl tensor as seen in Eq.~\eqref{eq:2.71.a} through Eq.~\eqref{eq:2.71.k}.
The second formulas may be computed more rapidly by a computer as they require fewer summations.
The drawback of the NP indices is that they require two inputs, the metric from the spacetime manifold, Eq.~\eqref{eq:2.1}, and a null tetrad, Eq.~\eqref{eq:2.8.a} through Eq.~\eqref{eq:2.8.d}.
The CM invariants for the warp drive metrics in Chapters \ref{Chapter4} and \ref{Chapter5} were computed by finding the NP indices.
All CM invariants contained in this dissertation were computed using Wolfram Mathematica $10.4$\textsuperscript{\textregistered}.
As a test, the output of the program matched exactly the known Riemann, Ricci, Weyl tensors and the Ricci scalar for the Schwarzschild metric, the MT wormhole, and the exponential metric.
In addition, derivations by hand of the four CM invariants for each wormhole in Chapter \ref{Chapter3} matched the program's output exactly.
The complete program is provided in Appendix \ref{ap:Program}. \\

%% file: ch3.tex
\catcode`\^=13\def^#1{\sp{#1}{}}
\catcode`\_=13\def_#1{\sb{#1}{}}

\chapter{Lorentzian Traversable Wormholes}
%In this chapter, Lorentzian Traversable Wormholes will be explored. 
%Based off of (but not copied) from my paper "Lorentzian Traversable Wormholes" and the various Drafts of it.
\label{Chapter3}
Before 1988, each identified wormhole line element precluded travel through it.
A traveler would encounter many crippling impossibilities such as a requirement of a trip duration lasting an infinite amount of time for an external observer, impassable throats on the order of the Planck length of $10^{-35}$m, naked singularities generating permanently destructive tidal forces, or curvature singularities lurking behind the event horizons.
Needless to say, scientists dismissed traversable wormholes as more science fiction than science fact. \\

Morris and Thorne demonstrated traversable wormholes by considering the converse situation to previous analysis \cite{Morris:1988A,Morris:1988B}.
Instead of investigating known spacetimes for wormholes, they generated spacetimes with wormholes that obeyed certain traversability requirements.
The traversable wormhole should allow a human (or any amount of matter of a similar size) to travel through it without damage and return in a reasonable amount of time.
To satisfy this main requirement, the wormhole spacetime should be both free of event horizons and naked singularities.
Wormhole spacetimes that contain no curvature singularities satisfy these two requirements.
Deriving the main curvature invariants using the process in the previous chapter and then plotting them will reveal any curvature singularities.

In this chapter, four basic wormholes will be presented:  the TS Flat-Face wormhole (\ref{chp3:TSFF}),  the MT wormhole (Section~\ref{chp3:MT}), the TS Schwarzschild wormhole (Section~\ref{Chp3:TSS}), and the exponential metric (\ref{Chp3:Exp}).
The non-trivial CM invariant functions for each wormhole will be presented and certain demonstrative plots for each will be displayed.
The plots will be inspected for places of great curvature and general traversability.
Appropriately, the wormholes will be compared and contrasted for similar features. \\

\section{Thin-Shell Flat-Face Wormhole}
\label{chp3:TSFF}
    The TS Flat-Face wormhole is one of the simplest wormhole solutions.
    It consists of two separate regions of Minkowski spacetime. 
    A small portion of each region is sliced out and connected with the other region using the TS Formalism.
    The formalism to compute the tensors for TS wormholes is outlined in \cite{Visser:1995}. 
    In brief, two copies of Minkowski flat space on either side of the wormhole's throat are assumed, identical regions from each space are removed, and then separate regions along the boundary are connected. 
    This formalism leads to a well-behaved wormhole, with the throat being located at the connecting boundary between the separate regions.
    
    In the TS formalism, the metric is modified to be: \begin{equation}
	g_{\mu\nu}(x)=\mathit{\Theta}\left(\eta(x)\right)g^+_{\mu\nu}(x)+\mathit{\Theta}\left(-\eta(x)\right)g^-_{\mu\nu}(x),
	\tag{3.1} \label{eq:3.1} 
	\end{equation} where $g^\pm_{\mu\nu}$ is the metric on the respective sides, $\mathit{\Theta}\left(\eta(x)\right)$ is the Heaviside-step function and $\eta(x)$ is the outward pointing normal from the wormhole’s throat. 
	The radius of the wormhole’s throat is located at the point the regions overlap, $x=a$ (that $x\geq a$ is important to note in regards to analyzing divergences). 
	This formalism requires the second fundamental form $K_{\mu\nu}^{\pm}$ for the analysis at the throat to be:
	\begin{equation}
	K_{\mu\nu}^{\pm}=
	\pm\begin{pmatrix}
	\ 0&\ \ 0&\ \ 0&\ \ 0\ \ \\[1mm]
	\ 0&\ \ \frac{1}{R_1}&\ \ 0&\ \ 0\ \ \\[1mm]
	\ 0&\ \ 0&\ \ \frac{1}{R_2}&\ \ 0\ \ \\[1mm]
	\ 0&\ \ 0&\ \ 0&\ \ 0\ \ 
	\end{pmatrix}, \tag{3.2}
	\label{eq:3.2}
	\end{equation}
	where $R_1$ and $R_2$ are the radii of curvature of the wormhole on either side. 
	The TS formalism modifies the Riemann tensor to become:
	\begin{equation}
	R_{\kappa \lambda \mu \nu}=-\delta(\eta)\left[k_{\kappa \mu}n_{\lambda}n_{\nu}+k_{\lambda \nu}n_{\kappa}n_{\mu}-k_{\kappa \nu}n_{\lambda} n_{\mu}-k_{\lambda \mu}n_{\kappa}n_{\nu}\right]+\mathit{\Theta}\left(\eta\right)R_{\kappa \lambda \mu \nu}^{+}+\mathit{\Theta}\left(-\eta\right)R_{\kappa \lambda \mu \nu}^{-}. \tag{3.3} \label{eq:3.3}
	\end{equation}
	where $\delta(\eta)$ is the delta function, $k_{\mu\nu}=K_{\mu\nu}^+-K_{\mu\nu}^-$ is the discontinuity in the second fundamental form, and $n_{\lambda}$ is the unit normal to the boundary of the shell.
	
	For the TS Flat-Face wormhole, the line element on either side of the throat is the Minkowski metric,
    \begin{equation}
        ds^2=-dt^2+dx^2+dy^2+dz^2. \tag{3.4} \label{eq:3.4}
    \end{equation}
    By computing the TS formalism, computing the TS Flat Face wormhole's CM invariants shows that all invariants vanish.
    The plot is the same as Minkowski space with zero curvature.
    The plot forms a single round disk with no divergences, singularities, discontinuities or other artifacts that might prevent travel through a TS wormhole.
    As the plot is incredibly simple, it was not included.
    Because the invariants are equivalent to flat space, a TS flat-face wormhole will be traversable.

\section{Morris-Thorne Wormhole}
\label{chp3:MT}
\begin{figure}[ht]
	\centering
	\begin{subfigure}{.48\linewidth}
		\includegraphics[scale=0.23]{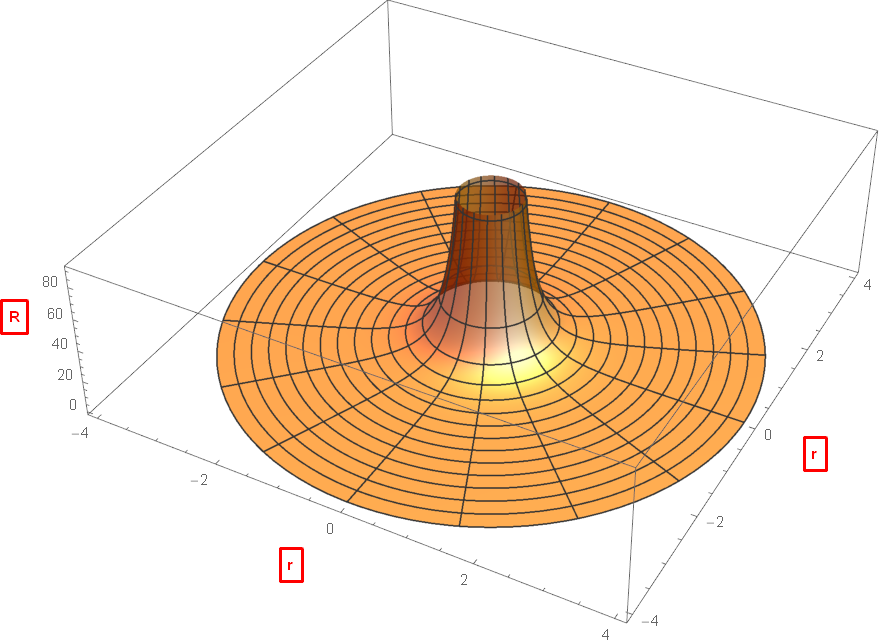}
		\caption{Plot of MT $R$}
		\label{RMTm}
	\end{subfigure}
	~
	\begin{subfigure}{.48\linewidth}
		\includegraphics[scale=0.23]{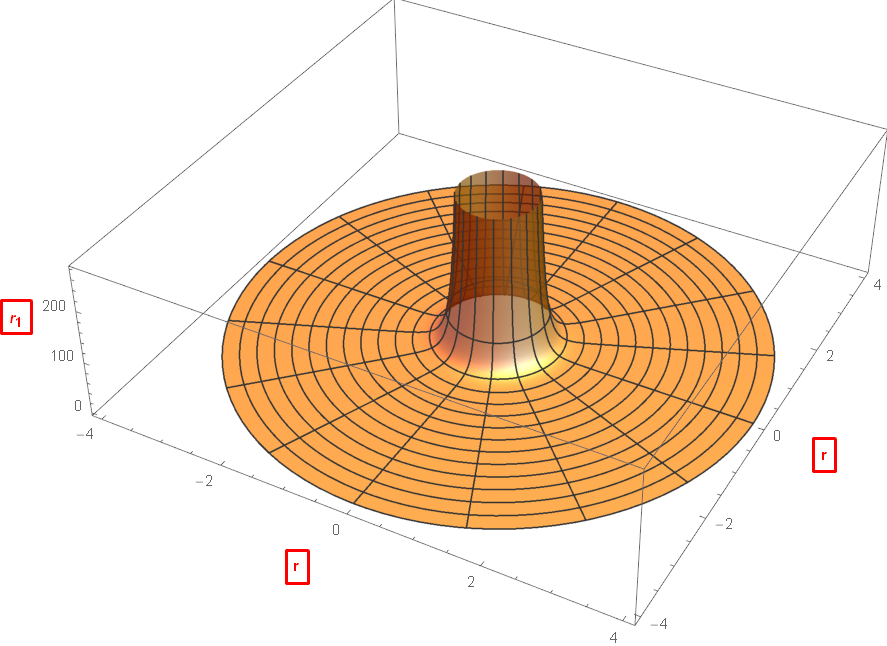}
		\caption{Plot of MT $r_1$}
		\label{r1MTm}
	\end{subfigure}
	\par \bigskip
	\begin{subfigure}{.48\linewidth}
		\includegraphics[scale=0.23]{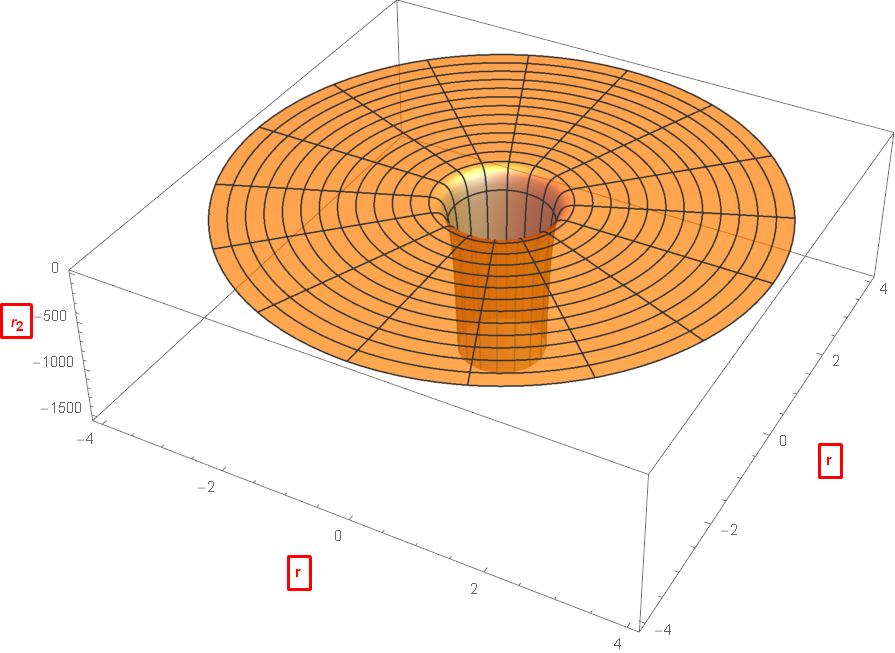}
		\caption{Plot of MT $r_2$}
		\label{r2MTm}
	\end{subfigure}
	~
	\begin{subfigure}{.48\linewidth}
		\includegraphics[scale=0.23]{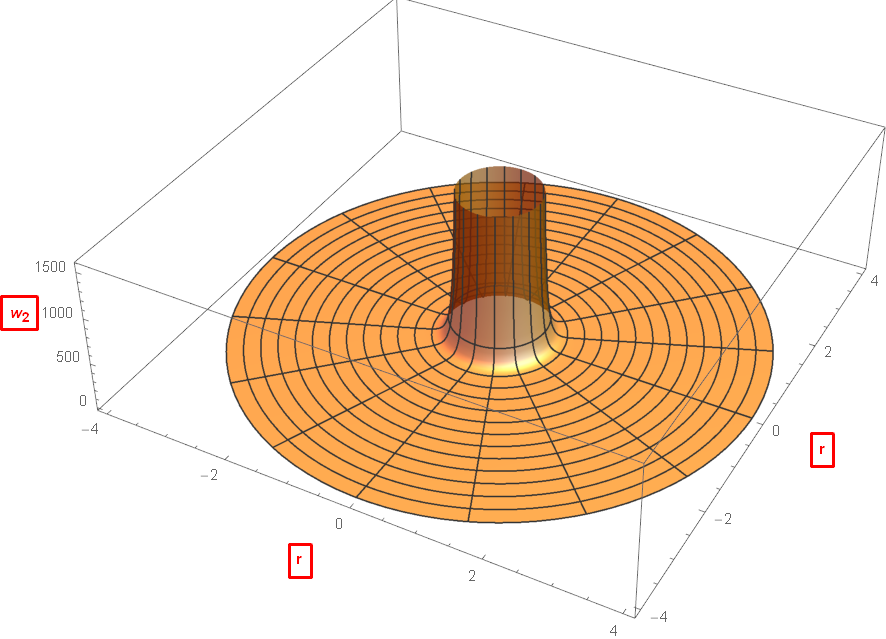}
		\caption{Plot of MT $w_2$}
		\label{w2MTm}
	\end{subfigure}
	\caption{Plots of the non-zero invariants for the MT wormhole. 
	The plots are in radial coordinates with $r\in\{0,4\}$. Each radial mesh line represents a radial distance of $r=0.2\bar{6}$. $G=M=1$ were normalized for simplicity and $r_0=2$ was chosen as the throat. 
	Notice the divergence at the center of each plot is completely inside the $r=2=r_0$ radial line. 
	This does not affect the traversability of the wormhole. } \label{fig:3.1}
\end{figure}

    The MT wormhole is a spherically symmetric and Lorentzian spacetime. 
    In the standard Schwarzschild coordinates \cite{Morris:1988A,Morris:1988B}, its line element is:
    \begin{equation}
    ds^2=-e^{2\phi^\pm(r)}dt^2+\frac{dr^2}{\left(1-\frac{b^\pm(r)}{r}\right)}+r^2(d\theta^2+\sin^2\theta  \ d\varphi^2). \tag{3.5} \label{eq:MT}
    \end{equation}
    The tetrad for the MT line element uses the spherical coordinates $(r$: with circumference $=2\pi r;\ 0\leq\theta < \pi;\ 0\leq\varphi < 2\pi)$, and $ (-\infty< t<\infty)$ is the proper time of a static observer.
    $\phi^\pm(r)$ is the freely specifiable redshift function that defines the proper time lapse through the wormhole throat. 
    $b^\pm(r)$ is the freely specifiable shape function that defines the wormhole throat’s spatial (hypersurface) geometry. 
    The $\pm$ indicates the side of the wormhole. 
    The throat described by Eq.~\eqref{eq:MT} is spherical.
    A fixed constant, $r_0$, is chosen to define the radius of the wormhole throat such that $b^\pm(r_0) = r_0$, which is an isolated minimum. 
    Two coordinate patches of the manifold are then joined at $r_0$. 
    Each patch represents either a different part of the same universe or another universe, and the patches range from $r_0\leq r<\infty$. 
    The condition that the wormhole is horizon-free requires that $g_{tt}=-e^{2\phi^\pm(r)} \neq 0$. This statement implies that $|\phi^\pm(r)|$ must be finite everywhere \cite{Lobo:2017,Visser:1995}.
    The use of Schwarzschild coordinates in Eq.~\eqref{eq:MT} leads to more efficient computations of the Riemann and Ricci curvature tensors, the Ricci scalar, and all four invariants.
    The four CM invariants for the MT Wormhole are
    \begin{align*}
    R &= \frac{1}{r^2}(b'(r \Phi'+2)+2r (b-r )\Phi''-2r (r-b)\Phi^{'2}+(3b -4r )\Phi'), \tag{3.6} \label{eq:3.6}
    \end{align*}
    ~
    \newpage
    ~
    \begin{align*}
    r_1 &= \frac{1}{16r^6}\bigl(r^2 \bigl(b^{'2} \left(r^2 \Phi^{'2}+2\right)-4r b'\Phi' \left(r^2 \Phi''+r^2 \Phi^{'2}-2 \right) \\ 
    & \ \ \ \ \ \ \ \ \ \ \ +4r^2\bigl(r^2 \Phi^{''2}+r^2\Phi^{'4}+2\Phi^{'2}\left(r^2\Phi''+1\right)\bigr)\bigr) \\ 
    & \ \ \ \ \ \ \ \ \ \ \ -2 r b \bigl(b'\left(-2r^3\Phi^{'3}+\Phi'\left(6r-2r^3\Phi ''\right)+r^2\Phi^{'2}+2\right) \\ 
    & \ \ \ \ \ \ \ \ \ \ \ + 2r\bigl(2r^3\Phi^{'4}+2\Phi^{'2} \left(2r^3\Phi''+r\right) \\
    & \ \ \ \ \ \ \ \ \ \ \ +2r\Phi'' \left(r^2\Phi''-1\right)-r^2\Phi^{'3}+\Phi'\left(2-r^2 \Phi''\right)\bigr)\bigr) \\
    & \ \ \ \ \ \ \ \ \ \ \
    + b^2\bigl(4 r^4\Phi^{''2}+4r^4\Phi^{'4} -4r^3\Phi^{'3}-8r^2\Phi''-4r\Phi'\left(r^2 \Phi''-3\right) \\
    & \ \ \ \ \ \ \ \ \ \ \ \ \ \ \ \ \ \ +\Phi^{'2}\left(8r^4\Phi''+r^2\right)+6\bigr)\bigr), \tag{3.7} \label{eq:3.7}
    \end{align*}
    ~
    \begin{align*}
    r_2 &= -\frac{3}{64r^9}\bigl(b\left(2r\Phi'+1\right)-r\left(b+2r\Phi'\right)\bigr)^2 \\
    & \ \ \ \ \ \ \ \ \ \ \times \bigl(r^2\left(b'\Phi'-2r\left(\Phi''+\Phi^{'2}\right)\right)+b\left(2r^2\Phi''+2r^2\Phi^{'2}-r\Phi'-2\right)\bigr), \tag{3.8} \label{eq:3.8} \\ \\
    w_2 &= \frac{1}{144 r^9}\bigl(r \left(b' \left(1-r \Phi'\right)+2r\left(r\Phi''+r\Phi^{'2}-\Phi'\right)\right)\\
    & \ \ \ \ \ \ \ \ \ \ \ \ \ -b\left(2 r^2 \Phi''+2r^2 \Phi^{'2}-3r\Phi'+3\right)\bigr)^3. \tag{3.9} \label{eq:3.9}
    \end{align*}
    All the invariants are non-zero and depend only on the radial coordinate, $r$, implying they are spherically symmetric. 
    The invariants are plotted in Fig.~\ref{fig:3.1} after selecting  a redshift function of $\phi(r)=0$ and the shape function of
    \begin{equation}
    b(r)=2GM\left(1-e^{r_0-r}\right)+r_0e^{r_0-r}. \tag{3.10} \label{eq:SF}
    \end{equation} 
    These functions satisfy the constraints on the asymptotic behavior and continuity at the wormhole's throat \cite{Visser:1995}. 
    At a distant greater than $0.5 \ r_0$, all the figures are asymptotically flat. 
    For $r\to0$, the figures diverge to infinity. 
    The intrinsic singularity at $r=0$ is not pathological as the radial coordinate $r$ has a minimum $r_0>0$ at the wormhole’s throat. 
    Thus, a traveler passing through the wormhole would not experience the divergence. 
    Any tidal forces on the traveler would be minimal.
    Consequently, the MT wormhole would be traversable as indicated by the included invariant plots.
    
\FloatBarrier

\subsection{Variation of Shape Function}
\label{chp3.2:Shape}
    The MT Wormhole given in Eq.~\eqref{eq:MT} has two freely specifiable functions, i.e., the redshift function, $\phi_\pm(r)$, and the shape function, $b_\pm(r)$. 
    These two functions satisfy the consistency requirements as \cite{Visser:1995} entails.
    However, additional constraints can be imposed by choice of the shape function for ease of calculations and temporal and spatial symmetry.
    The shape function must have:
    \begin{enumerate}
    	\item existence and finiteness at both limits, $\lim_{r\to\pm\infty} b(r)=b_\pm$,
	    \item the masses of the wormhole $M_\pm$ on the two sides are given by $b_\pm=2GM_\pm$,
    	\item $\exists r_*\|\forall r\in(r_0,r*), \ b'(r)<\frac{b(r)}{r}$,
    	\item $b_+(r_0)=b_-(r_0)$ and $b'_+(r_0)=b'_-(r_0)$ at the throat. 
    \end{enumerate}  
    
    The shape function chosen in Eq.~\eqref{eq:SF} obeys the listed conditions, with \\ $\lim_{r\to\pm\infty} b(r)=2GM$. 
    Thus, the shape function at the throat exists, is finite, and is continuous, which satisfies the second and fourth conditions.
    At the throat (i.e. $r\to r_0$), $b(r)=r_0$. 
    The derivative, $b'(r)=(2GM-r_0)e^{r_0-r}=b(r)+2GM$, is in agreement with the third condition mentioned above.\footnote{Since $G>0$ and $M>0$, $b'(r)<\frac{b(r)}{r}$ for all $r_0\le r\le \infty$ instead of a specific range of $r_*$ as necessitated by the third condition.}
    The freely specified shape function can have a significant impact on the form of the invariant functions.
    By applying the second derivative test, it is seen that a shape function with a term an $r^n$ term with $n\ge3$ will not have a discontinuity at $r=0$. 
    To test this, consider a series of shape functions 
    \begin{equation}
        b(r)=\frac{r^n}{r_0^3}\left(e^{r_0-r}\right), \tag{3.11} \label{eq:3.11}
    \end{equation}
    for $n=(0,1,2,3)$.
    Shape functions of this nature satisfy the conditions above.
    
\begin{figure}[ht]
	\centering
	\begin{subfigure}{.48\linewidth}
		\includegraphics[scale=0.35]{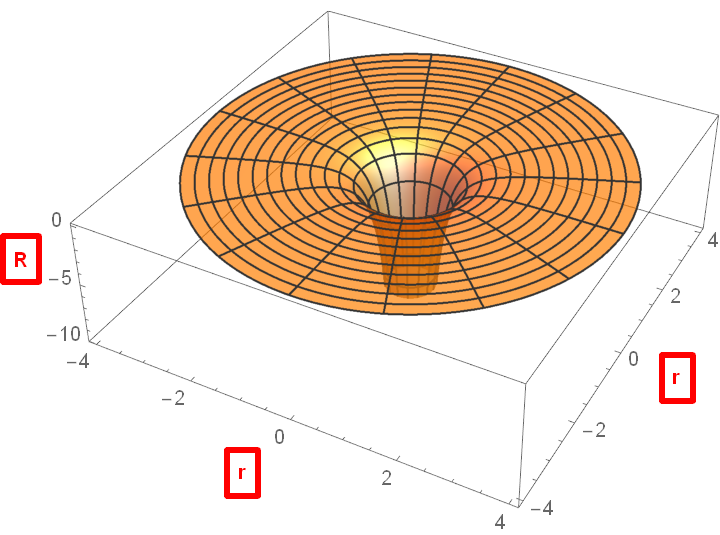}
		\caption{Plot of MT $R$ with $n=0$ in Eq.~\eqref{eq:3.11}}
		\label{vMTR0}
	\end{subfigure}
	~
	\begin{subfigure}{.48\linewidth}
		\includegraphics[scale=0.35]{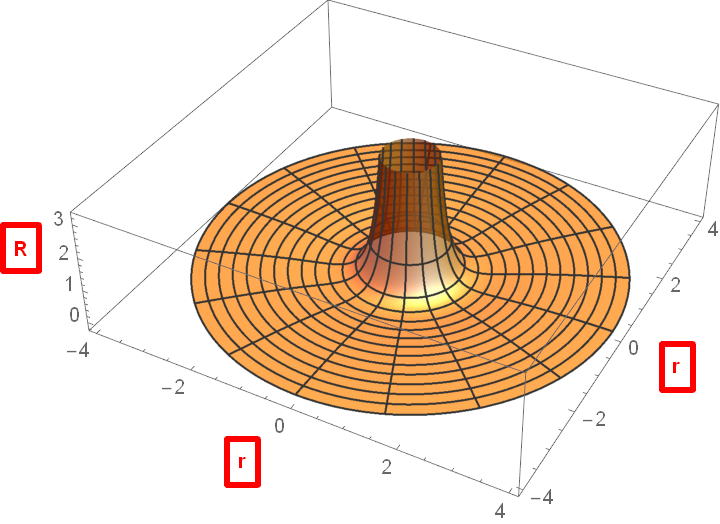}
		\caption{Plot of MT $R$ with $n=1$ in Eq.~\eqref{eq:3.11}}
		\label{vMTR1}
	\end{subfigure}
	\par \bigskip
	\begin{subfigure}{.5\linewidth}
		\includegraphics[scale=0.3]{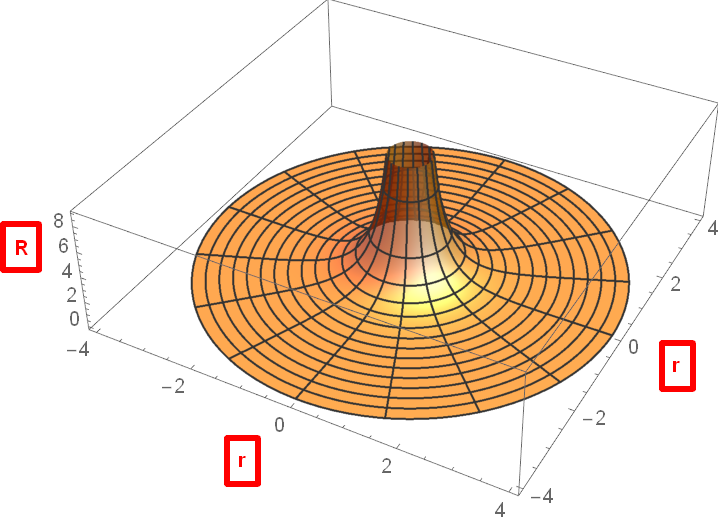}
		\caption{Plot of MT $R$ with $n=2$ in Eq.~\eqref{eq:3.11}}
		\label{vMTR2}
	\end{subfigure}
	~
	\begin{subfigure}{.52\linewidth}
		\includegraphics[scale=0.3]{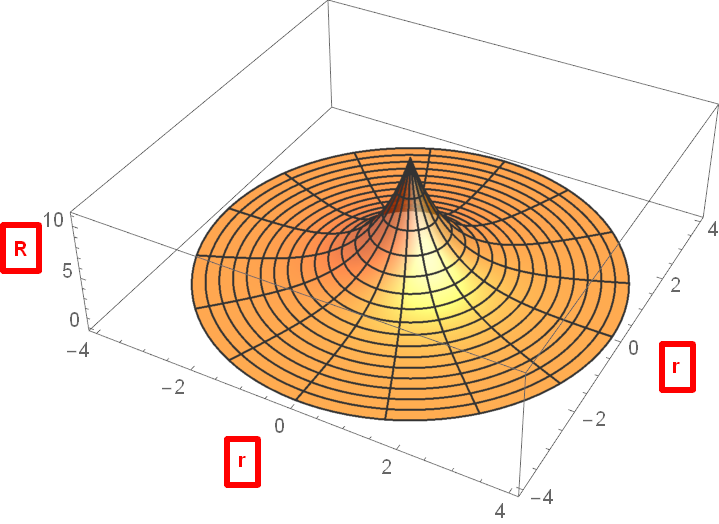}
		\caption{Plot of MT $R$ with $n=3$ in Eq.~\eqref{eq:3.11}}
		\label{vMTR3}
	\end{subfigure}
	\caption{Successive plots of the shape function for different powers of $r$ in the Ricci scalar for the MT wormhole. 
	The plots are in radial coordinates with $r\in\{0,4\}$. Each radial mesh line represents a radial distance of $r=0.2\bar{6}$. $G=M=1$ were normalized for simplicity and $r_0=2$ was chosen as the throat.  } \label{fig:3.2}
\end{figure}

    The successive shape functions for the Ricci scalar are displayed in Fig.~\ref{fig:3.2}.
    While the discontinuity at the center exists for the lower values of $n$, it  disappears at $r=0$ and the shape resembles that of an inverted cone with a finite depth. 
    While this is outside the domain of $r$ as discussed above, it demonstrates the effect that the shape function holds over the invariants and how controlling the shape function allows the removal of any singularities as a wormhole is engineered.
    
\begin{figure}[ht]
	\centering
	\includegraphics[scale=0.3]{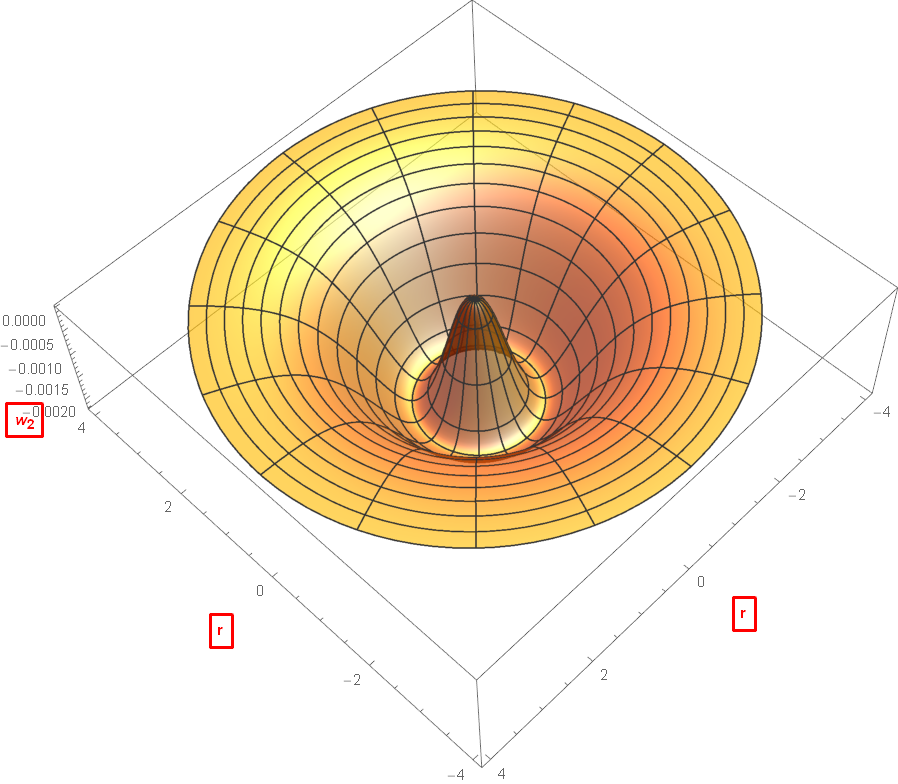}
	\caption{Plot of MT $w_2$ for the shape function given in Eq.~\eqref{eq:3.11}. The plots are in radial coordinates with $r\in\{0,4\}$ with $G=M=1$ normalized and $r_0=2$ chosen.}
	\label{ewMTm}
\end{figure}

    In fact, the power of the shape function extends further.
    Fig.~\ref{ewMTm} is very similar to the Ricci scalar from the exponential metric discussed in Section \ref{Chp3:Exp}.
    This suggests that the exponential metric can be obtained from the MT metric by a suitable choice of the shape function.
    This is an interesting research topic and is a consequence of the CM curvature invariant's ability to distinguish spacetime metrics by solving the equivalence problem.
    Potentially, a different choice of invariants like the Cartan set may be able to answer whether the MT and exponential metrics describe the same spacetime.

~
    
%\subsection{Variation of the RedShift Function}
%\label{chp3.2:Red}
    %The second freely specifiable function in the MT Wormhole given in \eqref{eq:MT} is the redshift function, $\phi_\pm(r)$. 
    %The consistency requiremenets for the redshift function are:
    %\begin{enumerate}
 	 %   \item continuity of the $t$ coordinate across the throat, $\phi_+(r_0)=\phi_-(r_0)$,
 	  %  \item existence and finiteness of both the limits, $\lim_{r\to\pm\infty} \phi(r)=\phi_\pm$.
    %\end{enumerate} 
    %These two aforementioned constraints are the minimum required for a traversable wormhole. 
    
\FloatBarrier

\section{Thin-Shell Schwarzschild Wormhole}
\label{Chp3:TSS}
    The next wormhole to investigate is the Thin-Shell Schwarzschild wormhole.
    It is two Schwarzschild black holes connected at their event horizons to make a wormhole.
    The Schwarzschild geometry in natural units is given by the line element: 
    \begin{equation}
        ds^2=-\left(1-\frac{2M}{r}\right)dt^2+\frac{dr^2}{\left(1-\frac{2M}{r}\right)}+r^2\left(d\theta^2+\sin^2\theta \ d\varphi^2\right),\tag{3.12} \label{eq:3.12}
    \end{equation}
    where $M$ is the mass of the wormhole. 
    The tetrad of the line element is the set of spherical coordinates. 
    The TS formalism developed in \cite{Visser:1995} and used in Section \ref{chp3:TSFF} is used to connect the two sides of the wormhole. 
    Each side is described by Eq.~\eqref{eq:3.12}. 
    The thin-shell formalism is applied with a unit normal $n_i=\left(0, \sqrt{1-\frac{2M}{r}}, 0, 0\right)$. 
    Regions described by $\Omega_{1,2}\equiv\left\{r_{1,2}\leq a\ |\ a>\frac{3M}{2} \right\}$ are removed from the two spacetimes leaving two separate and incomplete regions with boundaries given by the timelike hypersurfaces $\partial\Omega_{1,2}\equiv\left\{r_{1,2}=a\ |\ a>\frac{3M}{2} \right\}$. 
    The boundaries $\partial\Omega_{1}=\partial\Omega_{2}$ at the wormhole throat of $r=a$ are identified and connected.  
    The boundary at $a=\frac{3M}{2}$ is chosen to satisfy the Einstein equations and equation of state in \cite{Visser:1995}; however, an event horizon is expected.
    The resulting spacetime manifold is geodesically complete and contains two asymptotically flat regions connected by the wormhole.

\begin{figure}[ht]
    	\centering
    	\includegraphics[scale=0.3]{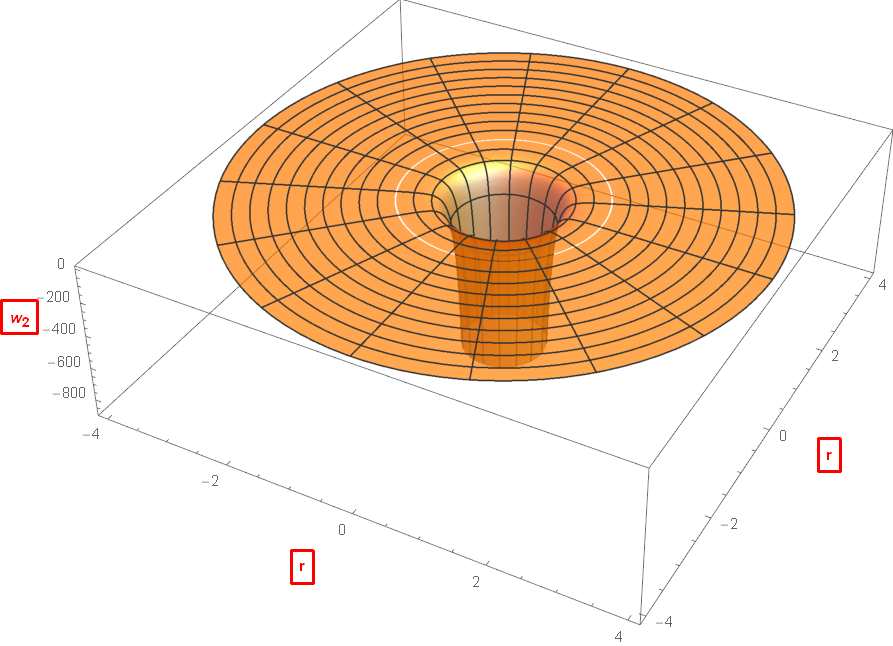}
    	\caption{Plot of Schwarzschild $w_2$. The plot is in radial coordinates with $r\in\{0,4\}$. Each mesh line represents a radial distance of $0.5$.  The $\delta$-function can be seen as a thin discontinuity at $r=\frac{3}{2}$ and its value is recorded in Eq.~\eqref{eq:3.14}.}
    	\label{w2tss}
    \end{figure}

    Of the four CM invariants computed for the Schwarzschild wormhole, three invariants $R$, $r_1$, and $r_2$ equal zero.
    The remaining invariant is
    \begin{align*}
    w_2 &= -\frac{12 M^3}{r^9}+\frac{6M^2}{a^2r^9}\sqrt{1-\frac{2M}{a}}\left(a \left(r-2 M\right)+2M\left(2M+2r^3-r\right)\right)\delta\left(r-a\right) \\
    & \ \ \ \ \ +\frac{12M}{a^5r^9}\left(a-2M\right)\bigl(4M^2\left(a-2M\right)^2+r^2\left(a-2M\right)^2 \\
    & \ \ \ \ \ \ \ \ \ \ \ \ \ \ \ \ \ \ \ \ \ \ \ \ \ \ \ \ \ \  -4Mr\left(a-2M\right)^2-2M^2r^6\bigr) \delta\left(r-a\right)^2 \\
    & \ \ \ \ \ + \frac{8}{a^6r^9}\left(1-\frac{2 M}{a}\right)^{3/2} \left((a-2 M)^3 \left(r-2M\right)^3+M^3r^9\right)\delta\left(r-a\right)^3. \tag{3.13} \label{eq:3.13}
    \end{align*}
    The $w_2$ invariant is broken into two main portions. 
    The leading term of $-\frac{12M^3}{r^9}$ is the Schwarzschild black hole's $w_2$ invariant. 
    The remaining portions of the function are all proportional to different powers of $\delta\left(r-a\right)$. 
    Consequently, they contribute to the throat of the wormhole. 
    Evaluating $w_2$ at the throat gives 
    \begin{align*}
    w_2|_{r=a} &= \frac{2}{a^{14}} \bigl( -6a^5M^3+4a^8 \left(\frac{\left(a-2M\right)^6}{a^9}+M^3\right)\left(1-\frac{2 M}{a}\right)^{\frac{3}{2}} \\
    & \ \ \ \ \ \ \ \ \ \ +3a^3M^2\left(4a^3M+a^2-4aM+4 M^2\right)\sqrt{1-\frac{2 M}{a}} \\
    & \ \ \ \ \ \ \ \ \ \ -6M\left(a-2M\right)\left(2a^6 M^2-a^4+8a^3M-24a^2M^2+32aM^3-16M^4\right) \bigr). \tag{3.14}
    \label{eq:3.14}
    \end{align*}
    Since $w_2\propto\frac{1}{a^{14}}$, the throat will experience greater curvature the smaller the radius is and vice versa. 
    The only nonzero invariant, $w_2$, is plotted in Fig.~\ref{w2tss}. 
    The mass and radius of the throat are normalized to $M=1$ and $a=\frac{3}{2}$ in the plot.
    Its plot has one divergence and one discontinuity. 
    The divergence occurs at $r=0$, which is outside the manifold of $\Omega_{1,2}$. 
    By the same argument for the apparent MT divergence, the first Schwarzschild divergence would not impede the traversability of the wormhole. 
    The discontinuity occurs at $r=a=\frac{3M}{2}$ and is located at the throat where the horizons are connected by the Schwarzschild wormholes. 
    In these invariants, it is represented by a discontinuous jump to the value in Eq.~\eqref{eq:3.14}. 
    Since the invariants at the horizon are inversely proportional to $a^{-14}$, the tidal forces on a traveler is benign at the horizon, and the Thin-Shell Schwarzschild wormhole would be traversable. \\
    
    \vspace{3mm}

\section{Exponential Metric}
\label{Chp3:Exp}
The exponential metric was demonstrated recently in \cite{Boonserm} to contain a traversable wormhole throat. 
    In natural units, its line element is
    \begin{equation}
        ds^2=-e^{-\frac{2M}{r}}dt^2+e^{+\frac{2M}{r}}\{dr^2+r^2\left(d\theta^2+\sin^2\theta d\varphi^2\right)\},\tag{3.15} \label{eq:3.15}
    \end{equation}
    where $M$ is the mass of the wormhole. 
    The tetrad utilizes the spherical coordinates. 
    It has a traversable wormhole throat at $r=M$. 
    The area of the wormhole is a concave function with a minimum at the throat where it satisfies the ``flare out'' condition.
    It does not have a horizon since $g_{tt}\neq0$ for all $r\geq0$. 
    The region $r<M$ on the other side of the wormhole is an infinite volume ``other universe'' that exhibits an ``underhill effect'' where time runs slower since $e^{-\frac{2M}{r}}>0$ in this region.
    The four curvature invariants for the exponential metric are 
    \begin{align*}
    R &= -\frac{2M^2}{r^4} e^{-\frac{2 M}{r}}, \tag{3.16} \label{eq:3.16} \\
    r_1 &= \frac{3M^4}{4 r^8} e^{-\frac{4 M}{r}}, \tag{3.17} \label{eq:3.17} \\
    r_2 &= \frac{3M^6}{8r^{12}} e^{-\frac{6 M}{r}}, \tag{3.18} \label{eq:3.18} \\
    w_2 &= -\frac{32 M^3(2 M-3 r)^3}{9 r^{12}} e^{-\frac{6 M}{r}}. \tag{3.19} \label{eq:3.19}
    \end{align*}
    Each invariant is nonzero and depends only the radial coordinate $r$ implying spherical symmetry. 
    In addition, they are finite at the throat $r=M$ and go to zero as $r\xrightarrow{}\infty$ in accordance with \cite{Boonserm}. 
    $w_2$ and $R$ have minima near the throat, while $r_1$ and $r_2$ have maxima. 
    The plots are finite everywhere and completely connected confirming the lack of a horizon.
    The encountered tidal forces would be minimal. 
    It can be concluded that the exponential metric represents a traversable wormhole.
    
    \begin{figure}[ht]
	\begin{subfigure}[ht]{.5\linewidth}
		\includegraphics[scale=0.5]{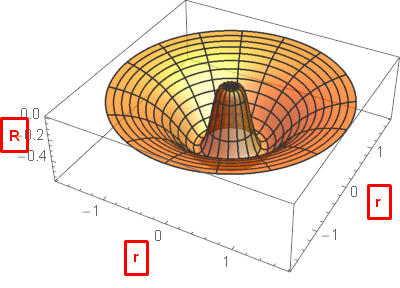}
		\caption{Plot of the exponential metric $R$}
		\label{Rem}
	\end{subfigure}
	~
	\begin{subfigure}[ht]{.5\linewidth}
		\includegraphics[scale=0.5]{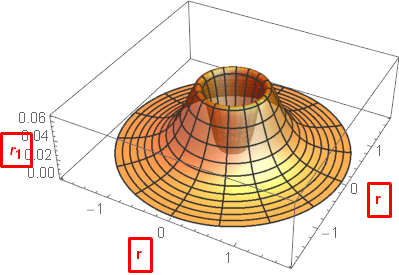}
		\caption{Plot of exponential metric $r_1$}
		\label{r1em}
	\end{subfigure}
	\par \bigskip
	\begin{subfigure}[hb]{.5\linewidth}
		\includegraphics[scale=0.5]{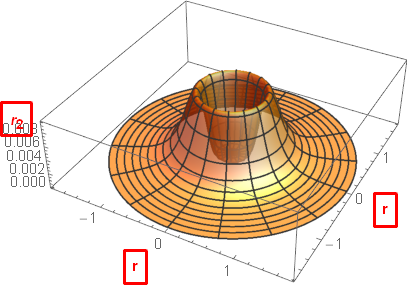}
		\caption{Plot of exponential metric $r_2$}
		\label{r2em}
	\end{subfigure}
	~
	\begin{subfigure}[hb]{.5\linewidth}
		\includegraphics[scale=0.5]{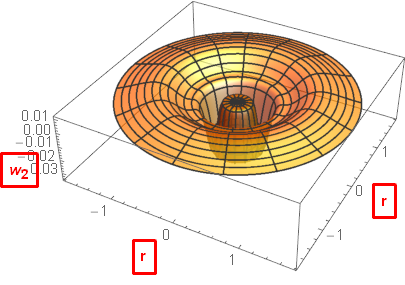}
		\caption{Plot of the exponential metric $w_2$}
		\label{w2em}
	\end{subfigure}
	\caption{Plots of the non-zero invariants for the exponential metric. The plots are in radial coordinates with $r\in\{0,1.8M\}$. Each mesh line represents a radial distance of $0.1M$. The throat begins at $r=M$.}
\end{figure}

%% file: ch4.tex
\chapter{Warp Drives Moving at a Constant Velocity}
\label{Chapter4}
%In this chapter, the curvature invariants of a warp drive moving at a constant velocity are presented and discussed.
The next set of spacetimes to be inspected are warp drive spacetimes moving at a constant velocity.
These examples of FTL spacetime share a recent yet rich history as seen in the following \cite{Alcubierre:1994,Krasnikov:1995ad,VanDenBroeck:1999sn,Natario:2001,Loup:2017,Davis,Marqu,Loup:2018}.
The key idea governing these theoretical spacetimes is that while motion through spacetime is limited by the speed of light, there is no limit to the rate spacetime itself can expand or contract.
The most famous FTL solution to the Einstein field equations demonstrates how a spaceship may make a trip to a distant star in an arbitrarily short proper time \cite{Alcubierre:1994}. 
A local contraction of spacetime in front of the spaceship paired with a local expansion of spacetime behind the ship allows the ship to surf along a warp bubble. 
While locally the spaceship remains within its own light cone as in Section \ref{chp1:TLC} and never exceeds $c$, globally the relative velocity\footnote{defined as proper spatial distance divided by proper time} may be much greater than $c$. 
FTL propulsion mechanisms based on this principle are called ``warp drives.''

In the following chapter, a procedure to plot warp drive spacetimes using curvature invariants will be presented.
First, the general metric for a warp drive spacetime in the $3+1$ ADM formalism will be demonstrated with its shift and lapse functions.
Then, the methods to calculate the Alcubierre and Nat\'ario warp drive metrics at constant velocity will be presented with their curvature invariants.
Each one will have their main variables and functions varied while maintaining all others as constants to see the individual affects on the invariant plots.
The Nat\'ario warp drive at changing velocities will be presented in the next chapter.

\section{Warp Drive Spacetimes}
\label{chp4.1:ds}
    Alcubierre and Nat\'ario developed warp drive theory using 3+1 ADM formalism \cite{Visser:1995,Marqu,ADM}.
    In a well defined coordinate patch, a spacetime may be decomposed into spacelike hypersurfaces, denoted as $\Sigma$, by use of an arbitrary time coordinate $dx^0$.
    Two nearby hypersurfaces, $\Sigma_t$ where $x^0=\mathrm{const}$ and $\Sigma_{t+dt}$ where $x+dx^0=\mathrm{const}$, are separated by the proper time $d\tau=N(x^\alpha, x^0)dx^0$.
    
    The ADM four-metric is 
    \begin{equation}
	g_{\mu\nu}=
	\begin{pmatrix}
	\ -N^2-N_i N_j g^{ij}&\ \ N_j \ \\
	\ N_i&\ \ g_{ij} \ \
	\end{pmatrix},
	\tag{4.1} \label{eq:4.1}
	\end{equation}
	where $N$ is the lapse function and $N_\alpha$ is the shift vector.
	These functions describe how to assemble the hypersurfaces to construct the entire spacetime.
	The lapse function gives the interval of proper time between nearby hypersurfaces as measured by Eulerian\footnote{These are observers whose four-velocity is normal to the hypersurfaces} observers.
	The shift vector relates the geometrical coordinates between different hypersurfaces.
	It relates the relative velocity between Eulerian observers and the lines of constant spatial coordinates.
	The 3-metric, $g_{ij}$, encodes the individual geometry of the 3-space for each hypersurface and measures the proper distance between two points inside each hypersurface.
	As long as the 3-metric is positive definite for all values of $t$, the spacetime is guaranteed to be globally hyperbolic.
	Consequently, the generic warp drive metric is a class B$_1$ spacetime.
    
    %The following paragraph comes heavily from the theorem's and definitions in Nat\'ario's Introduction.
    For a warp drive spacetime moving at a constant velocity, the lapse between infinitesimally displaced hypersurfaces, say $\Sigma_t$ and $\Sigma_{t+dt}$, remains constant.
    Thus, the lapse function is unitary, $N=1$.
    The shift function forms a time-dependent vector field in Euclidean 3-space given by the equation \cite{Lobo:2017,Alcubierre:1994,Natario:2001}: 
    \begin{equation}
        \textbf{X}=X^i\frac{\partial}{\partial x^i}=X\frac{\partial}{\partial x}+Y\frac{\partial}{\partial y}+Z\frac{\partial}{\partial z}. \label{eq:4.2} \tag{4.2}
    \end{equation} 
    Eq.~\eqref{eq:4.2} is important in defining the future pointing normal covector to the Cauchy surface as $n_\alpha=-dt\Leftrightarrow n^\alpha=\frac{\partial}{\partial t}+X^i\frac{\partial}{\partial x^i}=\frac{\partial}{\partial t}+\textbf{X}$.
    Any observer that travels along this covector is a Eulerian observer and a free-fall observer.
    A warp drive spacetime is flat wherever $\textbf{X}$ is spatially constant and a Killing vector field for the Euclidean metric.
    By assuming that the 3-metric is flat with Cartesian $(x,y,z)$ coordinates, Eq.~\eqref{eq:4.1} simplifies to
    \begin{equation*}
        ds^2=-dt^2+\sum_{i=1}^{3} (dx^i-X^idt)^2. \label{eq:4.3} \tag{4.3}
    \end{equation*}
    The warp drive spacetimes considered in this chapter correspond to specific choices for the shift function \textbf{X} in Eq.~\eqref{eq:4.3}. \\

\section{Alcubierre's Warp Drive}
\label{chp4:Alc}
    \begin{figure}[ht]
    \centering
	\includegraphics[scale=0.75]{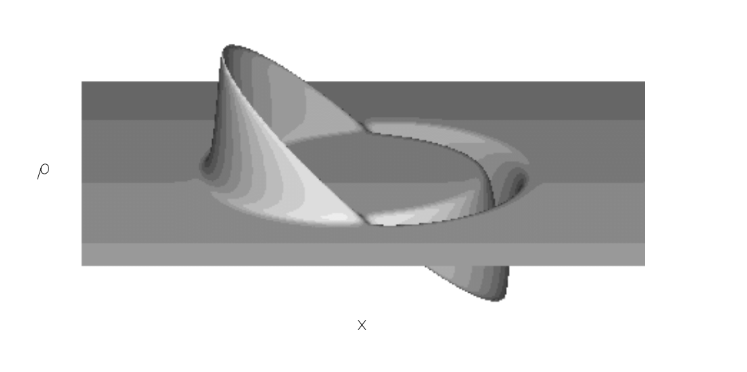}
	\caption{The expansion of the normal volume elements for the Alcubierre warp drive spacetime.} \label{fig:4.1}
    \end{figure}
    
    One of the simplest choices for the vector field \textbf{X} is for the warp bubble to be moving at a constant velocity in a single cardinal direction.
    Choosing the cardinal direction to be $dx^1=x$ corresponds mathematically to the vector field $(X,Y,Z)=(v_s f(r_s),0,0)$ \cite{Alcubierre:1994}.
    The velocity vector is $v_s(t)=\frac{dx_s(t)}{dt} = \mathrm{constant}$ represents the speed the warp bubble travels at to a distant observer.
    The shape function is $f(r_s)$, and Fig.~\ref{fig:4.1} is a plot of the volume elements of  Alcubierre's chosen shape function.
    It encodes the shape of the warp bubble in a similar manner to the shape functions for the wormholes discussed in Chapter \ref{Chapter3}.
    Then, Eq.~\eqref{eq:4.3} becomes
    \begin{equation}
        ds^2=-dt^2+(dx-v_s f(r_s)dt)^2+dy^2+dz^2. \label{eq:4.4} \tag{4.4}
    \end{equation}
    
    Eq.~\eqref{eq:4.4} is in traditional Cartesian coordinates $(-\infty<(x,y,z)<\infty)$ and its origin is at the beginning of its flight.
    $v_s(t)$ is the arbitrary speed Eulerian observers inside the warp bubble move in relation to Eulerian observers outside the warp bubble. 
    Thus, it is the speed of travel for the warp bubble itself.
    The radial distance is given by $r_s(t)=\sqrt{(x-x_s(t))^2+y^2+z^2}$.
    It is the path an Eulerian observer takes starting inside the warp bubble and traveling to the outside of the bubble. 
    Thus, it is the distance from the center of the warp bubble to any observers outside.
    The Alcubierre warp drive makes the choice for its continuous shape function, $f(r_s)$, as 
    \begin{equation}
        f(r_s)=\frac{\tanh{\sigma(r_s+\rho)}-\tanh{\sigma(r_s-\rho)}}{2\tanh{\sigma \rho}}. \label{eq:4.5} \tag{4.5}
    \end{equation}
    where $\sigma$ is the skin depth of the warp bubble and $\rho$ is the radius of the warp bubble.
    It is the shape of the warp bubble that can be represented by the expansion of the volume elements as in Fig.~\ref{fig:4.1}.
    The shape function is arbitrary other than $f=1$ in the interior of the bubble and $f=0$ in the exterior.
    The chosen shape function is spherically symmetric.
    The parameters, $\sigma>0$ and $\rho>0$, are arbitrary apart from being positive.
    
    The orthonormal tetrad may be read from the components in Eq.~\eqref{eq:4.4}. It is
    \begin{align}
        E_1 &=\begin{pmatrix}1\\0\\0\\0\end{pmatrix}, &
        E_2&=\begin{pmatrix}f(r_s) v_s\\-1\\0\\0\end{pmatrix},\nonumber \\
        E_3&=\begin{pmatrix}0\\0\\1\\0 \end{pmatrix}, & E_4 &=\begin{pmatrix}0\\0\\0\\1 \end{pmatrix}. \label{eq:4.6} \tag{4.6}
    \end{align}
    The null tetrad computed from Eq.~\eqref{eq:4.6} with Eq.~\eqref{eq:2.8.a} through Eq.~\eqref{eq:2.8.d} is 
    \begin{align}
        l_i &=\frac{1}{\sqrt{2}}\begin{pmatrix}1+f(r_s)v_s\\-1\\0\\0\end{pmatrix}, &
        k_i&=\frac{1}{\sqrt{2}}\begin{pmatrix}1-f(r_s)v_s\\1\\0\\0\end{pmatrix},\nonumber \\
        m_i&=\frac{1}{\sqrt{2}}\begin{pmatrix}0\\0\\1\\i \sin{\theta}\end{pmatrix}, & \bar{m}_i &=\frac{1}{\sqrt{2}}\begin{pmatrix}0\\0\\1\\-i \sin{\theta}\end{pmatrix}. \label{eq:4.7} \tag{4.7}
    \end{align}
    The comoving null tetrad describes light rays traveling parallel with the warp bubble. 
    Eqns. \eqref{eq:4.4} and \eqref{eq:4.7} may be applied to the equations in Chapter \ref{chapter2} to derive the four CM invariants.
    After making the specific choices detailed above, they are
    \begin{equation*} \label{eq:4.8}
        \begin{split}R &= \frac{1}{2} \sigma ^2 \text{v}_s^2 \coth (\rho  \sigma ) \\
        & \times (4 \tanh (\sigma  (\rho +\sqrt{(x-t \text{v}_s)^2})) \text{sech}^2(\sigma  (\rho +\sqrt{(x-t \text{v}_s)^2})) \\
        & -4 \tanh (\sigma  (\sqrt{(x-t \text{v}_s)^2}-\rho )) \text{sech}^2(\sigma  (\sqrt{(x-t \text{v}_s)^2}-\rho )) \\
        & -2 \sinh (\rho  \sigma ) \cosh ^3(\rho  \sigma ) (\cosh (2 \sigma  (\sqrt{(x-t \text{v}_s)^2}-\rho )) +\cosh (2 \sigma  (\rho +\sqrt{(x-t \text{v}_s)^2})) \\
        & -2 \cosh (4 \sigma  \sqrt{(x-t \text{v}_s)^2})+4) \text{sech}^4(\sigma  (\sqrt{(x-t \text{v}_s)^2}-\rho )) \text{sech}^4(\sigma  (\rho +\sqrt{(x-t \text{v}_s)^2})))\end{split} \tag{4.8} 
    \end{equation*}
    ~
    \begin{equation*} \label{eq:4.9}
        \begin{split}
            r_1 &= \frac{1}{16} \sigma^4 \text{v}_s^4 (\cosh ^4(\rho  \sigma ) \\
            & \times (\cosh (2 \sigma  (\sqrt{(x-t \text{v}_s)^2}-\rho ))+\cosh (2 \sigma  (\rho +\sqrt{(x-t \text{v}_s)^2})) \\
            & \ \ \ \ -2 \cosh (4 \sigma  \sqrt{(x-t \text{v}_s)^2})+4) \\
            & \times \text{sech}^4(\sigma  (\sqrt{(x-t \text{v}_s)^2}-\rho )) \text{sech}^4(\sigma  (\rho +\sqrt{(x-t \text{v}_s)^2})) \\
            & \ \ \ \ \ \ \ \ \ \ \ +2 \coth (\rho  \sigma ) (\tanh (\sigma  (\sqrt{(x-t \text{v}_s)^2}-\rho )) \text{sech}^2(\sigma  (\sqrt{(x-t \text{v}_s)^2}-\rho )) \\
            & \ \ \ \ \ \ \ \ \ \ \ -\tanh (\sigma  (\rho +\sqrt{(x-t \text{v}_s)^2})) \text{sech}^2(\sigma  (\rho +\sqrt{(x-t \text{v}_s)^2}))))^2
        \end{split} \tag{4.9} 
    \end{equation*}
    ~
    \begin{equation*}
        r_2 = 0 \tag{4.10} \label{eq:4.10}
    \end{equation*}
    ~
    \begin{equation*}\label{eq:4.11}
        \begin{split}
            w_2 &= -\frac{1}{288} \sigma ^6 \text{v}_s^6 \\
            & \times (2 \coth (\rho  \sigma ) (\tanh (\sigma  (\rho +\sqrt{(x-t \text{v}_s)^2})) \text{sech}^2(\sigma  (\rho +\sqrt{(x-t \text{v}_s)^2})) \\
            & \ \ \ \ \ \ \ \ \ \ \ \ \ \ \ \ -\tanh (\sigma  (\sqrt{(x-t \text{v}_s)^2}-\rho )) \text{sech}^2(\sigma  (\sqrt{(x-t \text{v}_s)^2}-\rho ))) \\
            & \ \ \ -\cosh ^4(\rho  \sigma ) (\cosh (2 \sigma  (\sqrt{(x-t \text{v}_s)^2}-\rho ))+\cosh (2 \sigma  (\rho +\sqrt{(x-t \text{v}_s)^2})) \\
            & \ \ \ -2 \cosh (4 \sigma  \sqrt{(x-t \text{v}_s)^2})+4) \text{sech}^4(\sigma  (\sqrt{(x-t \text{v}_s)^2}-\rho ))\\
            & \ \ \ \times \text{sech}^4(\sigma  (\rho +\sqrt{(x-t \text{v}_s)^2})))^3
        \end{split} \tag{4.11} 
    \end{equation*}
    
    While Eqs. \eqref{eq:4.8} through \eqref{eq:4.11} are very complicated functions, several features are apparent from inspecting them.
    First, $r_2$ is zero.
    It will not be plotted in the following subsections as its plots are flat disks showing no curvature.
    Next, each non-zero invariant depends only on the tetrad elements $t$ and $x$.
    The axes of the plots are chosen to be the tetrad components to see the effect of the free variables over time.
    In addition, each of the non-zero invariants is  proportional to both the skin depth $\sigma^n$ and the velocity, $\text{v}_s^n$.
    It should be expected that the magnitude of the invariants will then increase with the magnitude of both.
    Finally, each non-zero invariant does not have any recognizable singularities inside the spacetime manifold.
    In the next subsections, the non-zero CM invariants will be plotted to see any remaining individual affects of $v_s$, $\rho$ and $\sigma$.

    \subsection{Invariant Plots of Velocity for Alcubierre}
    \label{chp4.1:vel}
\begin{figure}[ht]
	\centering
	\begin{subfigure}{.48\linewidth}
		\includegraphics[scale=0.28]{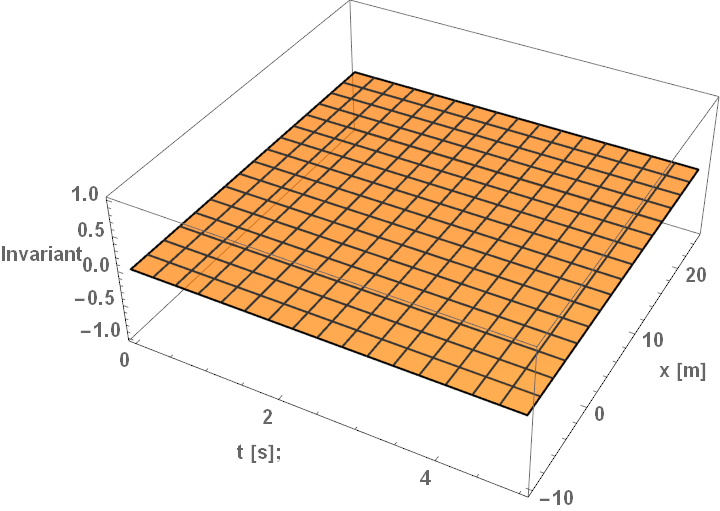}
		\caption{Plot of Alcubierre $R$ with and $v_s=0$}
		\label{ARs8r1v0}
	\end{subfigure}
	~
	\begin{subfigure}{.48\linewidth}
		\includegraphics[scale=0.28]{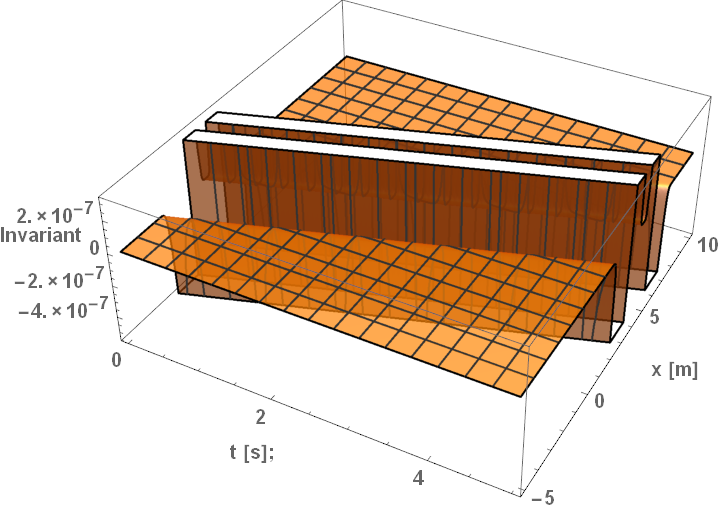}
		\caption{Plot of Alcubierre $R$ with and $v_s=1$}
		\label{ARs8r1v1 a}
	\end{subfigure}
	\par \bigskip
	\begin{subfigure}{.48\linewidth}
		\includegraphics[scale=0.28]{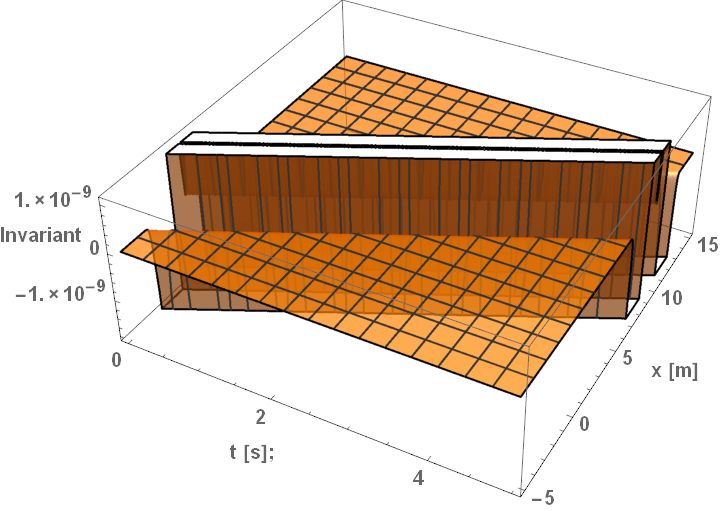}
		\caption{Plot of Alcubierre $R$ with and $v_s=2$}
		\label{ARs8r1v2}
	\end{subfigure}
	~
	\begin{subfigure}{.48\linewidth}
		\includegraphics[scale=0.28]{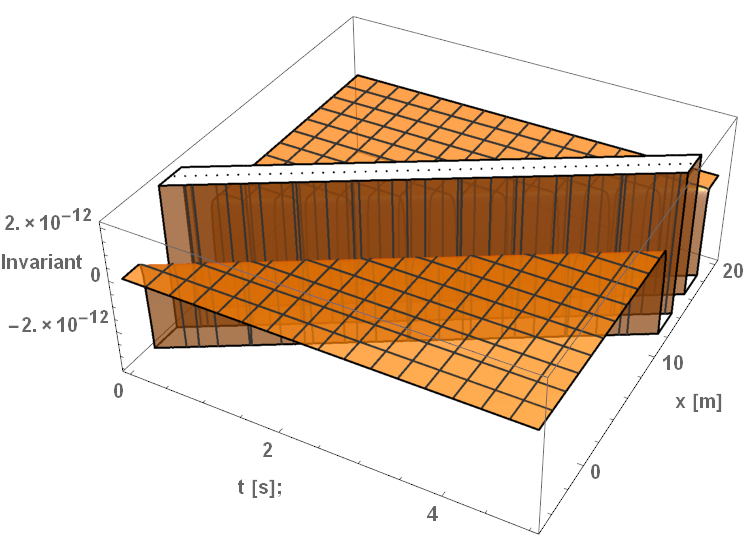}
		\caption{Plot of Alcubierre $R$ with $v_s=3$}
		\label{ARs8r1v3}
	\end{subfigure}
	~
	\begin{subfigure}{.48\linewidth}
		\includegraphics[scale=0.28]{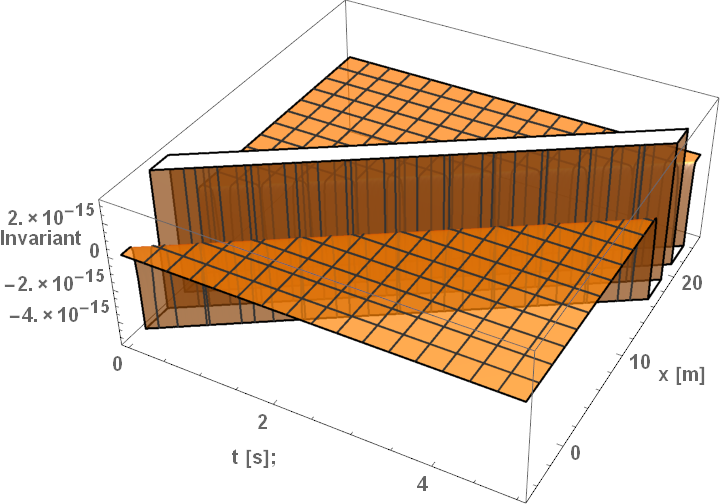}
		\caption{Plot of Alcubierre $R$ with $v_s=4$}
		\label{ARs8r1v4}
	\end{subfigure}
	~
	\begin{subfigure}{.48\linewidth}
		\includegraphics[scale=0.28]{Images/Chapter4/Alcubierre/Rs8r1v4.png}
		\caption{Plot of Alcubierre $R$ with $v_s=5$}
		\label{ARs8r1v5}
	\end{subfigure}
	\caption{Plots of the $R$ invariants for the Alcubierre warp drive while varying a velocity.
	$\sigma=8$ and $\rho=1$ as Alcubierre originally suggested in his paper \cite{Alcubierre:1994}.} 0\label{fig:4.2}
    \end{figure}

\begin{figure}[ht]
	\centering
	\begin{subfigure}{.48\linewidth}
		\includegraphics[scale=0.28]{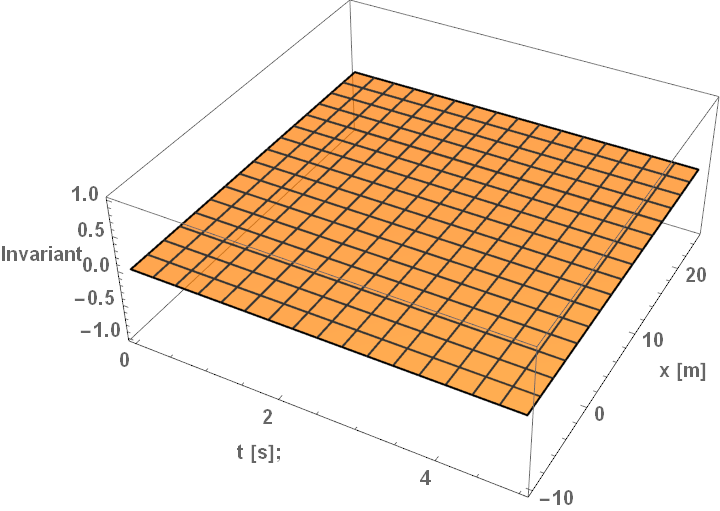}
		\caption{Plot of Alcubierre $r_1$ with and $v_s=0$}
		\label{Ar1s8r1v0}
	\end{subfigure}
	~
	\begin{subfigure}{.48\linewidth}
		\includegraphics[scale=0.28]{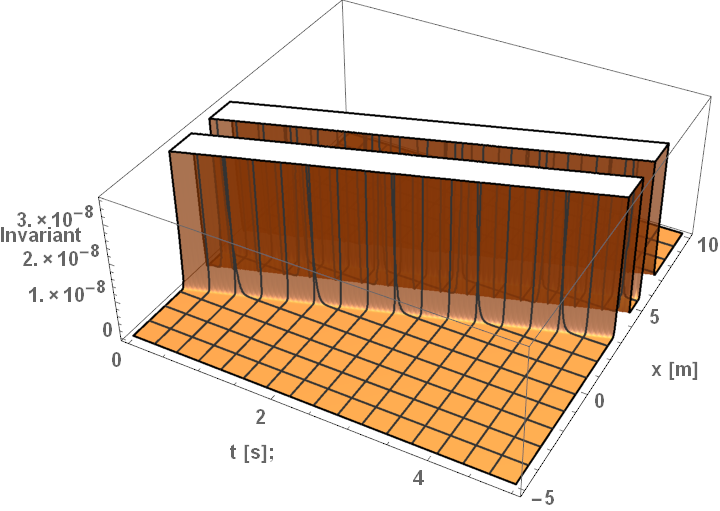}
		\caption{Plot of Alcubierre $r_1$ with and $v_s=1$}
		\label{Ar1s8r1v1 a}
	\end{subfigure}
	\par \bigskip
	\begin{subfigure}{.48\linewidth}
		\includegraphics[scale=0.28]{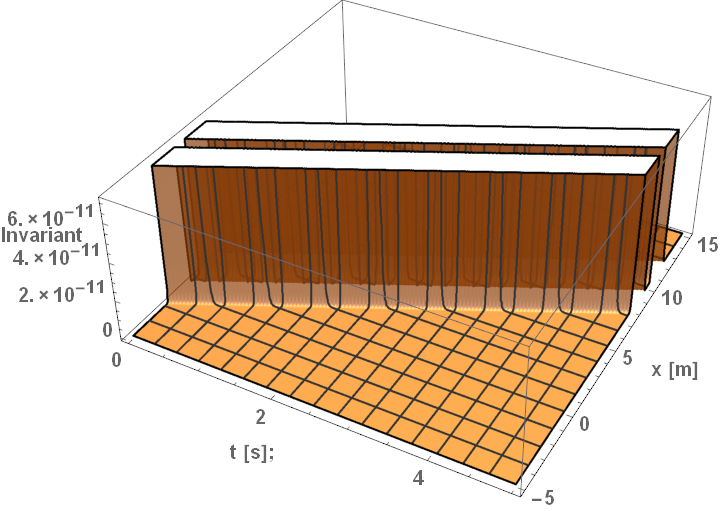}
		\caption{Plot of Alcubierre $r1$ with and $v_s=2$}
		\label{Ar1s8r1v2}
	\end{subfigure}
	~
	\begin{subfigure}{.48\linewidth}
		\includegraphics[scale=0.28]{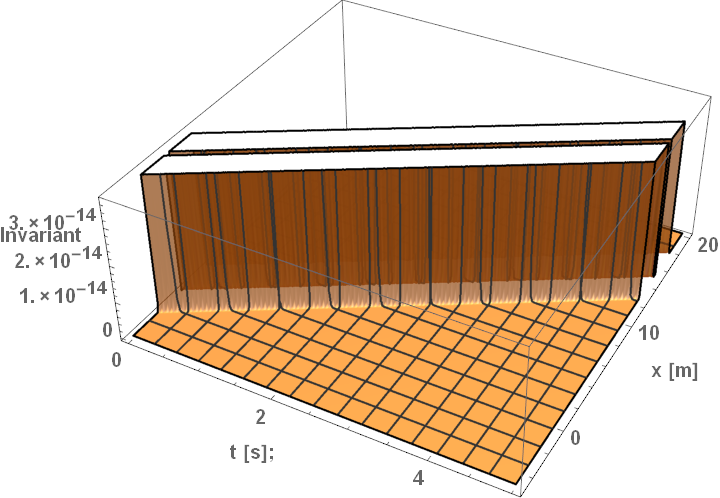}
		\caption{Plot of Alcubierre $r_1$ with $v_s=3$}
		\label{Ar1s8r1v3}
	\end{subfigure}
	~
	\begin{subfigure}{.48\linewidth}
		\includegraphics[scale=0.28]{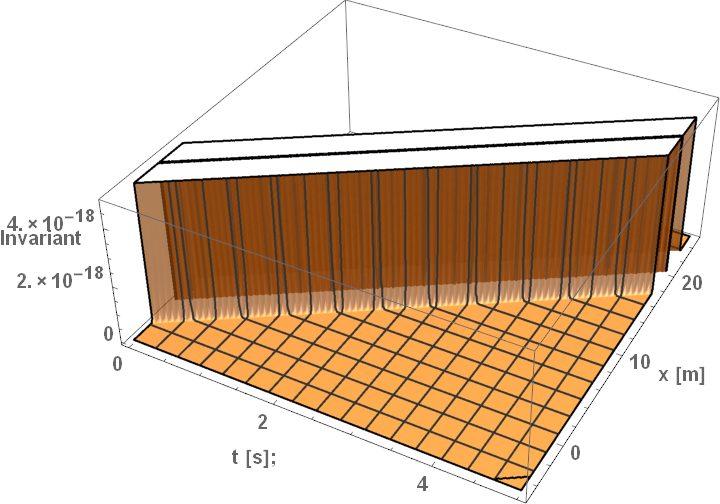}
		\caption{Plot of Alcubierre $r_1$ with $v_s=4$}
		\label{Ar1s8r1v4}
	\end{subfigure}
	~
	\begin{subfigure}{.48\linewidth}
		\includegraphics[scale=0.28]{Images/Chapter4/Alcubierre/r1s8r1v4.png}
		\caption{Plot of Alcubierre $r_1$ with $v_s=5$}
		\label{Ar1s8r1v5}
	\end{subfigure}
	\caption{Plots of the $r_1$ invariants for the Alcubierre warp drive while varying velocity.
	The other variables were chosen as $\sigma=8$ and $\rho=1$ to match the variables Alcubierre originally suggested in his paper \cite{Alcubierre:1994}.} \label{fig:4.3}
    \end{figure}

\begin{figure}[ht]
	\centering
	\begin{subfigure}{.48\linewidth}
		\includegraphics[scale=0.28]{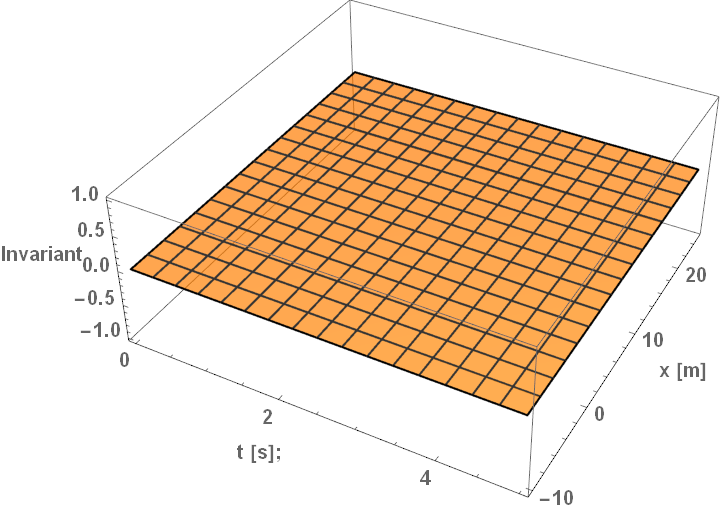}
		\caption{Plot of Alcubierre $w_2$ with and $v_s=0$}
		\label{Aw2s8r1v0}
	\end{subfigure}
	~
	\begin{subfigure}{.48\linewidth}
		\includegraphics[scale=0.28]{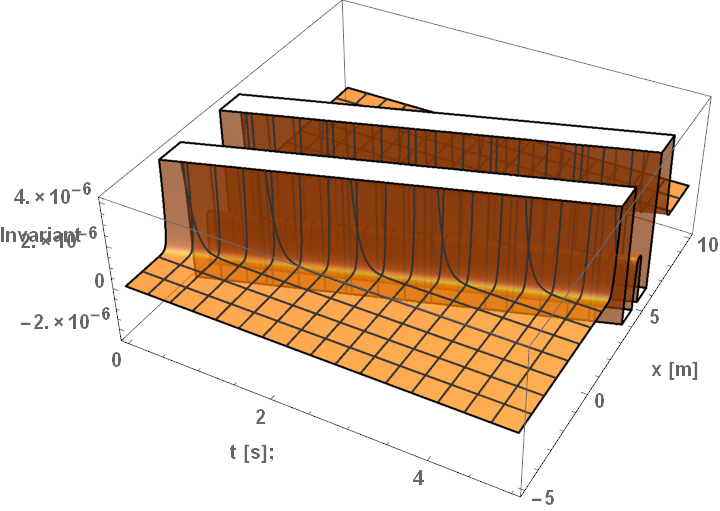}
		\caption{Plot of Alcubierre $w_2$ with and $v_s=1$}
		\label{Aw2s8r1v1 a}
	\end{subfigure}
	\par \bigskip
	\begin{subfigure}{.48\linewidth}
		\includegraphics[scale=0.28]{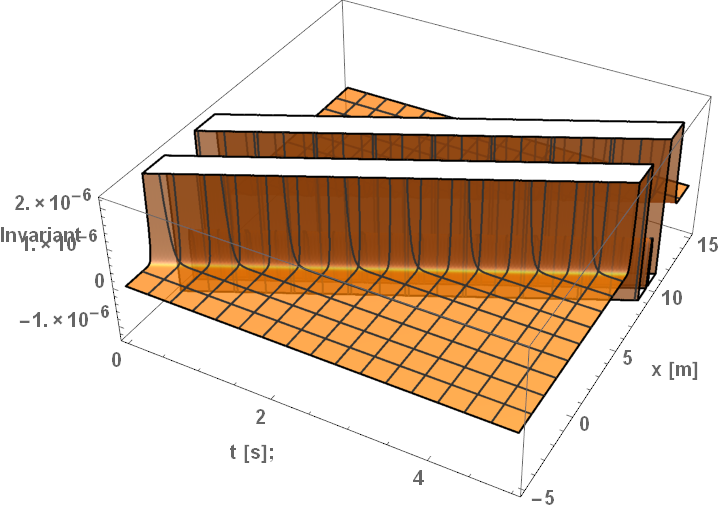}
		\caption{Plot of Alcubierre $w_2$ with and $v_s=2$}
		\label{Aw2s8r1v2}
	\end{subfigure}
	~
	\begin{subfigure}{.48\linewidth}
		\includegraphics[scale=0.28]{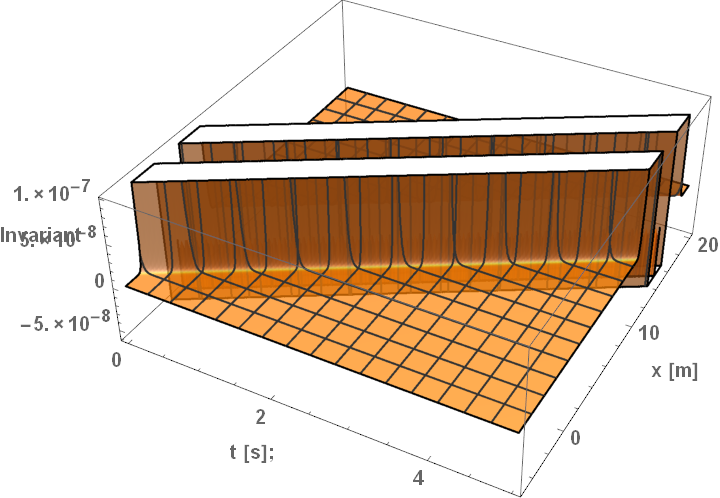}
		\caption{Plot of Alcubierre $w_2$ with $v_s=3$}
		\label{Aw2s8r1v3}
	\end{subfigure}
	~
	\begin{subfigure}{.48\linewidth}
		\includegraphics[scale=0.28]{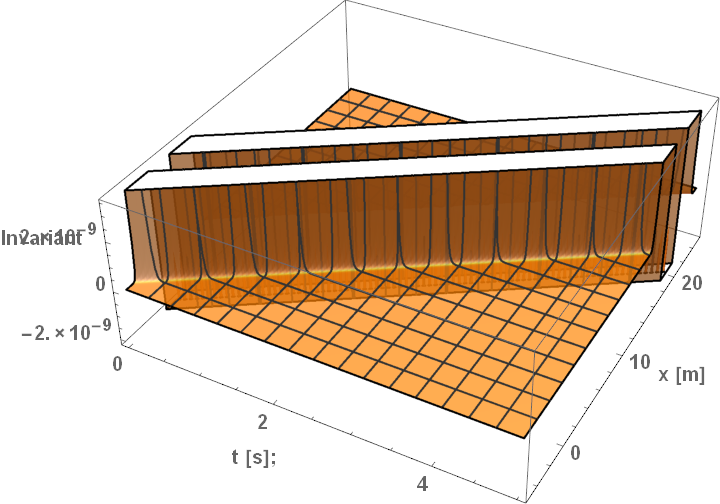}
		\caption{Plot of Alcubierre $w_2$ with $v_s=4$}
		\label{Aw2s8r1v4}
	\end{subfigure}
	~
	\begin{subfigure}{.48\linewidth}
		\includegraphics[scale=0.28]{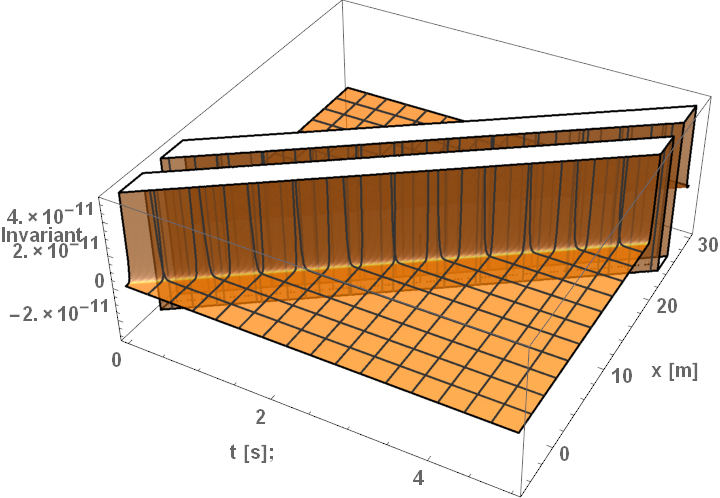}
		\caption{Plot of Alcubierre $w_2$ with $v_s=5$}
		\label{Aw2s8r1v5}
	\end{subfigure}
	\caption{Plots of the $w_2$ invariants for the Alcubierre warp drive while varying velocity.
	The radius was chosen as $\rho=1$ to match the variables Alcubierre originally suggested in his paper \cite{Alcubierre:1994}. The skin depth was chosen as $\sigma=2$ to keep the plots as machine size numbers. Both velocity and skin depth have an exponential affect on the magnitude of the invariants.} \label{fig:4.4}
    \end{figure}

        The plots for varying the velocities are included in Figures \ref{fig:4.2}, \ref{fig:4.3}, and \ref{fig:4.4} at the end of this subsection.
        The plots are 3D plots between the $x$ coordinate, $t$ coordinate, and the magnitude of the invariants along each axis.
        As natural units were selected, the plots have been normalized such that $c=1$ and a slope of $1$ in the $x$ vs. $t$ corresponds to the warp bubble traveling at light speed.
        The plots show a small range over the possible values of the variables to demonstrate many of the basic features of each invariant.
        First, the shape of all invariants resembles the ``top hat'' features of the shape function as discussed in \cite{Alcubierre:1994}.
        However, there are some minor variations between each invariant.
        The Ricci scalar $R$ oscillates from a trough, to a peak, to a flat area, to a peak and back to a trough.
        The $r_1$ invariant simply has a peak with a flat area followed by another peak.
        The $w_2$ follows the reverse pattern as the $R$ wavering from a peak, into a trough, into a flat area, into a trough and returning into a peak.
        In each invariant, a ship could safely surf along in the flat area, which is dubbed the harbor. 
        The central harbor disappears in Figures \ref{ARs8r1v3} through \ref{ARs8r1v5} because the plots lack precision.
        By plotting more points and consequently taking longer computational time, the central features will be recovered.
        The harbor's width is much less than the distance covered in these later plots.
        Inspecting the functions, the harbor remains, and choosing smaller time intervals allows it to reappear in the plots.
    
        Varying the velocity has several effects.
        First, it increases the amount of distance, the $x$ coordinate, covered per unit time.
        The warp bubble's velocity acts exactly like $v=d/t$ and is a good check that the program has been encoded correctly.
        The warp bubble will cover an increasing amount of distance over time as observed by an Eulerian observer.
        Second, the velocity causes the magnitude of the invariants to decrease exponentially.
        This observation is in contrast to what was predicted by inspecting the leading terms of the invariants.
        It can be concluded that the additional terms overpower the leading term.
        Next, the shape of the warp bubble remains constant throughout the flight.
        The only affect of time is the distance covered.
        Finally, the plots have no holes or discontinuities, agreeing with the inspection of the invariant functions that no intrinsic singularities exist.
        While not truly practical, the invariants reveal nothing that will prevent a spaceship to surf the center channel.
        
    \newpage
    
    \newpage
    
    \FloatBarrier

    \subsection{Invariant Plots of Skin Depth for Alcubierre}
    \label{chp4.1:skin}
\begin{figure}[ht]
	\centering
	\begin{subfigure}{.48\linewidth}
		\includegraphics[scale=0.28]{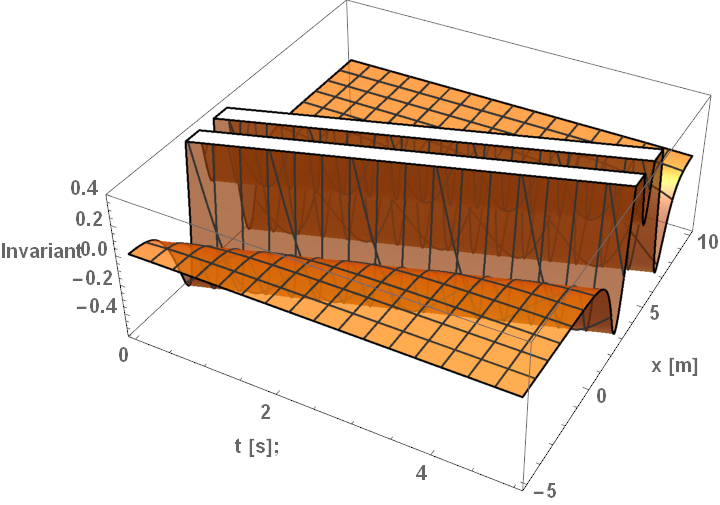}
		\caption{Plot of Alcubierre $R$ with and $\sigma=1$}
		\label{ARs1r1v1 b}
	\end{subfigure}
	~
	\begin{subfigure}{.48\linewidth}
		\includegraphics[scale=0.28]{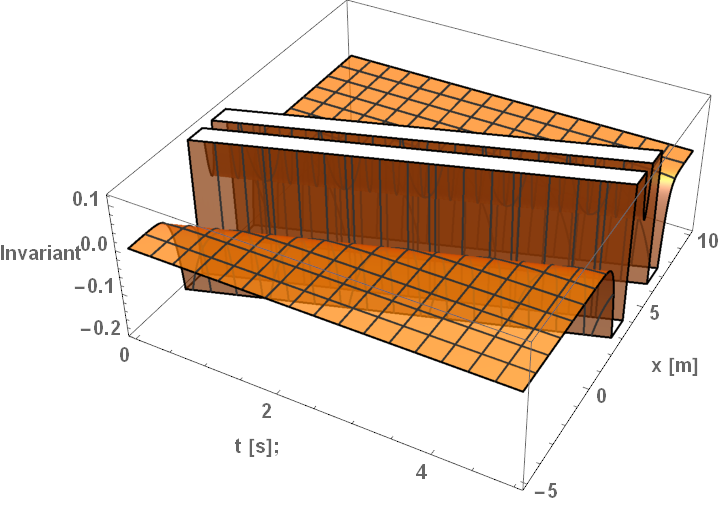}
		\caption{Plot of Alcubierre $R$ with and $\sigma=2$}
		\label{ARs2r1v1}
	\end{subfigure}
	\par \bigskip
	\begin{subfigure}{.48\linewidth}
		\includegraphics[scale=0.28]{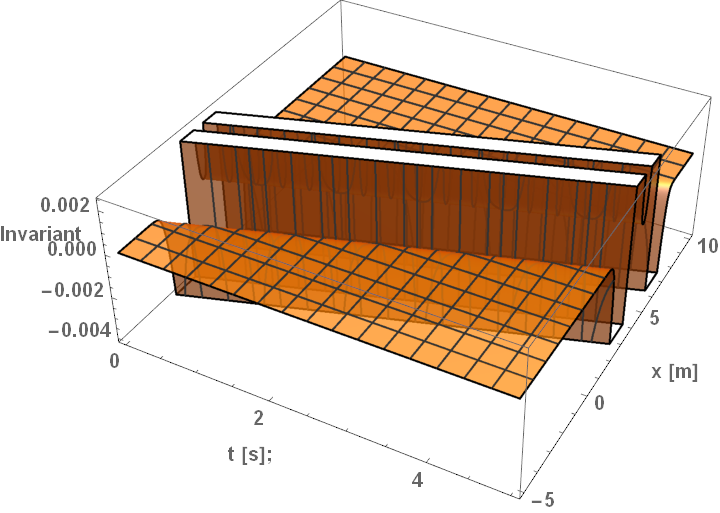}
		\caption{Plot of Alcubierre $R$ with and $\sigma=4$}
		\label{ARs4r1v1}
	\end{subfigure}
	~
	\begin{subfigure}{.48\linewidth}
		\includegraphics[scale=0.28]{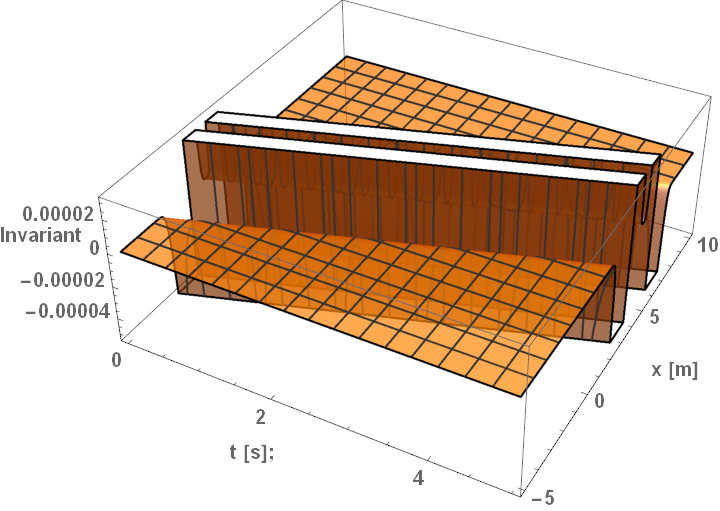}
		\caption{Plot of Alcubierre $R$ with $v_s=6$}
		\label{ARs6r1v1}
	\end{subfigure}
	~
	\begin{subfigure}{.48\linewidth}
		\includegraphics[scale=0.28]{Images/Chapter4/Alcubierre/Rs8r1v1.png}
		\caption{Plot of Alcubierre $R$ with $\sigma=8$}
		\label{ARs8r1v1}
	\end{subfigure}
	~
	\begin{subfigure}{.48\linewidth}
		\includegraphics[scale=0.28]{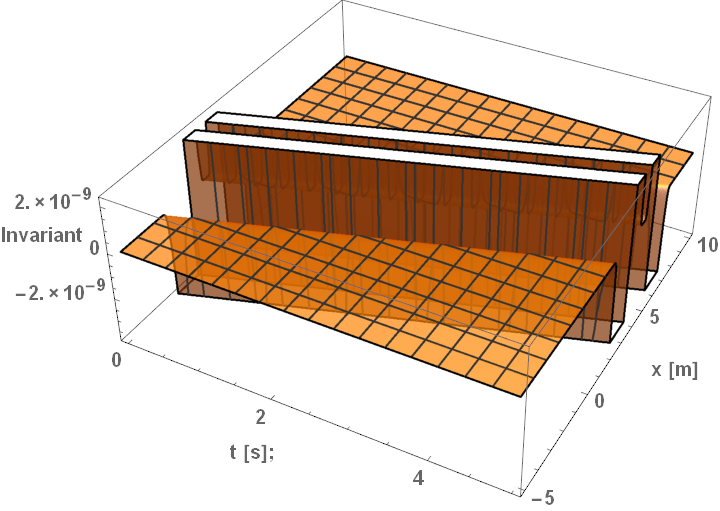}
		\caption{Plot of Alcubierre $R$ with $\sigma=10$}
		\label{ARs10r1v1}
	\end{subfigure}
	\caption{Plots of the $R$ invariants for the Alcubierre warp drive while varying skin depth.
	The variables were chosen as $v_s=8$ and $\rho=1$ to match the variables Alcubierre originally suggested in his paper \cite{Alcubierre:1994}.} \label{fig:4.5}
    \end{figure}

\begin{figure}[ht]
	\centering
	\begin{subfigure}{.48\linewidth}
		\includegraphics[scale=0.28]{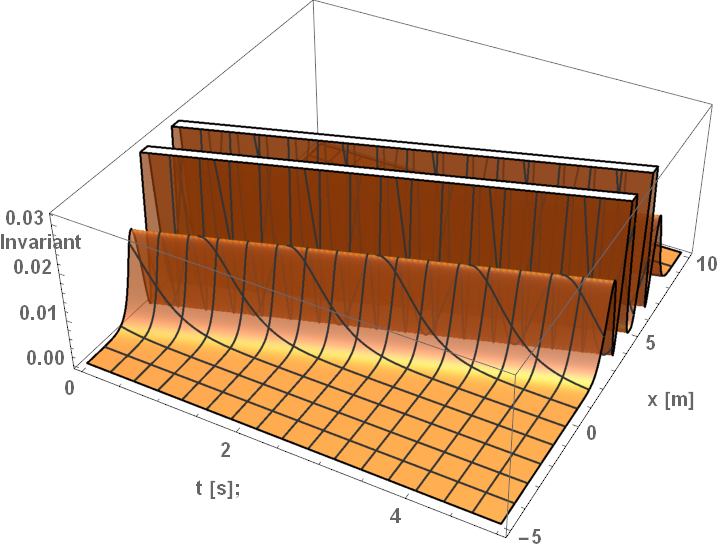}
		\caption{Plot of Alcubierre $r_1$ with and $\sigma=1$}
		\label{Ar1s1r1v1 b}
	\end{subfigure}
	~
	\begin{subfigure}{.48\linewidth}
		\includegraphics[scale=0.28]{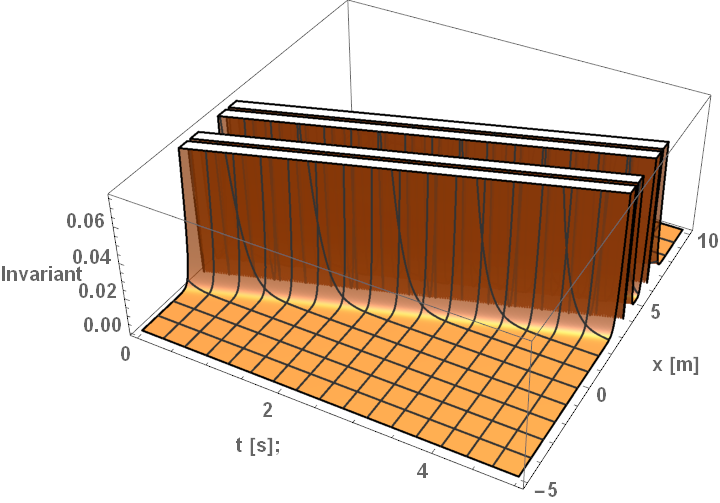}
		\caption{Plot of Alcubierre $r_1$ with and $\sigma=2$}
		\label{Ar1s2r1v1}
	\end{subfigure}
	\par \bigskip
	\begin{subfigure}{.48\linewidth}
		\includegraphics[scale=0.28]{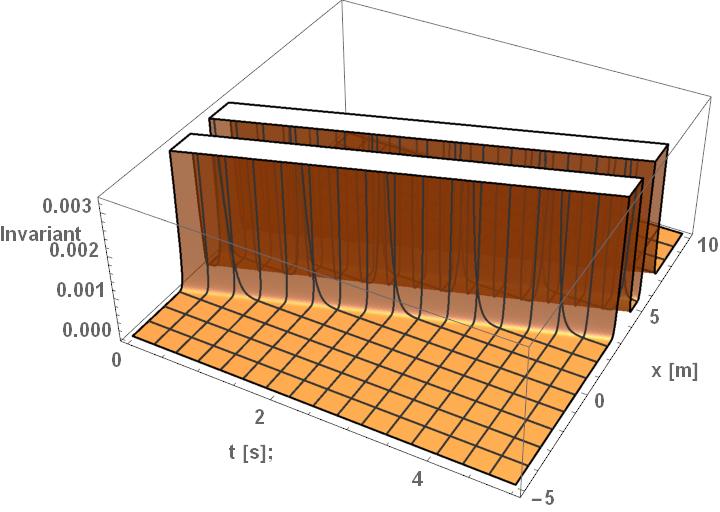}
		\caption{Plot of Alcubierre $r1$ with and $\sigma=4$}
		\label{Ar1s4r1v1}
	\end{subfigure}
	~
	\begin{subfigure}{.48\linewidth}
		\includegraphics[scale=0.28]{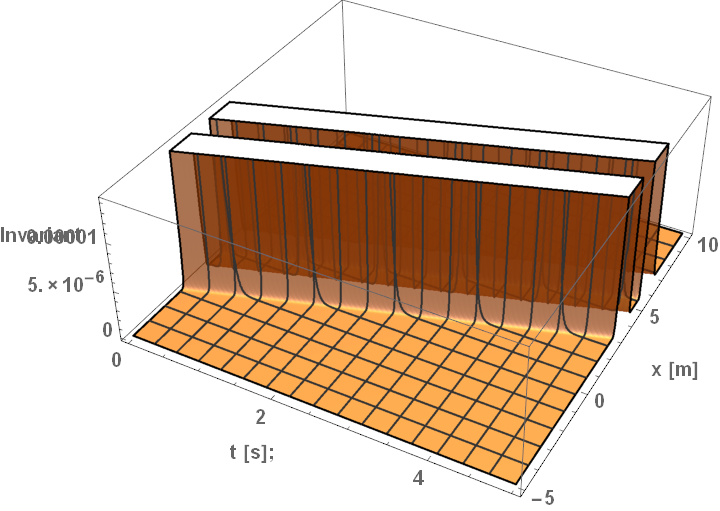}
		\caption{Plot of Alcubierre $r_1$ with $\sigma=6$}
		\label{Ar1s6r1v1}
	\end{subfigure}
	~
	\begin{subfigure}{.48\linewidth}
		\includegraphics[scale=0.28]{Images/Chapter4/Alcubierre/r1s8r1v1.png}
		\caption{Plot of Alcubierre $r_1$ with $\sigma=8$}
		\label{Ar1s8r1v1}
	\end{subfigure}
	~
	\begin{subfigure}{.48\linewidth}
		\includegraphics[scale=0.28]{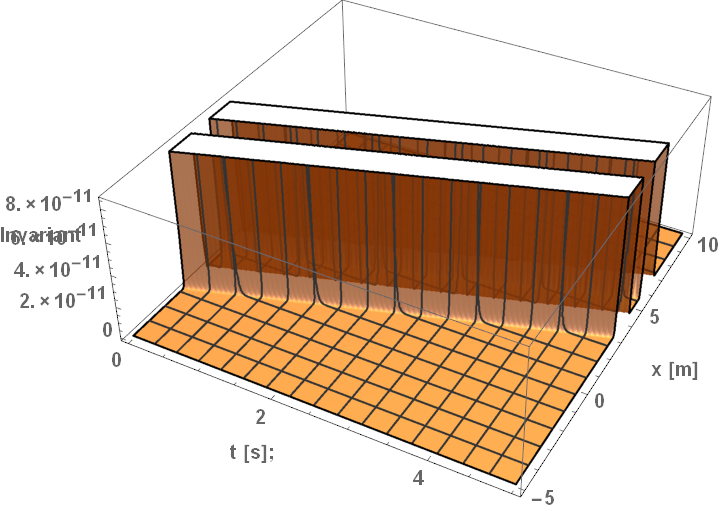}
		\caption{Plot of Alcubierre $r_1$ with $\sigma=10$}
		\label{Ar1s10r1v1}
	\end{subfigure}
	\caption{Plots of the $r_1$ invariants for the Alcubierre warp drive while varying skin depth.
	The other variables were chosen as $v_s=1$ and $\rho=1$ to match the variables Alcubierre originally suggested in his paper \cite{Alcubierre:1994}.} \label{fig:4.6}
    \end{figure}

\begin{figure}[ht]
	\centering
	\begin{subfigure}{.48\linewidth}
		\includegraphics[scale=0.28]{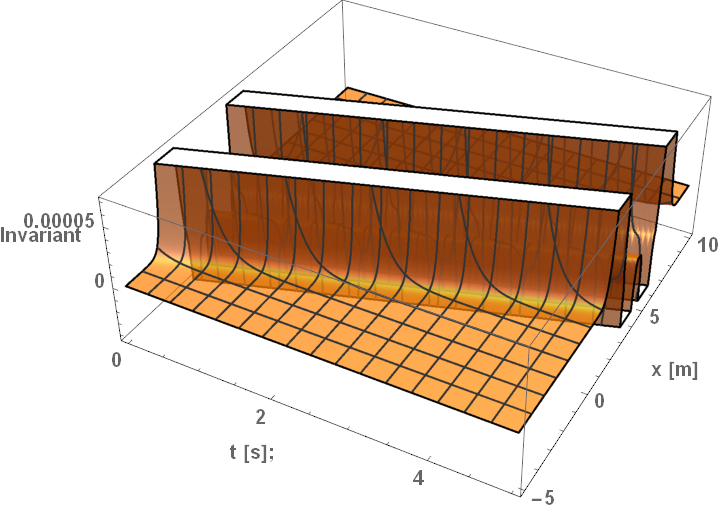}
		\caption{Plot of Alcubierre $w_2$ with and $\sigma=1$}
		\label{Aw2s1r1v1}
	\end{subfigure}
	~
	\begin{subfigure}{.48\linewidth}
		\includegraphics[scale=0.28]{Images/Chapter4/Alcubierre/w2s2r1v1.png}
		\caption{Plot of Alcubierre $w_2$ with and $\sigma=2$}
		\label{Aw2s2r1v1}
	\end{subfigure}
	\par \bigskip
	\begin{subfigure}{.48\linewidth}
		\includegraphics[scale=0.28]{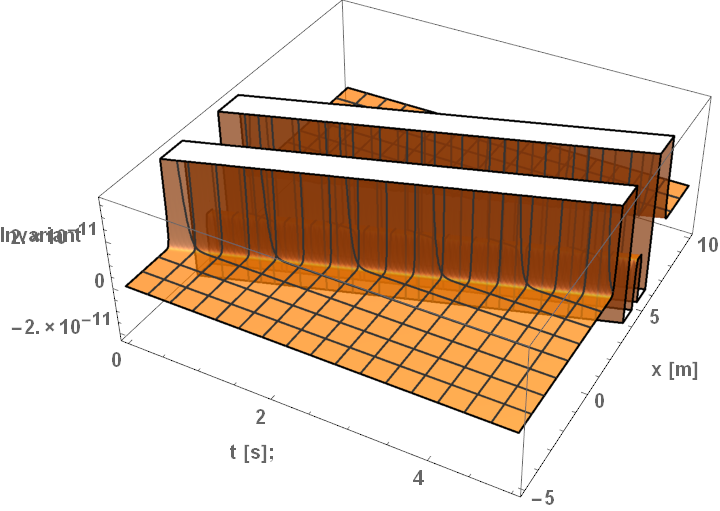}
		\caption{Plot of Alcubierre $w_2$ with and $\sigma=4$}
		\label{Aw2s4r1v1}
	\end{subfigure}
	~
	\begin{subfigure}{.48\linewidth}
		\includegraphics[scale=0.28]{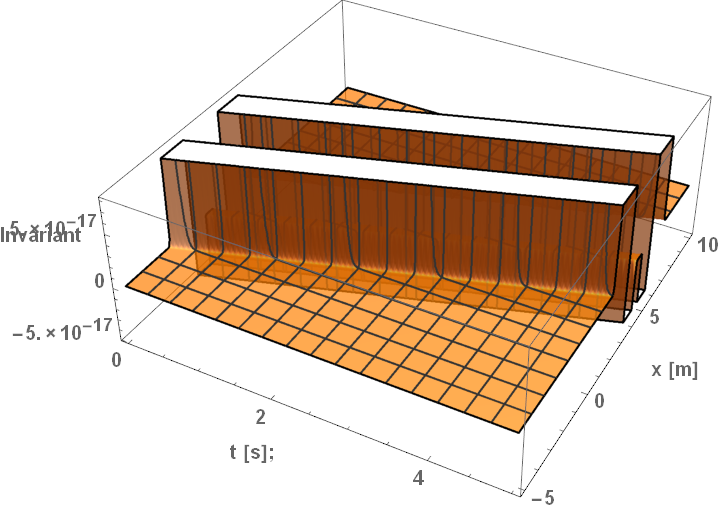}
		\caption{Plot of Alcubierre $w_2$ with $\sigma=6$}
		\label{Aw2s6r1v1}
	\end{subfigure}
	~
	\begin{subfigure}{.48\linewidth}
		\includegraphics[scale=0.28]{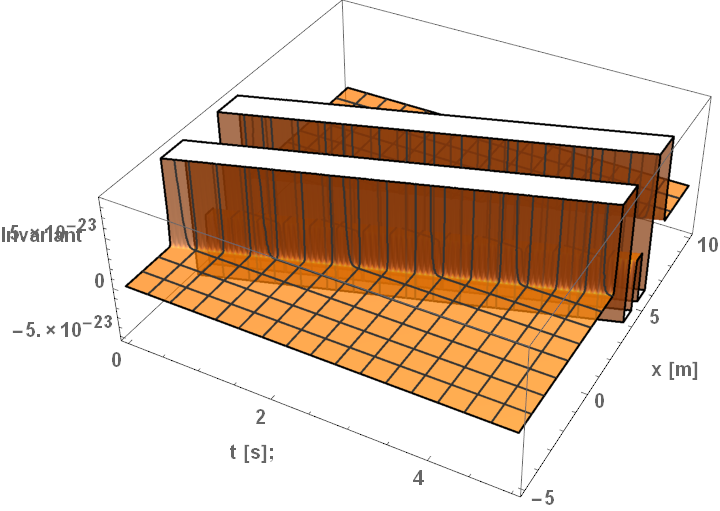}
		\caption{Plot of Alcubierre $w_2$ with $\sigma=8$}
		\label{Aw2s8r1v1}
	\end{subfigure}
	~
	\begin{subfigure}{.48\linewidth}
		\includegraphics[scale=0.28]{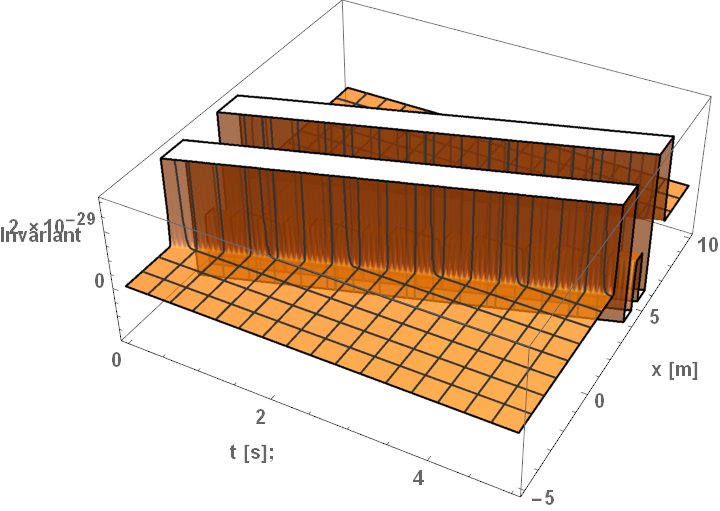}
		\caption{Plot of Alcubierre $w_2$ with $\sigma=10$}
		\label{Aw2s10r1v1}
	\end{subfigure}
	\caption{Plots of the $w_2$ invariants for the Alcubierre warp drive while varying skin-depth.
	The radius and velocity were chosen as $\rho=1$ and $v_s=1$ to match the variables Alcubierre originally suggested in his paper \cite{Alcubierre:1994}.} \label{fig:4.7}
    \end{figure}

    The plots for varying the skin depth are included in Figures \ref{fig:4.5}, \ref{fig:4.6}, and \ref{fig:4.7} at the end of this subsection.
    Repeating the procedure of Section \ref{chp4.1:vel}, the variable $\sigma$ has been varied between values of $1$ and $8$ while maintaining the other variables at the constant values of $\rho=1$ and $v_s=1$.
    Many of the features in these plots are the same as those discussed at the beginning of Section \ref{chp4.1:vel}; thus, the skin depth variation reveals two additional features.
    First, the plots advance towards the ``top hat'' function by slowly straightening out any dips.
    This feature is most notable in the plots of $r_1$ in Figures \ref{ARs1r1v1 b} and \ref{Ar1s1r1v1 b}.
    Multiple ripples occur in these two plots initially, but then  gradually smooth out as $\sigma$ increases.
    These unforeseen ripples could be the source of a rich internal structure inside of the warp bubbles and potentially affect its flight. 
    Second, the relative magnitude of the Ricci scalar and $r_1$ is several orders of magnitude greater than that of $w_2$.
    This can be seen as $\sigma\rightarrow 8$ the Ricci scalar goes to $10^{-9}$, $r_1$ goes to $10^{-11}$, and $w_2$ goes to the order of $10^{-28}$.
    Consequently, the trace terms of the Riemann tensor will have the greatest effect on the curvature as both the Ricci scalar and $r_1$ are members of the Ricci invariants in Eq.~\eqref{eq:2.22} and Eq.~\eqref{eq:2.41}.
    The terms of the Weyl tensor will have negligible effects since $w_2$ is a member of the Weyl tensor in Eq.~\eqref{eq:2.40}.
    This conclusion can help warp drive calculations by focusing on the effects in the easier to calculate Ricci tensor. 
    
    The main effects of varying the skin depth is to decrease the magnitude of the warp bubble's curvature exponentially.
    This can be seen in each of the invariants as the magnitude decreases from being on the order of $10^{-3}$ to $10^{-28}$.
    The exponential decrease implies that thinner values of the warp bubble's thin depth would propel itself at greater velocities due to the greater amount of curvature.
    This novel idea needs further exploration.
        
    \newpage
    
    \newpage
    
    \FloatBarrier

    \subsection{Invariant Plots of Radius for Alcubierre}
    \label{chp4.1:radius}
\begin{figure}[ht]
	\centering
	\begin{subfigure}{.48\linewidth}
		\includegraphics[scale=0.28]{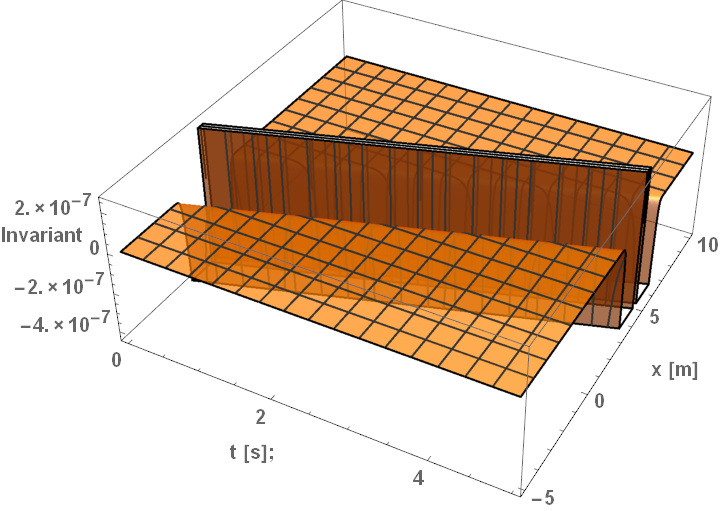}
		\caption{Plot of Alcubierre $R$ with and $\rho=0.1$}
		\label{ARs8rp1v1}
	\end{subfigure}
	~
	\begin{subfigure}{.48\linewidth}
		\includegraphics[scale=0.28]{Images/Chapter4/Alcubierre/Rs8r1v1.png}
		\caption{Plot of Alcubierre $R$ with and $\rho=1$}
		\label{ARs8r10v1 c}
	\end{subfigure}
	\par \bigskip
	\begin{subfigure}{.48\linewidth}
		\includegraphics[scale=0.28]{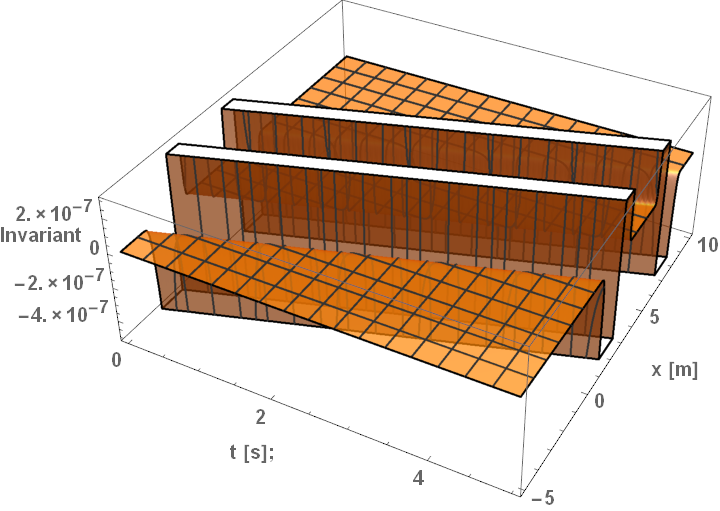}
		\caption{Plot of Alcubierre $R$ with and $\rho=2$}
		\label{ARs8r2v1}
	\end{subfigure}
	~
	\begin{subfigure}{.48\linewidth}
		\includegraphics[scale=0.28]{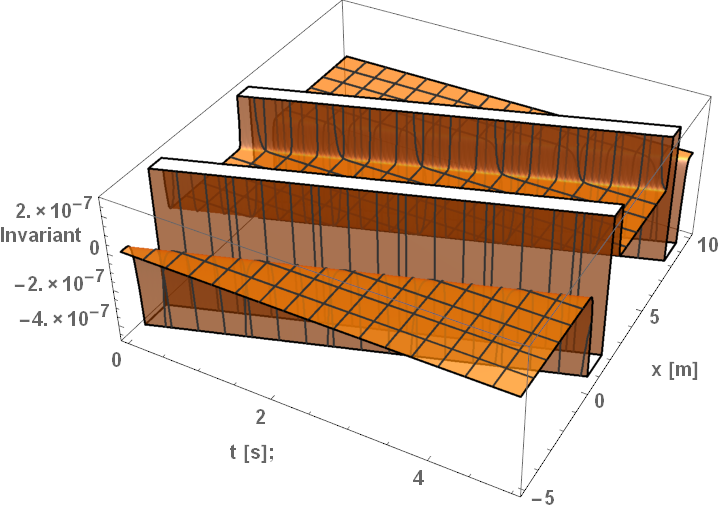}
		\caption{Plot of Alcubierre $R$ with $\rho=3$}
		\label{ARs8r3v1}
	\end{subfigure}
	~
	\begin{subfigure}{.48\linewidth}
		\includegraphics[scale=0.28]{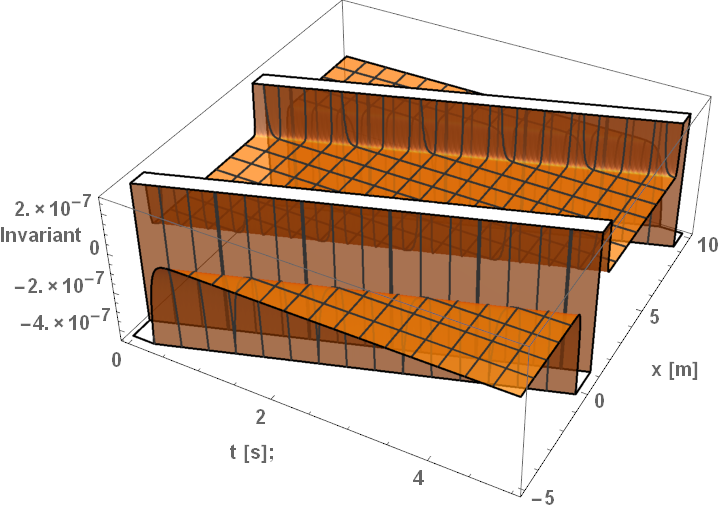}
		\caption{Plot of Alcubierre $R$ with $\rho=4$}
		\label{ARs8r4v1}
	\end{subfigure}
	~
	\begin{subfigure}{.48\linewidth}
		\includegraphics[scale=0.28]{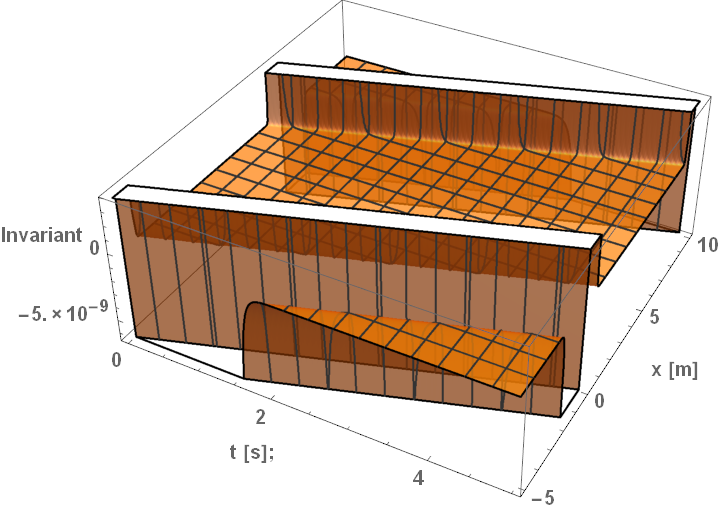}
		\caption{Plot of Alcubierre $R$ with $\rho=5$}
		\label{ARs8r5v1}
	\end{subfigure}
	\caption{Plots of the $R$ invariants for the Alcubierre warp drive while varying radius. The other variables were chosen as $\sigma=8$ and $v_s=1$ to match the variables Alcubierre originally suggested in his paper \cite{Alcubierre:1994}.} \label{fig:4.8}
    \end{figure}

\begin{figure}[ht]
	\centering
	\begin{subfigure}{.48\linewidth}
		\includegraphics[scale=0.28]{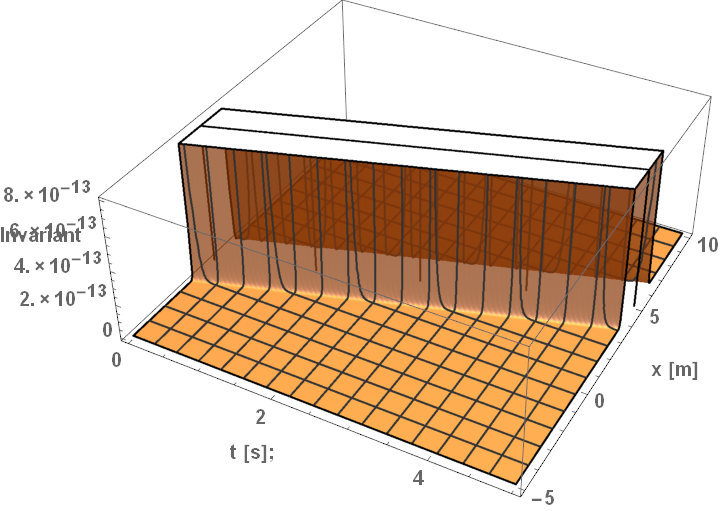}
		\caption{Plot of Alcubierre $r_1$ with and $\rho=0.1$}
		\label{Ar1s8rp1v1}
	\end{subfigure}
	~
	\begin{subfigure}{.48\linewidth}
		\includegraphics[scale=0.28]{Images/Chapter4/Alcubierre/r1s8r1v1.png}
		\caption{Plot of Alcubierre $r_1$ with and $\rho=1$}
		\label{Ar1s8r1v1 c}
	\end{subfigure}
	\par \bigskip
	\begin{subfigure}{.48\linewidth}
		\includegraphics[scale=0.28]{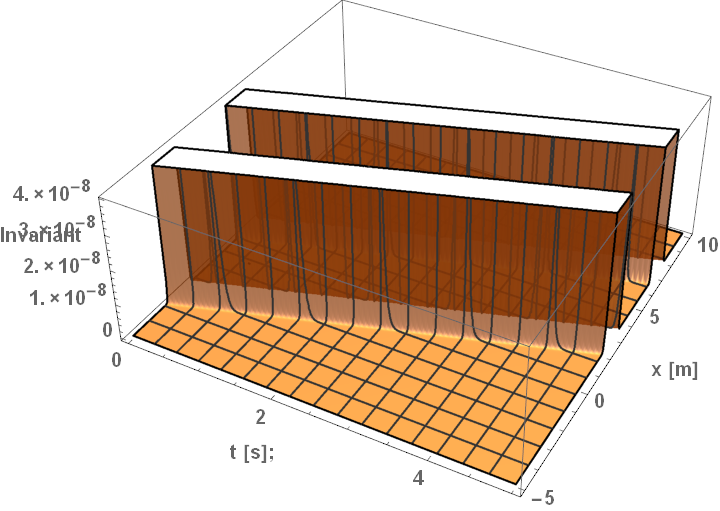}
		\caption{Plot of Alcubierre $r_1$ with and $\rho=3$}
		\label{Ar1s8r2v1}
	\end{subfigure}
	~
	\begin{subfigure}{.48\linewidth}
		\includegraphics[scale=0.28]{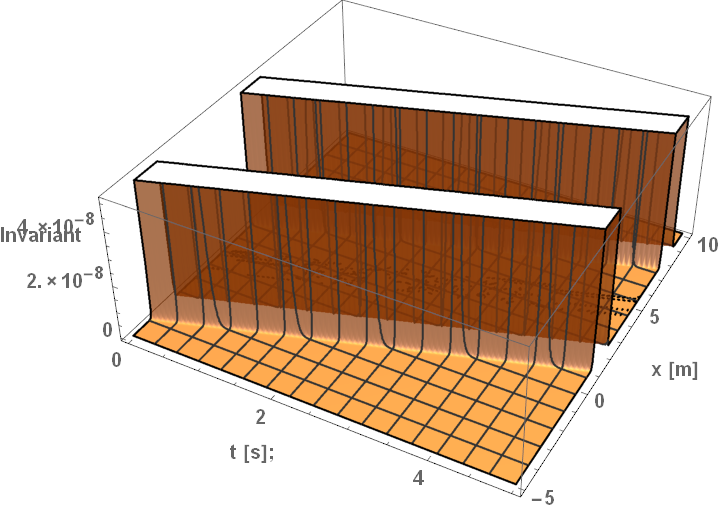}
		\caption{Plot of Alcubierre $r_1$ with $\rho=3$}
		\label{Ar1s8r3v1}
	\end{subfigure}
	~
	\begin{subfigure}{.48\linewidth}
		\includegraphics[scale=0.28]{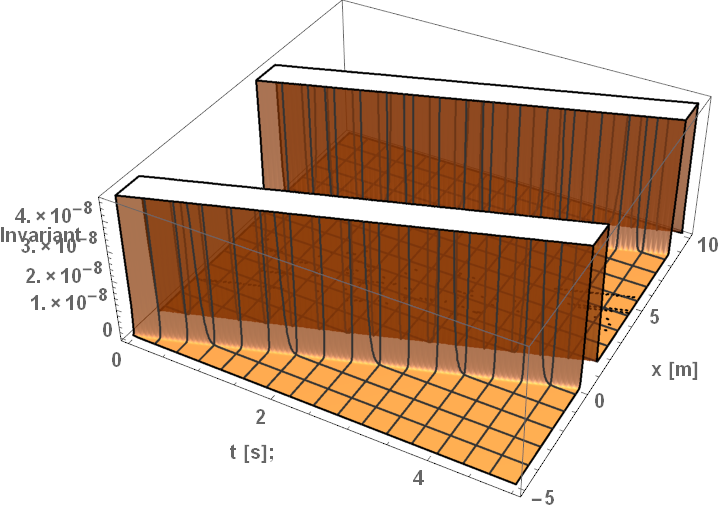}
		\caption{Plot of Alcubierre $r_1$ with $\rho=4$}
		\label{Ar1s8r4v1}
	\end{subfigure}
	~
	\begin{subfigure}{.48\linewidth}
		\includegraphics[scale=0.28]{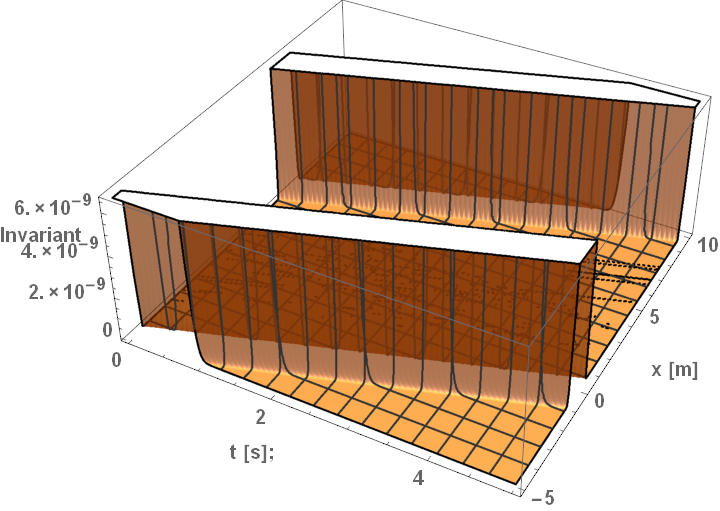}
		\caption{Plot of Alcubierre $r_1$ with $\rho=5$}
		\label{Ar1s8r5v1}
	\end{subfigure}
	\caption{Plots of the $r_1$ invariants for the Alcubierre warp drive while varying the radius. The other variables were chosen as $\sigma=8$ and $v_s=1$ to match the variables Alcubierre originally suggested in his paper \cite{Alcubierre:1994}.} \label{fig:4.9}
    \end{figure}

\begin{figure}[ht]
	\centering
	\begin{subfigure}{.48\linewidth}
		\includegraphics[scale=0.28]{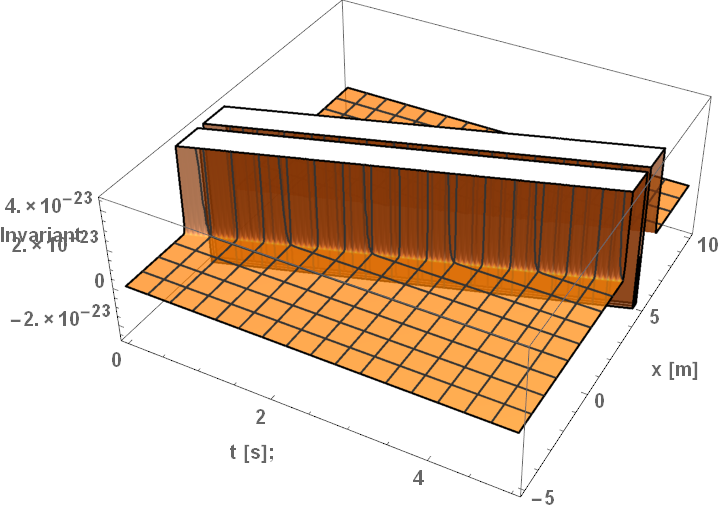}
		\caption{Plot of Alcubierre $w_2$ with and $\rho=0.1$}
		\label{Aw2s8rp1v1}
	\end{subfigure}
	~
	\begin{subfigure}{.48\linewidth}
		\includegraphics[scale=0.28]{Images/Chapter4/Alcubierre/w2s8r1v1.png}
		\caption{Plot of Alcubierre $w_2$ with and $\rho=1$}
		\label{Aw2s8r1v1 c}
	\end{subfigure}
	\par \bigskip
	\begin{subfigure}{.48\linewidth}
		\includegraphics[scale=0.28]{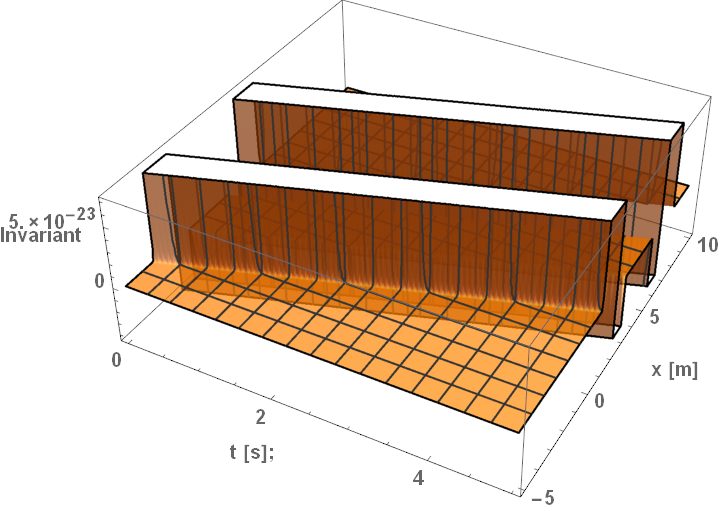}
		\caption{Plot of Alcubierre $w_2$ with and $\rho=2$}
		\label{Aw2s8r2v1}
	\end{subfigure}
	~
	\begin{subfigure}{.48\linewidth}
		\includegraphics[scale=0.28]{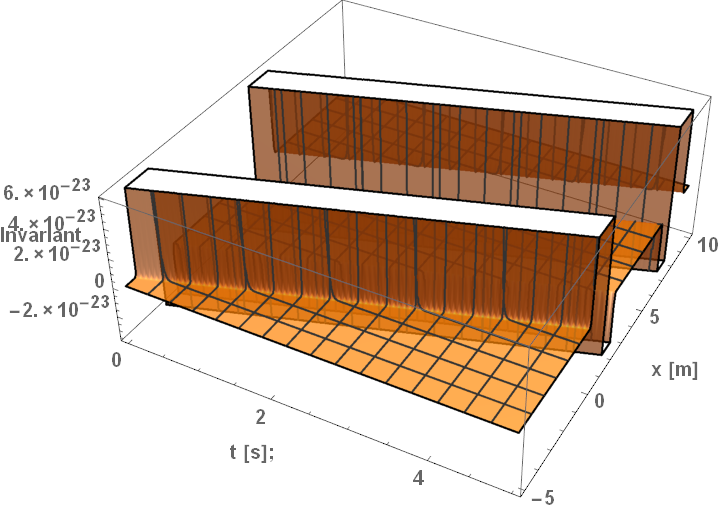}
		\caption{Plot of Alcubierre $w_2$ with $\rho=3$}
		\label{Aw2s8r3v1}
	\end{subfigure}
	~
	\begin{subfigure}{.48\linewidth}
		\includegraphics[scale=0.28]{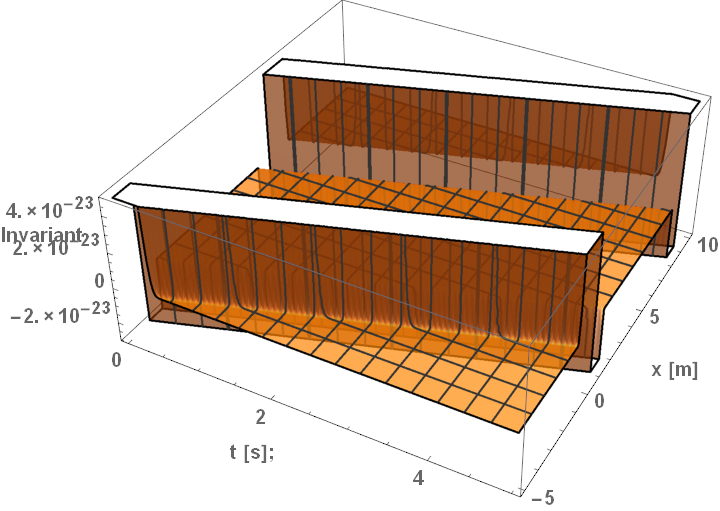}
		\caption{Plot of Alcubierre $w_2$ with $\rho=4$}
		\label{Aw2s8r4v1}
	\end{subfigure}
	~
	\begin{subfigure}{.48\linewidth}
		\includegraphics[scale=0.28]{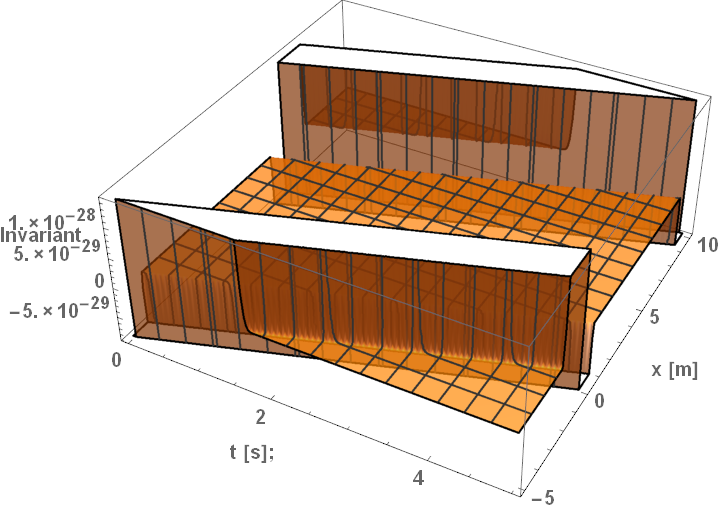}
		\caption{Plot of Alcubierre $w_2$ with $\rho=5$}
		\label{Aw2s8r5v1}
	\end{subfigure}
	\caption{Plots of the $w_2$ invariants for the Alcubierre warp drive while varying radius.
	The other variables were chosen as $\sigma=8$ and $v_s=1$ to match the variables Alcubierre originally suggested in his paper \cite{Alcubierre:1994}.} \label{fig:4.10}
    \end{figure}

    The plots for varying the radius $\rho$ of the Alcubierre warp bubble are included in Figures \ref{fig:4.8}, \ref{fig:4.9}, and \ref{fig:4.10} at the end of this subsection.
    Following the procedure of Section \ref{chp4.1:vel}, the variable $\rho$ has been varied between values of $0.1$~m and $5$~m while maintaining the other variables at the constant values of $\sigma=8$ and $v_s=1$.
    Many of the features in these plots are the same as those discussed at the beginning of Section \ref{chp4.1:vel}, but the variation of the radius does reveal an additional feature.
    The spatial size for a spaceship to harbor inside the warp bubble is directly affected by the value of $\rho$.
    By inspecting the $x$-axis of each plot, the size of the harbor is the same value of that of $\rho$.
    While expected of the radius, it is confirmation that the program is encoded correctly and that the invariant functions reveal spacetime's curvature.
    Of greater interest, the magnitude of the invariants does not have a clear correlation with $\rho$. 
    As an example, consider the $r_1$ plots in Fig.~\ref{fig:4.9}.
    When $\rho$ = 0~m, the $r_1$ invariants has its lowest magnitude order of $10^{-13}$.
    As the radius increases in the next four plots, the invariant increases to an order of $10^{-8}$.
    At the largest value $\rho$ = 5~m, the invariant decreases to an order of $10^{-9}$.
    Inspecting the invariant function itself in Eq.~\eqref{eq:4.9}, $\rho$ does not have a noticeable relationship that explains this behavior.
    It can be hypothesized that resonance values of the radius $\rho$ exist that could create massive warp bubbles. 
    In conclusion, the radius $\rho$ defines the size of the harbor and the warp bubble. 
    It must always be chosen large enough for the ship to be unaffected by the curvature of the warp bubble itself.
        
    \newpage
    
    \newpage
    
    \FloatBarrier

\section{Nat\'ario's Warp Drive}
\label{chp4:Nat}
    The Nat\'ario warp drive spacetime describes a spacetime that ``slides'' the warp bubble region through space \cite{Natario:2001}.
    A region in front of the bubble will be contracting and balanced by a region behind the bubble, which will be expanding.
    Their net effect may propel the warp bubble at arbitrary velocities potentially even greater than lightspeed. 
    Its line element is
    \begin{equation}
        ds^2=(1-X_{r_s}X^{r_s}-X_{\theta}X^{\theta})dt^2+2(X_{r_s}dr_s+X_{\theta} r d\theta)dt-dr_s^2-r_s^2(d\theta^2-\sin^2{\theta} \ d\varphi^2). \label{eq:4.12} \tag{4.12}
    \end{equation}
    The standard spherical coordinates are $(0\leq r_s<\infty$;\ $0\leq\theta < \pi$;$\ 0\leq\varphi < 2\pi)$, and $(-\infty< t<\infty)$  \cite{Natario:2001,Loup:2017}.  
    The vector field in Eq.~\eqref{eq:4.3} is converted to spherical coordinates and then set to 
    \begin{equation}
        \textbf{X}\sim -v_s(t) d[n(r_s)r_s^2\sin^2{\theta}d\phi]\sim-2v_sn(r_s)\cos{\theta}\textbf{e}_{r_s}+v_s(2n(r_s)+r_sn'(r_s))\sin{\theta}\textbf{e}_{\theta}. \label{eq:4.13} \tag{4.13}
    \end{equation}
    $v_s(t)$ is the constant speed of the warp bubble observed by Eulerian observers and $n(r_s)$ is the shape function of the warp bubble. 
    The shape function, $n(r_s)$, is arbitrary other than the conditions $n(r_s)=\frac{1}{2}$ for large r and $n(r_s)=0$ for small r.
    These conditions on the shape function match the conditions for the ``top-hat'' function from Section \ref{chp4:Alc}.
    The selected shape function is 
    \begin{equation}
        n(r_s)=\frac{1}{2}(1-(\frac{1}{2}(1-\tanh{\sigma(r-\rho)}))). \label{eq:4.14} \tag{4.14}
    \end{equation}
    where $\sigma$ is the skin depth of the bubble and $\rho$ is the radius of the bubble \cite{Loup:2018}.
    The front of the warp bubble corresponds to $\cos\theta>0$ and the back vice versa.
    At the front, there is a compression in the radial direction and an expansion in the perpendicular direction.
    The comoving null tetrad is:
    \begin{align}
        l_i&=\frac{1}{\sqrt{2}}\begin{pmatrix}1+X_{r_s}\\-1\\0\\0\end{pmatrix}, &
        k_i&=\frac{1}{\sqrt{2}}\begin{pmatrix}1-X_{r_s}\\1\\0\\0\end{pmatrix}, \nonumber \\
        m_i&=\frac{1}{\sqrt{2}}\begin{pmatrix}X_\theta\\0\\-r\\i r \sin{\theta}\end{pmatrix}, &
        \bar{m}_i&=\frac{1}{\sqrt{2}}\begin{pmatrix}X_\theta\\0\\-r\\-i r \sin{\theta}\end{pmatrix}. \label{eq:4.15} \tag{4.15}
    \end{align}
    Inputting Eqs.~\eqref{eq:4.12} and \eqref{eq:4.15} into the method from Chapter $2$, one may derive the four CM invariants.
    The Ricci scalar is
    \begin{equation*} \label{eq:4.16}
         \begin{split}
            R &=-\frac{1}{8} \sigma ^2 v_s^2 \text{sech}^4(\sigma  (r-\rho )) (\cos (2 \theta )+r^2 \sigma ^2 \sin ^2(\theta ) \tanh ^2(\sigma  (r-\rho )) \\
            & -2 r \sigma  \sin ^2(\theta ) \tanh (\sigma  (r-\rho ))+2)
        \end{split}, \tag{4.16} 
    \end{equation*}
    The Ricci scalar is included alone in this chapter as a demonstrative example.
    The remaining three are included in Appendix \ref{ap:vel}.
    
    Like the Alcubierre invariants, the Nat\'ario invariants are exceptionally complicated.
    By inspecting the functions, similar features may be observed.
    First, the Nat\'ario invariants do not change with time, but instead with $r$ and $\theta$.
    The warp bubble skims along the comoving null tetrad in Eq.~\eqref{eq:4.15}.
    As a consequence, the coordinates for the plots are chosen to be $r$ and $\theta$.
    They will show the shape of the bubble around the ship during flight.
    Second, each invariant is proportional to both $v_s^n$ and $\sigma^n$ like the Alcubierre invariants.
    The magnitude of the bubble's curvature then increases exponentially with both velocity and skin depth.
    In addition, the Nat\'ario invariants are proportional to $\cos^n(\frac{\theta}{2})$ and $\text{sech}^n(\sigma(r-\rho))$.
    The warp bubble is shaped such that the curvature is at a maximum in front of the ship around $\frac{\theta}{2}=0$ and a minimum behind the ship around $\frac{\theta}{2}=\frac{\pi}{2}$.
    It is at a maximum for $r=\rho$ along the center of the warp bubble, since there $\text{sech}(0)=1$.
    Outside these values, the curvature should then fall off and go asymptotically to $0$.
    These features match that of the ``top-hat'' function described in \cite{Alcubierre:1994}.
    Finally, there are no intrinsic singularities.
    The manifold is asymptotically flat and completely connected.
    The flight of such a warp bubble should be significantly less affected by any gravitational tidal forces as compared to the Alcubierre metric in Section \ref{chp4:Alc}.
    The CM curvature invariants confirm that the Nat\'ario warp drive is a more realistic alternative to Alcubierre's.
    
    Despite the complexity of the invariants, the shape of their plots is very simple.
    It forms a very narrow and jagged ring as in Fig.~\ref{fig:4.1}.
    Precisely at $r=\rho$, the CM curvature invariants spike to non-zero magnitudes depending on the invariant.
    The Ricci scalar $R$ takes the form of a smooth disc outside the the warp bubble.
    The shape of the $r_1$ invariant is that of a jagged disc at $r=\rho$.
    The disc has jagged edges in the negative direction, with sharp spikes at radial values $r=\rho$ and at polar angle values of $\theta=0$ and $\theta=\pi$.
    The shape of the $r_2$ invariant is that of a jagged disc at $r=\rho$.
    Its edges vary between positive and negative values depending on the polar angle $\theta$.
    The shape of the $w_2$ invariant is that of a jagged disc at $r=\rho$.
    In front of the harbor ($\theta>0$), the invariant has rapidly changing negative values between $0$ and $1$.
    Behind the harbor ($\theta<0$), the invariant has positive values rapidly changing.
    The jagged edges of the plots must mean that the $r_1$, $r_2$ and $w_2$ invariants oscillate rapidly between $0$ and $1$ along the circumference of the warp bubble.
    The oscillations are more rapid than the plotted precision.
    Potentially, the $r_1$ invariant could be replotted at a greater precision but at an extreme loss of computational speed.
    Outside, the CM invariants of the Nat\'ario warp bubble are asymptotically flat.
    Inside, the CM invariants of the Nat\'ario warp bubble are asymptotically flat, implying there is a safe harbor for a ship to reside.
    The simplicity of the plots in comparison to the complexity of the invariant functions implies that only a single term (or a very small subset of all the terms) in the invariant functions dominates the size of the warp bubble magnitude.
    In the remainder of this section, the effect of the velocity $v_s$, the skin depth, $\sigma$, and the radius $\rho$ is analyzed.

\subsection{Invariant Plots of Velocity for Nat\'ario}
\label{chp4.2:vel}
\begin{figure}[htb]
    	\begin{subfigure}{.45\linewidth}
    	    \centering
    		\includegraphics[scale=0.2]{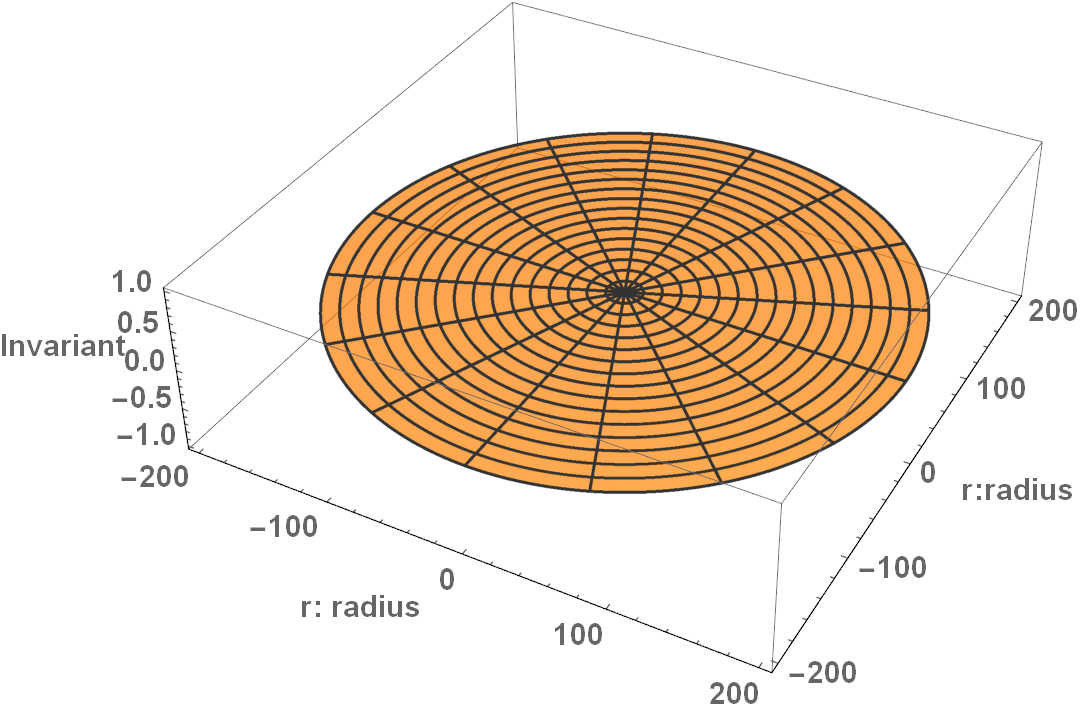}
    		\caption{$v=0.0 \frac{m}{s}$}
    		\label{fig:4.11a}
    	\end{subfigure}
    	~
    	\begin{subfigure}{.55\linewidth}
    	    \centering
    		\includegraphics[scale=0.2]{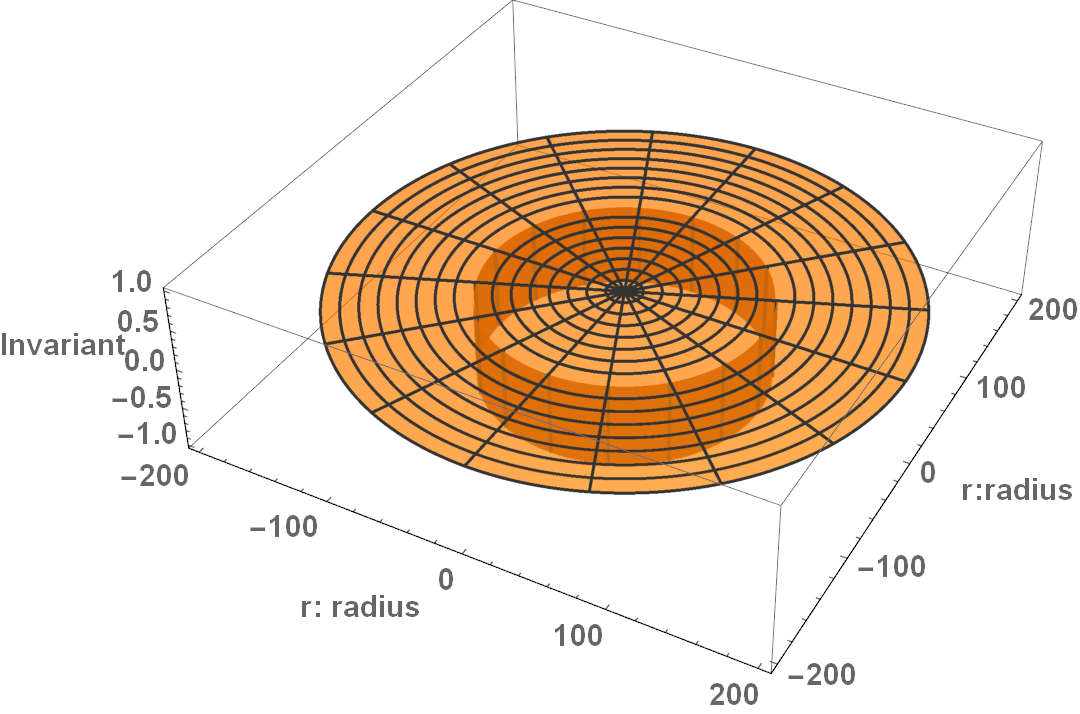}
    		\caption{$v=0.01 \frac{m}{s}$}
    		\label{fig:4.11b}
    	\end{subfigure}
    	~
    	\begin{subfigure}{.45\linewidth}
    	    \centering
    		\includegraphics[scale=0.2]{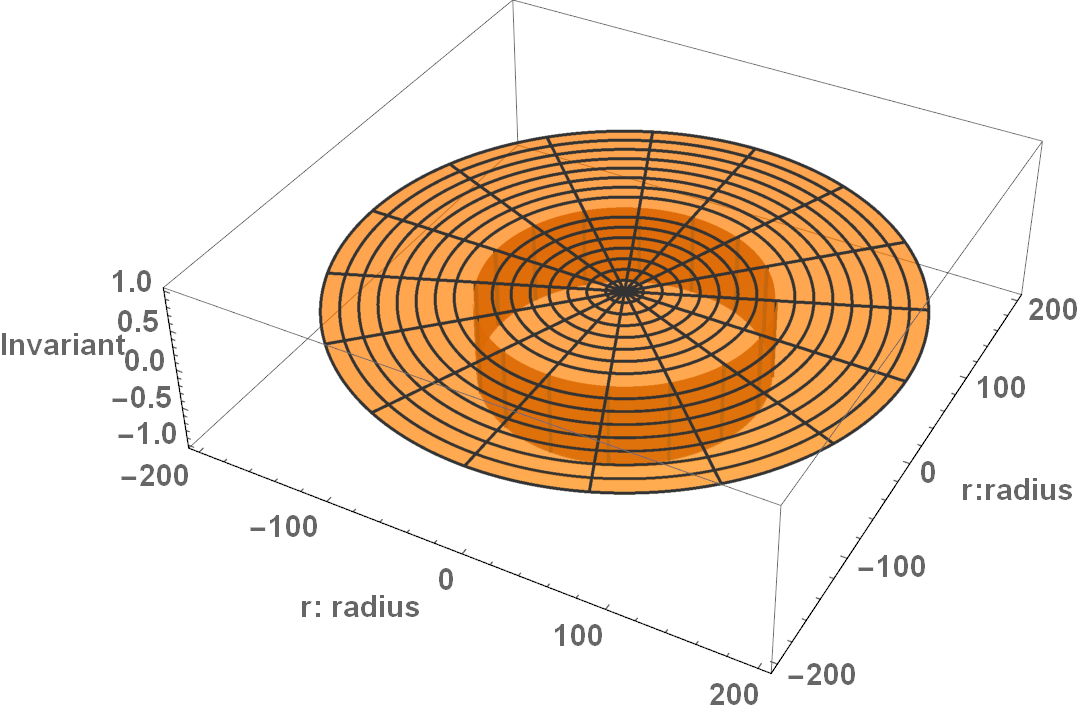}
    		\caption{$v=0.1 \frac{m}{s}$}
    		\label{fig:4.11c}
    	\end{subfigure}
    	~
    	\begin{subfigure}{.55\linewidth}
    	    \centering
    		\includegraphics[scale=0.2]{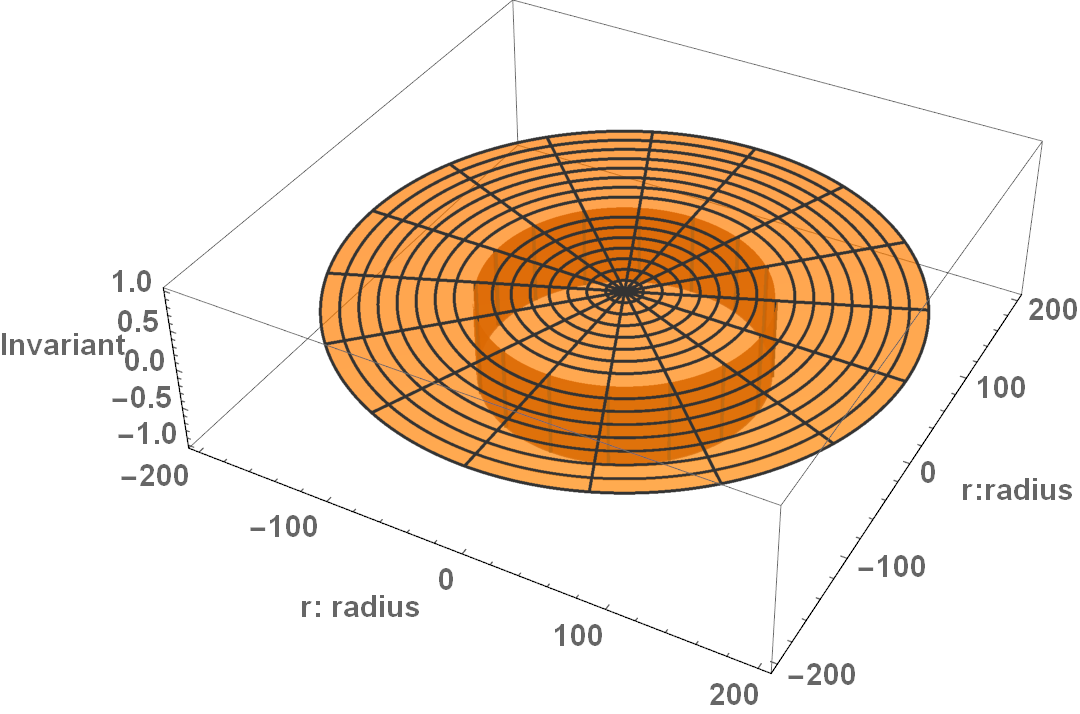}
    		\caption{$v=1 \frac{m}{s}$}
    		\label{fig:4.11d}
    	\end{subfigure}
    	~
    	\begin{subfigure}{.45\linewidth}
    	    \centering
    		\includegraphics[scale=0.2]{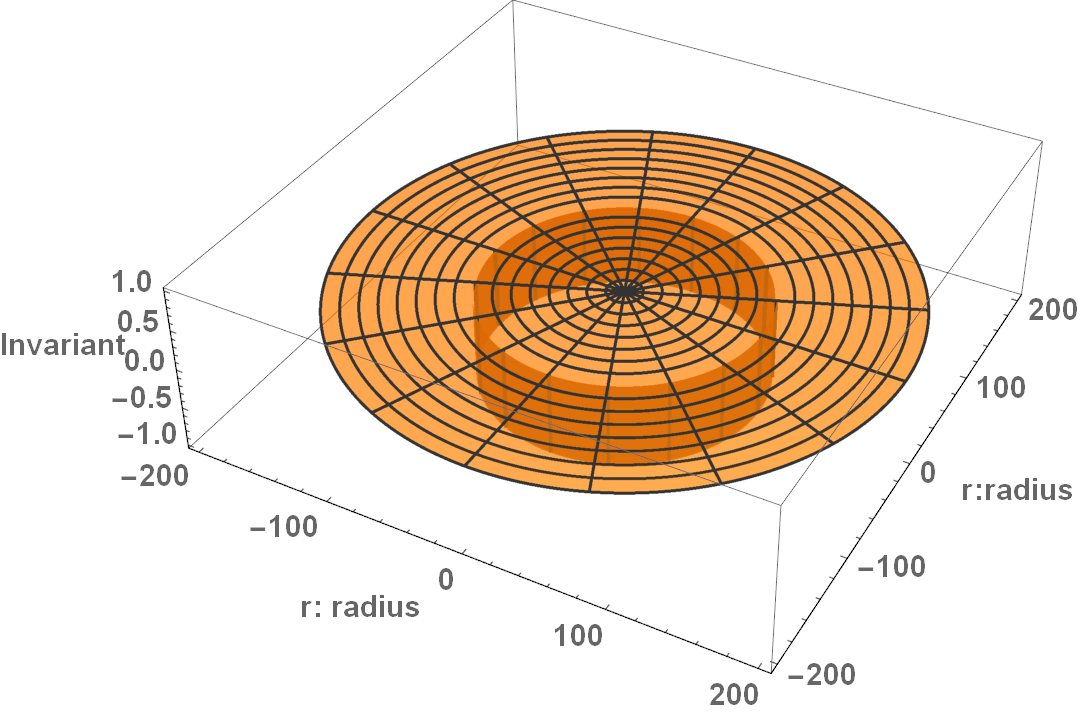}
    		\caption{$v=10 \frac{m}{s}$}
    		\label{fig:4.11e}
    	\end{subfigure}
    	~
    	\begin{subfigure}{.55\linewidth}
    	    \centering
    		\includegraphics[scale=0.2]{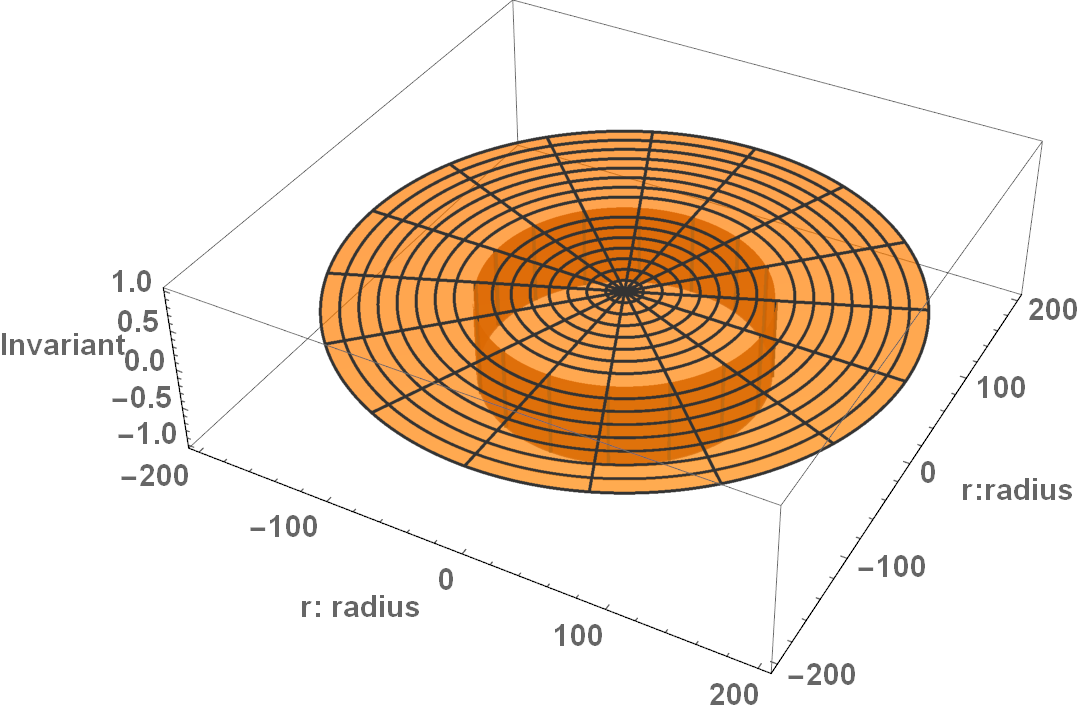}
    		\caption{$v=100 \frac{m}{s}$}
    		\label{fig:4.11f}
    	\end{subfigure}
	\caption{The velocity evolution of R, the Ricci scalar for the Nat\'ario warp drive at a constant velocity.
	It is understood that $v_s$ is multiplied by $c$.
	The other variables are set to $\sigma$ = 50,000~$\frac{1}{\mathrm{m}}$ and $\rho$ = 100~m.} \label{fig:4.11}
    \end{figure}

\begin{figure}[htb]
	\begin{subfigure}{.45\linewidth}
	    \centering
		\includegraphics[scale=0.2]{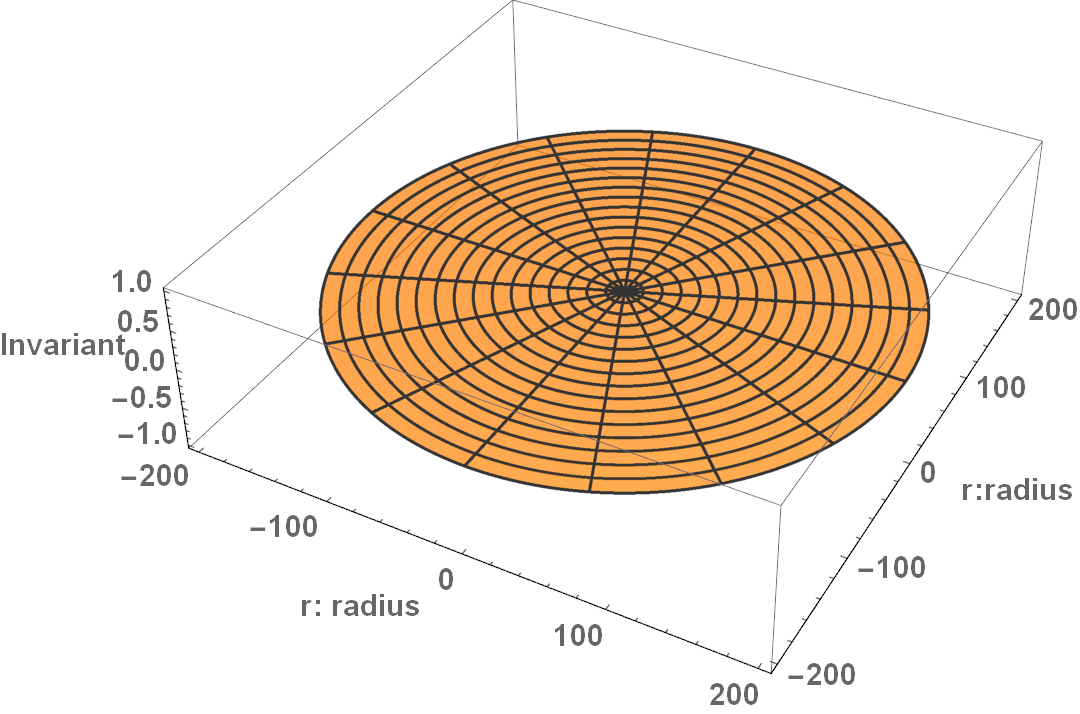}
		\caption{$v=0.0 \frac{m}{s}$}
		\label{fig:4.12a}
	\end{subfigure}
	~
	\begin{subfigure}{.55\linewidth}
	    \centering
		\includegraphics[scale=0.2]{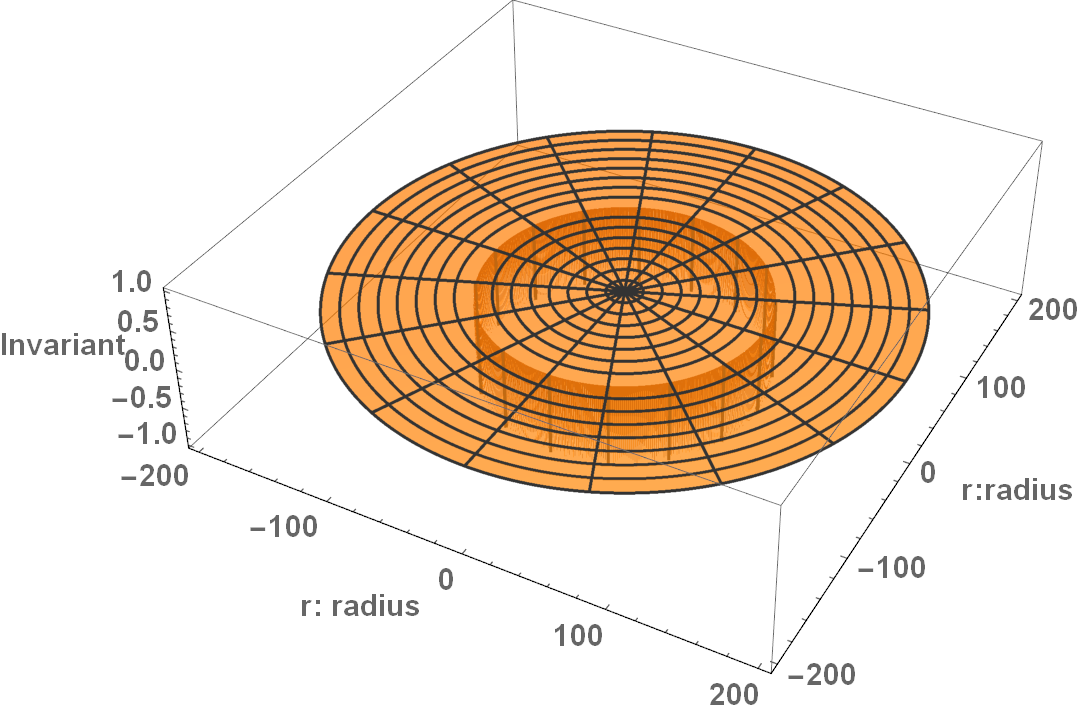}
		\caption{$v=0.01 \frac{m}{s}$}
		\label{fig:4.12b}
	\end{subfigure}
	~
	\begin{subfigure}{.45\linewidth}
	    \centering
		\includegraphics[scale=0.2]{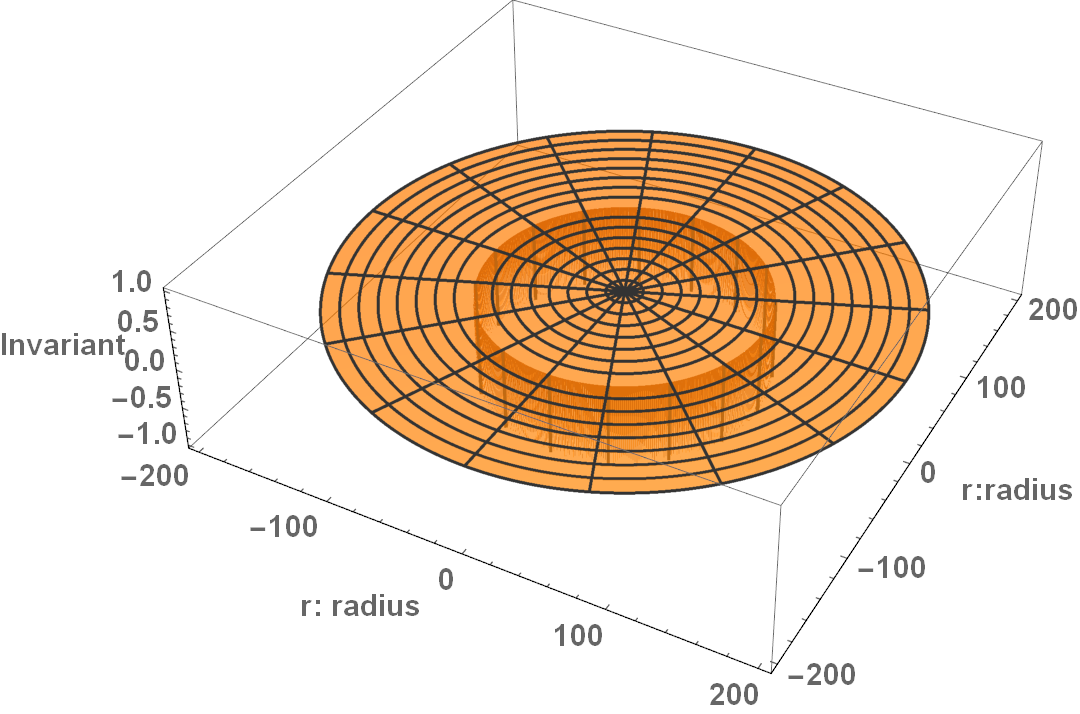}
		\caption{$v=0.1 \frac{m}{s}$}
		\label{fig:4.12c}
	\end{subfigure}
	~
	\begin{subfigure}{.55\linewidth}
	    \centering
		\includegraphics[scale=0.2]{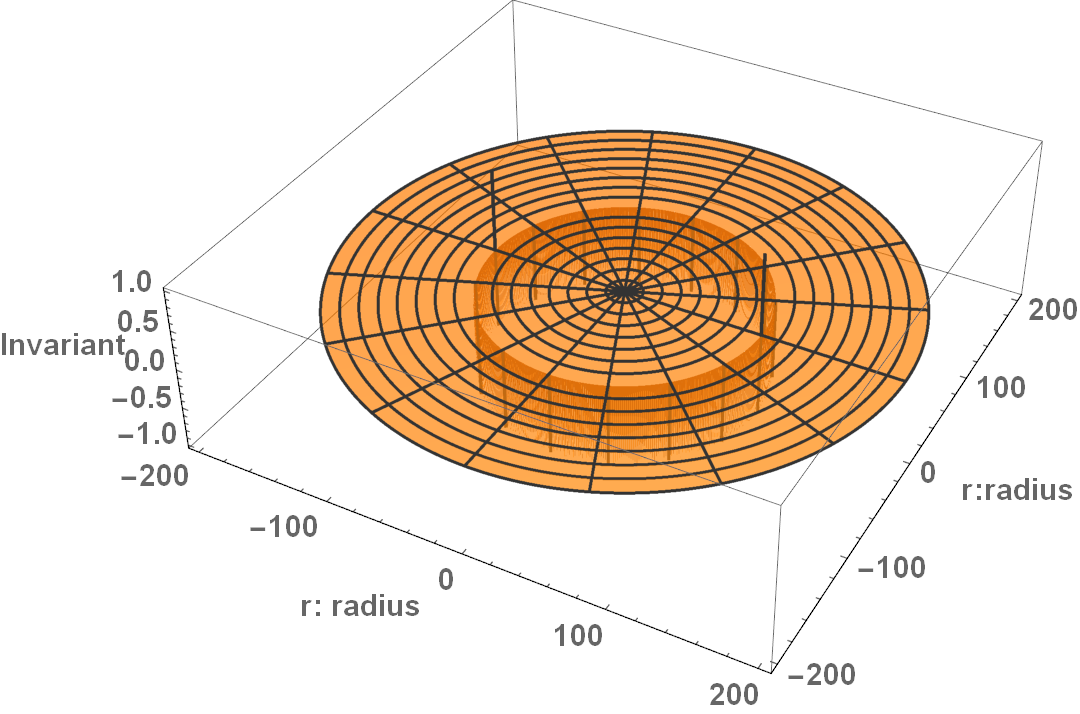}
		\caption{$v=1 \frac{m}{s}$}
		\label{fig:4.12d}
	\end{subfigure}
	~
	\begin{subfigure}{.45\linewidth}
	    \centering
		\includegraphics[scale=0.2]{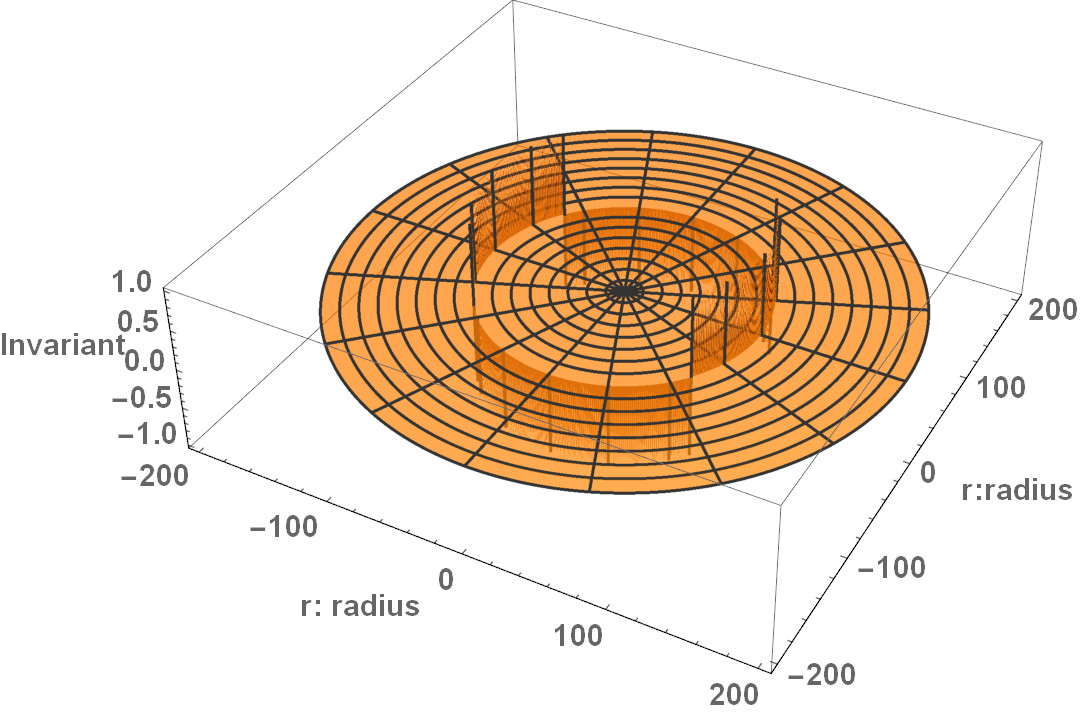}
		\caption{$v=10 \frac{m}{s}$}
		\label{fig:4.12e}
	\end{subfigure}
	~
	\begin{subfigure}{.55\linewidth}
	    \centering
		\includegraphics[scale=0.2]{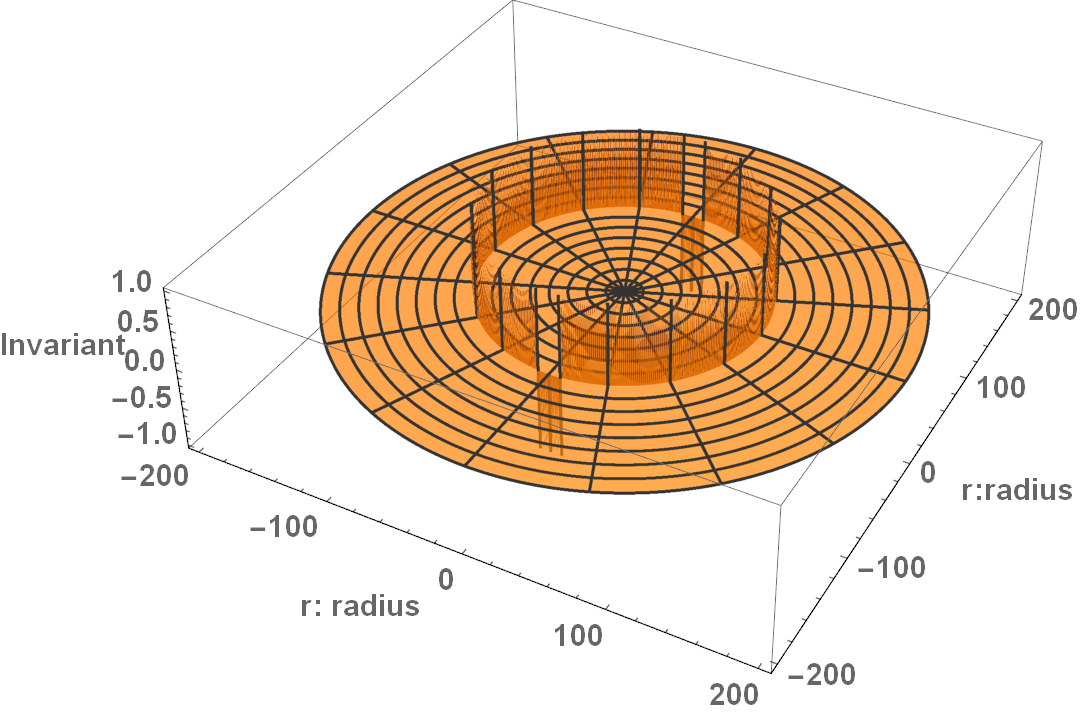}
		\caption{$v=100 \frac{m}{s}$}
		\label{fig:4.12f}
	\end{subfigure}
	\caption{The velocity evolution of the $r_1$ invariant for the Nat\'ario warp drive at a constant velocity.
	It is understood that $v_s$ is multiplied by $c$.
	The other variables are set to $\sigma$ = 50,000~$\frac{1}{\mathrm{m}}$ and $\rho$ = 100~m.} \label{fig:4.12}
    \end{figure}

\begin{figure}[htb]
	\begin{subfigure}{.45\linewidth}
	    \centering
		\includegraphics[scale=0.2]{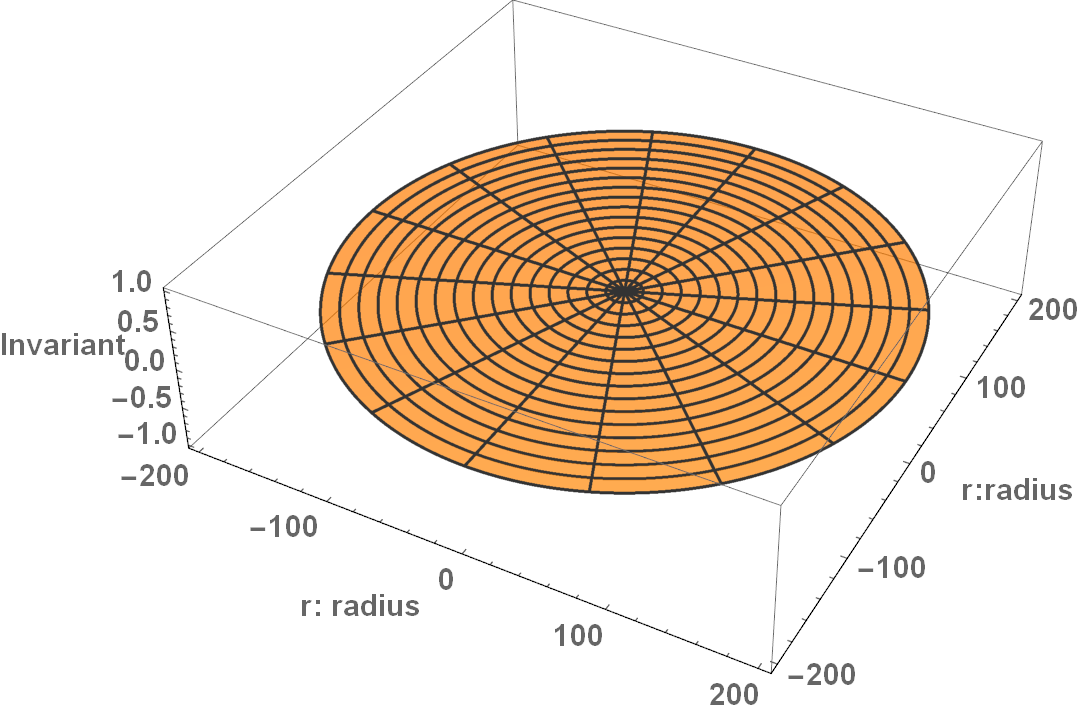}
		\caption{$v=0.0 \frac{m}{s}$}
		\label{fig:4.13a}
	\end{subfigure}
	~
	\begin{subfigure}{.55\linewidth}
	    \centering
		\includegraphics[scale=0.2]{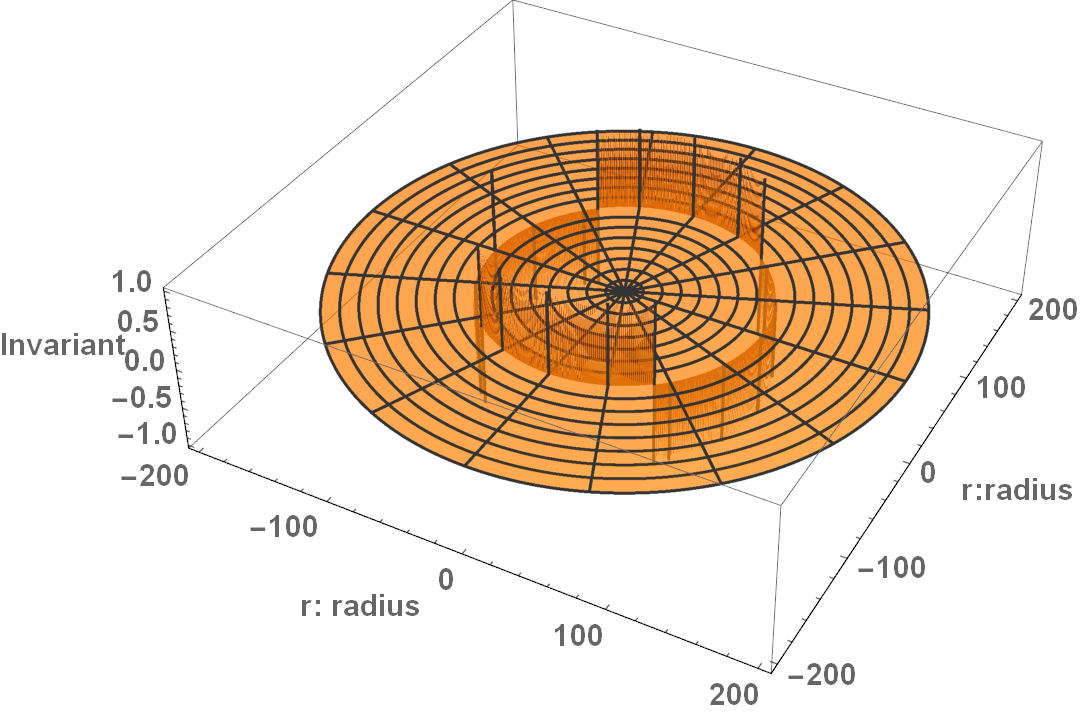}
		\caption{$v=0.01 \frac{m}{s}$}
		\label{fig:4.13b}
	\end{subfigure}
	~
	\begin{subfigure}{.45\linewidth}
	    \centering
		\includegraphics[scale=0.2]{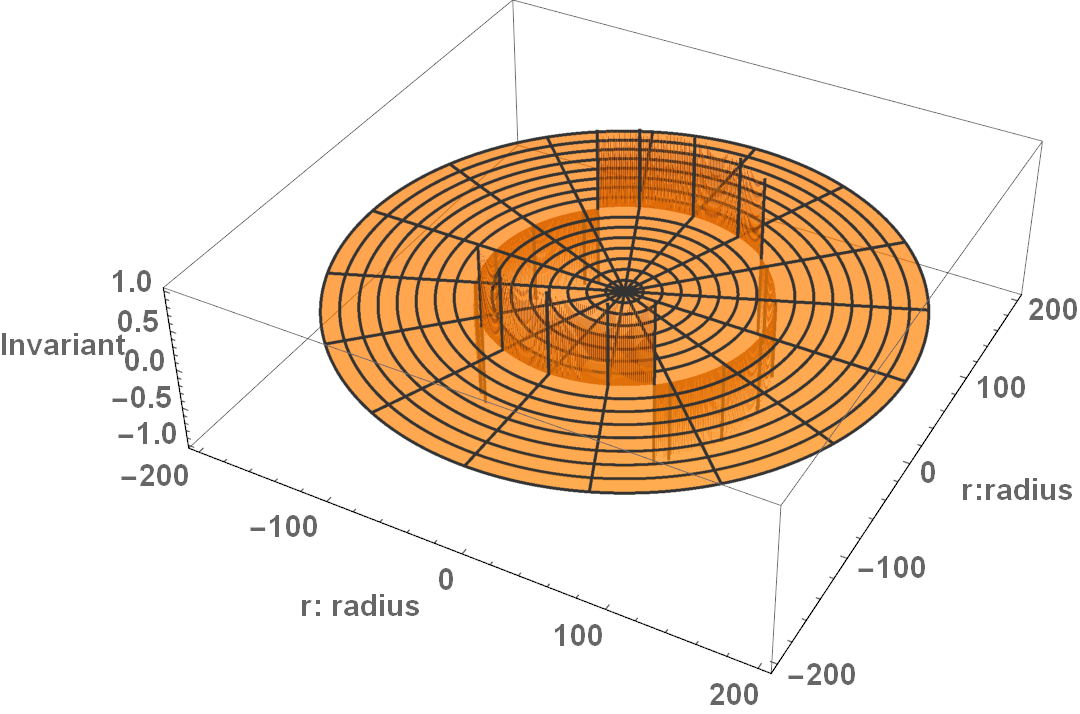}
		\caption{$v=0.1 \frac{m}{s}$}
		\label{fig:4.13c}
	\end{subfigure}
	~
	\begin{subfigure}{.55\linewidth}
	    \centering
		\includegraphics[scale=0.2]{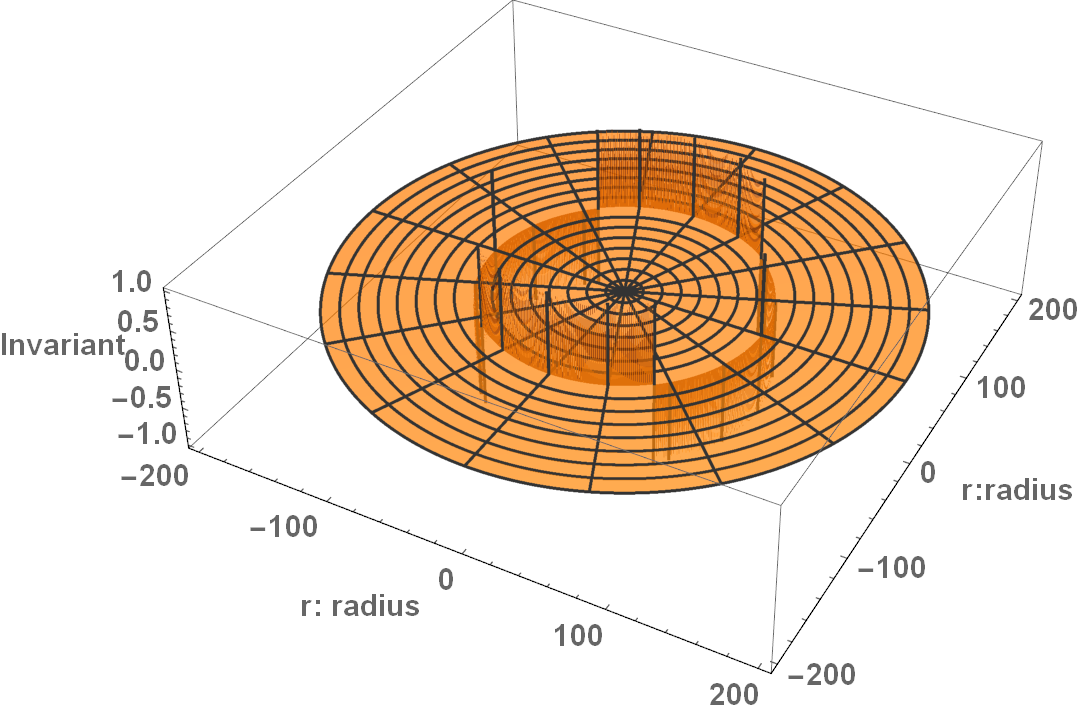}
		\caption{$v=1 \frac{m}{s}$}
		\label{fig:4.13d}
	\end{subfigure}
	~
	\begin{subfigure}{.45\linewidth}
	    \centering
		\includegraphics[scale=0.2]{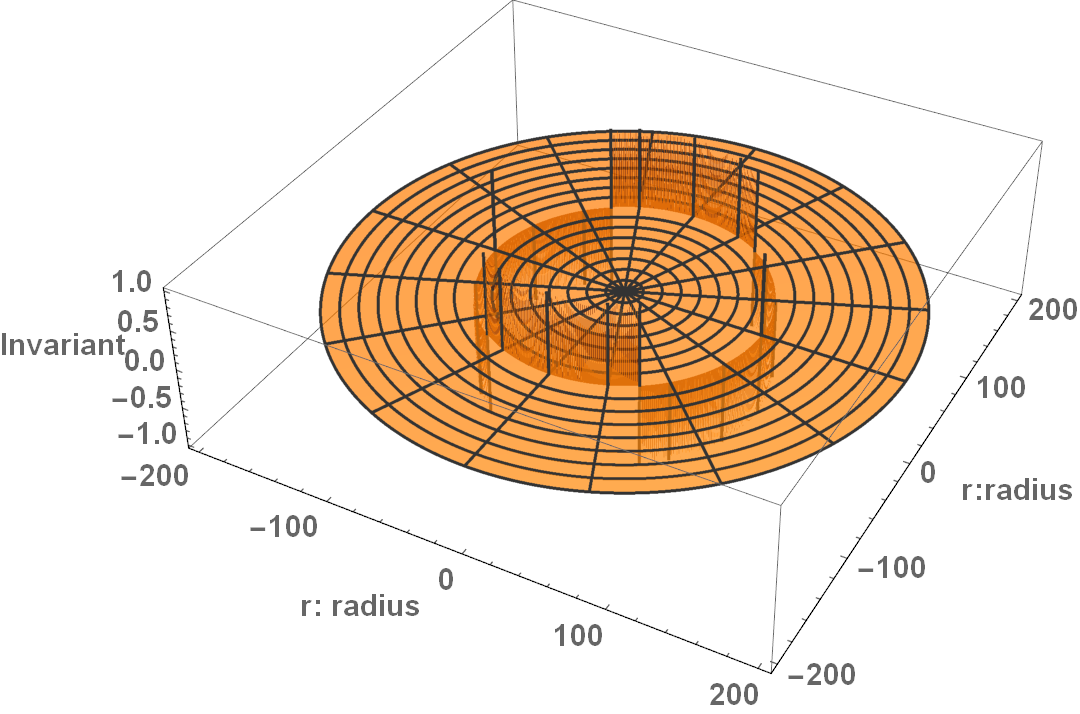}
		\caption{$v=10 \frac{m}{s}$}
		\label{fig:4.13e}
	\end{subfigure}
	~
	\begin{subfigure}{.55\linewidth}
	    \centering
		\includegraphics[scale=0.2]{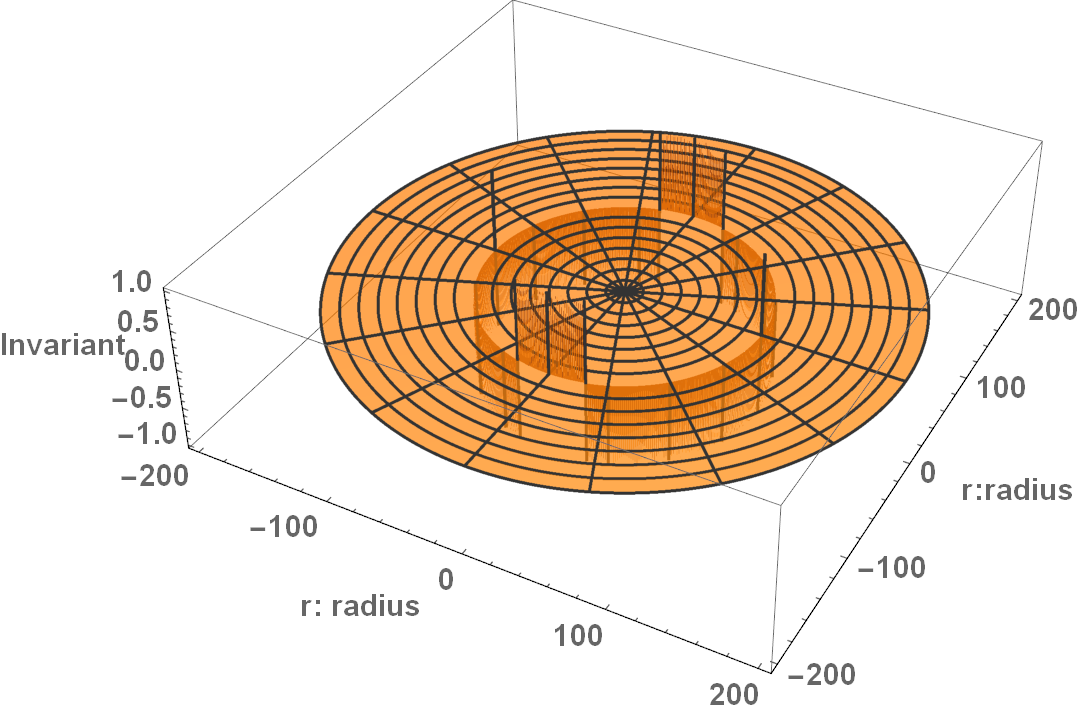}
		\caption{$v=100 \frac{m}{s}$}
		\label{fig:4.13f}
	\end{subfigure}
	\caption{The velocity Evolution of the $r_2$ invariant for the Nat\'ario warp drive at a constant velocity.
	It is understood that $v_s$ is multiplied by $c$.
	The other variables are set to $\sigma$ = 50,000~$\frac{1}{\mathrm{m}}$ and $\rho$ = 100~m.} \label{fig:4.13}
    \end{figure}

\begin{figure}[htb]
	\begin{subfigure}{.45\linewidth}
	    \centering
		\includegraphics[scale=0.2]{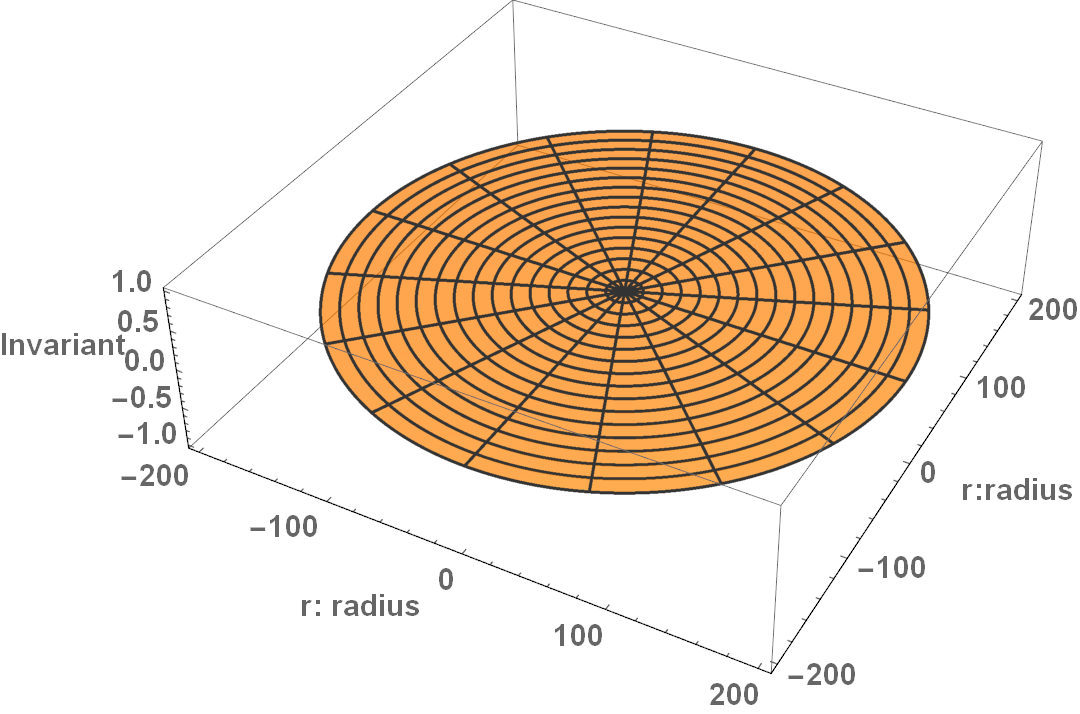}
		\caption{$v=0.0 \frac{m}{s}$}
		\label{fig:4.14a}
	\end{subfigure}
	~
	\begin{subfigure}{.55\linewidth}
	    \centering
		\includegraphics[scale=0.2]{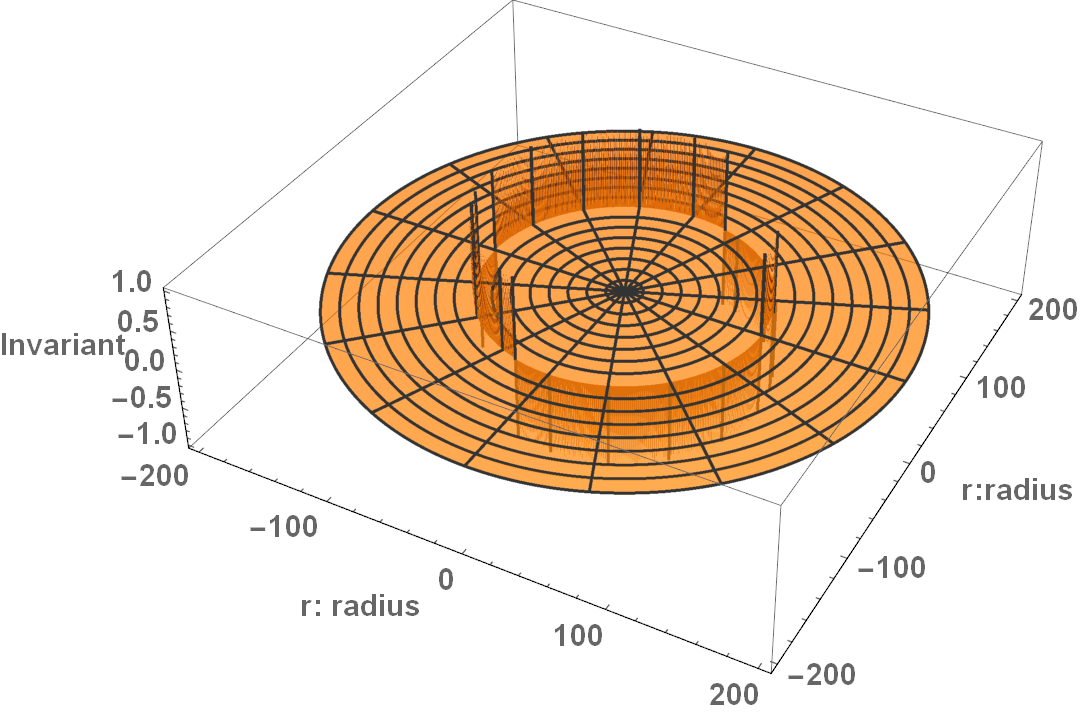}
		\caption{$v=0.01 \frac{m}{s}$}
		\label{fig:4.14b}
	\end{subfigure}
	~
	\begin{subfigure}{.45\linewidth}
	    \centering
		\includegraphics[scale=0.2]{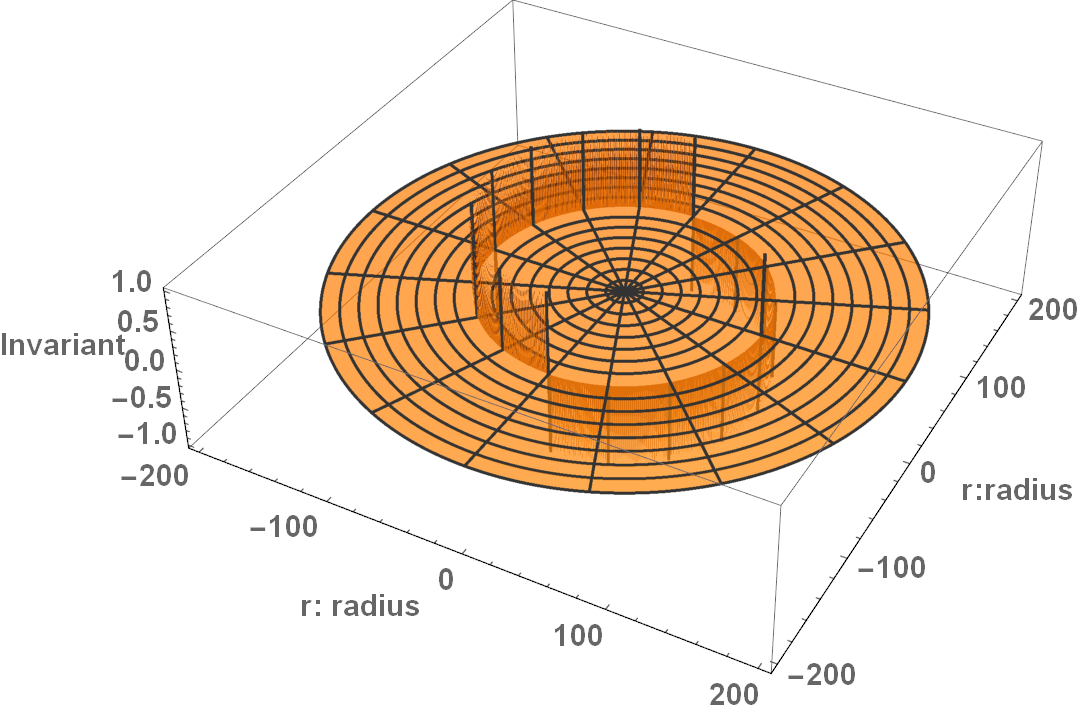}
		\caption{$v=0.1 \frac{m}{s}$}
		\label{fig:4.14c}
	\end{subfigure}
	~
	\begin{subfigure}{.55\linewidth}
	    \centering
		\includegraphics[scale=0.2]{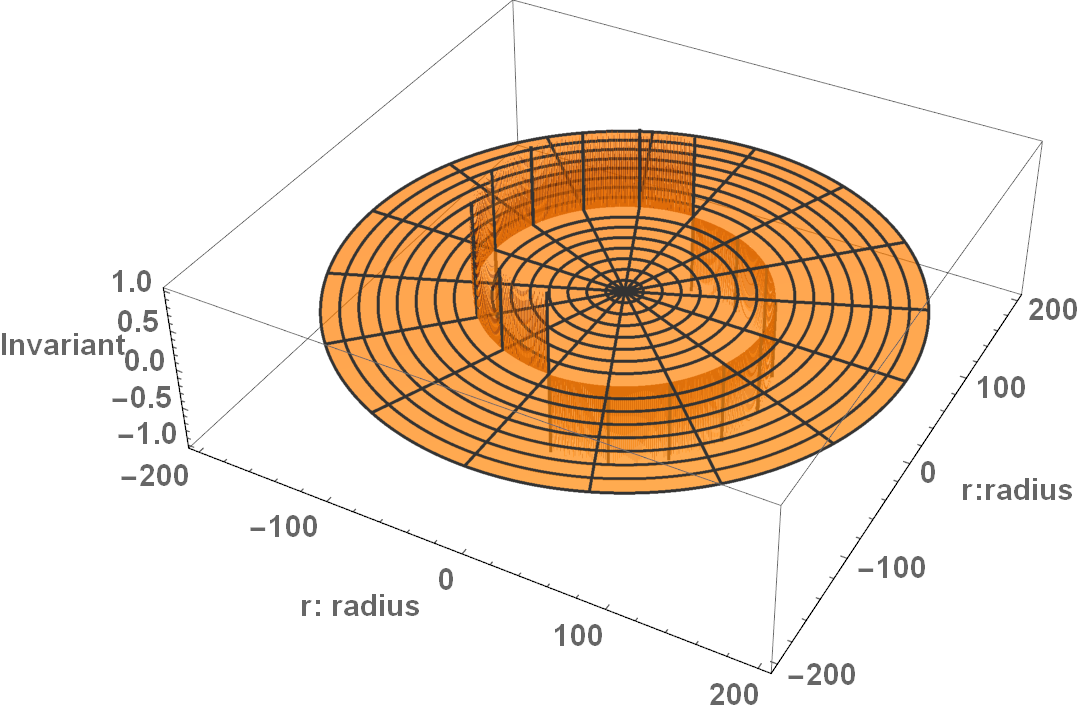}
		\caption{$v=1 \frac{m}{s}$}
		\label{fig:4.14d}
	\end{subfigure}
	~
	\begin{subfigure}{.45\linewidth}
	    \centering
		\includegraphics[scale=0.2]{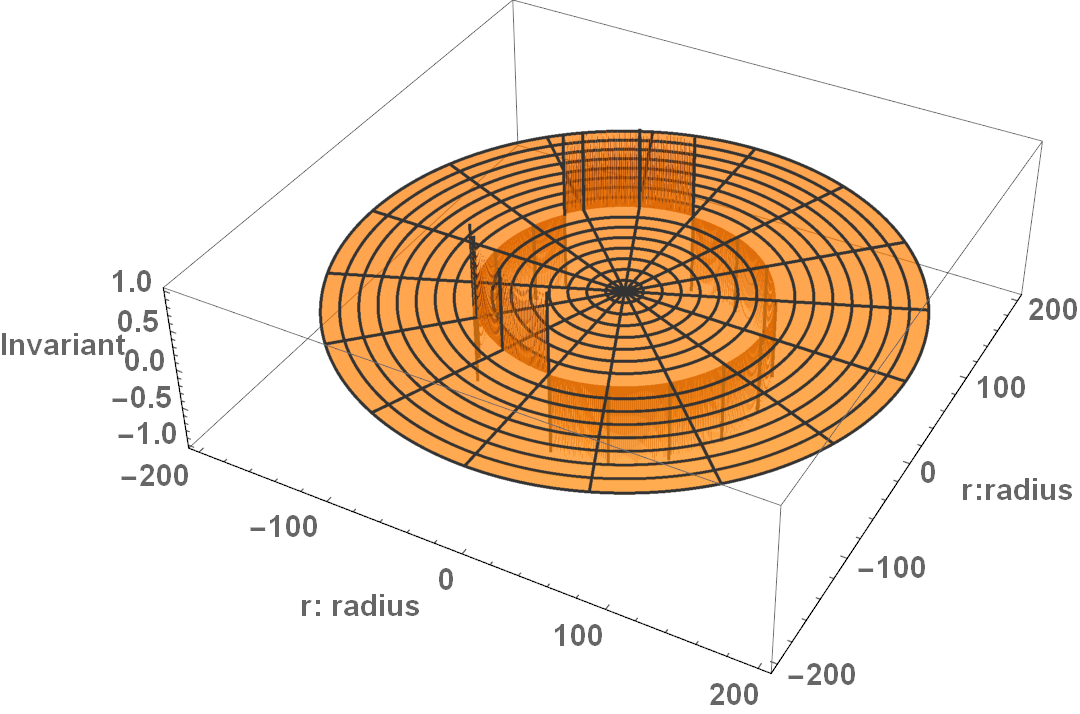}
		\caption{$v=10 \frac{m}{s}$}
		\label{fig:4.14e}
	\end{subfigure}
	~
	\begin{subfigure}{.55\linewidth}
	    \centering
		\includegraphics[scale=0.2]{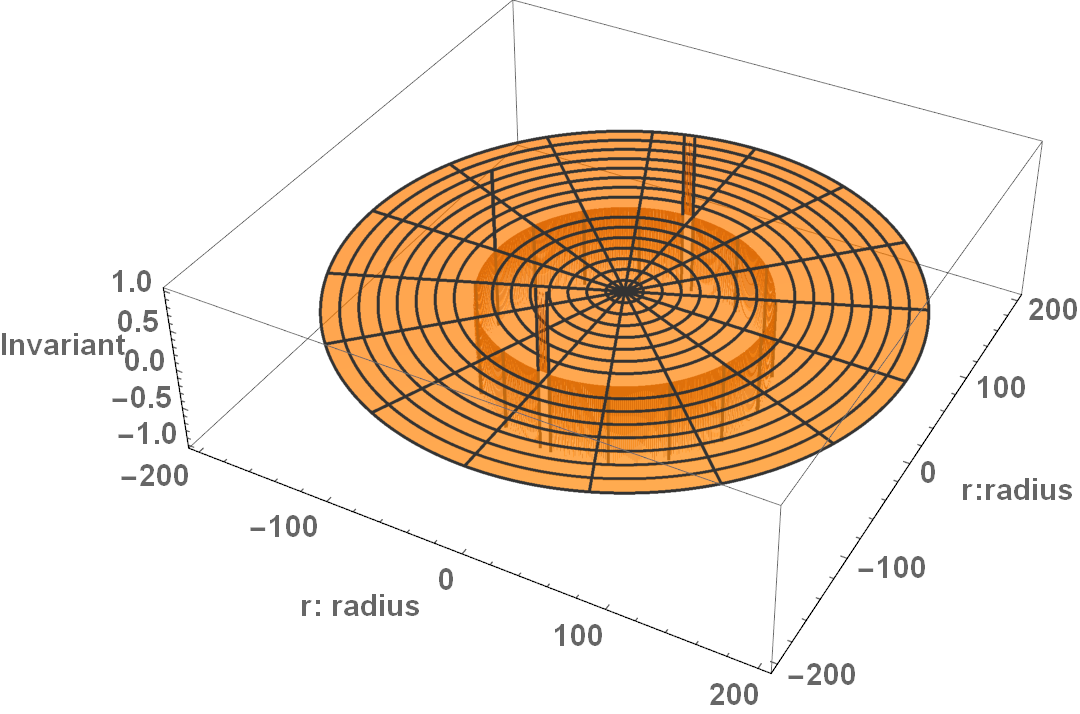}
		\caption{$v=100 \frac{m}{s}$}
		\label{fig:4.14f}
	\end{subfigure}
	\caption{The velocity evolution of the $w_2$ invariant for the Nat\'ario warp drive at a constant velocity.
	The other variables are set to $\sigma$ = 50,000~$\frac{1}{\mathrm{m}}$ and $\rho$ = 100~m.} \label{fig:4.14}
    \end{figure}

    Figures \ref{fig:4.11} through \ref{fig:4.14} plot the Nat\'ario invariants while changing the velocity.
    First, the manifold is completely flat when $v_s=0$ for each invariant.
    The warp bubble must be turned off at this velocity, which confirms that the program encoded the invariant functions correctly.
    For the Ricci scalar, a non-zero velocity causes the invariant's magnitude to jump to a small negative value at $r=\rho$.
    For $r_1$, an increase in velocity causes the magnitude of the invariant to swap from negative values to positive values as the velocity increases along the circumferences of the circle with a radius of $r=\rho$.
    For $r_2$, an increase in velocity causes the magnitude of the invariant to swap from positive values to negative values along the circumferences of the circle with a radius of $r=\rho$.
    For $w _2$, an increase in velocity swap the magnitude of the invariant between positive values to negative values as the velocity increases just along the semicircle of radius $r=\rho$ behind the harbor.
    But in front of the harbor, the $w_2$ invariant function remains negative regardless of the velocity.
        
    Our prediction of an exponential increase in the invariants due to the velocity is not consistent with the invariants' plots.
    A potential reason for this discrepancy is a dominant term inside of each CM invariant that overcomes the exponential increase in $v_s$.
    The dominant term in the invariant functions must either not depend on $v_s$ or the values for $\sigma$ are the dominant factor.
    The research in this dissertation may be extended either greater values of $v_s$ or lower values of the other variables to further investigate this anomaly.
    If the CM invariants can be manipulated by different choices of $v_s$, then a theoretical ship could control its navigation.
    ~
    %New plots for r_1
    ~
    ~
    %New plots for r_2
    ~
    ~
    %New plots for w_2
    ~
    
    \FloatBarrier
    
    \subsection{Invariant Plots of Skin Depth for Nat\'ario}
    \label{chp4.2:skin}
\begin{figure}[hb] \label{fig:4.16p1}
    	\begin{subfigure}{.45\linewidth}
    	    \centering
    		\includegraphics[scale=0.2]{Images/Chapter4/Natario/NcV-Rs50000p100v1.png}
    		\caption{The invariant $R$ with $\sigma$ = 50,000 $\frac{1}{\mathrm{m}}$}
    		\label{fig:4.15a}
    	\end{subfigure}
	~
    	\begin{subfigure}{.55\linewidth}
        	\centering
    		\includegraphics[scale=0.2]{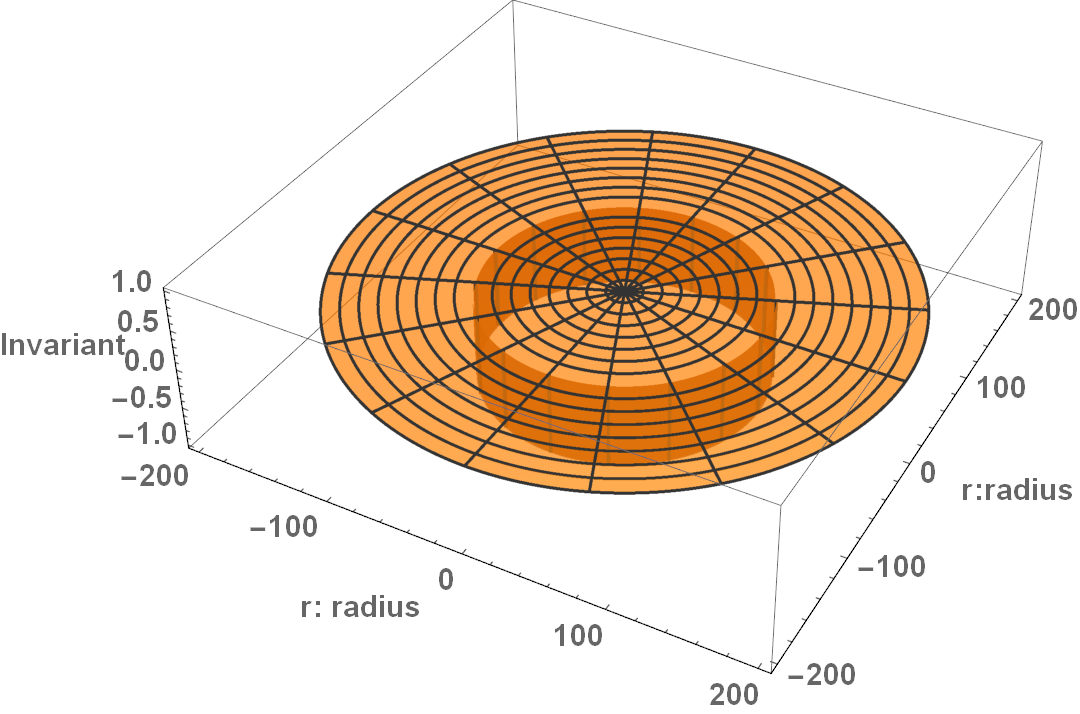}
    		\caption{The invariant $R$ with $\sigma$ = 100,000 $\frac{1}{\mathrm{m}}$}
    		\label{fig:4.15b}
    	\end{subfigure}
	~
    	\begin{subfigure}{.45\linewidth}
    		\includegraphics[scale=0.2]{Images/Chapter4/Natario/NcV-r1s50000p100v1.png}
    		\caption{The invariant $r_1$ with $\sigma$ = 50,000 $\frac{1}{\mathrm{m}}$}
    		\label{fig:4.15c}
    	\end{subfigure}
	~
    	\begin{subfigure}{.55\linewidth}
    	    \centering
    		\includegraphics[scale=0.2]{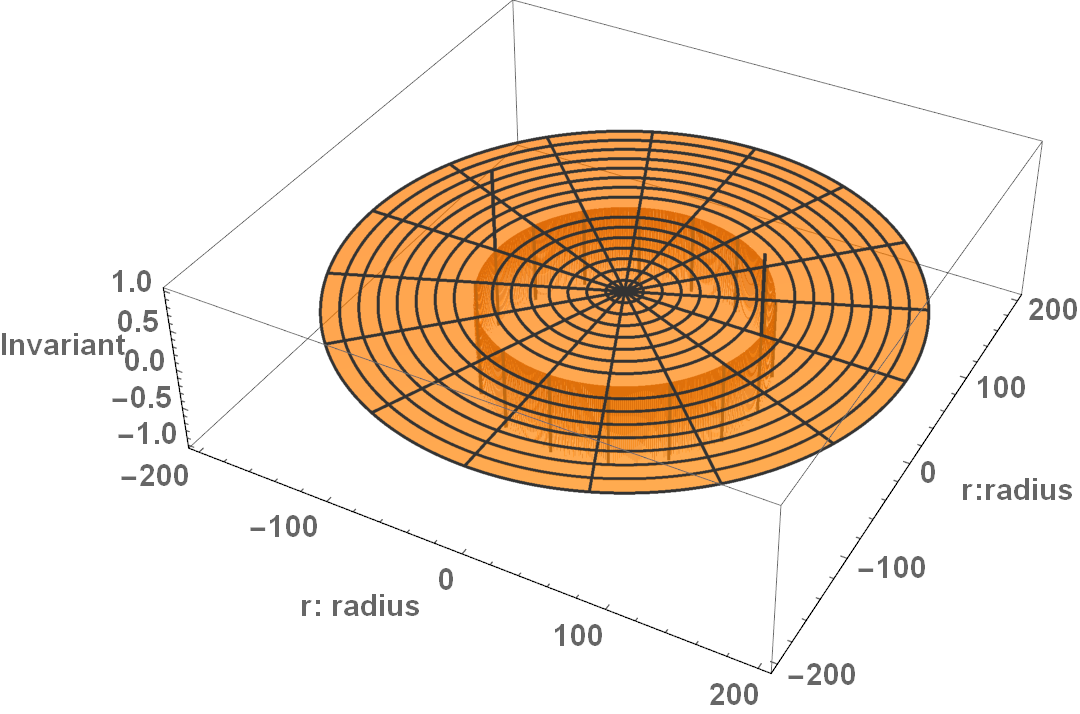}
    		\caption{The invariant $r_1$ with $\sigma$ = 100,000 $\frac{1}{\mathrm{m}}$}
    		\label{fig:4.15d}
    	\end{subfigure}
    	\caption{The warp bubble skin depth for the Ricci scalar and $r_1$ for the Nat\'ario warp drive at a constant velocity.
    	The other variables were chosen to be $v=1 \frac{\mathrm{m}}{\mathrm{s}}$, and $\rho$ = 100~m in natural units.} \label{fig:4.15}
	\end{figure}

\begin{figure}[ht]
    	\begin{subfigure}{.45\linewidth}
    	    \centering
    		\includegraphics[scale=0.2]{Images/Chapter4/Natario/NcV-r2s50000p100v1.png}
    		\caption{The invariant $r_2$ with $\sigma$ = 50,000 $\frac{1}{\mathrm{m}}$}
    		\label{fig:4.16a}
    	\end{subfigure}
	~
    	\begin{subfigure}{.55\linewidth}
    	    \centering
    		\includegraphics[scale=0.2]{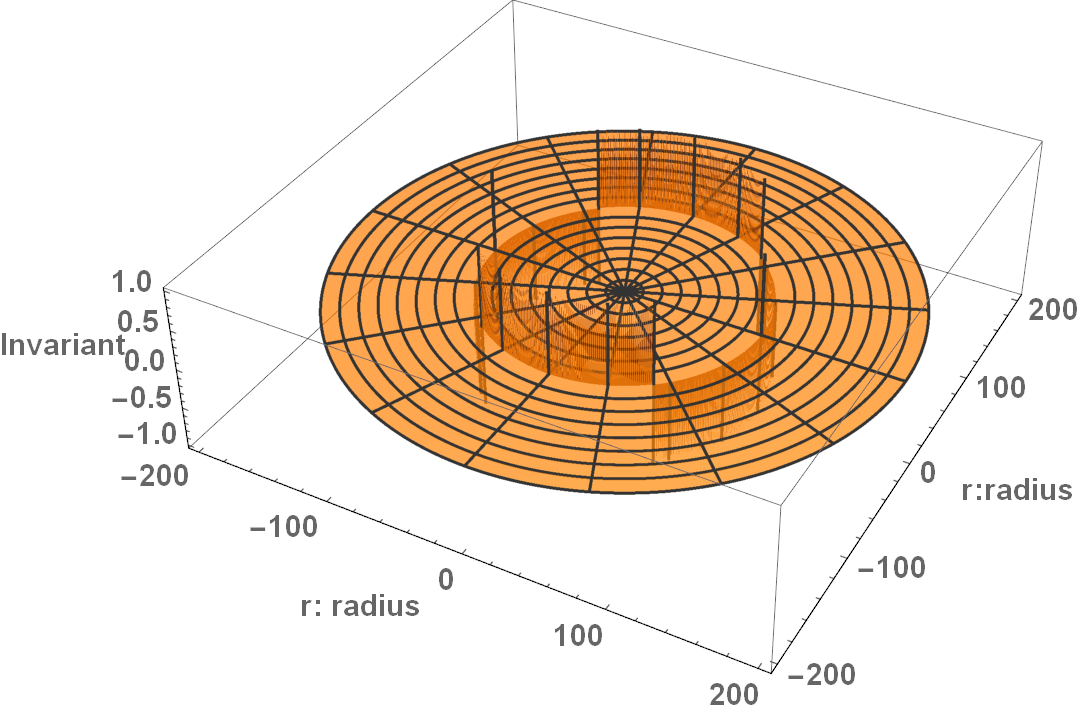}
    		\caption{The invariant $r_2$ with $\sigma$ = 100,000 $\frac{1}{\mathrm{m}}$}
    		\label{fig:4.16b}
    	\end{subfigure}
	~
    	\begin{subfigure}{.45\linewidth}
    		\includegraphics[scale=0.2]{Images/Chapter4/Natario/NcV-w2s50000p100v1.png}
    		\caption{The invariant $w_2$ with $\sigma$ = 50,000 $\frac{1}{\mathrm{m}}$}
    		\label{fig:4.16c}
    	\end{subfigure}
	~
    	\begin{subfigure}{0.55\linewidth}
    	    \centering
    		\includegraphics[scale=0.2]{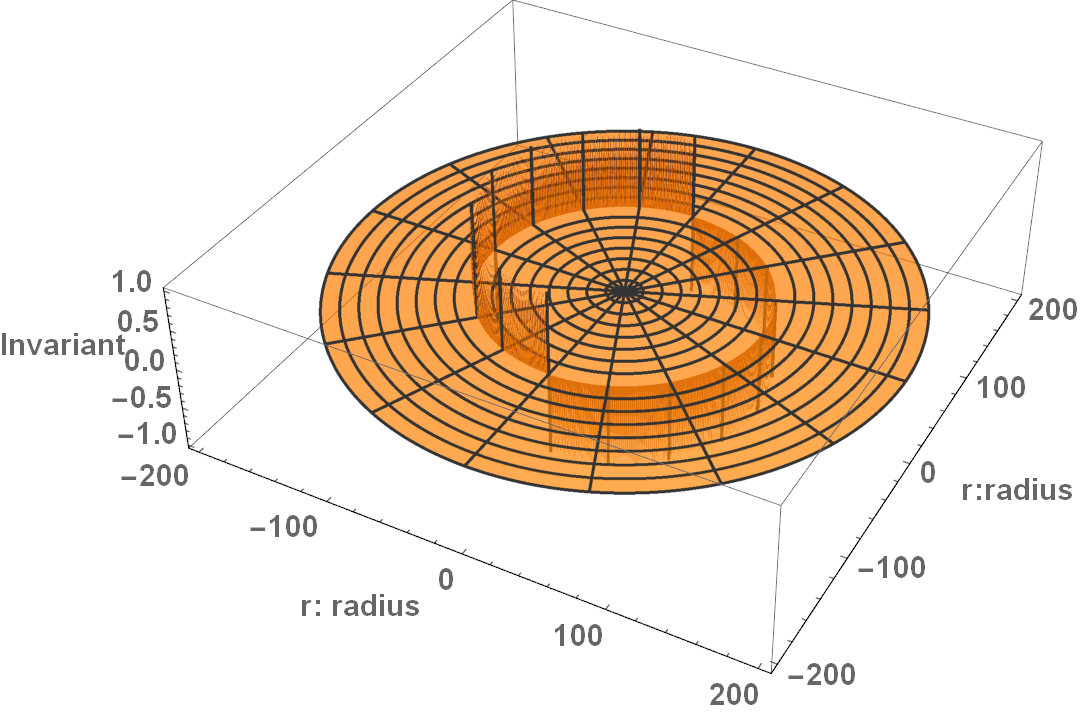}
    		\caption{The invariant $w_2$ with $\sigma$ = 100,000 $\frac{1}{\mathrm{m}}$}
    		\label{fig:4.16d}
    	\end{subfigure}
    	\caption{The warp bubble skin depth for $r_2$ and $w_2$ for the Nat\'ario warp drive at a constant velocity.
    	The other variables were chosen to be $v=1 \frac{\mathrm{m}}{\mathrm{s}}$, and $\rho$ = 100~m in natural units.} \label{fig:4.16}
    \end{figure}

    Figs.~\ref{fig:4.15} and \ref{fig:4.16}, plot the Nat\'ario invariants while changing the skin depth from $\sigma=500000$ to $\sigma=100000$.
    Unlike the shape expected from inspecting the invariant functions, the shape of the invariants remains the same.
    Since $\text{sech}(r-\rho) \rightarrow 1$ as $(r-\rho \rightarrow0$), the spike in the invariant functions match the values of the limit of the $\text{sech}$ and Eq.~\eqref{eq:4.14}.
    The dominant term(s) in the invariant must then be proportional to $\text{sech}(r-\rho)$. 
    The $\sigma$ plots add further evidence that the shape of the CM invariants is a consequence of the ``top-hat'' shape function. \\
        
	~
    
    \FloatBarrier
    
    \newpage
        
    \subsection{Invariant Plots of Radius for Nat\'ario}
    \label{chp4.2:radius}
\begin{figure}[h]
    	\begin{subfigure}{.45\linewidth}
    	    \centering
    		\includegraphics[scale=0.2]{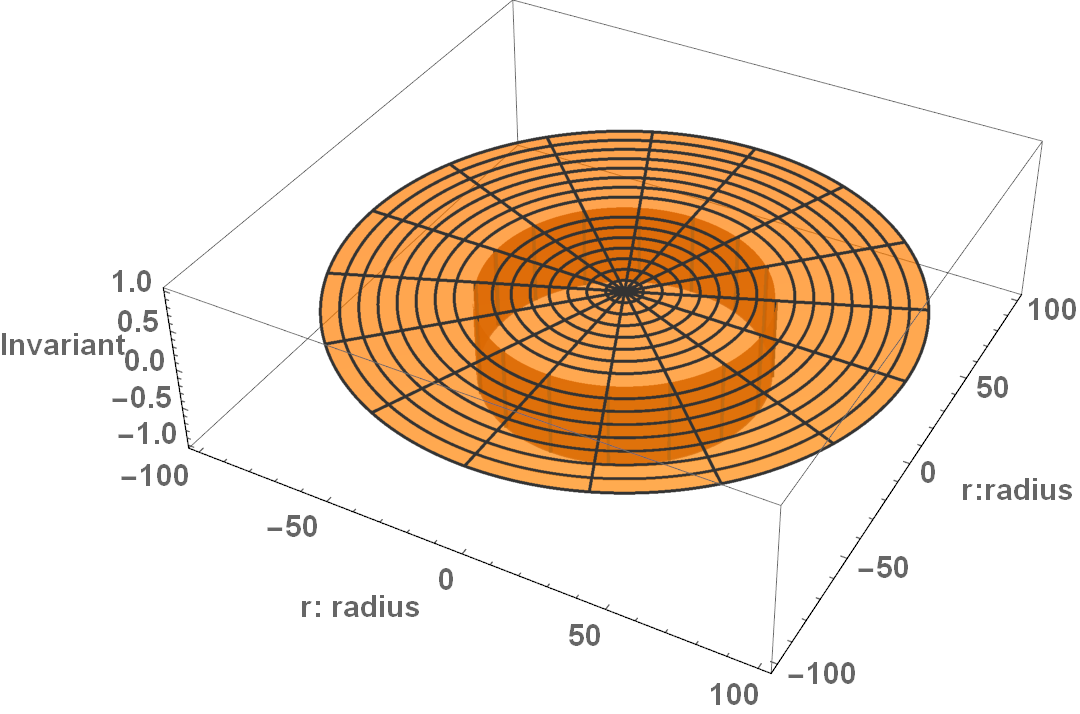}
    		\caption{The invariant $w_2$ with $\rho$ = 50~m}
    		\label{fig:4.17.a}
    	\end{subfigure}
	~
    	\begin{subfigure}{.56\linewidth}
        	\centering
    		\includegraphics[scale=0.2]{Images/Chapter4/Natario/NcV-Rs50000p100v1.png}
    		\caption{The invariant $R$ with $\rho$ = 100~m}
    		\label{fig:4.17.b}
    	\end{subfigure}
	~
    	\begin{subfigure}{.45\linewidth}
    		\includegraphics[scale=0.2]{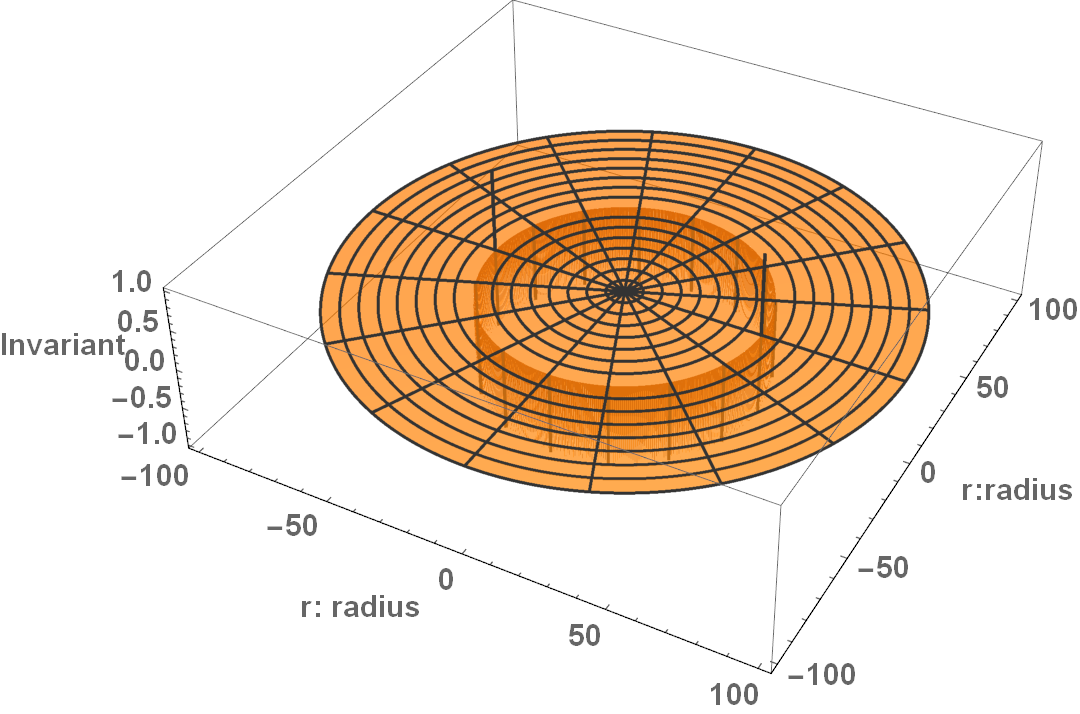}
    		\caption{The invariant $r_1$ with $\rho$ = 50~m}
    		\label{fig:4.17.c}
    	\end{subfigure}
	~
    	\begin{subfigure}{.45\linewidth}
    	    \centering
    		\includegraphics[scale=0.2]{Images/Chapter4/Natario/NcV-r1s50000p100v1.png}
    		\caption{The invariant $r_1$ with $\rho$ = 100~m}
    		\label{fig:4.17.d}
    	\end{subfigure}
    	\caption{The warp bubble radius for the Ricci scalar and $r_1$ for the Nat\'ario warp drive at a constant velocity.
    	The other variables were chosen to be $v=1 \frac{\mathrm{m}}{\mathrm{s}}$, and $\sigma=50,000~\frac{1}{\mathrm{m}}$ in natural units.} \label{fig:4.17}
	\end{figure}

\begin{figure}[h]
    	\begin{subfigure}{.45\linewidth}
    	    \centering
    		\includegraphics[scale=0.2]{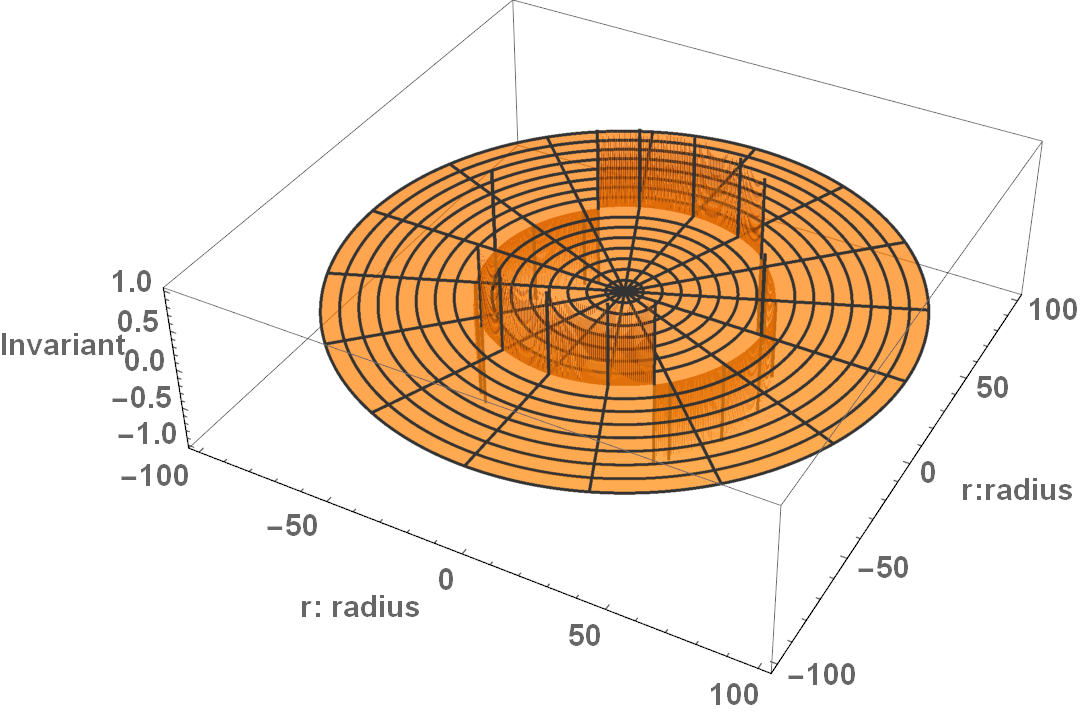}
    		\caption{The invariant $r_2$ with $\rho$ = 50~m}
    		\label{fig:4.18.a}
    	\end{subfigure}
	~
    	\begin{subfigure}{.55\linewidth}
    	    \centering
    		\includegraphics[scale=0.2]{Images/Chapter4/Natario/NcV-r2s50000p100v1.png}
    		\caption{The invariant $r_2$ with $\rho$ = 100~m}
    		\label{fig:4.18.b}
    	\end{subfigure}
	~
    	\begin{subfigure}{.45\linewidth}
    		\includegraphics[scale=0.2]{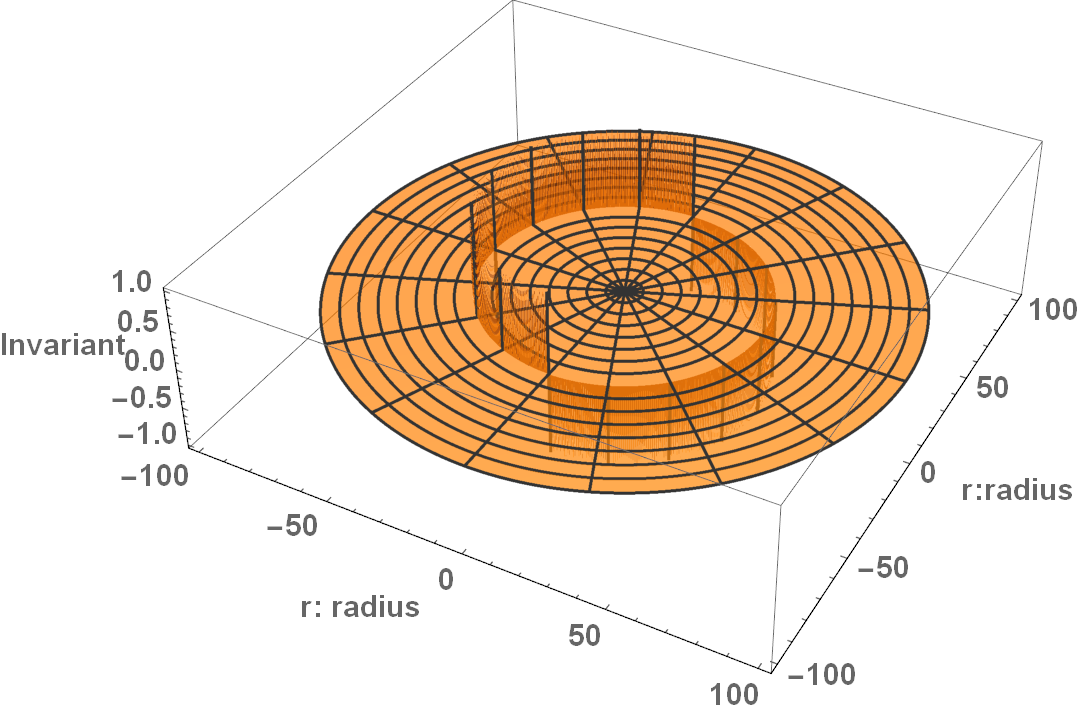}
    		\caption{The invariant $w_2$ with $\rho$ = 50~m}
    		\label{fig:4.18.c}
    	\end{subfigure}
	~
    	\begin{subfigure}{0.55\linewidth}
    	    \centering
    		\includegraphics[scale=0.2]{Images/Chapter4/Natario/NcV-w2s50000p100v1.png}
    		\caption{The invariant $w_2$ with $\rho$ = 100~m}
    		\label{fig:4.18.d}
    	\end{subfigure}
    	\caption{The warp bubble radius for the $r_2$, and $w_2$ for the Nat\'ario warp drive at a constant velocity.
    	The other variables were chosen to be $v=1 \frac{\mathrm{m}}{\mathrm{s}}$, and $\sigma=50,000~\frac{1}{\mathrm{m}}$ in natural units.} \label{fig:4.18}
    \end{figure}

    Like the Alcubierre plots in Section \ref{chp4.1:radius}, the main effect of changing $\rho$ is to increase the size of the Nat\'ario safe harbor.
    As $\rho$ increases from $\rho=50$ to $\rho=100$ in Figures \ref{fig:4.17} through \ref{fig:4.18}, the sizes of the safe harbor and the bubble double.
    The plots confirm that the radius $\rho$ moderates the size of the bubble in the same fashion as Section \ref{chp4.1:radius}.
    The spacing between the fringes in $r_1$, $r_2$, and $w_2$ is not affected by changing $\rho$.
    The nature of the internal structure in the warp bubble itself remains a topic for further research.
        
	~
        
    \FloatBarrier

%% file: ch5.tex
\chapter{Warp Drives Moving at a Constant Acceleration}
\label{Chapter5}
%In this chapter, the curvature invariants of Nat\'ario's warp drive moving at a constant acceleration are presented and discussed.
The final set of spacetimes to be investigated in this dissertation are Nat\'ario warp drive spacetimes moving at a constant acceleration. 
The earliest papers acknowledged the need for an accelerating bubble to transport a ship from a state of rest in its dock to its desired FTL velocity.
However, no papers initially considered accelerating solutions of the ADM metric in Eq.~\eqref{eq:4.1}.
Loup considered non-unity values for the lapse function $N$ and showed that these choices led to a warp bubble transporting at a constant acceleration \cite{Loup:2017}.
Later, he expanded his work to six line elements for the Nat\'ario spacetime with constant acceleration for contravariant, covariant, and mixed spacetimes \cite{Loup:2018}.
Deriving the appropriate lapse functions for the other potential warp drives remains an ongoing area of research.

In this chapter, the work of Loup will be presented to expand the warp drive line solutions to ADM spacetimes moving at a constant acceleration.
Then, the CM invariants for the covariant form of the accelerating Nat\'ario metric will be plotted.
The size of the invariant functions for these spacetimes is prohibitively large, so they will be included as a link to a .pdf document.
However, the invariant functions' main features such as singularities will be discussed.
The different free variables $v_s$, $t$, $\sigma$, and $\rho$ will be varied individually to see their effect on the CM invariant plots.
Finally, the effect of each of these free variables will be discussed.

\section{The Accelerating Nat\'ario Spacetime}
\label{chp5.1:ds}
\begin{figure}[ht]
    	\centering
    	\includegraphics[scale=0.3]{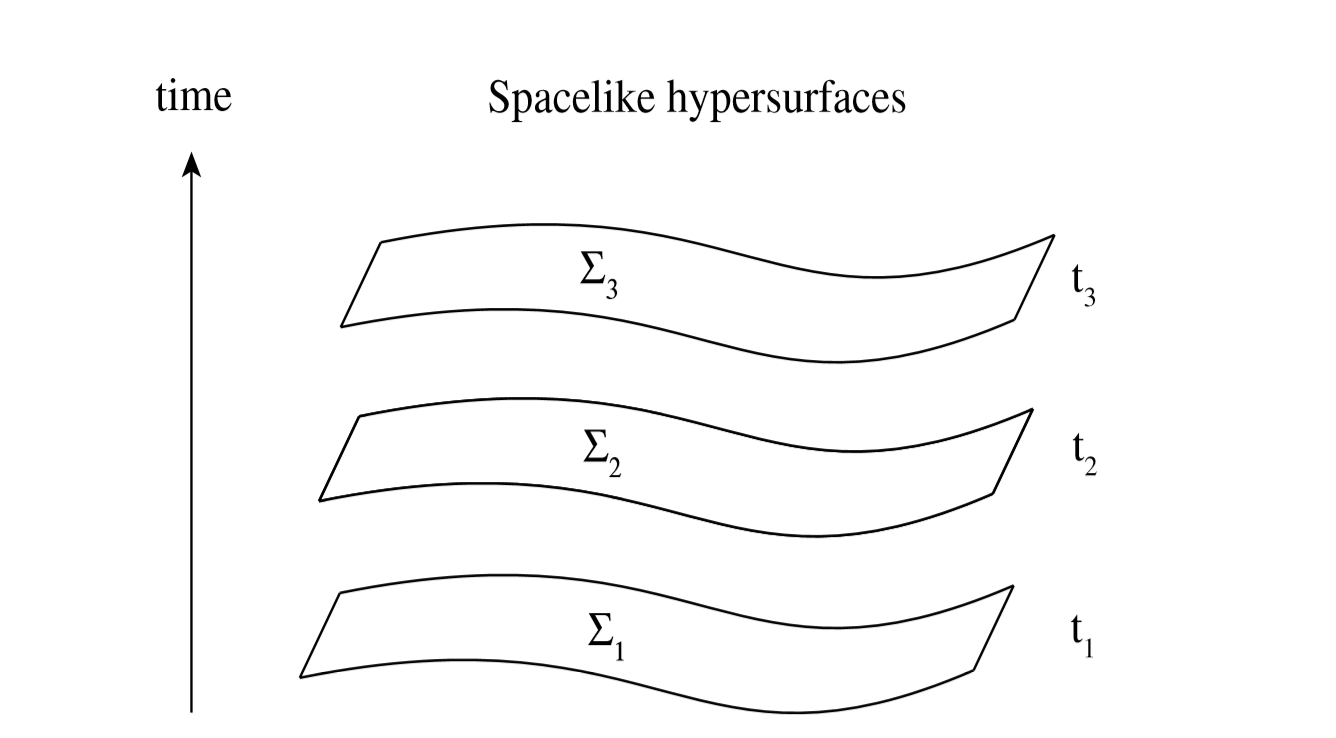}
    	\caption{Foliation of spacetime into 3D spacelike hypersurfaces at three different times from \cite{Frauendiener:2011}.}
    	\label{fig:5.1}
    \end{figure}

    \vspace{3mm}
    Consider the three spacelike hypersurfaces in Fig.~\ref{fig:5.1} for a Nat\'ario warp drive undergoing a constant acceleration.
    Then, the lapse of time $t_f=t_3-t_2$ for the warp to evolve to the hypersurface $\Sigma_3$ from the hypersurface $\Sigma_2$ must be less than the time lapse $t_i=t_2-t_1$ since the constant acceleration requires that $v_f>v_i$.
    The lapse function $N$ in Eq.~\eqref{eq:4.1} measures this lapse of proper time between different hypersurfaces.
    Choosing $c=1$ and $G_{tt}=1$, the lapse function will be 
    \begin{equation*}
        N^2=(1+X_t+X_t X^t), \tag{5.1} \label{eq:5.1} 
    \end{equation*}
    where $X_t$ is the time component of a covariant vector field for Nat\'ario's warp drive.
    
    An accelerating Nat\'ario warp drive spacetime must exhibit the same properties as the constant velocity Nat\'ario spacetime in Section \ref{chp4:Nat}.
    Namely, $n\textbf{X}=0$ and $X=v_s=0$ for the values of the radial coordinate $r$ inside the warp bubble's harbor and $n\textbf{X}=v_s(t)dx$ and $X=v_s=v_s$ for the values of the radial coordinate $r$ outside of the warp bubble.
    A natural choice for the variable velocity of the warp bubble is 
    \begin{equation*}
        v_s=2 n(r_s) a t. \tag{5.2} \label{eq:5.2}
    \end{equation*}
    where $n(r_s)$ is the Nat\'ario shape function, $a$ is the constant acceleration of the warp bubble, and $t$ is the time variable.
    Since time between the Cauchy surfaces decreases due to the lapse, the shift vector $N^i=\textbf{X}$ describing the relative velocity of Eulerian observers inside of each hypersurface must also be modified to $n\textbf{X}=v_s \star (dx) + x \star (dv_s)$, where $\star$ is the Hodge star product.
    For constant velocity, $x \star (dv_s)$ vanishes.
    But for a constant acceleration, the total differential is \begin{equation}
        dv_s=2 \left[a t \frac{d n(r_s)}{dr_s} dr_s + n(r_s) a dt \right] \tag{5.3} \label{eq:5.3}
    \end{equation}
    in spherical polar coordinates.
    Combining Eq.~\eqref{eq:5.1}, Eq.~\eqref{eq:5.2} and Eq.~\eqref{eq:4.13}, the Nat\'ario vector field $n\textbf{X}$ for variable velocities is
    \begin{equation*}
        n\textbf{X}=X^tdt+X^{r_s} dr_s + X^{\theta}r_s d\theta \tag{5.4} \label{eq:5.4}
    \end{equation*}
    with the contravariant shift vector components given by 
    \begin{align}
        X_t&=2n(r_s) a r_s \cos{\theta}, \tag{5.5} \label{eq:5.5} \\
        X_{r_s}&=2\left[2n(r_s)^2+r_sn'(r_s) \right]at\cos{\theta}, \tag{5.6} \label{eq:5.6} \\
        X_{\theta}&=-2n(r_s)at \left[2n(r_s)+r_sn'(r_s)\right]\sin{\theta}. \tag{5.7} \label{eq:5.7}
    \end{align}
    
    The specific equation for the Nat\'ario warp drive line element in the parallel covariant $3+1$ ADM is
    \begin{equation}
        ds^2=(1-2 X_t+(X_t)^2-(X_{r_s})^2-(X_{\theta})^2)dt^2+2(X_{r_s}dr_s+X_{\theta} r_s d\theta)dt-dr_s^2-r_s^2d\theta^2-r_s^2\sin^2{\theta} \ d\varphi^2 \tag{5.8} \label{eq:5.8}
    \end{equation}
    in spherical coordinates\footnote{$(0\le r_s <\infty;\ 0\leq\theta\leq\pi;\ 0\leq\varphi\leq 2\pi)$, and $ (-\infty< t<\infty)$} and where $a$ is the constant acceleration.
    The covariant shift vector components can be obtained by raising Eq.~\eqref{eq:5.5} through Eq.~\eqref{eq:5.7} by the metric to get 
    \begin{align}
        X_t&=2n(r_s) a r_s \cos{\theta}, \tag{5.9} \label{eq:5.9} \\
        X_{r_s}&=2\left[2n(r_s)^2+r_sn'(r_s)\right]at\cos{\theta}, \tag{5.10} \label{eq:5.10} \\
        X_{\theta}&=-2n(r_s)at\left[2n(r_s)+r_sn'(r_s)\right]r_s^2\sin{\theta}. \tag{5.11} \label{eq:5.11}
    \end{align}
    The Nat\'ario warp drive continuous shape function is Eq.~\eqref{eq:4.14}.
    The comoving null tetrad for Eq.~\eqref{eq:5.3} is
    \begin{align}
        l_i&=\begin{pmatrix}1-X_t+X_{r_s}\\-1\\0\\0\end{pmatrix}, &
        k_i&=\begin{pmatrix}1-X_t-X_{r_s}\\1\\0\\0\end{pmatrix}, \nonumber\\
        m_i&=\begin{pmatrix}X_\theta\\0\\-r\\i r \sin{\theta}\end{pmatrix}, &  
        \bar{m}_i&=\begin{pmatrix}X_\theta\\0\\-r\\-i r \sin{\theta}\end{pmatrix}.  \tag{5.12} \label{eq:5.12}
    \end{align}
    The comoving null tetrad describes light rays traveling parallel with the warp bubble. 
    Substituting Eqs.~\eqref{eq:5.8} and \eqref{eq:5.12} into the procedure from Chapter \ref{chapter2}, one may derive the four CM invariants.
    The Ricci scalar is included in Appendix \ref{ap:acc} as Eq.~\eqref{eq:C.1} as a demonstrative example.
    Due to the remaining three invariants' massive size, a link to a .pdf file for each is included in Appendix \ref{ap:acc}.
    Each of the four invariants has two singularities as can be seen in the Ricci scalar in Appendix \ref{ap:acc}.
    The first is at $r=0$ and the second is at
    \begin{equation}
        0=a r \cos (\theta )+a r \tanh ((r-\rho ) \sigma) \cos(\theta) - 2. \tag{5.13} \label{eq:5.13}
    \end{equation}
    Each invariant has the second singularity to increasing powers with $w_2$ maxing out at $n=12$. 
    The effect of these singularities will be analyzed in the plots in the next section.

\section{Invariant Plots of Time for Nat\'ario}
\label{chp5:time}
\begin{figure}[ht]
	\begin{subfigure}{.45\linewidth}
	    \centering
		\includegraphics[scale=0.29]{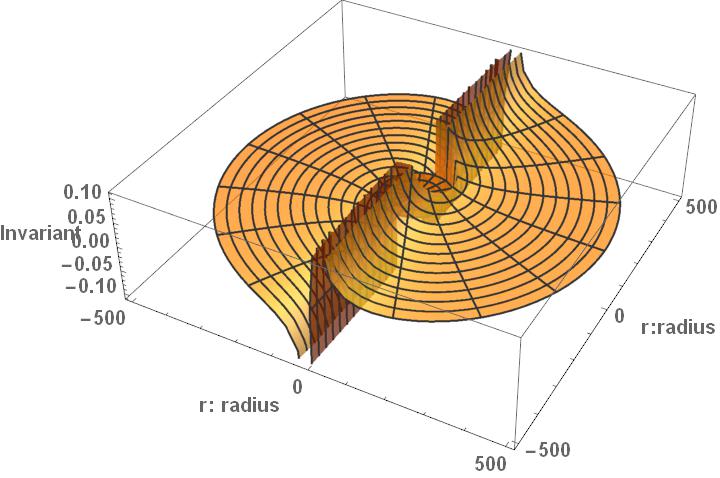}
		\caption{$t=0.0$~s}
		\label{fig:5.2a}
	\end{subfigure}
	~
	\begin{subfigure}{.55\linewidth}
	    \centering
		\includegraphics[scale=0.29]{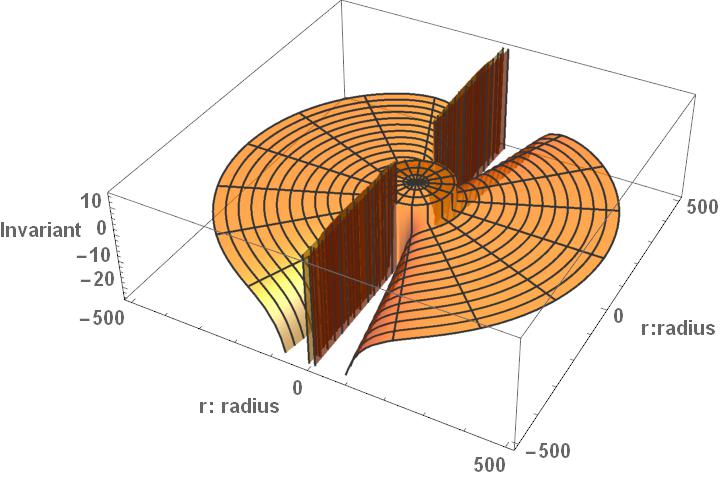}
		\caption{$t=1.0$~s}
		\label{fig:5.2b}
	\end{subfigure}
	~
	\begin{subfigure}{.45\linewidth}
	    \centering
		\includegraphics[scale=0.29]{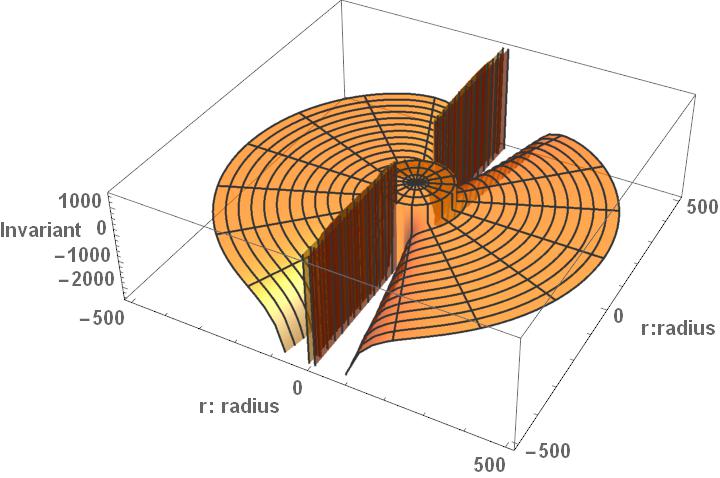}
		\caption{$t=10.0$~s}
		\label{fig:5.2c}
	\end{subfigure}
	~
	\begin{subfigure}{.55\linewidth}
	    \centering
		\includegraphics[scale=0.29]{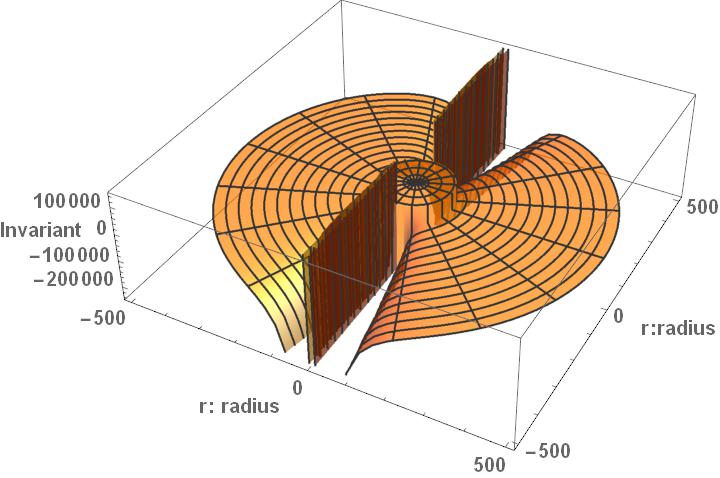}
		\caption{$t=100.0$~s}
		\label{fig:5.2d}
	\end{subfigure}
	\caption{The time evolution of $R$, the Ricci scalar.
	The other variables were chosen to be $a=1~\frac{\textrm{m}}{\textrm{s}^2}$, $\sigma$ = 50,000~$\frac{1}{\mathrm{m}}$, and $\rho=100$~m in natural units.} \label{fig:5.2}
    \end{figure}

    The invariants evolve dynamically over time.
    Figure~\ref{fig:5.2} shows how the Ricci scalar evolves from $t = 0$~s to $t = 100$~s, while setting $\rho = 100$~m, $\sigma$~= 50,000~$\frac{1}{\mathrm{m}}$, and $a =1.0 \  \frac{\mathrm{m}}{\mathrm{s}^2}$. 
    The figures provide rich details of the features in and around the warp bubble. 
    Each plot has a safe harbor within $r\leq100$ where the curvature plots are flat.
    Consequently, the singularity identified in the invariant functions must not be an intrinsic singularity.
    A spaceship riding in the interior of the harbor would experience only flat space throughout the entire time evolution. 
    The warp bubble travels perpendicular to the wake along the $\theta=(\frac{\pi}{2},\frac{3\pi}{2})$ line.
    The wake is a volume of large curvature as shown on the plots and is the primary effect of the singularity noted in the previous section.
    While the form of the wake quickly reaches a constant shape, the Ricci scalar's magnitude increases approximately proportional to time.
    The linear increase in the warp bubble's curvature suggests an engineering constraint on a maximum  achievable global velocity.
    The wake's shape also shows that there is internal structure to the warp bubble. 
    Finally, it can be seen that far in front and behind the warp bubble that the invariants are zero; thus, the space is asymptotically flat.
    
\begin{figure}[htb]
	\begin{subfigure}[t]{.45\linewidth}
	    \centering
		\includegraphics[scale=0.29]{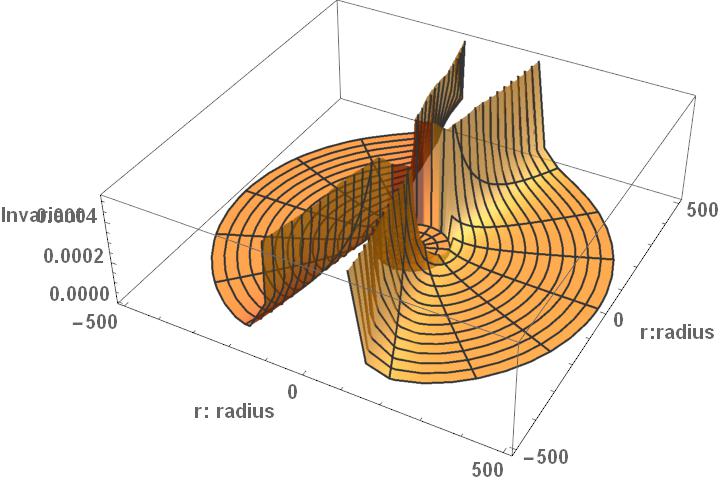}
		\caption{$r_1$ and $t=0.0$~s}
		\label{fig:5.3a}
	\end{subfigure}
	~
	\begin{subfigure}[t]{.55\linewidth}
	    \centering
		\includegraphics[scale=0.29]{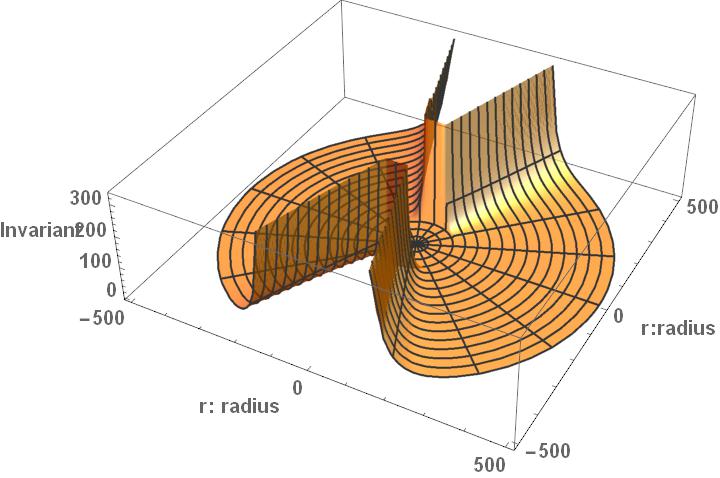}
		\caption{$r_1$ and $t=1.0$~s}
		\label{fig:5.3b}
	\end{subfigure}
	~
	\begin{subfigure}[t]{.45\linewidth}
	    \centering
		\includegraphics[scale=0.29]{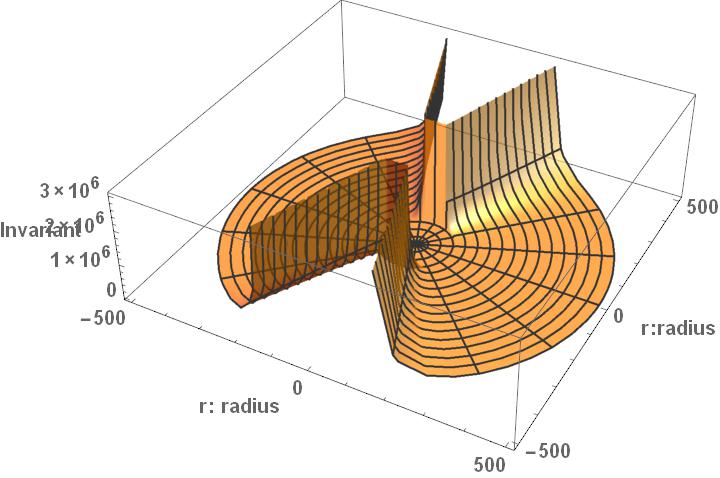}
		\caption{$r_1$ and $t=10.0$~s}
		\label{fig:5.3c}
	\end{subfigure}
	~
	\begin{subfigure}[t]{.55\linewidth}
	    \centering
		\includegraphics[scale=0.29]{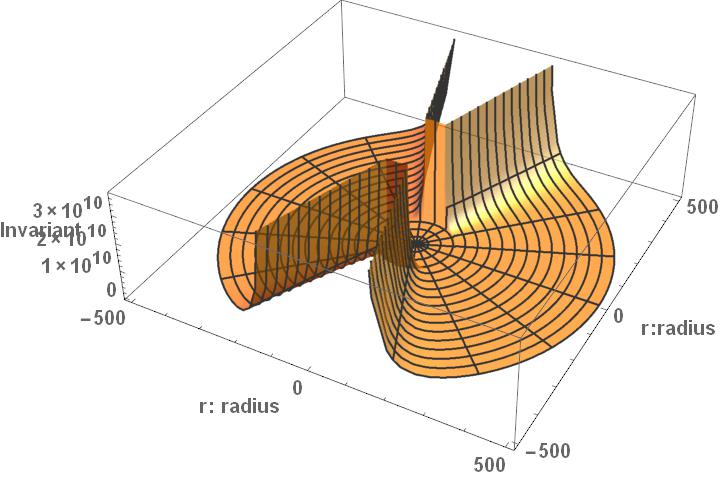}
		\caption{$r_1$ and $t=100.0$~s}
		\label{fig:5.3d}
	\end{subfigure}
	\caption{The time evolution of the invariant $r_1$.
	The other variables were chosen to be $a=1~\frac{\textrm{m}}{\textrm{s}^2}$, $\sigma$ = 50,000~$\frac{1}{\mathrm{m}}$, and $\rho=100$~m in natural units.} \label{fig:5.3}
    \end{figure}

    Choosing the same values for $\rho$, $\sigma$, and $a$ as the Ricci scalar and varying the time, Fig.~\ref{fig:5.3} shows the time evolution for the $r_1$ invariant. 
    It has many similar features to the Ricci scalar. 
    It contains the safe harbor, a wake running perpendicular to the direction of motion, is asymptotically flat in front and behind the bubble, and increases approximately linearly with time. 
    The first apparent difference is the positive magnitude and lack of internal structure in the wakes. 
    Subtly, the wake increases in angular size as time increases.
    
    The invariant $r_2$ shares the same basic properties of the Ricci scalar and $r_1$. 
    It contains the safe harbor, a wake running perpendicular to the direction of motion, is asymptotically flat in front and behind the bubble, and increases approximately linearly with time. 
    It is similar in shape to $r_1$ and has the same internal structure as the Ricci scalar, but it increases in magnitude more drastically.
    
    The invariant $w_2$ shares the same basic properties of the Ricci scalar, $r_1$, and $r_2$. It contains the safe harbor, a wake running perpendicular to the direction of motion, is asymptotically flat in front and behind the bubble, and increases approximately linearly with time. 
    The invariant's wake contains a small amount of internal structure at lower time values, but as time reaches 100~s, the internal structure begins to form crenellations. 
    As time continues to evolve, the crenellations travel out parallel to the length of the wake from the center of the warp bubble. 
    It can be speculated that this would cause an erratic flight path of the bubble since the crenellations are not symmetric. 
    It can also be speculated that the Ricci scalar and $r_2$ would exhibit crenellations at higher time values.
    
    The wake along the $\theta=(\frac{\pi}{2},\frac{3\pi}{2})$ line appears to propagate to infinity instantaneously.
    A distant Eulerian observer along that axis should be able to detect it the split second the warp drive accelerates.
    Due to the size of the invariant functions, the wake would also impact the observer such as with a gravitational wave.
    Consequently, the wake would causally effect an observer at infinity and violate the laws of causality.
    Moreover, the location of the wake matches the singularity identified in the invariant functions; thus, it must be an intrinsic curvature singularity.
    If the wake is an inherent feature of any accelerating warp drive spacetime and not due to the specific choices in the design of the warp drive such as the shape function, then the wake would be evidence that an accelerating warp drive is not possible in our universe.
    The wake and the identified intrinsic curvature singularity is under continued investigation for its effect on accelerating warp drive spacetimes.

    %New Figure
    
    %New Figure%
    
    \begin{figure}[htb]
	\begin{subfigure}[t]{.45\linewidth}
	    \centering
		\includegraphics[scale=0.29]{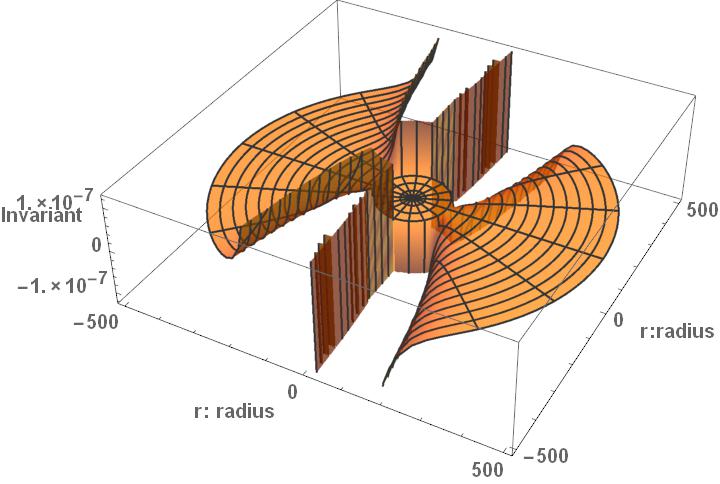}
		\caption{$r_2$ and $t=0.0$~s}
		\label{fig:5.4a}
	\end{subfigure}
	~
	\begin{subfigure}[t]{.55\linewidth}
	    \centering
		\includegraphics[scale=0.29]{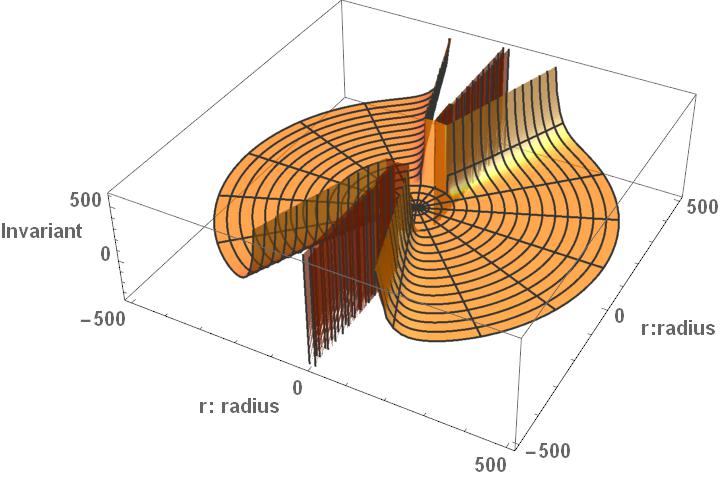}
		\caption{$r_2$ and $t=1.0$~s}
		\label{fig:5.4b}
	\end{subfigure}
	~
	\begin{subfigure}[t]{.45\linewidth}
	    \centering
		\includegraphics[scale=0.29]{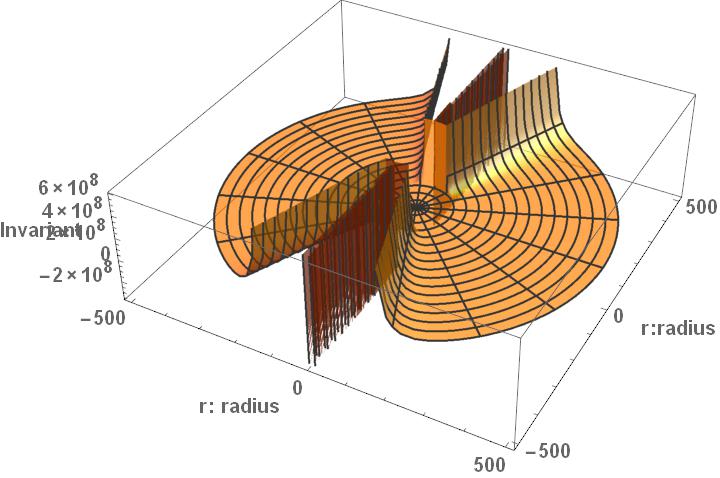}
		\caption{$r_2$ and $t=10.0$~s}
		\label{fig:5.4c}
	\end{subfigure}
	~
	\begin{subfigure}[t]{.55\linewidth}
	    \centering
		\includegraphics[scale=0.29]{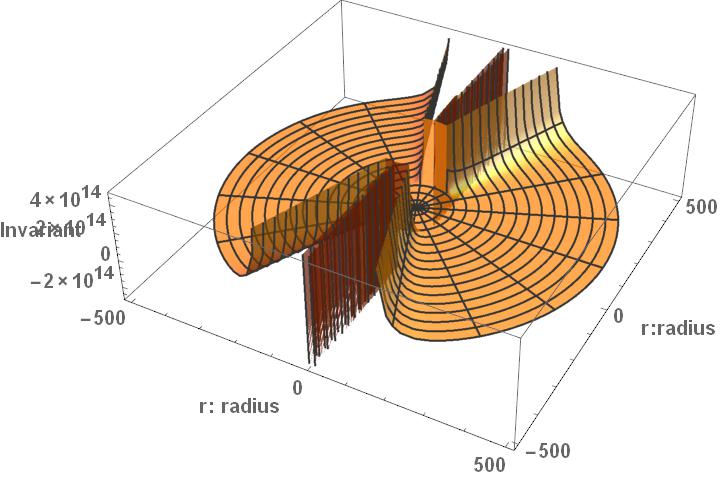}
		\caption{$r_2$ and $t=100.0$~s}
		\label{fig:5.4d}
	\end{subfigure}
	\caption{The time evolution of the invariant $r_2$.
	The other variables were chosen to be $a=1~\frac{\textrm{m}}{\textrm{s}^2}$, $\sigma$ = 50,000~$\frac{1}{\mathrm{m}}$, and $\rho=100$~m in natural units.} \label{fig:5.4}
    \end{figure}
    
    %New Figure%
    
    \begin{figure}[htb]
    	\begin{subfigure}[t]{.45\linewidth}
    	    \centering
    		\includegraphics[scale=0.29]{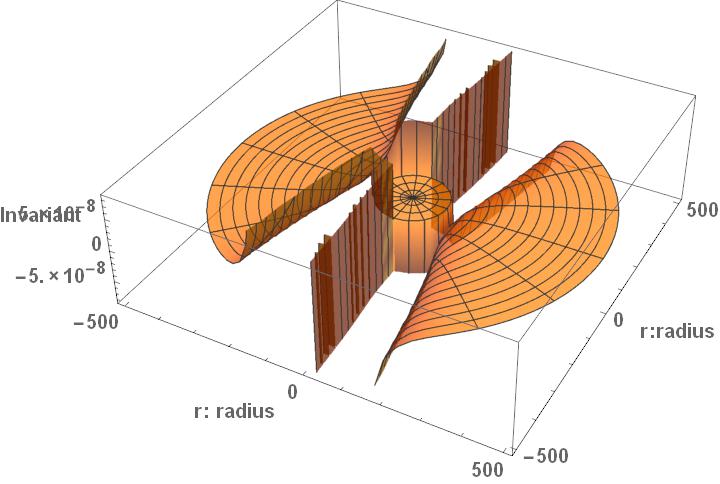}
    		\caption{$w_2$ and $t=0.0$~s}
    		\label{fig:5.5a}
    	\end{subfigure}
    	~
    	\begin{subfigure}[t]{.55\linewidth}
    	    \centering
    		\includegraphics[scale=0.29]{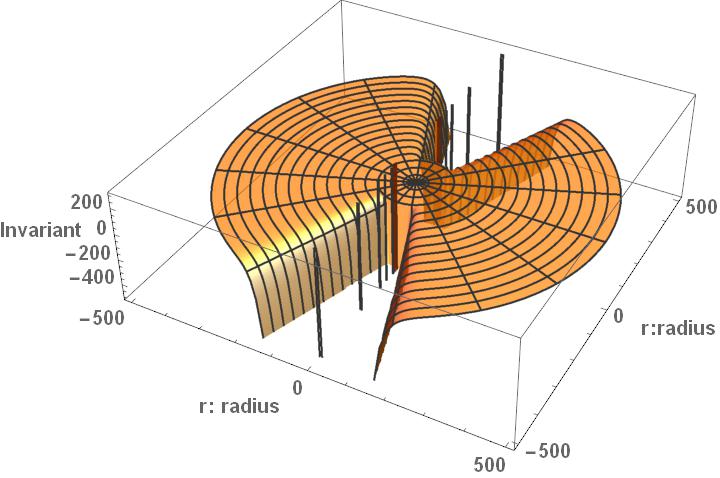}
    		\caption{$w_2$ and $t=1.0$~s}
    		\label{fig:5.5b}
    	\end{subfigure}
    	~
    	\begin{subfigure}[t]{.45\linewidth}
    	    \centering
    		\includegraphics[scale=0.29]{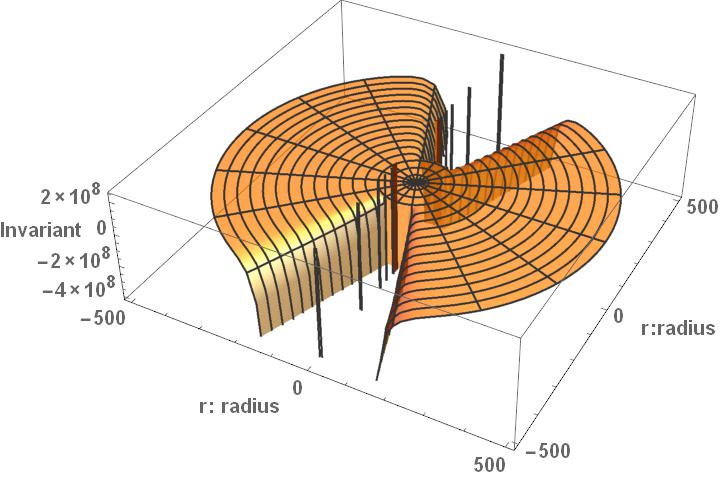}
    		\caption{$w_2$ and $t=10.0$~s}
    		\label{fig:5.5c}
    	\end{subfigure}
    	~
    	\begin{subfigure}[t]{.55\linewidth}
    	    \centering
    		\includegraphics[scale=0.29]{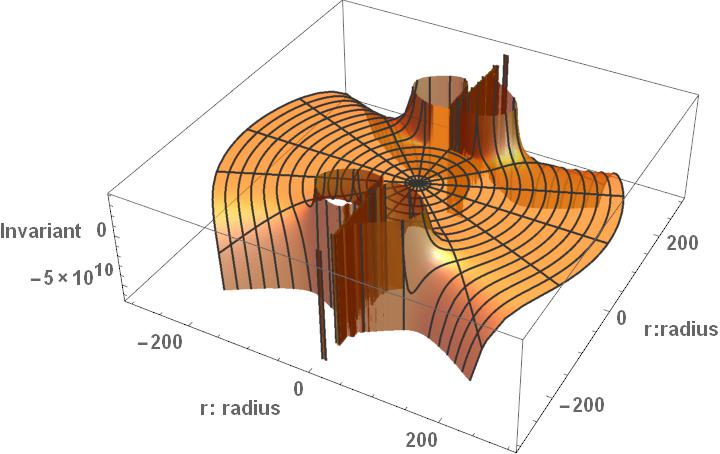}
    		\caption{$w_2$ and $t=100.0$~s}
    		\label{fig:5.5d}
    	\end{subfigure}
    	~
    	\begin{subfigure}[t]{.45\linewidth}
    	    \centering
    		\includegraphics[scale=0.29]{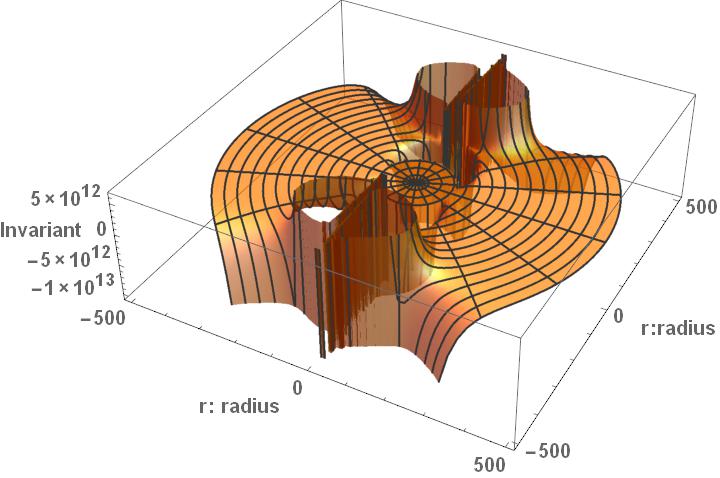}
    		\caption{$w_2$ and $t=200.0$~s}
    		\label{fig:5.5e}
    	\end{subfigure}
    	~
    	\begin{subfigure}[t]{.55\linewidth}
    	    \centering
    		\includegraphics[scale=0.29]{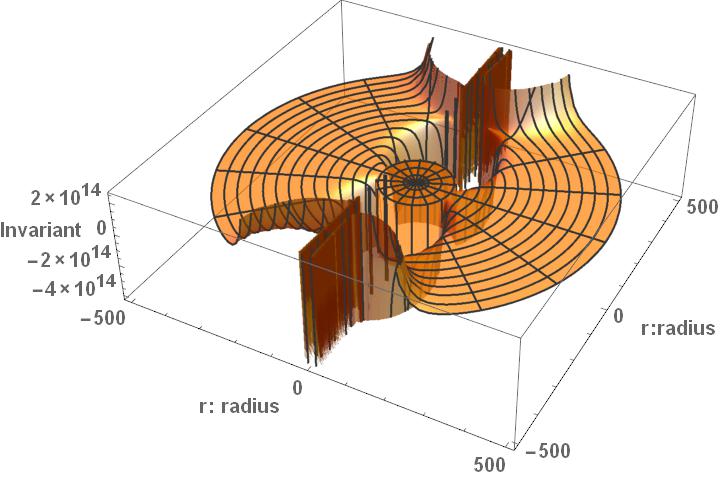}
    		\caption{$w_2$ and $t=300.0$~s}
    		\label{fig:5.5f}
    	\end{subfigure}
        \caption{The time evolution for the invariant $w_2$.
        The other variables were chosen to be $a=1~\frac{\textrm{m}}{\textrm{s}^2}$, $\sigma$ = 50,000~$\frac{1}{\mathrm{m}}$, and $\rho=100$~m in natural units.} \label{fig:5.5}
    \end{figure}
    
    \FloatBarrier

\section{Invariant Plots of Acceleration for Nat\'ario}
\label{chp5:accel}
\begin{figure}[ht]
	\centering
    	\begin{subfigure}[t]{.47\linewidth}
    	    \centering
    		\includegraphics[scale=0.29]{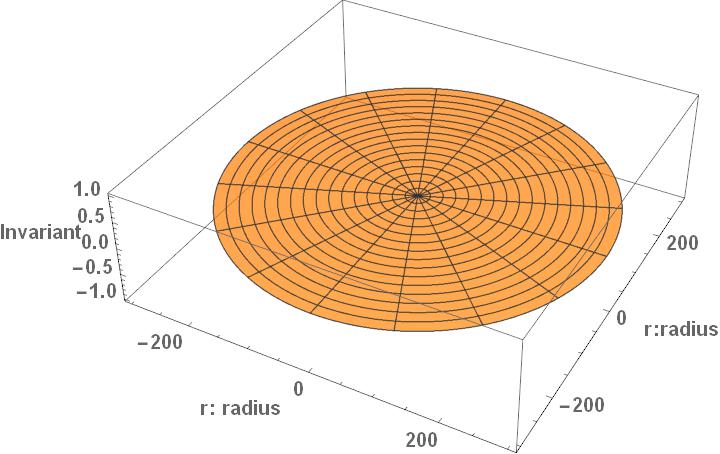}
    		\caption{$a=0.0~\frac{\mathrm{m}}{\mathrm{s}^2}$}
    		\label{fig:5.6a}
    	\end{subfigure}
	~
    	\begin{subfigure}[t]{.47\linewidth}
    	    \centering
    		\includegraphics[scale=0.29]{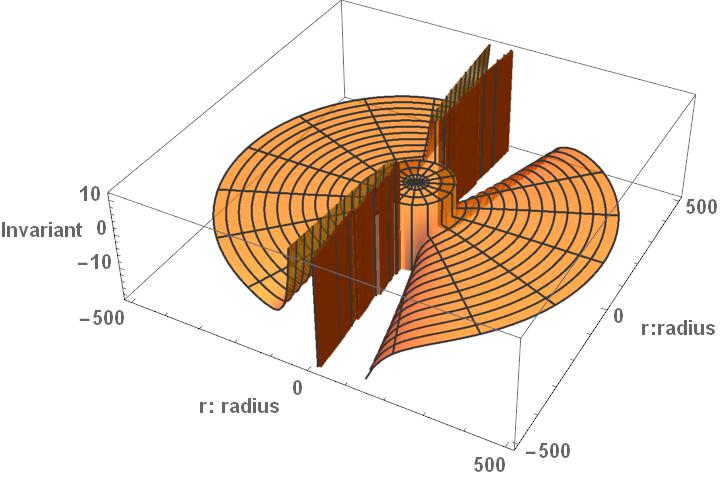}
    		\caption{$a=0.1~\frac{\mathrm{m}}{\mathrm{s}^2}$}
    		\label{fig:5.6b}
    	\end{subfigure}
	~
    	\begin{subfigure}[t]{.47\linewidth}
    	    \centering
    		\includegraphics[scale=0.29]{Images/Chapter5/R/Rs50000p100a1t1.jpg}
    		\caption{$a=1.0~\frac{\mathrm{m}}{\mathrm{s}^2}$}
    		\label{fig:5.6c}
    	\end{subfigure}
	~
    	\begin{subfigure}[t]{.47\linewidth}
    	    \centering
    		\includegraphics[scale=0.29]{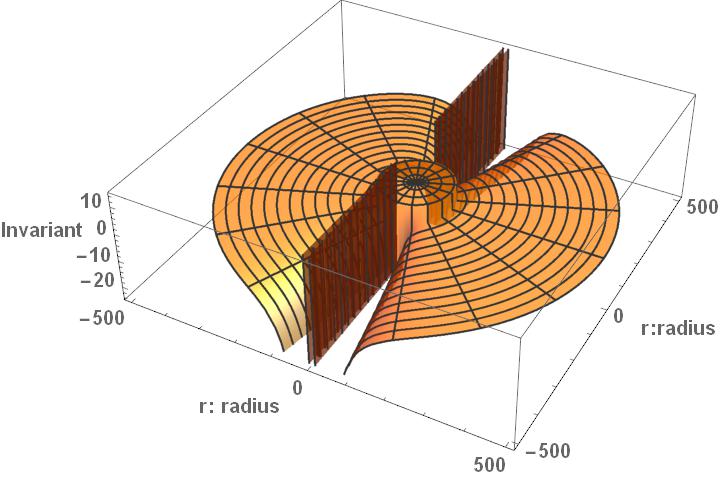}
    		\caption{$a=10.0~\frac{\mathrm{m}}{\mathrm{s}^2}$}
    		\label{fig:5.6d}
    	\end{subfigure}
	\caption{Varying the acceleration of $R$, the Ricci scalar.
	The other variables were chosen to be $a=1~\frac{\textrm{m}}{\textrm{s}^2}$, $\sigma$ = 50,000~$\frac{1}{\mathrm{m}}$, and $\rho=100$~m in natural units.} \label{fig:5.6}
    \end{figure}

    Varying the acceleration of the invariants speeds up or slows down the time evolution of the previous section. 
    Setting $\rho=100$~m, $\sigma$~= 50,000~$\frac{1}{\mathrm{m}}$, and $t=1.0$~s, Fig.~\ref{fig:5.6} shows the acceleration's variation for the Ricci scalar.
    The first plot of $a=0$ is consistent with the lapse function being zero for all time. 
    The distance between hypersurfaces remains constant. 
    The space is flat and no warp bubble forms. 
    The second plot corresponds to a time slice between Figs.~\ref{fig:5.2a} and \ref{fig:5.2b}. 
    Similarly, the third plot is identical to Fig.~\ref{fig:5.2b}, and the fourth plot is also identical to Fig.~\ref{fig:5.2c}. 
    It can be concluded that modifying the acceleration parameter corresponds with modifying the rate of change of the hypersurfaces.
    The analysis in the previous section holds for this case as well.
    
\begin{figure}[ht]
    	\begin{subfigure}[t]{1.0\linewidth}
    	    \centering
    		\includegraphics[scale=0.33]{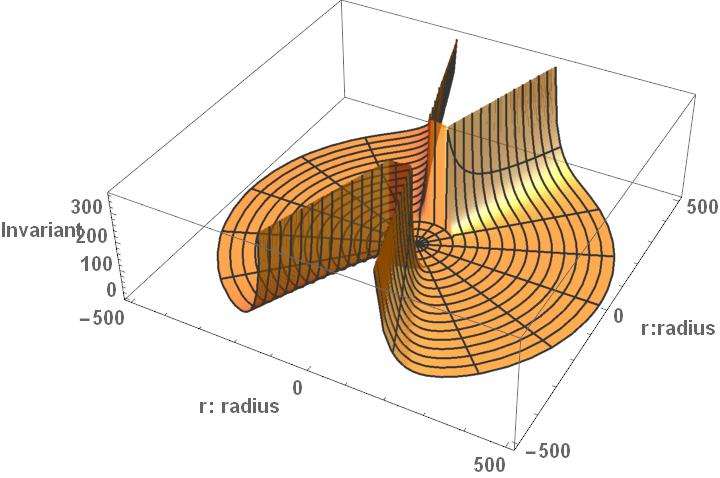}
    		\caption{$r_1$ and $a=0.1~\frac{\mathrm{m}}{\mathrm{s}^2}$}
    		\label{fig:5.7a}
    	\end{subfigure}
	~
    	\begin{subfigure}[t]{1.0\linewidth}
    	    \centering
    		\includegraphics[scale=0.33]{Images/Chapter5/r1/r1s50000p100a1t1.jpg}
    		\caption{$r_1$ and $a=1.0~\frac{\mathrm{m}}{\mathrm{s}^2}$}
    		\label{fig:5.7b} 
    	\end{subfigure} \\
	~
    	\begin{subfigure}[t]{1.0\linewidth}
    	    \centering
    		\includegraphics[scale=0.33]{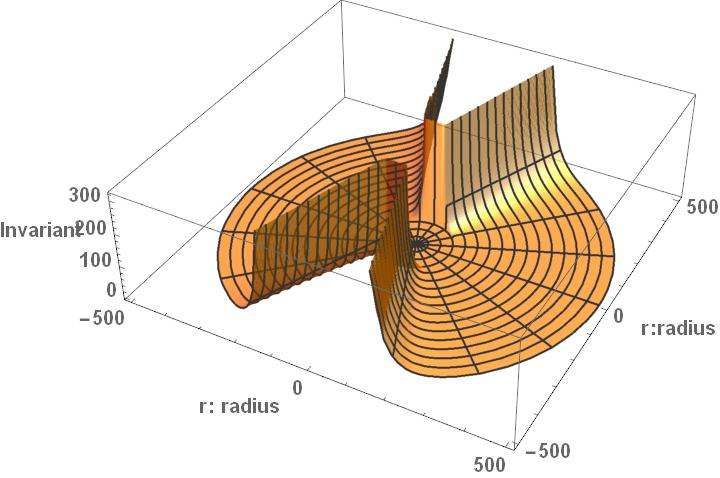}
    		\caption{$r_1$ and $a=10.0~\frac{\mathrm{m}}{\mathrm{s}^2}$}
    		\label{fig:5.7c}
    	\end{subfigure}
	\caption{Varying acceleration for the invariant $r_1$.
	The other variables were chosen to be $a=1~\frac{\textrm{m}}{\textrm{s}^2}$, $\sigma$ = 50,000~$\frac{1}{\mathrm{m}}$, and $\rho=100$~m in natural units.} \label{fig:5.7}
    \end{figure}

\begin{figure}[ht]
    	\begin{subfigure}{1.0\linewidth}
    	    \centering
    		\includegraphics[scale=0.33]{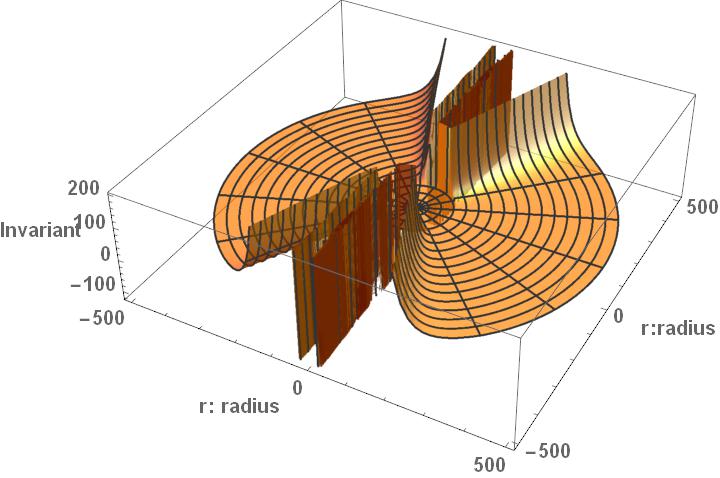}
    		\caption{$r_2$ and $a=0.1~\frac{\mathrm{m}}{\mathrm{s}^2}$}
    		\label{fig:5.8a}
    	\end{subfigure} \\
	~
    	\begin{subfigure}{1.0\linewidth}
    	    \centering
    		\includegraphics[scale=0.33]{Images/Chapter5/r2/r2s50000p100a1t1.jpg}
    		\caption{$r_2$ and $a=1.0~\frac{\mathrm{m}}{\mathrm{s}^2}$}
    		\label{fig:5.8b}
    	\end{subfigure} \\
	~
    	\begin{subfigure}{1.0\linewidth}
    	    \centering
    		\includegraphics[scale=0.33]{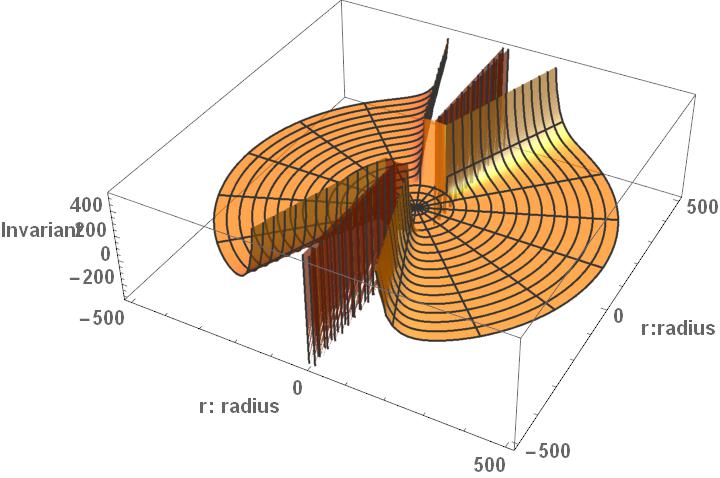}
    		\caption{$r_2$ and $a=10.0~\frac{\mathrm{m}}{\mathrm{s}^2}$}
    		\label{fig:5.8c}
    	\end{subfigure}
    \caption{Varying acceleration for the invariant $r_2$.
    The other variables were chosen to be $a=1~\frac{\textrm{m}}{\textrm{s}^2}$, $\sigma$ = 50,000~$\frac{1}{\mathrm{m}}$, and $\rho=100$~m in natural units.} \label{fig:5.8}
    \end{figure}

    The plots of the invariants $r_1$, $r_2$ and $w_2$ follow a similar process to the Ricci scalar. 
    They are plotted in Figs.~\ref{fig:5.7} and \ref{fig:5.8}. When $a=0$, their plots are identical to Fig.~\ref{fig:5.6a}. 
    Their next plots can be seen as additional time slices between the plots shown in Figs.~\ref{fig:5.3a} and \ref{fig:5.3d} for $r_1$ and Figs.~\ref{fig:5.4a} and \ref{fig:5.5d} for $r_2$. 
    Some additional features are present in the plots. 
    The curvature invariants warp much more significantly and non-symmetrically than their counterparts for the Ricci scalar, as can be seen in Fig.~\ref{fig:5.8c}. 
    The nonzero invariant inside the spaceship found in the previous time slices is also present, but due to the low magnitude of the $r_1$, $r_2$, and $w_2$ invariants, this can be safely ignored.
    
    %New Figure%
    
    %New Figure%
    
    %New Figure %
    
    \begin{figure}[ht]
    	\begin{subfigure}{1.0\linewidth}
    	    \centering
    		\includegraphics[scale=0.33]{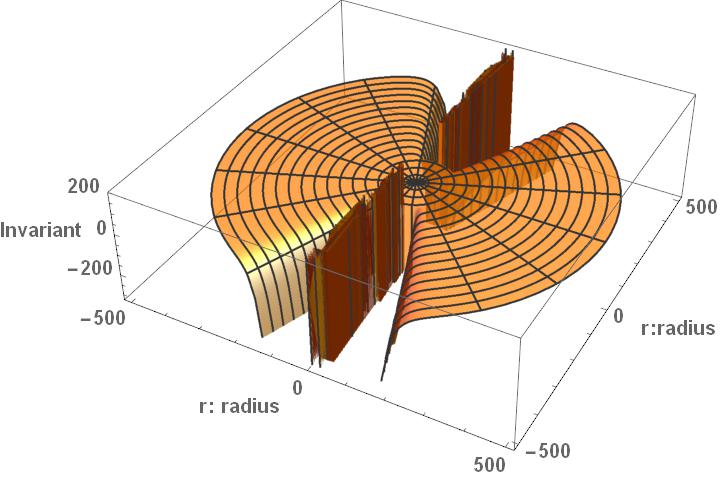}
    		\caption{$w_2$ and $a=0.1~\frac{\mathrm{m}}{\mathrm{s}^2}$}
    		\label{fig:5.9a}
    	\end{subfigure} \\
	~
    	\begin{subfigure}{1.0\linewidth}		            
    	    \centering
    	    \includegraphics[scale=0.33]{Images/Chapter5/w2/w2s50000p100a1t1.jpg}
    		\caption{$w_2$ and $a=1.0~\frac{\mathrm{m}}{\mathrm{s}^2}$}
    		\label{fig:5.9b}
    	\end{subfigure} \\
	~
    	\begin{subfigure}{1.0\linewidth}
    	    \centering
    		\includegraphics[scale=0.33]{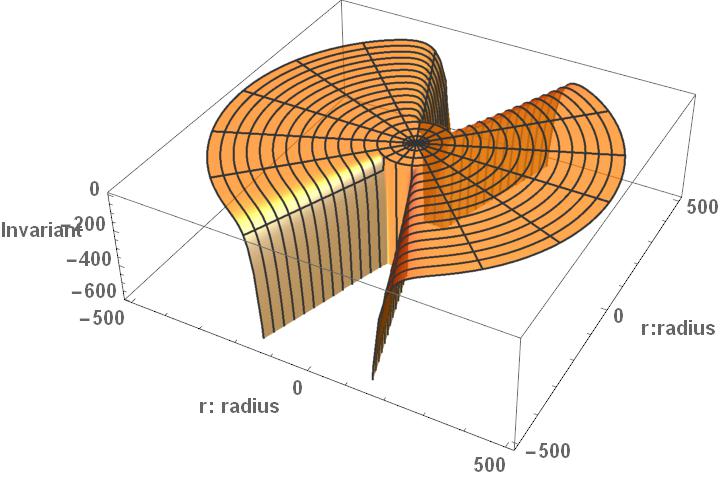}
    		\caption{$w_2$ and $a=10.0~\frac{\mathrm{m}}{\mathrm{s}^2}$}
    		\label{fig:5.9c}
    	\end{subfigure}
        \caption{Varying acceleration for the invariant $w_2$.
        The other variables were chosen to be $a=1~\frac{\textrm{m}}{\textrm{s}^2}$, $\sigma$ = 50,000~$\frac{1}{\mathrm{m}}$, and $\rho=100$~m in natural units.}
    \end{figure}
    
    \FloatBarrier

\section{Invariant Plots of Skin Depth for Nat\'ario}
\label{chp5:skin}
\begin{figure}[hb]
    	\begin{subfigure}{.45\linewidth}
    	    \centering
    		\includegraphics[scale=0.3]{Images/Chapter5/R/Rs50000p100a1t1.jpg}
    		\caption{The invariant $R$ with $\sigma$~= 50,000~$\frac{1}{\textrm{m}}$}
    		\label{fig:5.10a}
    	\end{subfigure}
	~
    	\begin{subfigure}{.55\linewidth}
        	\centering
    		\includegraphics[scale=0.3]{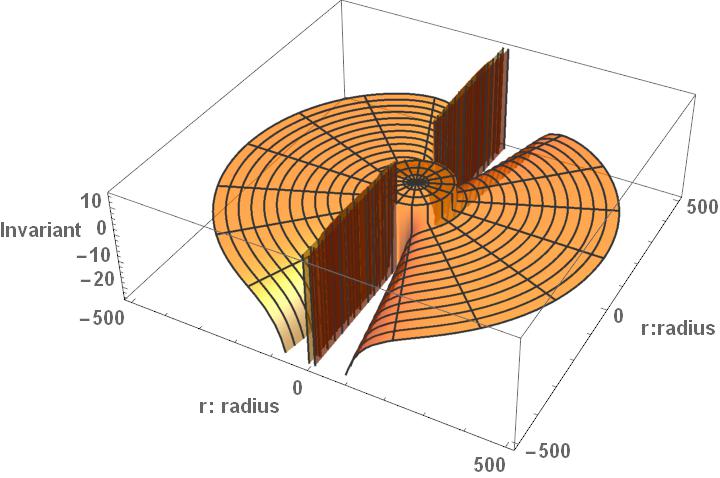}
    		\caption{The invariant $R$ with $\sigma$~= 100,000~$\frac{1}{\textrm{m}}$}
    		\label{fig:5.10b}
    	\end{subfigure}
	~
    	\begin{subfigure}{.45\linewidth}
    		\includegraphics[scale=0.3]{Images/Chapter5/r1/r1s50000p100a1t1.jpg}
    		\caption{The invariant $r_1$ with $\sigma$~= 50,000~$\frac{1}{\textrm{m}}$}
    		\label{fig:5.10c}
    	\end{subfigure}
	~
    	\begin{subfigure}{.55\linewidth}
    	    \centering
    		\includegraphics[scale=0.3]{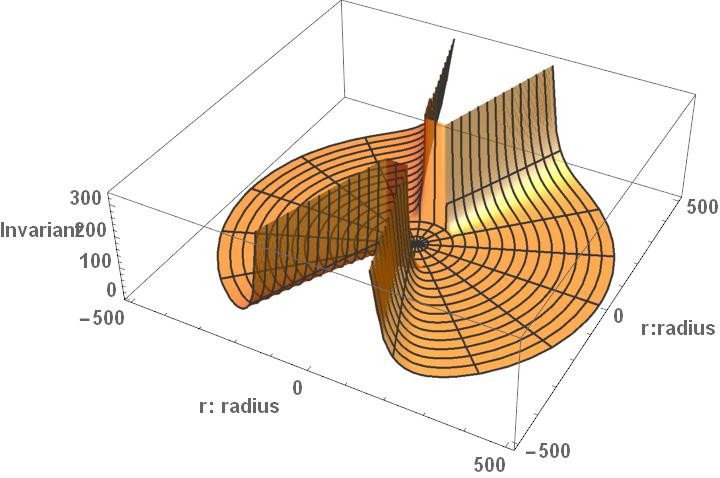}
    		\caption{The invariant $r_1$ with $\sigma$~= 100,000~$\frac{1}{\textrm{m}}$}
    		\label{fig:5.10d}
    	\end{subfigure}
    	\caption{The warp bubble skin depth for the Ricci scalar and $r_1$.
    	The other variables were chosen to be $t=1$~s, $a=1~\frac{\mathrm{m}}{\mathrm{s}^2}$, and $\rho=100$~m in natural units.} \label{fig:5.10}
    \end{figure}

\begin{figure}[ht]
    	\begin{subfigure}{.45\linewidth}
    	    \centering
    		\includegraphics[scale=0.3]{Images/Chapter5/r2/r2s50000p100a1t1.jpg}
    		\caption{The invariant $r_2$ with $\sigma$~= 50,000~$\frac{1}{\textrm{m}}$}
    		\label{fig:5.10e}
    	\end{subfigure}
	~
    	\begin{subfigure}{.55\linewidth}
    	    \centering
    		\includegraphics[scale=0.3]{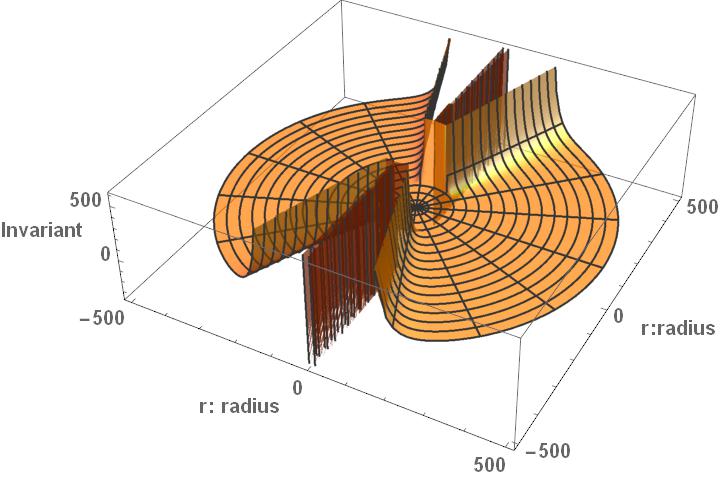}
    		\caption{The invariant $r_2$ with $\sigma$~= 100,000~$\frac{1}{\textrm{m}}$}
    		\label{fig:5.10f}
    	\end{subfigure}
	~
    	\begin{subfigure}{.45\linewidth}
    		\includegraphics[scale=0.3]{Images/Chapter5/w2/w2s50000p100a1t1.jpg}
    		\caption{The invariant $w_2$ with $\sigma$~= 50,000~$\frac{1}{\textrm{m}}$}
    		\label{fig:5.10g}
    	\end{subfigure}
	~
    	\begin{subfigure}{0.55\linewidth}
    	    \centering
    		\includegraphics[scale=0.3]{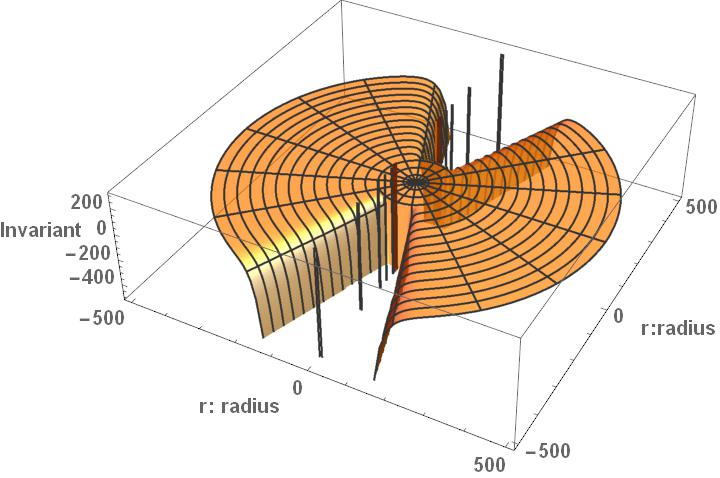}
    		\caption{The invariant $w_2$ with $\sigma$~= 100,000~$\frac{1}{\textrm{m}}$}
    		\label{fig:5.10h}
    	\end{subfigure}
    	\caption{The warp bubble skin depth for $r_2$ and $w_2$.
    	The other variables were chosen to be $t=1$~s, $a=1~\frac{\mathrm{m}}{\mathrm{s}^2}$, and $\rho=100$~m in natural units.} \label{fig:5.11}
    \end{figure}

    Varying the skin depth of the warp bubble, $\sigma$, does not noticeably affect the invariant plots reaching the same conclusion as Section \ref{chp4.2:skin}.  
    Figures \ref{fig:5.10} and \ref{fig:5.11} present the plots for each of the four invariants while doubling the skin depth and setting $\sigma$~= 100,000~$\frac{1}{\textrm{m}}$, $t=1.0$~s, and $a=1.0~\frac{\mathrm{m}}{\mathrm{s}^2}$. 
    The shape of the invariant does not change in the figures after the doubling.
    It can be speculated that either the invariants are independent of skin depth or that the impact of skin depth is minimal compared to the other variables. \\
    
	~

    \FloatBarrier
    
\section{Invariant Plots of Radius for Nat\'ario}
\label{chp5:radius}
\begin{figure}[ht]
	\begin{subfigure}{.45\linewidth}
	    \centering
		\includegraphics[scale=0.3]{Images/Chapter5/R/Rs50000p100a1t1.jpg}
		\caption{Ricci scalar with $\rho=100$~m}
		\label{fig:5.11a}
	\end{subfigure}
	~
	\begin{subfigure}{.55\linewidth}
    	\centering
		\includegraphics[scale=0.3]{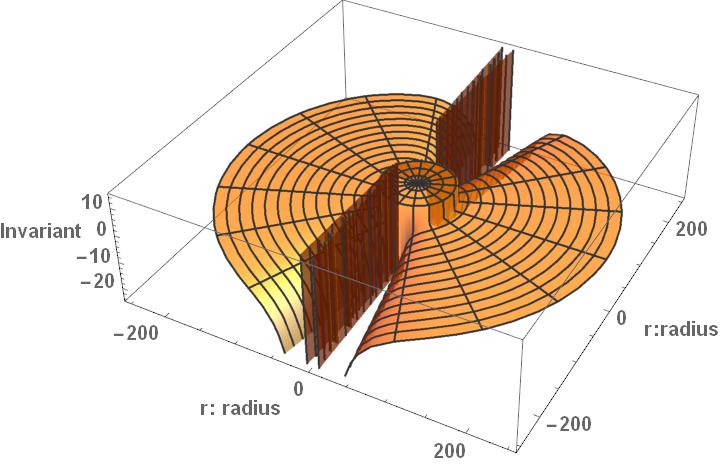}
		\caption{Ricci scalar with $\rho=200$~m}
		\label{fig:5.11b}
	\end{subfigure}
	~
	\begin{subfigure}{.45\linewidth}
		\includegraphics[scale=0.3]{Images/Chapter5/r1/r1s50000p100a1t1.jpg}
		\caption{The invariant $r_1$ with $\rho=100$~m}
		\label{fig:5.11c}
	\end{subfigure}
	~
	\begin{subfigure}{.55\linewidth}
	    \centering
		\includegraphics[scale=0.3]{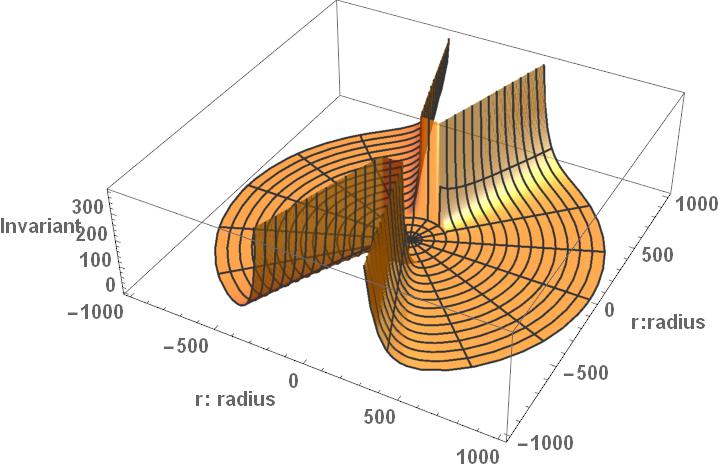}
		\caption{The invariant $r_1$ with $\rho=200$~m}
		\label{fig:5.11d}
	\end{subfigure}
	\caption{The warp bubble radius for the Ricci scalar and $r_1$.
    	The other variables were chosen to be $t=1$~s, $a=1~\frac{\mathrm{m}}{\mathrm{s}^2}$, and $\sigma$ = 50,000~$\frac{1}{\mathrm{m}}$ in natural units.} \label{fig:5.12}
	\end{figure}

\begin{figure}[ht]
	\begin{subfigure}{.45\linewidth}
	    \centering
		\includegraphics[scale=0.3]{Images/Chapter5/r2/r2s50000p100a1t1.jpg}
		\caption{The invariant $r_2$ with $\rho=100$~m}
		\label{fig:5.11e}
	\end{subfigure}
	~
    	\begin{subfigure}{.55\linewidth}
    	    \centering
    		\includegraphics[scale=0.3]{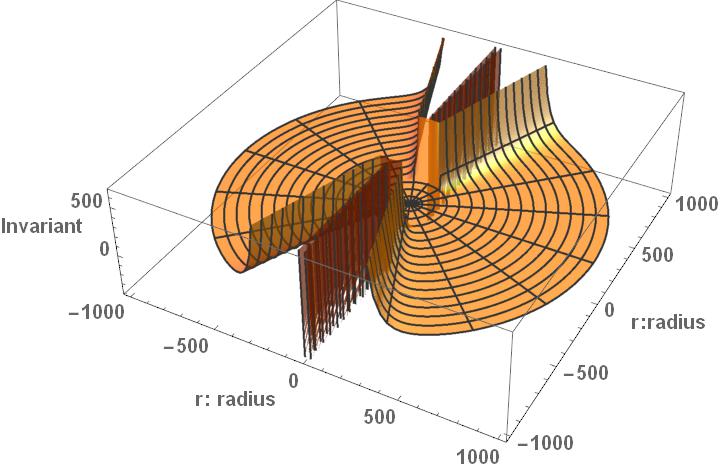}
    		\caption{The invariant $r_2$ with $\rho=200$~m}
    		\label{fig:5.11f}
    	\end{subfigure}
	~
    	\begin{subfigure}{.45\linewidth}
    		\includegraphics[scale=0.3]{Images/Chapter5/w2/w2s50000p100a1t1.jpg}
    		\caption{The invariant $w_2$ with $\rho=100$~m}
    		\label{fig:5.11g}
    	\end{subfigure}
	~
    	\begin{subfigure}{0.55\linewidth}
    	    \centering
    		\includegraphics[scale=0.3]{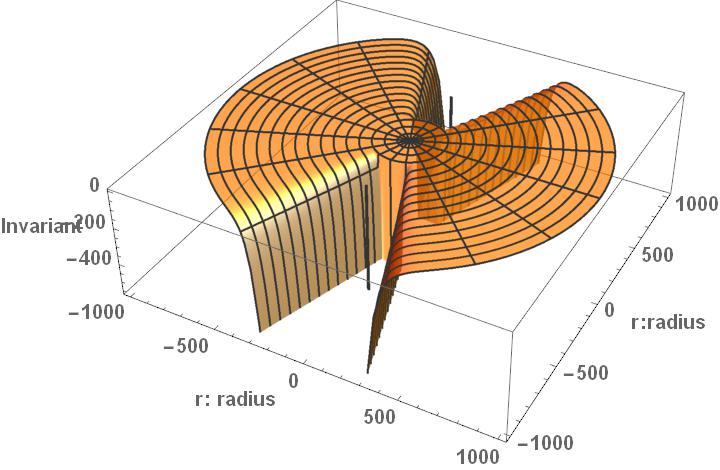}
    		\caption{The invariant $w_2$ with $\rho=200$~m}
    		\label{fig:5.11h}
    	\end{subfigure}
    	\caption{The warp bubble radius for $r_2$ and $w_2$.
    	The other variables were chosen to be $t=1$~s, $a=1~\frac{\mathrm{m}}{\mathrm{s}^2}$, and $\sigma$ = 50,000~$\frac{1}{\mathrm{m}}$ in natural units.} \label{fig:5.13}
    \end{figure}

    Varying the radius of the warp bubble, $\rho$, increases the size of the safe harbor inside the invariants, reaching the same conclusion as Section \ref{chp4.2:radius}.  
    Setting $\rho=100$~m, $t=1.0$~s, and $a=1.0~\frac{\mathrm{m}}{\mathrm{s}^2}$, the following plots for the invariants result in Figs.~\ref{fig:5.12} and \ref{fig:5.13}. 
    In the figure, the radial coordinate clearly doubles in each invariant without affecting the shape of the plots. 
    The safe harbor of $\rho\leq100$ in the left hand column also doubles in size to $\rho\leq200$. 
    The only other pertinent feature is in the internal structure of $w_2$. 
    The structures are reduced implying that they cluster near the center.
    
	~
    
    \FloatBarrier

%\lipsum[2-3]

%% file: ch6.tex
\chapter{Conclusion}
\label{Chapter6}
%In this conclusion, I will talk about the main impacts of my research on our research group.
The spacetime elements considered in this research can at best be described as speculative and at worst as physical impossibilities.
No experimental evidence exists of either a warp drive or a wormhole.
It is well known that wormholes and warp drives violate the null energy conditions (NEC) and require exotic matter to stabilize \cite{Lobo:2017,Visser:1995}.
Moreover, it appears excessively difficult to engineer a local spacetime in a way that a human could use for travel.
Wormholes either appear at galactic scales as a possibility lying beyond a black hole's event horizon or as a solution to how spacetime warps on the Planck scale. 
Dark energy seems to be the only possibility for constructing the Spacetime expansion and contraction required to propel a warp drive.
Dark energy acts on Hubble volume scales and its energy density is on the scale of $7\times10^{-30} \frac{g}{cm^3}$.
Exponentially increasing its energy density to propel a spacecraft capable of carrying a human being is an engineering impossibility for the foreseeable future.
The most dangerous potential consequence is how easy it is to create a CTC by adjusting the relative velocities of the wormhole throats or the warp-drive bubbles.
CTCs create time machines, which implies that both spacetimes violate causality.
Causality is fundamental to our understanding of physics, so both types of spacetime should be prohibited from existing in the known and possible universe by a short proof of contradiction.

Then, the important question is why should these spacetimes be researched at all.
First, all spacetimes considered herein are solutions to Einstein's equations.
They appear to be fully possible inside the known frameworks or semi-classical approximations of GR.
If causality prohibits these spacetimes, then how and why they cannot exist needs to be understood.
So by researching them, additional constraints on possible spacetimes inside GR can be deduced.
Second, simple extensions of GR containing QM and even classical scalar field theories show physical violations of the NECs \cite{Barcelo:2002}.
The energy conditions in GR provide a precise definition of local energy densities.
They lay the basis for the amount a light cone may be tipped over as discussed in Section \ref{chp1:TLC}.
Probing mathematically realistic but physically unlikely theories of wormholes and warp drives will help our understanding of the breadth of NEC violations.
Finally, assume for a moment that wormholes and warp drives are both mathematically possible and physically realistic.
Then, methods and techniques must be developed to identify them when they occur in nature.
For example, the jump discontinuity found in Fig.~\ref{fig:2.1.2} can lead to a redshift that would distinguish a wormhole from a black hole.
Second, the wakes observed for each warp drive spacetime should generate gravitational waves.
Potentially, these waves could be detected by significantly more sensitive LIGO-like experiments.
If either of these were detected, this would provide significant evidence for their physical existence.

% Next paragraph transitions from these general reasons to study the spacetime metrics to the specific use of invariants.
The research contained in this dissertation and in the program in Appendix \ref{ap:Program} forms a foundation for the computation of any set of invariants.
The program computes everything needed to find any SP invariant from a given line element.
It computes each tensor and all NP indices in Section \ref{chp2:GR}.
While not discussed in this dissertation, the program may also solve the Einstein equations and obtain other tensors such as $E_{ijkl}$.
The real power of the program is its modality.
It can be easily adapted for any given set of invariants in addition to the CM invariants discussed herein.
In the next section, how the program may be adapted to other sets of invariants or GR calculations will be presented. \\

\section{Invariant Research}
\label{chp6.1:Inv}
%In this section, I will discuss the use of my program for solving general invariant problems. \\
    The program in Appendix \ref{ap:Program} forms the basis for this research, and it can be used to find the invariants of any given basis for any given metric.
    It takes a metric and a null tetrad as inputs and outputs the needed set of invariants.
    In this research, the question was whether any curvature singularities existed in the considered spacetimes.
    If a singularity was discovered, its nature would be revealed by computing and plotting only the CM invariants for the considered spacetimes.
    However, sets of invariants can also be used in studying the Segre and Petrov type of the line element and the equivalence problem.\footnote{The question of whether two spacetime metrics are equivalent.}
    To a limited extent for the wormhole metrics, the research shows that the MT wormhole and the exponential metric may be equivalent with appropriate choices of shape function and redshift function. 
    As it stands, the research and program can be adapted to study these problems for any given set of invariants or line element.
    In this section, several of the ongoing uses of the program will be presented.
    
    Computing and plotting the invariant functions has significant advantages for the inspection of these exotic spacetimes. 
    As mentioned previously, the advantage of plotting the invariants is that they are free from coordinate mapping distortions and other artifacts of the chosen coordinates.
    The resulting invariants properly illustrate the entire underlying spacetime independent of the coordinate system chosen.
    Plotting the invariants exposes the presence of any artifacts, divergences, or discontinuities anywhere on the manifold. 
    Once the artifacts are revealed by the invariants, they can be related mathematically to the textbook tensors. 
    Any artifact's effect can then be analyzed based on where the curvature invariant locates it, what type of artifact the curvature invariant reveals it to be, and how the artifact may affect the object's motion based on the plots of the curvature invariant at that location.
    
    A second advantage is the relative ease with which the invariants can be plotted. 
    Software packages exist or can be developed to calculate the textbook tensors such as the one provided in Appendix \ref{ap:Program}. 
    Then, the CM invariants can be derived from the textbook using minimal edits to the software packages.
    The CM invariants were chosen to be computed and plotted in this paper because they had general independence, were of lowest possible degree, and a minimal independent set for any Petrov Type and for any specific choice of the Ricci tensor \cite{CM}.
    It is of research interest that other choices for the set of invariants exist, such as the Cartan invariants previously discussed, as wells as the Witten and Petrov invariants \cite{Zakhary:1997,McNutt:2018}.
    These additional sets of invariants may answer several, unconsidered questions, such as whether the underlying spacetime is unique.
    These sets may also be computed and plotted without difficulty.
    They are related to the CM invariants by polynomial functions. 
    Since the invariants are either scalars or pseudoscalars, they can be straightforwardly plotted and visually interpreted.
    
    The first extension of this research in this paper is to include the non-independent CM invariants presented in Eqs.~\eqref{eq:2.37} through \eqref{eq:2.46}.
    While the syzygies for Class B spacetimes reduced this group to the four in Eq.~\eqref{eq:2.67}, other classes of spacetimes will have different sets of syzygies and different sets of independent invariants.
    The remaining invariants may then be computed.
    The program in Appendix \ref{ap:Program} currently computes the tetrad and the tetrad components of the Ricci tensor and the Weyl tensor.
    The functional relationships with the non-independent CM invariants are the next expansion to the program's code.

    Another extension of this work is to expand the program to other sets of invariants such as the Cartan invariants.
    An alternate question to the one of completeness for a set of invariants is how to construct a set of invariants from the Riemann tensor, and/or its derivatives that will sufficiently cover a spacetime \cite{Stephani:2003}.
    Such a set may be constructed from the SP invariants; however, the SP invariants will not uniquely cover the spacetime.
    A better alternative considers the Cartan set of invariants, which provide a unique coordinate-independent characterization.
    The coordinate components of different metrics may then be compared given the coordinate components for curvature and the derivatives of the Riemann tensor. \\
    
    The Cartan invariants are the non-zero components of the Riemann tensor $R_{\alpha\beta \gamma \delta}$ and its covariant derivatives \cite{Stephani:2003,McNutt:2018}.
    Invariants that are constructed from or are equal to the Cartan invariants of any order are called extended invariants.
    The algorithm to compute the Cartan invariants in any dimension is as follows:
    \begin{enumerate}
        \item Set the order of differentiation $q$ to 0.
        \item Calculate the derivatives of the Riemann tensor up to the $q$th order.
        \item Find the canonical form of the Riemann tensor and its derivatives.
        \item Fix the frame as far as possible by this canonical form and note the residual frame freedom (the group of allowed transformations is the linear isotropy group $H_q$). The dimension $H_q$ is the dimension of the remaining vertical freedom of the frame bundle.
        \item Find the number $t_q$ of independent functions of spacetime position  in the components of the Riemann tensor and its covariant derivatives, in the canonical form. This tells the remaining horizontal freedom.
        \item If the isotropy group and number of independent functions are the same as in the previous step, let $p+1=q$, and the algorithm terminates, if they differ (or if $q=0$), increase $q$ by 1 and return to step 2.
    \end{enumerate}
    As the algorithm follows a simple set of finite steps, it can be automated and added to the computer program in Appendix \ref{ap:Program}.
    Upon completing this algorithm, sufficiently smooth metrics may be compared as a test of the equivalence.
    
    The research in this dissertation has already hinted that two of the metrics, the MT wormhole and the exponential metric, rely on the same underlying spacetime.
    A possibility exists that every type of wormhole is a specific choice of either the shape function $b(r)$ and/or the redshift function $\Phi(r)$ of the MT metric.
    Even more tantalizing is that wormholes and warp drives have the same topological structure.
    If true, then the two will be connected by a suitable choice of shape functions $b(r)$, $f(r_s)$, and/or $n(r_s)$.
    Computing the Cartan invariants may confirm or deny this hypothesis.
    Research is underway to adapt the program in this manner to construct the Cartan Invariants based on the procedure outlined in this section. 
    Its primary purpose is to answer the question of equivalence among the previous wormhole and warp drive metrics among others to be considered below. \\

\section{Wormhole Research}
\label{chp6.2:Wormhole}
    Chapter \ref{Chapter3} demonstrates how to compute and plot the curvature invariants of various wormhole line elements.
    The CM curvature invariants reveal the entire wormhole spacetime manifold and whether the wormhole is traversable or not.  
    As examples, plotting the curvature invariants of the $\left(i\right)$ spherically symmetric MT, $\left(ii\right)$ TS Schwarzschild and $\left(iii\right)$ Exponential metric wormholes showed they are traversable in agreement with \cite{Visser:1995,Boonserm,Morris:1988A,Morris:1988B}. 
    The invariants of the MT wormhole were found to be non-zero and are plotted in Figs.~\ref{RMTm}--\ref{w2MTm}.
    A divergence is found in all four.
    The divergence does not affect the wormhole's traversability as it is outside the physical range of the radial coordinate, $r\in{\left(r_0,\infty\right)}$. 
    For the TS Schwarzschild wormhole, $w_2$ is found to be the single non-zero invariant. 
    As plotted in Fig.~\ref{w2tss}, it has a divergence at the center and a ring discontinuity. 
    The divergence is outside the physical radial coordinate and can be safely ignored. 
    The ring discontinuity represents a jump due to the $\delta$-function from the TS formalism. 
    It is shown to be inversely proportional to ${a^{-14}}$; thus, it will not affect any transport through the wormhole. 
    The SP invariants of the exponential metric were found to be non-zero and were plotted in Figs.~\ref{Rem}--\ref{w2em}. 
    The plots are continuous across the entire manifold and traversable.

    Potentially, the ring discontinuity in the TS Schwarzschild wormhole may lead to a redshift of light rays that pass through the wormhole. 
    The redshift could be used to distinguish a TS Schwarzschild wormhole from the famous Schwarzschild black holes.
    The TS Schwarzschild wormhole is the most common example of a large class of wormholes. 
    The class includes wormholes with different radii of curvature, $R$, masses, $M$ and/or different charge, $Q$, on either side of their throat, and time-dependent wormholes. 
    For charged wormholes, a second ring artifact at $r=Q$ is likely to exist since the metric has a singularity at that point.
    Similar shifts could be expected also from their more complicated line elements such as the rotating traversable wormholes \cite{Teo:1998}.
    Significant research using the methods and programs in this dissertation is underway to compute the invariants of the rotating traversable wormholes, which will aid in their potential identification.
    
    One future application of studying curvature invariants of wormholes is an investigation of the rotating traversable wormhole in \cite{Teo:1998}.
    The rotating wormhole is the most general extension of the MT wormhole discussed in Section~\ref{chp3:MT}. 
    Its metric is
    \begin{equation*}
        ds^2=-N^2dt^2 +e^\mu dr^2 +r^2K^2 \left[d\theta^2+\sin^2\theta(d\phi-\omega dt)^2\right], \tag{6.1} \label{eq:6.1}
    \end{equation*}
    where $N$ is the redshift function, $\mu$ is the shape function, $K$ determines proper radial distance, and $\omega$ is the wormhole's angular velocity.
    It should be expected that the polar symmetry observed in the MT plots will be broken by the angular velocity.
    Several interesting features of rotating wormholes include specific geodesics that do not encounter exotic matter and an ergoregion surrounding the throat.
    Consequently, a hypothetical interstellar traveler would choose the rotating traversable wormhole for transport over the ones discussed previously.
    Research is ongoing into this wormhole metric.

    Another prospective future application of this work is an investigation of wormholes with throats that change dynamically over time as in \cite{Visser:1995}.
    The invariants for these wormholes should have the size of the central discontinuity change over time.
    The ring discontinuity in the invariant functions will change as a function of time as a primary consequence. 
    Hence, dynamic wormholes are more technically demanding to study as compared to static wormholes.
    Consequently, it can be expected that the computation of a dynamic wormhole's invariants and their plots increase in difficulty and computational runtime.
    The program in Appendix \ref{ap:Program} should use the NP indices to compute the invariants for this class of wormhole.
    Like rotating wormholes, dynamic wormholes are a more physically realistic class of wormholes, and they can be related to primordial black holes. \\

\section{Warp Drive Research}
\label{chp6.3:WarpDrive}
    The research in Chapters \ref{Chapter4} and \ref{Chapter5} demonstrate how computing and plotting the curvature invariants for various variables of warp drive spacetimes can reveal their underlying curvature. 
    While the individual functions are mammoth in size and may take days, weeks, or even months to calculate, their plots can be quickly scanned and understood. 
    The plots give the magnitude of curvature at each point around the ship. 
    Where the curvature invariant's magnitudes are large, space is greatly warped, and vice versa.
    Also by observing the changes in slopes on the plots, the rate at which spacetime is being folded can be analyzed.
    Using this information can help map the spacetime around the ship and aid potential navigation.

    The curvature invariant functions and plots were displayed for the Alcubierre warp drive at constant velocity, the Nat\'ario warp drive at constant velocity, and the accelerating Nat\'ario warp drive metric.
    Different choices of the free variables were varied to see the individual effect on each invariant.
    The curvature invariants reveal a safe harbor for a ship to travel inside the warp bubble and an asymptotically flat space outside the bubble in all cases.
    At the radial position $r=\rho$ of the warp bubble(s), the curvature invariants have local maxima implying that $\rho$ is the location where spacetime is warped the most.
    For the constant velocity Alcubierre warp drive, the warp bubbles resemble two troughs with simple internal structures.
    For the constant velocity Nat\'ario warp drive, the warp bubbles peak around $r=\rho$ and display rich internal structure.
    For the accelerating Nat\'ario warp drive, two wakes radiate out from the warp bubble along the $\theta=(\frac{\pi}{2},\frac{3\pi}{2})$ axis.
    The wakes have rich internal structures that ripple and bubble as time advances.
    The internal structures of the warp bubble found in this paper are novel and require more diligent research to discover their effects on the warp drive's flight.
    
    For the accelerating Nat\'ario warp drive, a variation of time shows that each invariant's plots experience a sudden jump from positive curvature in the direction of motion to negative curvature. 
    As time progresses, the shapes of the $R$, $r_1$, and $r_2$ invariants remain constant, but the magnitude of the invariants increases roughly linearly in time and the angular arc of the wake subtlety along the polar axis. 
    The $w_2$ invariant begins to exhibit crenellations in the interior of the warp bubble after $100 s$.
    By varying the acceleration, the invariant plots skip through the time slices and internal structures become more prominent.
    Changing the skin depth did not change either the shape or magnitude of the invariant plots.
    Doubling the radius did double the size of the warp bubble and safe harbor without affecting the shape of the invariants.
    The invariant plots give a rich and detailed understanding of the curvature of spacetime surrounding a warp drive.

    In addition to the research presented in this dissertation on inspecting the different invariants, further work can be done in mapping warp drive spacetimes.
    The work in Chapter \ref{Chapter5} can be further expanded by considering time slices greater than 300~s.
    Potentially, crenellations like the ones observed for the $w_2$ invariant exist within the warp bubbles for the $R$, $r_1$ and $r_2$ invariants.
    The crenellations would make controlling the direction of the warp bubble challenging.
    As the crenellations bubble out along the polar axis, the warp bubble would be pulled chaotically by the crenellation's large curvature.
    The crenellations imply that high frequency gravitational waves would be produced by an accelerating warp drive.
    Potentially, these waves could be detected by an extremely sensitive detector.
    A detector like the one proposed recently would be suitable \cite{Woods}.
    Mapping the effect on the warp bubble's path is of critical importance to plotting a complete journey to a distant star.
    
    The linear increase in the magnitude of the curvature implies that the warp drive requires a linear increase in the total amount of energy needed to accelerate the warp bubble.
    The invariant plots in Figs.~\ref{fig:5.1} to \ref{fig:5.4} show that there is a linear increase in the magnitude of the curvature the longer the warp drive accelerates.
    The invariant plots in Figs.~\ref{fig:5.5} to \ref{fig:5.8} reveal how a greater magnitude of the acceleration causes the magnitude of the invariants to increase.
    From the stress-energy tensor for the warp drive, these observations indicate that the energy requirements for an accelerating warp drive increase proportionally over time \cite{Alcubierre:1994,Natario:2001,Loup:2018}. 
    Therefore, a realistic warp drive will be able to accelerate to some finite $v_s$ that is potentially greater than $c$.
    The superluminal censorship theorem should be revisited in light of this new evidence  \cite{Barcelo:2002}.
    Further research is needed to establish the maximum achievable $v_s$ for a warp drive.
    
    The next logical extension of the research on warp drives is to combine the constant velocity Nat\'ario and the constant acceleration Nat\'ario warp drives into a complete trip to a distant star such as our closest star, Alpha Centauri.
    It would consist of four stages.
    The first stage would be under traditional rocket propulsion as the spaceship flies a safe distance away from its origin.
    Second, the warp bubble would be activated for a period of time to accelerate it at $v_s=2n(r_s) at$ until a maximum velocity of $v_s=200c$ is reached.
    The third stage would be a period of constant velocity at $v_s=200c$ as the ship travels between the stars. 
    Fourth, the warp bubble would decelerate at $v_s=-2n(r_s) at$ as the warp bubble is slowly deactivated and the ship returned to standard propulsion.
    It should be noted that Eqs.~\eqref{eq:5.8} through \eqref{eq:5.11} would need to be modified for infinitesimal hypersurface lapses increasing in proper time for the deceleration.
    Finally, the ship would continue under rocket propulsion as it begins its orbit of the destination star or planet.
    
    The flight could answer several questions about warp drive propulsion. 
    The effect of the wake and crenellations of the accelerating Nat\'ario warp drive should be identified and studied to see their effect on the origin and destination.
    If the effect is disastrous, the distance the ship would have to travel under traditional rocket propulsion before the warp drive is turned on would be known.
    The duration of each individual stage of the FTL trip could then be found.
    Once the duration is known, the amount of ``exotic matter'' and negative energy could be calculated using the methods in \cite{Alcubierre:1994,Natario:2001,Loup:2018}.
    Finally if the crenellations do cause gravitational waves, methods to detect approaching or departing warp bubbles could be developed. \\
    
    The technique of plotting the invariants can be applied to the other warp drive space times such as Alcubierre's at a constant acceleration, Krasnikov's at either constant velocity or constant acceleration, or Van Den Broeck's at either constant velocity or constant acceleration \cite{Krasnikov:1995ad,VanDenBroeck:1999sn,Loup:2018}.
    In addition, the lapse functions for the Krasnikov and Van Den Broeck's warp drives would need to be identified and then their accelerating line elements could be derived.
    After plotting their line elements for the invariants, each proposed warp drive could be compared and contrasted to their corresponding invariants at a constant velocity contained in Chapter \ref{Chapter4}.
    The Cartan invariants from Section \ref{chp6.1:Inv} computed for each warp drive element.
    The warp drive's Cartan invariants may then be compared with the wormholes and other spacetimes for equivalence. \\

\section{Closing Thoughts}
\label{Chp6.4:Close}
%Section going over key findings.
    The study of wormholes and warp drives is a rich vein that contains many nuggets to be mined.
    The research in this dissertation adds significantly to their study.
    We have investigated the curvature invariants and confirmed that wormholes may be traversable.
    We discovered that there is a significant redshift in the TS Schwarzschild wormhole that may be used to distinguish it from the classic Schwarzschild black hole.
    We compared the exponential metric to the MT metric and saw that their shape functions are potentially identical.
    The curvature invariants contained no singularities for the Alcubierre and Nat\'ario warp drives moving at a constant velocity.
    Their plots show how warp bubbles evolve over the different variables $v_s$, $\sigma$ and $\rho$.
    They reveal the never-before-seen internal structures of the warp bubbles.
    Under every condition, a safe harbor existed in the warp bubbles that allows a ship to travel peacefully.
    The Nat\'ario warp drive at a constant acceleration shared many of these properties.
    It also contains a sizable wake with rich internal structures.
    Unfortunately, it also contained a curvature singularity along the polar angle, which requires additional study.
    
    %Paragraph on the place of this dissertation in history.
    None of these findings would have been possible without developing and vigorously testing the Mathematica\textsuperscript{\textregistered} program contained in Appendix \ref{ap:Program}.
    It is extensive in its applications as it contains every tensor in GR, each NP index, and the independent CM curvature invariants.
    While this research focused on wormholes and warp drives, the program may compute the CM invariants for any known spacetime.
    The program takes as its input a metric and a tetrad, then outputs the CM curvature invariants.
    Including these inputs is standard practice for any research paper on novel spacetimes in GR.
    The program can then easily identify any curvature singularities in the spacetimes by following the same procedure as in this research.
    The program may be easily adapted to any basis set of invariants.
    The basis for computing any element in a set of invariants is already contained in Chapter \ref{chapter2} and included in the program.
    A future researcher only needs to modify the ``Computing the Invariants'' section of the program for the particular set of invariants of interest such as the Cartan invariants discussed in Section \ref{chp6.1:Inv}.
    The program lays the foundation for many branches of future research in Numerical GR.
    
    %Where does this leave us?
    While wormholes and warp drives capture our imaginations about what is possible, a realistic engineering of one is many years beyond what humanity can currently achieve.
    Taking an affect that is as visible as Dark energy on a multi-galactic scale and shrinking it to the scale of a human is beyond any construction of GR currently known.
    However, these spacetimes are prime examples of the boundary between what is possible in the mathematics of GR and what is observed experimentally.
    They provide an opportunity to test and analyze where our understanding of GR ends.
    The intention of the research contained in this dissertation is to explore this boundary intensely.
    It plays with our notions of possibility.
    The aim is to ask what is possible and why is the possibility not observed.
    This research does just that and may act as a stepping stone to further advances in our knowledge and our imagination.  \\

%% file: Appendix.tex
%\chapter{IRB Approval}

%\chapter{Additional Figures}

\chapter{Mathematica Program} \label{ap:Program}
%In this Appendix, the program for computing the CM invariants for the Natario metric at a constant velocity will be presented as a demonstrative.
The individual Mathematica programs for each spacetime metric contained in this dissertation may be found at the following website locations: 
\begin{enumerate}
    \item Morris-Thorne metric --- \\ 
    \url{https://baylor.box.com/s/fj3bhd8trjnbg6tc1zqt867oaqzmqld1}
    \item Thin-Shell Schwarzschild metric --- \\ 
    \url{https://baylor.box.com/s/fm1bt2asy3ulyyaqv9ro0f3o2p3k4fc1}
    \item Exponential metric --- \\
    \url{https://baylor.box.com/s/rfgfmlajr637g3aetd3zv28n8o70b1n5}
    \item Alcubierre metric at a constant velocity --- \\
    \url{https://baylor.box.com/s/3rllwrve15ut31kj5cruwcbr66slta9s}
    \item Nat\'ario Metric at a constant velocity --- \\
    \url{https://baylor.box.com/s/lba8r6cr2uukfxveji9foejh5a9su8fc}
    \item The invariants, $R$, for the Nat\'ario metric at a constant acceleration --- \\
    \url{https://baylor.box.com/s/lyod1zkjlmfc961ny5vosm0rxwbruxbb}
    \item The invariants, $r_1$, for the Nat\'ario metric at a constant acceleration --- \\
    \url{https://baylor.box.com/s/7y9nkx4pq2q4qnzxesyz0knxai3qoj9y}
    \item The invariants, $r_2$, for the Nat\'ario metric at a constant acceleration --- \\
    \url{https://baylor.box.com/s/tbxgl1gwsxeup2y27d8icqde737f3lv3}
    \item The invariants, $w_2$, for the Nat\'ario metric at a constant acceleration --- \\
    \url{https://baylor.box.com/s/b794s0enxthvaxt97he48f2wv0xm446p}
\end{enumerate}

\begin{figure}[ht]
    \centering
    \includegraphics[page=1,scale=0.75]{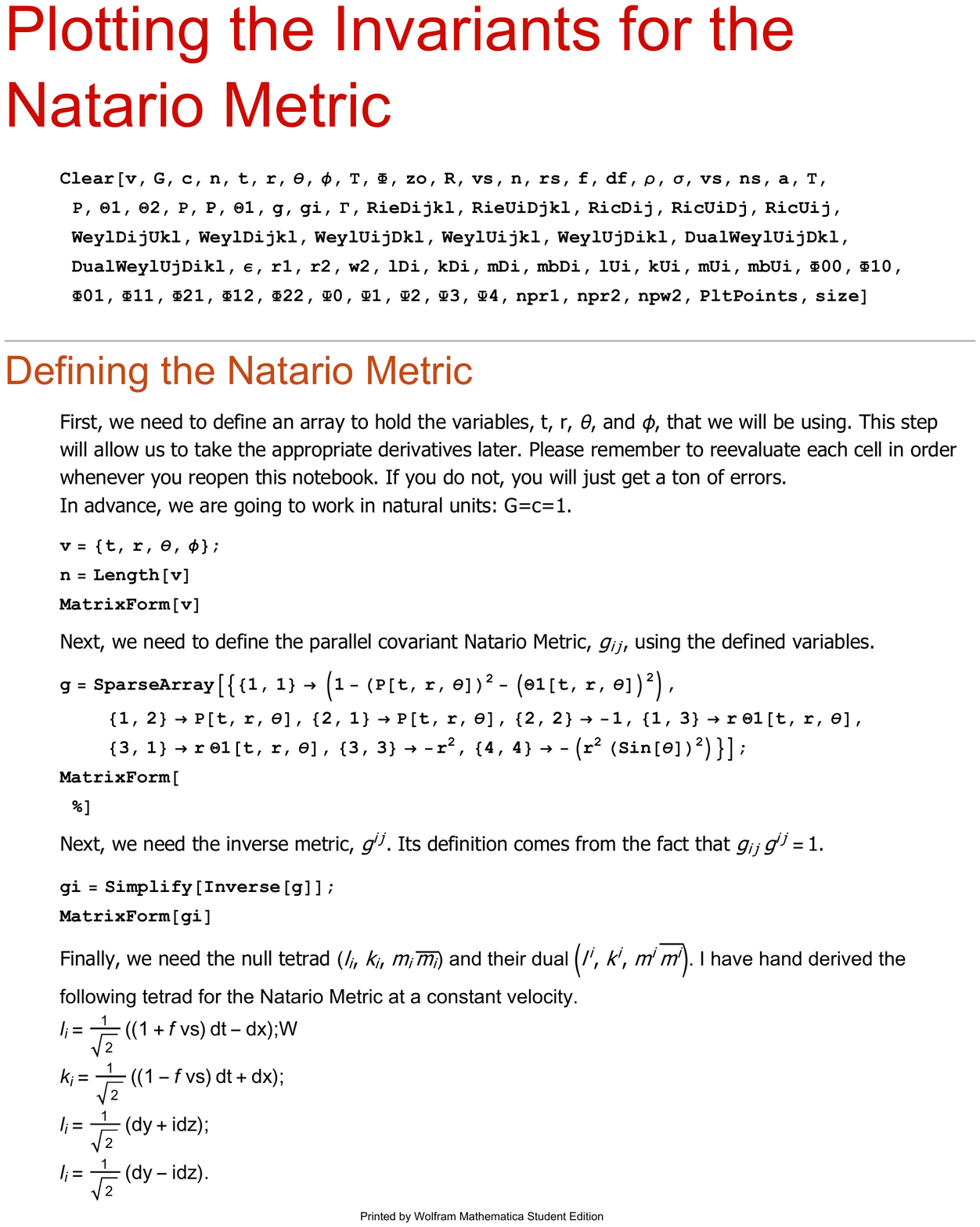}
    \label{fig:Math.p1}
\end{figure}
~
\begin{figure}[ht]
    \centering
    \includegraphics[page=2,scale=0.75]{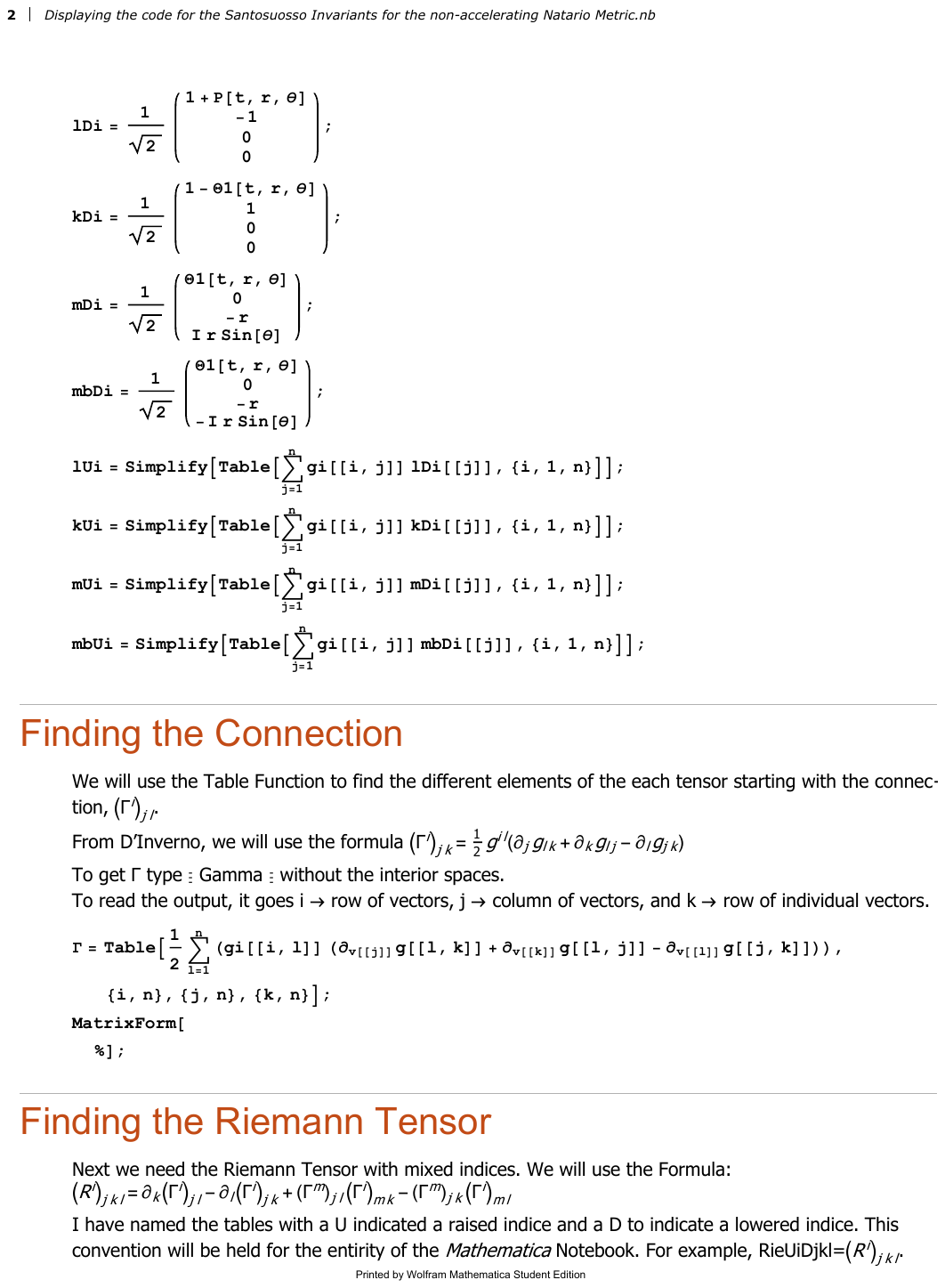}
    \label{fig:Math.p2}
\end{figure}
~
\begin{figure}[ht]
    \centering
    \includegraphics[page=3,scale=0.75]{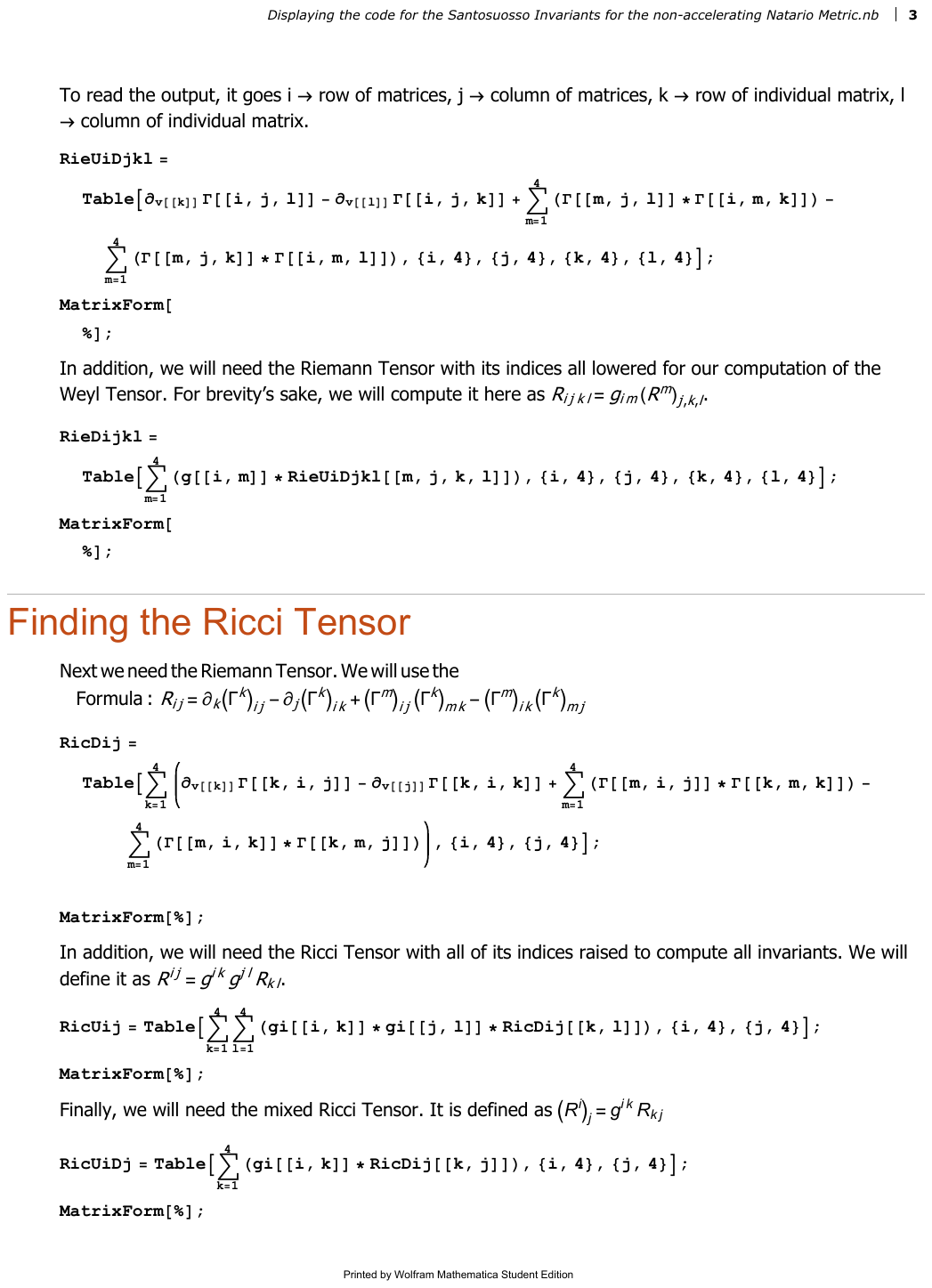}
    \label{fig:Math.p3}
\end{figure}
~
\begin{figure}[ht]
    \centering
    \includegraphics[page=4,scale=0.75]{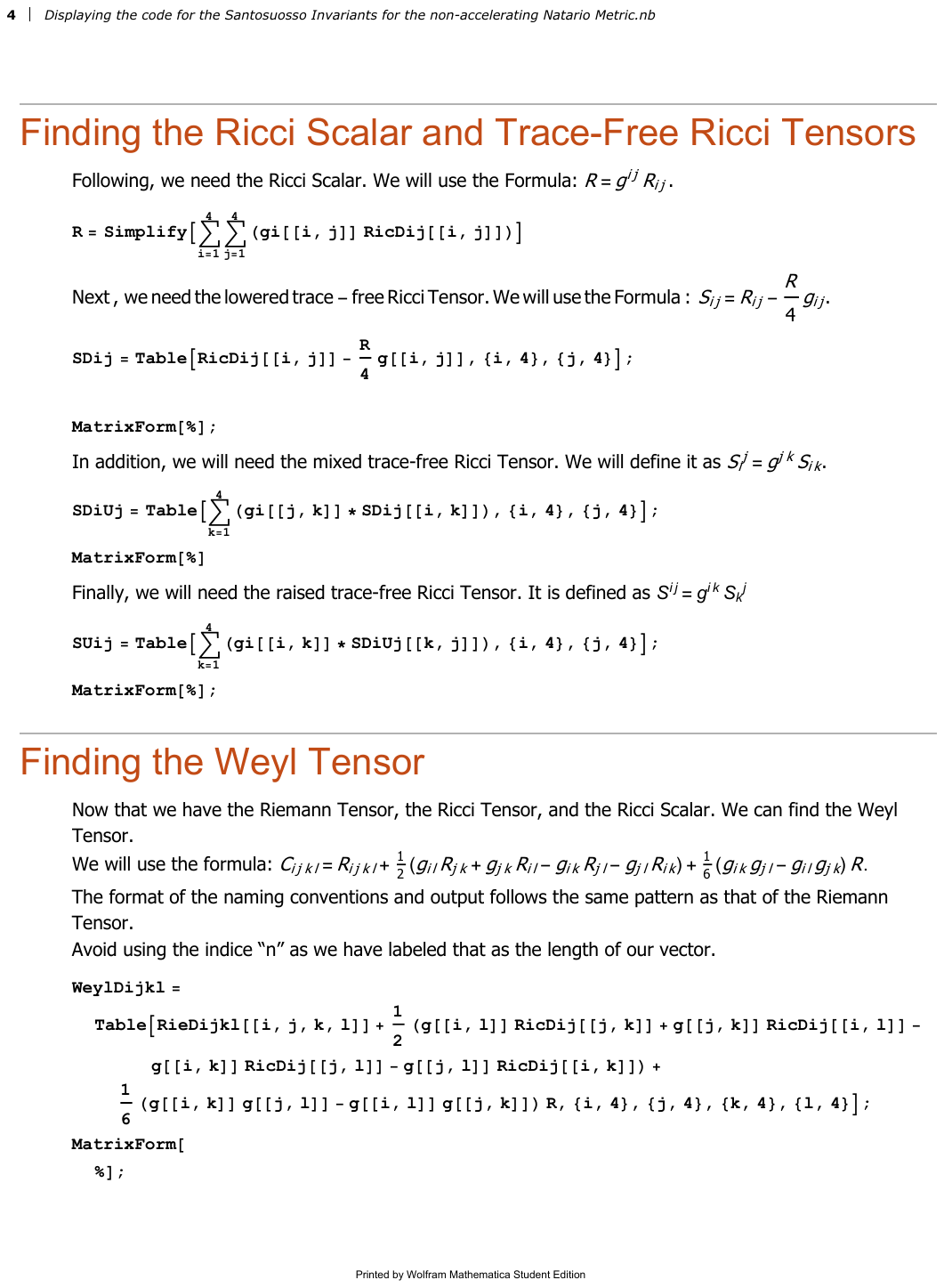}
    \label{fig:Math.p4}
\end{figure}
~
\begin{figure}[ht]
    \centering
    \includegraphics[page=5,scale=0.75]{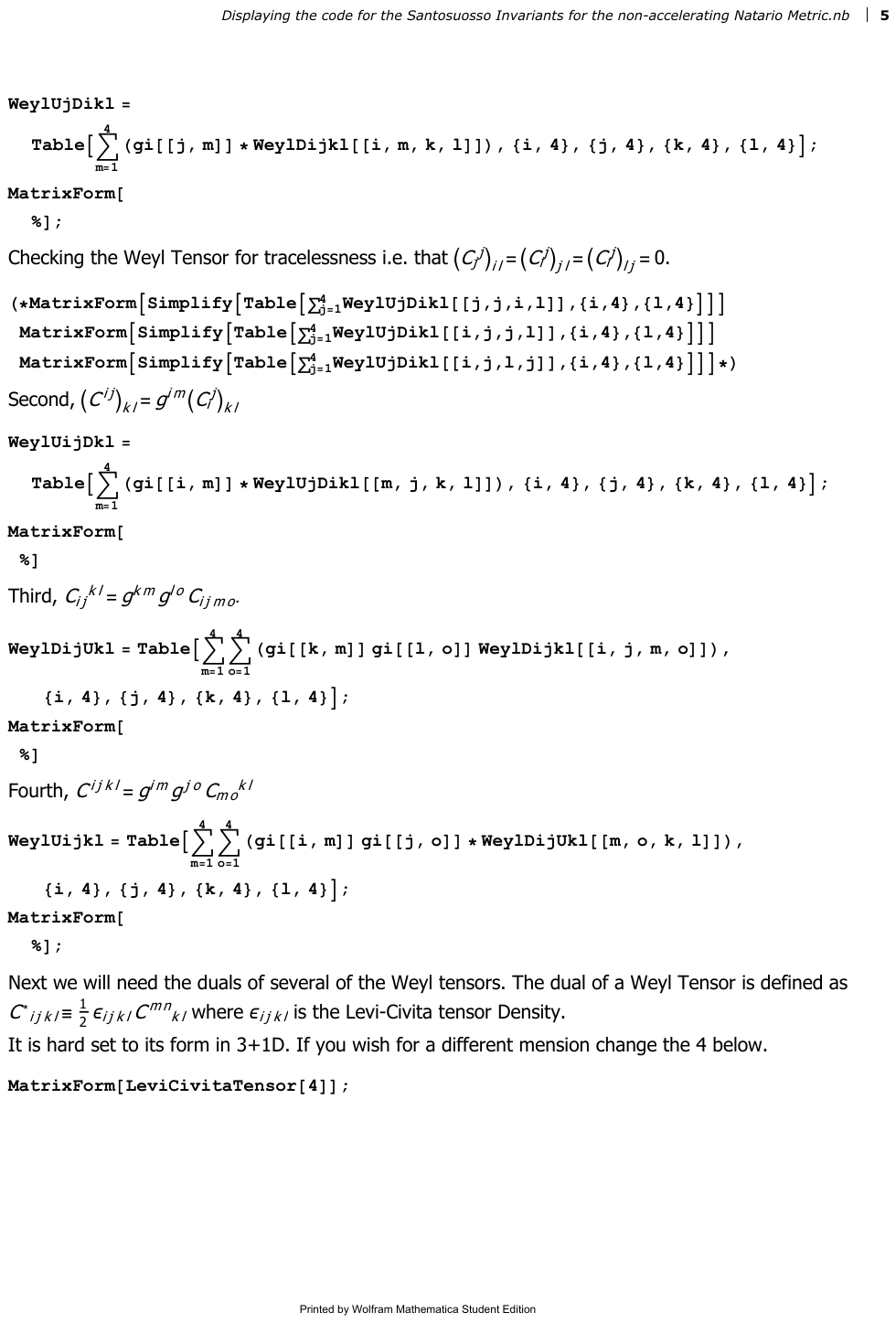}
    \label{fig:Math.p5}
\end{figure}
~
\begin{figure}[ht]
    \centering
    \includegraphics[page=6,scale=0.75]{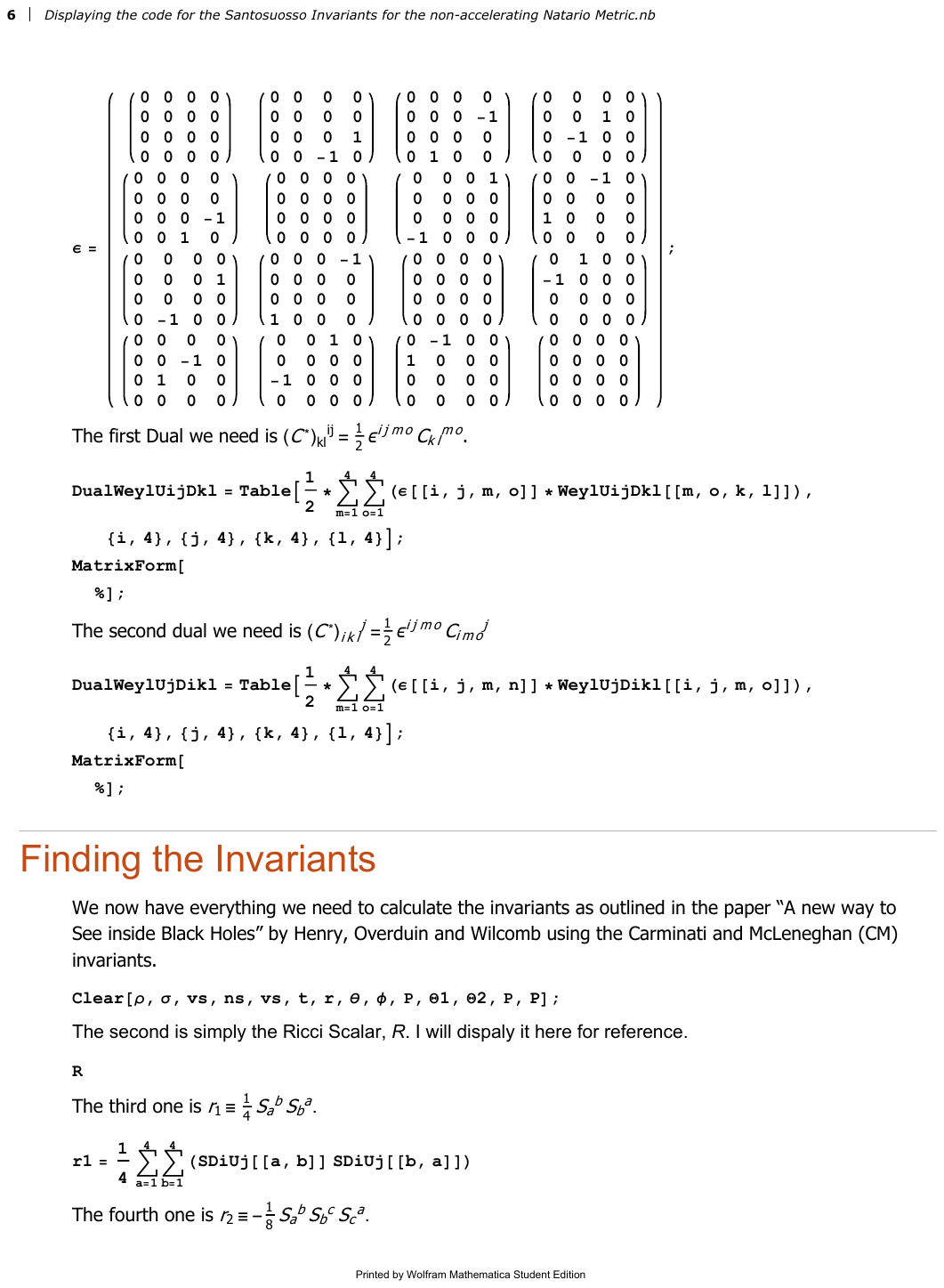}
    \label{fig:Math.p6}
\end{figure}
~
\begin{figure}[ht]
    \centering
    \includegraphics[page=7,scale=0.75]{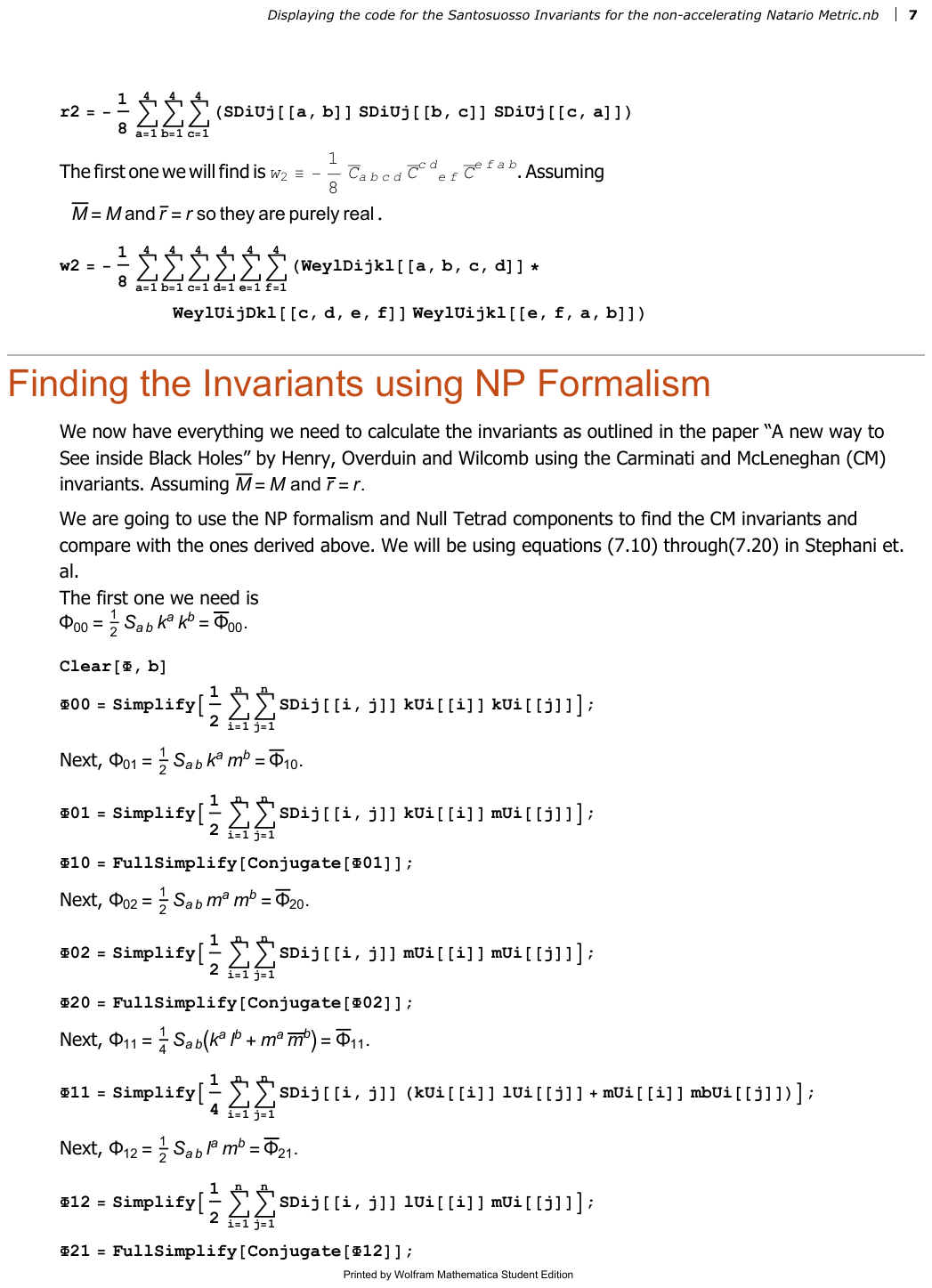}
    \label{fig:Math.p7}
\end{figure}
~
\begin{figure}[ht]
    \centering
    \includegraphics[page=8,scale=0.75]{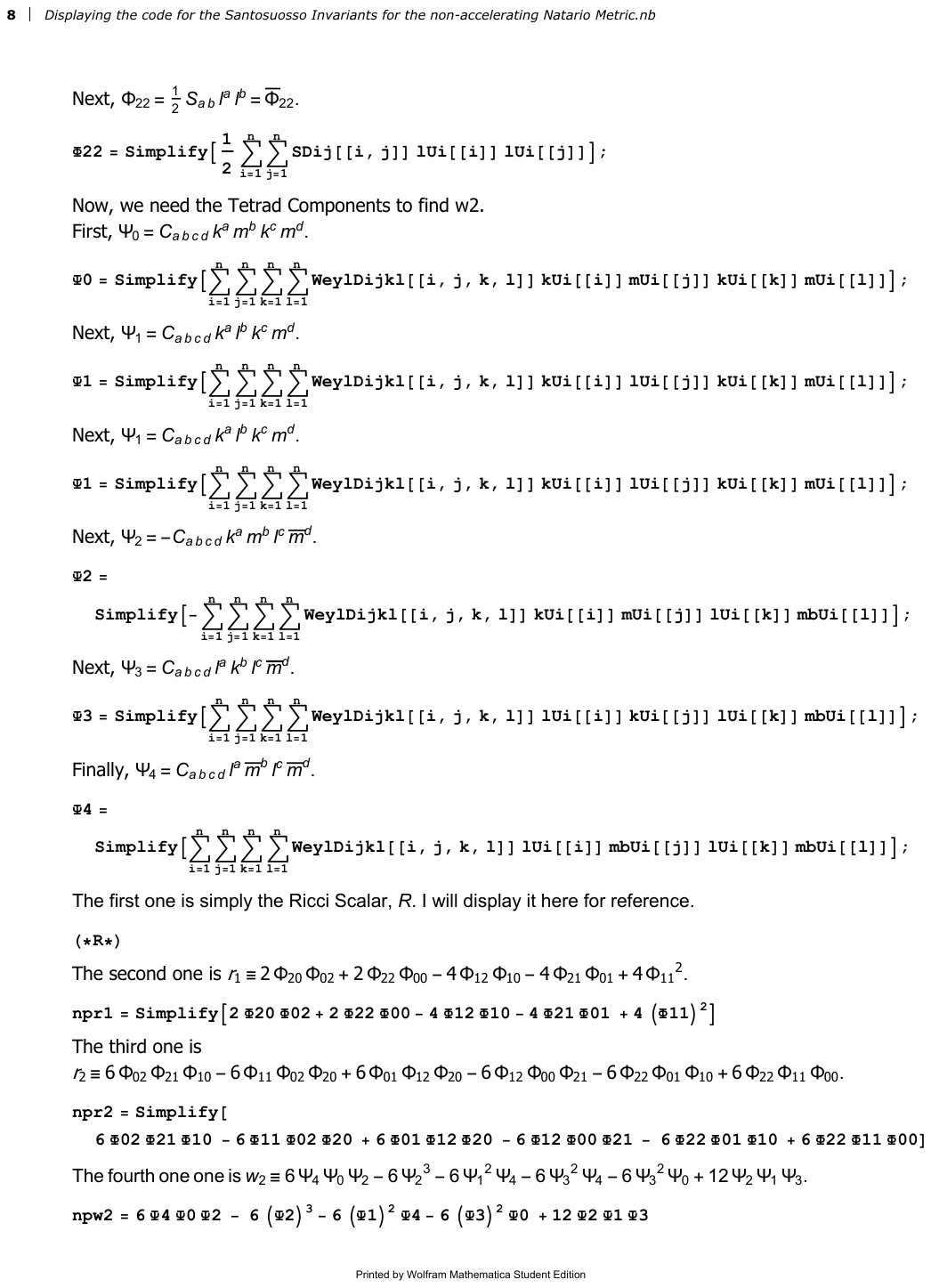}
    \label{fig:Math.p8}
\end{figure}
~
\begin{figure}[ht]
    \centering
    \includegraphics[page=9,scale=0.75]{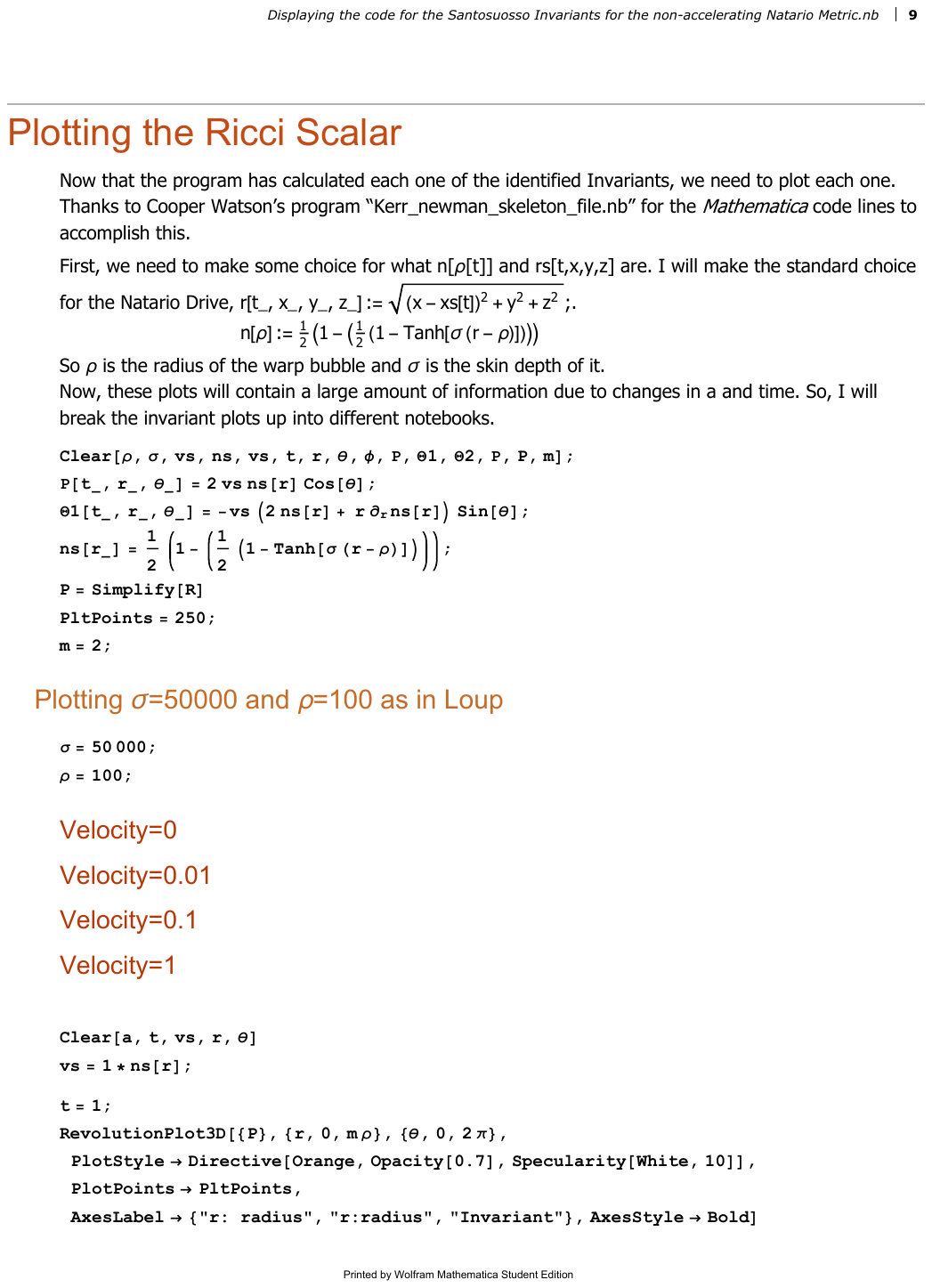}
    \label{fig:Math.p9}
\end{figure}
~
\begin{figure}[ht]
    \centering
    \includegraphics[page=10,scale=0.75]{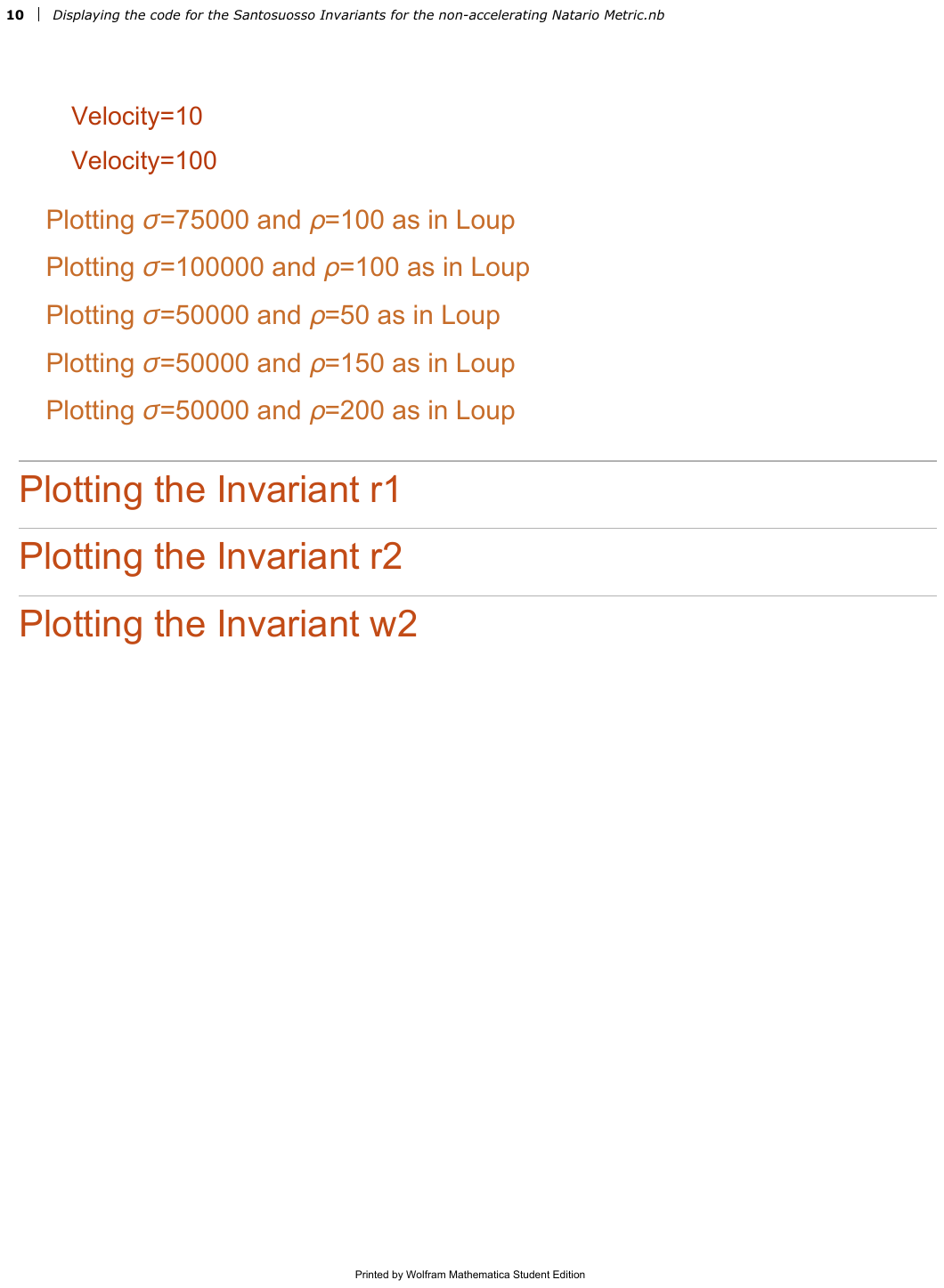}
    \label{fig:Math.p10}
\end{figure}

\newpage

\chapter{Invariants for the Nat\'ario Metric at Constant Velocity} \label{ap:vel}
    \begin{align*} 
        r_1 &= \frac{1}{1024 r^2}v_s^2 \sigma ^2 \cos ^8(\frac{\theta }{2}) \text{sech}^4((r-\rho ) \sigma ) ((32 r^4 v_s^2 \sigma ^4 (\sin (\frac{\theta }{2}) \\
        &
        -\sin (\frac{3 \theta }{2}))^2 \tanh ^6((r-\rho ) \sigma ) +64 r^3 v_s^2 \sigma ^3 (r \sigma -1) (\sin (\frac{\theta }{2}) \\
        & -\sin (\frac{3 \theta }{2}))^2 \tanh ^5((r-\rho ) \sigma ) \\
        &-4 r^2 \sigma ^2 (-r^2 \sigma ^2 v_s^2+4 r \sigma  v_s^2+(r^2 \sigma ^2 -4 r \sigma -4) \cos (4 \theta ) v_s^2-12 v_s^2+16 r^2 \sigma ^2 \\
        & -8 (v_s^2+2 r^2 \sigma ^2) \cos (2 \theta )) \sec ^2(\frac{\theta }{2}) \tanh ^4((r-\rho ) \sigma )+2 r \sigma  (-4 r^2 \sigma ^2 v_s^2+48 r \sigma  v_s^2 \\
        & +(4 r^2 \sigma ^2+16 r \sigma +3) \cos (4 \theta ) v_s^2-7 v_s^2+192 r^2 \sigma ^2 \\
        & +4 (v_s^2 (8 r \sigma +1)-48 r^2 \sigma ^2) \cos (2 \theta )) \sec ^2(\frac{\theta }{2}) \tanh ^3((r-\rho ) \sigma ) \\
        & +(48 r^2 \sigma ^2 v_s^2-28 r \sigma  v_s^2+(16 r^2 \sigma ^2+12 r \sigma -3) \cos (4 \theta ) v_s^2+7 v_s^2-768 r^2 \sigma ^2 \\
        & +4 ((8 r^2 \sigma ^2+4 r \sigma -1) v_s^2+160 r^2 \sigma ^2) \cos (2 \theta )) \sec ^2(\frac{\theta }{2}) \tanh ^2((r-\rho ) \sigma ) \\
        & +2 (-7 r \sigma  v_s^2+3 (r \sigma -1) \cos (4 \theta ) v_s^2+7 v_s^2+320 r \sigma \\
        & + 4 (v_s^2 (r \sigma -1)-16 r \sigma ) \cos (2 \theta )) \sec ^2(\frac{\theta }{2}) \tanh ((r-\rho ) \sigma ) \\
        & -2 (3 \cos (2 \theta )+5) (\cos (2 \theta ) v_s^2-v_s^2+32) \sec ^2(\frac{\theta }{2})) \sec ^6(\frac{\theta }{2}) \\
        & -16 r \sigma  \text{sech}^2((r-\rho ) \sigma ) \tan ^2(\frac{\theta }{2}) (2 r^3 v_s^2 \sigma ^3 (3 \cos (2 \theta )-1) \tanh ^4((r-\rho ) \sigma ) \\
        & +r^2 v_s^2 \sigma ^2 (2 r \sigma +(10 r \sigma -9) \cos (2 \theta )+9) \tanh ^3((r-\rho ) \sigma ) \\
        & +r \sigma  ((4 r^2 \sigma ^2+5 r \sigma -9) v_s^2 \\
        & +(4 r^2 \sigma ^2-13 r \sigma +9) \cos (2 \theta ) v_s^2-32 r^2 \sigma ^2) \tanh ^2((r-\rho ) \sigma ) \\
        & +((-4 r^2 \sigma ^2-5 r \sigma +9) v_s^2-(4 r^2 \sigma ^2-13 r \sigma -7) \cos (2 \theta ) v_s^2+96 r^2 \sigma ^2) \tanh ((r-\rho ) \sigma ) \\
        & +9 v_s^2+4 r v_s^2 \sigma -32 r \sigma +7 v_s^2 \cos (2 \theta ) +4 r v_s^2 \sigma  \cos (2 \theta )) \\
        &\sec ^4(\frac{\theta }{2})+\frac{1}{4} r^2 \sigma ^2 \text{sech}^4((r-\rho ) \sigma ) ((4 r^2 \sigma ^2 v_s^2+16 r \sigma  v_s^2 -(4 r^2 \sigma ^2+16 r \sigma -3) \cos (4 \theta ) v_s^2\\
        &+109 v_s^2-64 r^2 \sigma ^2-4 (v_s^2-16 r^2 \sigma ^2) \cos (2 \theta )) \sec ^8(\frac{\theta }{2}) \\
        & -416 r^2 v_s^2 \sigma ^2 (\cos (2 \theta )-2) \tan ^2(\frac{\theta }{2}) \tanh ^2((r-\rho ) \sigma ) \sec ^4(\frac{\theta }{2}) \\
        & -64 r v_s^2 \sigma  (2 r \sigma +2 r \cos (2 \theta ) \sigma +17) \tan ^2(\frac{\theta }{2}) \tanh ((r-\rho ) \sigma ) \sec ^4(\frac{\theta }{2}) \\
        & +704 r^4 v_s^2 \sigma ^4 \tan ^4(\frac{\theta }{2}) \tanh ^4((r-\rho ) \sigma )-2816 r^3 v_s^2 \sigma ^3 \tan ^4(\frac{\theta }{2}) \tanh ^3((r-\rho ) \sigma ))), 
        \tag{B.1} \label{eq:B.1}
    \end{align*}
    ~
    \newpage
    ~
    \begin{align*}  
        r_2 & = -\frac{1}{32768 r^3}3 v_s^4 \sigma ^3 \cos ^{12}(\frac{\theta }{2}) \text{sech}^6((r-\rho ) \sigma ) (\frac{1}{2} (\tanh ((r-\rho ) \sigma )+1) \\
        & \times (512 r^5 v_s^2 \sigma ^5 \cos ^4(\theta ) \tan ^2(\frac{\theta }{2}) \tanh ^7((r-\rho ) \sigma )+32 r^4 v_s^2 \sigma ^4 (4 r \sigma + \\
        & \times (4 r \sigma -3) \cos (2 \theta )-5) \sec ^2(\frac{\theta }{2}) (\sin (\frac{\theta }{2})-\sin (\frac{3 \theta }{2}))^2 \tanh ^6((r-\rho ) \sigma ) \\
        & +4 r^3 \sigma ^3 (2 r^2 \sigma ^2 v_s^2-10 r \sigma  v_s^2-2 (r^2 \sigma ^2-5 r \sigma -4) \cos (4 \theta ) v_s^2-r^2 \sigma ^2 \cos (6 \theta ) v_s^2 \\
        & +3 r \sigma  \cos (6 \theta ) v_s^2+4 \cos (6 \theta ) v_s^2+24 v_s^2-64 r^2 \sigma ^2 \\ 
        & +((r^2 \sigma ^2-3 r \sigma +28) v_s^2+64 r^2 \sigma ^2) \cos (2 \theta )) \sec ^4(\frac{\theta }{2}) \tanh ^5((r-\rho ) \sigma ) \\
        & +2 r^2 \sigma ^2 (-10 r^2 \sigma ^2 v_s^2+96 r \sigma  v_s^2+3 r^2 \sigma ^2 \cos (6 \theta ) v_s^2+16 r \sigma  \cos (6 \theta ) v_s^2 \\
        & +4 \cos (6 \theta ) v_s^2-20 v_s^2+816 r^2 \sigma ^2+(v_s^2 (-3 r^2 \sigma ^2+112 r \sigma +12)-832 r^2 \sigma ^2) \cos (2 \theta ) \\
        & +2 ((5 r^2 \sigma ^2+16 r \sigma +2) v_s^2+8 r^2 \sigma ^2) \cos (4 \theta )) \sec ^4(\frac{\theta }{2}) \tanh ^4((r-\rho ) \sigma ) \\
        & + r \sigma  (96 r^2 \sigma ^2 v_s^2-80 r \sigma  v_s^2+16 r^2 \sigma ^2 \cos (6 \theta ) v_s^2+16 r \sigma  \cos (6 \theta ) v_s^2-3 \cos (6 \theta ) v_s^2\\
        & +22 v_s^2 -3968 r^2 \sigma ^2+((112 r^2 \sigma ^2+48 r \sigma -13) v_s^2+3072 r^2 \sigma ^2) \cos (2 \theta )\\
        & +2 (v_s^2 (16 r^2 \sigma ^2+8 r \sigma -3) -64 r^2 \sigma ^2) \cos (4 \theta )) \sec ^4(\frac{\theta }{2}) \tanh ^3((r-\rho ) \sigma ) \\
        & +2 (-20 r^2 \sigma ^2 v_s^2+22 r \sigma  v_s^2+4 r^2 \sigma ^2 \cos (6 \theta ) v_s^2
         -3 r \sigma  \cos (6 \theta ) v_s^2-\cos (6 \theta ) v_s^2 \\
        & -2 v_s^2+2448 r^2 \sigma ^2+(v_s^2 (12 r^2 \sigma ^2-13 r \sigma +1)-448 r^2 \sigma ^2) \cos (2 \theta ) \\
        & +2 ((2 r^2 \sigma ^2-3 r \sigma +1) v_s^2+24 r^2 \sigma ^2) \cos (4 \theta )) \sec ^4(\frac{\theta }{2}) \tanh ^2((r-\rho ) \sigma ) \\
        & -(-22 r \sigma  v_s^2+3 r \sigma  \cos (6 \theta ) v_s^2+4 \cos (6 \theta ) v_s^2+8 v_s^2+3136 r \sigma +((13 r \sigma -4) v_s^2 \\
        & +1024 r \sigma ) \cos (2 \theta ) +(v_s^2 (6 r \sigma -8)-64 r \sigma ) \cos (4 \theta )) \sec ^4(\frac{\theta }{2}) \tanh ((r-\rho ) \sigma ) \\
        & +32 (\cos (4 \theta ) v_s^2-v_s^2+112 \cos (2 \theta )+144) \tan ^2(\frac{\theta }{2})) \sec ^8(\frac{\theta }{2}) \\
        & +2 r^2 \sigma ^2 \text{sech}^4((r-\rho ) \sigma ) \tan ^2(\frac{\theta }{2}) (-32 r^5 v_s^2 \sigma ^5 (\cos (2 \theta )-3) \tan ^2(\frac{\theta }{2}) \tanh ^6((r-\rho ) \sigma ) \\
        & -16 r^4 v_s^2 \sigma ^4 (-2 r \sigma +(6 r \sigma -7) \cos (2 \theta )+31) \tan ^2(\frac{\theta }{2}) \tanh ^5((r-\rho ) \sigma ) \\
        & +r^3 \sigma ^3 (-4 r^2 \sigma ^2 v_s^2-57 r \sigma  v_s^2+(4 r^2 \sigma ^2-19 r \sigma -9) \cos (4 \theta ) v_s^2+141 v_s^2+64 r^2 \sigma ^2 \\
        & +4 (v_s^2 (19 r \sigma -37)-16 r^2 \sigma ^2) \cos (2 \theta )) \sec ^4(\frac{\theta }{2}) \tanh ^4((r-\rho ) \sigma ) \\
        & -r^2 \sigma ^2 (-12 r^2 \sigma ^2 v_s^2-69 r \sigma  v_s^2+(12 r^2 \sigma ^2+17 r \sigma -15) \cos (4 \theta ) v_s^2+215 v_s^2+320 r^2 \sigma ^2 \\
        & +4 (3 v_s^2 (19 r \sigma -6)-80 r^2 \sigma ^2) \cos (2 \theta )) \sec ^4(\frac{\theta }{2}) \tanh ^3((r-\rho ) \sigma )\\
        & -r \sigma  (60 r^2 \sigma ^2 v_s^2+119 r \sigma  v_s^2 +(4 r^2 \sigma ^2-31 r \sigma +29) \cos (4 \theta ) v_s^2-157 v_s^2-512 r^2 \sigma ^2 \\
        & +8 r \sigma  ((8 r \sigma -23) v_s^2+64 r \sigma ) \cos (2 \theta )) \sec ^4(\frac{\theta }{2}) \tanh ^2((r-\rho ) \sigma ) \\
        & +(-128 r^3 \sigma ^3+12 r^3 v_s^2 \sigma ^3-416 r^2 \sigma ^2+94 r^2 v_s^2 \sigma ^2+113 r v_s^2 \sigma -98 v_s^2 \\
        & + 4 ((4 r^3 \sigma ^3+28 r^2 \sigma ^2-12 r \sigma -19) v_s^2+136 r^2 \sigma ^2) \cos (2 \theta ) \\
        & +v_s^2 (4 r^3 \sigma ^3+18 r^2 \sigma ^2-33 r \sigma -18) \cos (4 \theta )) \sec ^4(\frac{\theta }{2}) \tanh ((r-\rho ) \sigma ) \\
        & +2 (-r^2 \sigma ^2 v_s^2-22 r \sigma  v_s^2+(r^2 \sigma ^2-2 r \sigma -9) \cos (4 \theta ) v_s^2-49 v_s^2+144 r^2 \sigma ^2+224 r \sigma \\
        & +(16 r \sigma  (7 r \sigma +6)-2 v_s^2 (12 r \sigma +19)) \cos (2 \theta )) \sec ^4(\frac{\theta }{2})) \sec ^4(\frac{\theta }{2}) \\
        & +\frac{1}{8} r \sigma  \text{sech}^2((r-\rho ) \sigma ) (r^2 \sigma ^2 (72 r^2 \sigma ^2 v_s^2-664 r \sigma  v_s^2+44 r^2 \sigma ^2 \cos (6 \theta ) v_s^2\\
        & +44 r \sigma  \cos (6 \theta ) v_s^2 +69 \cos (6 \theta ) v_s^2+750 v_s^2-3904 r^2 \sigma ^2 \\
        &+((-108 r^2 \sigma ^2+532 r \sigma +139) v_s^2+5376 r^2 \sigma ^2) \cos (2 \theta ) \\
        & -2 ((4 r^2 \sigma ^2-44 r \sigma -97) v_s^2 +736 r^2 \sigma ^2) \cos (4 \theta )) \tanh ^4((r-\rho ) \sigma ) \sec ^8(\frac{\theta }{2}) \\
        & +2 r \sigma  (512 r^3 \sigma ^3-16 r^3 v_s^2 \sigma ^3+8 r^3 v_s^2 \cos (6 \theta ) \sigma ^3+2240 r^2 \sigma ^2-130 r^2 v_s^2 \sigma ^2 \\
        & +5 r^2 v_s^2 \cos (6 \theta ) \sigma ^2 +726 r v_s^2 \sigma +73 r v_s^2 \cos (6 \theta ) \sigma -186 v_s^2 \\
        & +(v_s^2 (-8 r^3 \sigma ^3+139 r^2 \sigma ^2+135 r \sigma +137) -256 r^2 \sigma ^2 (2 r \sigma +13)) \cos (2 \theta ) \\
        & +2 ((8 r^3 \sigma ^3-7 r^2 \sigma ^2+109 r \sigma +13) v_s^2+544 r^2 \sigma ^2) \cos (4 \theta ) \\
        & +23 v_s^2 \cos (6 \theta )) \tanh ^3((r-\rho ) \sigma ) \sec ^8(\frac{\theta }{2}) \\
        & +(-3968 r^3 \sigma ^3+48 r^3 v_s^2 \sigma ^3-8 r^3 v_s^2 \cos (6 \theta ) \sigma ^3 -3328 r^2 \sigma ^2+654 r^2 v_s^2 \sigma ^2 \\
        & +85 r^2 v_s^2 \cos (6 \theta ) \sigma ^2-728 r v_s^2 \sigma +100 r v_s^2 \cos (6 \theta ) \sigma +112 v_s^2 \\
        & +((8 r^3 \sigma ^3+123 r^2 \sigma ^2+540 r \sigma -82) v_s^2+3584 r^2 \sigma ^2 (r \sigma +2)) \cos (2 \theta ) \\
        & +((-48 r^3 \sigma ^3+290 r^2 \sigma ^2+88 r \sigma -32) v_s^2+384 r^2 \sigma ^2 (r \sigma -6)) \cos (4 \theta ) \\
        & +2 v_s^2 \cos (6 \theta )) \tanh ^2((r-\rho ) \sigma ) \sec ^8(\frac{\theta }{2})+2 (8 r \sigma  v_s^2+4 r \sigma  \cos (6 \theta ) v_s^2 \\
        & +\cos (6 \theta ) v_s^2+56 v_s^2 -704 r \sigma +(256 (r \sigma +7)-v_s^2 (4 r \sigma +41)) \cos (2 \theta ) \\
        & -8 (v_s^2 (r \sigma +2) -4 (14 r \sigma +3)) \cos (4 \theta )+1184) \sec ^8(\frac{\theta }{2}) \\
        & +2 (-24 r^2 \sigma ^2 v_s^2-170 r \sigma  v_s^2+4 r^2 \sigma ^2 \cos (6 \theta ) v_s^2 \\
        & +31 r \sigma  \cos (6 \theta ) v_s^2+2 \cos (6 \theta ) v_s^2+112 v_s^2+2112 r^2 \sigma ^2-192 r \sigma \\
        & -((4 r^2 \sigma ^2-129 r \sigma +82) v_s^2+256 r \sigma  (5 r \sigma +12)) \cos (2 \theta )+2 ((12 r^2 \sigma ^2+5 r \sigma -16) v_s^2 \\
        & +32 r \sigma  (3-13 r \sigma )) \cos (4 \theta )) \tanh ((r-\rho ) \sigma ) \sec ^8(\frac{\theta }{2}) \\
        & -64 r^4 \sigma ^4 (-r^2 \sigma ^2 v_s^2+8 r \sigma  v_s^2+(r^2 \sigma ^2-8 r \sigma +3) \cos (4 \theta ) v_s^2-23 v_s^2+16 r^2 \sigma ^2 \\
        & -4 (3 v_s^2+4 r^2 \sigma ^2) \cos (2 \theta )) \tan ^2(\frac{\theta }{2}) \tanh ^6((r-\rho ) \sigma ) \sec ^4(\frac{\theta }{2}) \\
        & +32 r^3 \sigma ^3 (-8 r^2 \sigma ^2 v_s^2+68 r \sigma  v_s^2+(8 r^2 \sigma ^2-20 r \sigma -13) \cos (4 \theta ) v_s^2-119 v_s^2+256 r^2 \sigma ^2 \\
        & + 4 (v_s^2 (4 r \sigma -15)-64 r^2 \sigma ^2) \cos (2 \theta )) \tan ^2(\frac{\theta }{2}) \tanh ^5((r-\rho ) \sigma ) \sec ^4(\frac{\theta }{2}) \\
        & +2048 r^6 v_s^2 \sigma ^6 \cos ^2(\theta ) \tan ^4(\frac{\theta }{2}) \tanh ^8((r-\rho ) \sigma ) \\
        & +4096 r^5 v_s^2 \sigma ^5 (r \sigma -2) \cos ^2(\theta ) \tan ^4(\frac{\theta }{2}) \tanh ^7((r-\rho ) \sigma )) \sec ^4(\frac{\theta }{2}) \\
        & +r^3 \sigma ^3 \text{sech}^6((r-\rho ) \sigma ) (-\frac{1}{16} (-(3 (2 r^2 \sigma ^2+8 r \sigma -69) v_s^2+256 r^2 \sigma ^2) \cos (2 \theta ) \\
        & +4 (3 (r^2 \sigma ^2+4 r \sigma +11) v_s^2+40 r^2 \sigma ^2) \cos (4 \theta )+3 (v_s^2 (2 r^2 \sigma ^2+8 r \sigma +3) \cos (6 \theta ) \\
        & -4 (v_s^2 (r^2 \sigma ^2+4 r \sigma -7)-8 r^2 \sigma ^2))) \sec ^{12}(\frac{\theta }{2})-\frac{1}{2} r^2 \sigma ^2 (-4 r^2 \sigma ^2 v_s^2-32 r \sigma  v_s^2 \\
        & +(4 r^2 \sigma ^2+32 r \sigma +45) \cos (4 \theta ) v_s^2-201 v_s^2+64 r^2 \sigma ^2 \\
        & +4 (69 v_s^2-16 r^2 \sigma ^2) \cos (2 \theta )) \tan ^2(\frac{\theta }{2}) \tanh ^2((r-\rho ) \sigma ) \sec ^8(\frac{\theta }{2}) \\
        & +r \sigma  (-4 r^2 \sigma ^2 v_s^2-52 r \sigma  v_s^2+(4 r^2 \sigma ^2+4 r \sigma +33) \cos (4 \theta ) v_s^2-61 v_s^2+64 r^2 \sigma ^2 \\
        & + (v_s^2 (52-48 r \sigma )-64 r^2 \sigma ^2) \cos (2 \theta )) \tan ^2(\frac{\theta }{2}) \tanh ((r-\rho ) \sigma ) \sec ^8(\frac{\theta }{2}) \\
        & -16 r^4 v_s^2 \sigma ^4 (20 \cos (2 \theta )-27) \tan ^4(\frac{\theta }{2}) \tanh ^4((r-\rho ) \sigma ) \sec ^4(\frac{\theta }{2}) \\
        & -64 r^3 v_s^2 \sigma ^3 (r \sigma +(r \sigma -4) \cos (2 \theta )+13) \tan ^4(\frac{\theta }{2}) \tanh ^3((r-\rho ) \sigma ) \sec ^4(\frac{\theta }{2}) \\
        & +192 r^6 v_s^2 \sigma ^6 \tan ^6(\frac{\theta }{2}) \tanh ^6((r-\rho ) \sigma )-1152 r^5 v_s^2 \sigma ^5 \tan ^6(\frac{\theta }{2}) \tanh ^5((r-\rho ) \sigma ))),
        \tag{B.2} \label{eq:B.2}
    \end{align*}
    ~
    \begin{align*} 
        w_2 &= \frac{1}{2415919104}v_s^4 \sigma ^3 \text{sech}^6((r-\rho ) \sigma ) (-32 v_s^2 \sigma ^3 (r v_s \sigma  \sin (\theta ) \text{sech}^2((r-\rho ) \sigma ) \\
        & +2 (v_s \cos (\theta )+v_s \sin (\theta )+v_s (\cos (\theta )+\sin (\theta )) \tanh ((r-\rho ) \sigma )+4))^3 \\
        & \times (\text{sech}^2((r-\rho ) \sigma ) (6 \cos ^2(\theta ) +7 \sin ^2(\theta )+4 r^2 \sigma ^2 \sin ^2(\theta ) \tanh ^2((r-\rho ) \sigma ) \\
        & -8 r \sigma  \sin ^2(\theta ) \tanh ((r-\rho ) \sigma )) \\
        & -3 (3 \cos (2 \theta )+1) \tanh ((r-\rho ) \sigma ) (\tanh ((r-\rho ) \sigma )+1))^3 \\
        & +\frac{27 i}{r^3} \text{sech}^8((r-\rho ) \sigma ) \sin ^4(\theta ) (\cosh (2 (r-\rho ) \sigma ) (4+4 i)+(4+4 i)+(1+i) v_s \cos (\theta ) \\
        & +i v_s \cos (\theta +2 i (r-\rho ) \sigma )+v_s \cos (\theta -2 i (r-\rho ) \sigma )+(1+i) v_s \sin (\theta ) \\
        & +(1+i) r v_s \sigma  \sin (\theta ) +v_s \sin (\theta +2 i (r-\rho ) \sigma )\\
        &+i v_s \sin (\theta -2 i (r-\rho ) \sigma ))^2 (-r^2 \sigma ^2 v_s^2-2 r \sigma  v_s^2+r^2 \sigma ^2 \cos (2 \theta ) v_s^2 \\
        & +2 r \sigma  \cos (2 \theta ) v_s^2-2 i \cos (2 \theta +2 i (r-\rho ) \sigma ) v_s^2+(1-i) r \sigma  \cos (2 \theta +2 i (r-\rho ) \sigma ) v_s^2 \\
        & -i \cos (2 \theta +4 i (r-\rho ) \sigma ) v_s^2+2 i \cos (2 \theta -2 i (r-\rho ) \sigma ) v_s^2\\
        & +(1+i) r \sigma  \cos (2 \theta -2 i (r-\rho ) \sigma ) v_s^2 +i \cos (2 \theta -4 i (r-\rho ) \sigma ) v_s^2-2 r \sigma  \sin (2 \theta ) v_s^2\\
        &-2 \sin (2 \theta ) v_s^2-(1+i) r \sigma  \sin (2 \theta +2 i (r-\rho ) \sigma ) v_s^2 -2 \sin (2 \theta +2 i (r-\rho ) \sigma ) v_s^2\\
        &-\sin (2 \theta +4 i (r-\rho ) \sigma ) v_s^2-(1-i) r \sigma  \sin (2 \theta -2 i (r-\rho ) \sigma ) v_s^2 \\
        & -2 \sin (2 \theta -2 i (r-\rho ) \sigma ) v_s^2-\sin (2 \theta -4 i (r-\rho ) \sigma ) v_s^2-2 r \sigma  \sinh (2 (r-\rho ) \sigma ) v_s^2 \\
        & -4 \sinh (2 (r-\rho ) \sigma ) v_s^2-2 \sinh (4 (r-\rho ) \sigma ) v_s^2-2 v_s^2-12 \cos (\theta ) v_s\\
        &-(8+4 i) \cos (\theta +2 i (r-\rho ) \sigma ) v_s-(2+2 i) \cos (\theta +4 i (r-\rho ) \sigma ) v_s \\
        & -(8-4 i) \cos (\theta -2 i (r-\rho ) \sigma ) v_s-(2-2 i) \cos (\theta -4 i (r-\rho ) \sigma ) v_s -8 r \sigma  \sin (\theta ) v_s \\
        &-12 \sin (\theta ) v_s -(8-4 i) \sin (\theta +2 i (r-\rho ) \sigma ) v_s-4 r \sigma  \sin (\theta +2 i (r-\rho ) \sigma ) v_s\\
        &-(2-2 i) \sin (\theta +4 i (r-\rho ) \sigma ) v_s -(8+4 i) \sin (\theta -2 i (r-\rho ) \sigma ) v_s\\
        &-4 r \sigma  \sin (\theta -2 i (r-\rho ) \sigma ) v_s-(2+2 i) \sin (\theta -4 i (r-\rho ) \sigma ) v_s \\
        & -2 ((r \sigma +2) v_s^2+16) \cosh (2 (r-\rho ) \sigma )\\
        &-2 (v_s^2+4) \cosh (4 (r-\rho ) \sigma )-24) (\tanh ((r-\rho ) \sigma )+1) \\
        & \times (r \sigma  \tanh ((r-\rho ) \sigma )-1) ((4 r \sigma -v_s (r^2 \sigma ^2-r \sigma +1) \cos (\theta )) \tanh ^3((r-\rho ) \sigma ) \\
        & -3 (2 r^2 \sigma ^2+v_s (r^2 \sigma ^2-r \sigma +1) \cos (\theta )-2) \tanh ^2((r-\rho ) \sigma ) \\
        & -3 (v_s (r^2 \sigma ^2-r \sigma +1) \cos (\theta )-4 r \sigma ) \tanh ((r-\rho ) \sigma )-2 r^2 \sigma ^2-r^2 v_s \sigma ^2 \cos (\theta )\\
        & -v_s \cos (\theta ) +r v_s \sigma  \cos (\theta )+\text{sech}^2((r-\rho ) \sigma ) (6 r^2 \sigma ^2+v_s (3 r^2 \sigma ^2+r \sigma -3) \cos (\theta ) \\
        & + (4 r \sigma +v_s (3 r^2 \sigma ^2+r \sigma -1) \cos (\theta )) \tanh ((r-\rho ) \sigma )+6)+2)^2\\
        & + \frac{2304}{r^2}v_s^2(\sigma  \sin ^4(\theta ) (\tanh ((r-\rho ) \sigma )+1)^2 (r \sigma  \tanh ((r-\rho ) \sigma )-1)^2\\
        & \times (r v_s \sigma  \sin (\theta ) \text{sech}^2((r-\rho ) \sigma ) \\
        & +2 (v_s \cos (\theta )+v_s \sin (\theta )+v_s (\cos (\theta )+\sin (\theta )) \tanh ((r-\rho ) \sigma )+4)) \\
        &\times (\text{sech}^2((r-\rho ) \sigma ) (6 \cos ^2(\theta ) +7 \sin ^2(\theta )+4 r^2 \sigma ^2 \sin ^2(\theta ) \tanh ^2((r-\rho ) \sigma )\\
        &-8 r \sigma  \sin ^2(\theta ) \tanh ((r-\rho ) \sigma ))\\
        & -3 (3 \cos (2 \theta )+1) \tanh ((r-\rho ) \sigma ) (\tanh ((r-\rho ) \sigma )+1))\\
        &\times (r^2 v_s^2 \sigma ^2 \sin ^2(\theta ) \text{sech}^4((r-\rho ) \sigma ) \\
        & +2 r v_s \sigma  (2 \sin (\theta ) (v_s \cos (\theta )+v_s \sin (\theta )+2)+v_s (2 \sin ^2(\theta )+\sin (2 \theta ))\\
        &\times \tanh ((r-\rho ) \sigma )) \text{sech}^2((r-\rho ) \sigma ) \\
        & +2 (-\text{sech}^2((r-\rho ) \sigma ) (\cos (\theta )+\sin (\theta ))^2 v_s^2+(\cos (\theta )+\sin (\theta ))^2 \tanh ^2((r-\rho ) \sigma ) v_s^2 \\
        & +6 \cos (\theta ) \sin (\theta ) v_s^2+3 v_s^2+8 \cos (\theta ) v_s+8 \sin (\theta ) v_s \\
        & +4 (\cos (\theta )+\sin (\theta )) (v_s \cos (\theta )+v_s \sin (\theta )+2) \tanh ((r-\rho ) \sigma ) v_s+16))) \\
        & -\frac{144 \sigma  \sin ^2(\theta )}{r^2} (r v_s \sigma  \sin (\theta ) \text{sech}^2((r-\rho ) \sigma )+2 (v_s \cos (\theta ) +v_s \sin (\theta ) \\
        & +v_s (\cos (\theta )+\sin (\theta )) \tanh ((r-\rho ) \sigma )+4))^3 (\text{sech}^2((r-\rho ) \sigma ) (6 \cos ^2(\theta )+7 \sin ^2(\theta ) \\
        & +4 r^2 \sigma ^2 \sin ^2(\theta ) \tanh ^2((r-\rho ) \sigma )-8 r \sigma  \sin ^2(\theta ) \tanh ((r-\rho ) \sigma )) \\ 
        & -3 (3 \cos (2 \theta )+1) \tanh ((r-\rho ) \sigma ) (\tanh ((r-\rho ) \sigma )+1)) \\
        & \times (r \sigma  (2 r \sigma +v_s (r \sigma +1) \cos (\theta )) \text{sech}^2((r-\rho ) \sigma ) \\
        & -2 (r^2 v_s \sigma ^2 \cos (\theta ) \tanh ^3((r-\rho ) \sigma )+r \sigma  (2 r \sigma +v_s (r \sigma -1) \cos (\theta )) \tanh ^2((r-\rho ) \sigma ) \\
        & +((v_s-r v_s \sigma ) \cos (\theta )-4 r \sigma ) \tanh ((r-\rho ) \sigma )+v_s \cos (\theta )-2)) \\
        & \times (\frac{1}{2} r^2 v_s^2 \sigma ^2 (r \sigma +1) \sin (2 \theta ) \text{sech}^4((r-\rho ) \sigma )+r \sigma  (-r^2 v_s^2 \sigma ^2 \sin (2 \theta ) \tanh ^3((r-\rho ) \sigma ) \\
        & -r v_s^2 \sigma  (r \sigma -1) \sin (2 \theta ) \tanh ^2((r-\rho ) \sigma )+2 v_s^2 \cos (\theta ) (r \sigma  \cos (\theta )+\cos (\theta ) \\
        & + 2 r \sigma  \sin (\theta )) \tanh ((r-\rho ) \sigma )+2 v_s^2 (r \sigma +1) \cos ^2(\theta )+4 v_s (r \sigma +1) \cos (\theta ) \\
        & + r \sigma  (v_s^2 \sin (2 \theta )-8)) \text{sech}^2((r-\rho ) \sigma )\\
        &-2 (r^2 v_s^2 \sigma ^2 (\cos (2 \theta )+\sin (2 \theta )+1) \tanh ^4((r-\rho ) \sigma ) \\
        & +2 r v_s \sigma  \cos (\theta ) (2 r \sigma +v_s (2 r \sigma -1) \cos (\theta )+v_s (2 r \sigma -1) \sin (\theta )) \tanh ^3((r-\rho ) \sigma ) \\
        & + (-8 r^2 \sigma ^2+4 r v_s (r \sigma -1) \cos (\theta ) \sigma +2 v_s^2 (r \sigma -1)^2 \cos ^2(\theta ) \\
        & + v_s^2 (r \sigma -1)^2 \sin (2 \theta )) \tanh ^2((r-\rho ) \sigma )+(-2 (r \sigma -2) \cos ^2(\theta ) v_s^2 \\
        & -(r \sigma -2) \sin (2 \theta ) v_s^2-4 (r \sigma -1) \cos (\theta ) v_s+16 r \sigma ) \tanh ((r-\rho ) \sigma )+2 v_s^2 \cos ^2(\theta ) \\
        & +4 v_s \cos (\theta )+v_s^2 \sin (2 \theta )+8))\\
        & +\frac{864}{r^3} \sin ^4(\theta ) (\tanh ((r-\rho ) \sigma )+1) (r \sigma  \tanh ((r-\rho ) \sigma )-1) (r v_s \sigma  \sin (\theta ) \text{sech}^2((r-\rho ) \sigma )\\
        &+2 (v_s \cos (\theta )+v_s \sin (\theta )+v_s (\cos (\theta )+\sin (\theta )) \tanh ((r-\rho ) \sigma )+4))^2 \\
        & \times (\frac{1}{2} r^2 v_s^2 \sigma ^2 (r \sigma +1) \sin (2 \theta ) \text{sech}^4((r-\rho ) \sigma )+r \sigma  (-r^2 v_s^2 \sigma ^2 \sin (2 \theta ) \tanh ^3((r-\rho ) \sigma ) \\
        & -r v_s^2 \sigma  (r \sigma -1) \sin (2 \theta ) \tanh ^2((r-\rho ) \sigma )+2 v_s^2 \cos (\theta ) (r \sigma  \cos (\theta )+\cos (\theta )+2 r \sigma  \sin (\theta )) \\
        & \times \tanh ((r-\rho ) \sigma )+2 v_s^2 (r \sigma +1) \cos ^2(\theta )+4 v_s (r \sigma +1) \cos (\theta )\\
        & +r \sigma  (v_s^2 \sin (2 \theta )-8)) \text{sech}^2((r-\rho ) \sigma ) \\
        & -2 (r^2 v_s^2 \sigma ^2 (\cos (2 \theta )+\sin (2 \theta )+1) \tanh ^4((r-\rho ) \sigma )\\
        &+2 r v_s \sigma  \cos (\theta ) (2 r \sigma +v_s (2 r \sigma -1) \cos (\theta ) +v_s (2 r \sigma -1) \sin (\theta )) \tanh ^3((r-\rho ) \sigma )\\
        &+(-8 r^2 \sigma ^2+4 r v_s (r \sigma -1) \cos (\theta ) \sigma +2 v_s^2 (r \sigma -1)^2 \cos ^2(\theta ) \\
        & +v_s^2 (r \sigma -1)^2 \sin (2 \theta )) \tanh ^2((r-\rho ) \sigma )\\
        & +(-2 (r \sigma -2) \cos ^2(\theta ) v_s^2-(r \sigma -2) \sin (2 \theta ) v_s^2 -4 (r \sigma -1) \cos (\theta ) v_s\\
        & +16 r \sigma ) \tanh ((r-\rho ) \sigma )+2 v_s^2 \cos ^2(\theta )+4 v_s \cos (\theta )+v_s^2 \sin (2 \theta )+8))^2).
        \tag{B.3} \label{eq:B.3}
    \end{align*}
    
    \newpage
    
\chapter{Invariants for the Nat\'ario Metric at Constant Acceleration} \label{ap:acc}
    \begin{align*} \label{eq:C.1}
        R &= -\frac{a}{32 r^2 (a r \cos (\theta )+a r \tanh ((r-\rho ) \sigma ) \cos (\theta )-2)^3} \\
        & \times (a r^8 t^2 \sigma ^4 \sin ^2(\theta ) (a r \cos (\theta )+a r \tanh ((r-\rho ) \sigma ) \cos (\theta )-2) \text{sech}^8((r-\rho ) \sigma ) \\
        & -4 a r^3 t^2 \sigma ^3 (a \sigma  \cos (\theta ) \sin ^2(\theta ) \tanh ^3((r-\rho ) \sigma ) r^6 \\
        & +(a (2 r \sigma -3) \cos (\theta )-2 \sigma ) \sin ^2(\theta ) \tanh ^2((r-\rho ) \sigma ) r^5 \\
        & -a (3 r^2+2) \cos (\theta ) \sin ^2(\theta ) r^3+(a (\sigma  r^3-6 r^2-2) \cos (\theta ) \\
        & -2 r (r \sigma -3)) \sin ^2(\theta ) \tanh ((r-\rho ) \sigma ) r^3 +2 (3 r^2+2) \sin ^2(\theta ) r^2\\
        & +16 a (2 r \sigma +3) \cos ^3(\theta ) r-16 (r^2+4 \sigma  r-2) \cos ^2(\theta )) \text{sech}^6((r-\rho ) \sigma ) \\
        & +4 a r^2 t^2 \sigma ^2 (a \sigma ^2 \cos (\theta ) \sin ^2(\theta ) \tanh ^5((r-\rho ) \sigma ) r^7 \\
        & +\sigma  (3 a (r \sigma -2) \cos (\theta )-2 \sigma ) \sin ^2(\theta ) \tanh ^4((r-\rho ) \sigma ) r^6+3 a \cos ^3(\theta ) r^5\\
        & +5 a \sin (\theta ) \sin (2 \theta ) r^5 -6 \cos ^2(\theta ) r^4-16 \sin ^2(\theta ) r^4-64 a \cos ^3(\theta ) r^3\\
        & +13 a \cos (\theta ) \sin ^2(\theta ) r^3-16 a \sigma  \cos ^3(\theta ) r^2 \\
        & +184 \cos ^2(\theta ) r^2-58 \sin ^2(\theta ) r^2+104 a \cos ^3(\theta ) r+(-4 \sigma  (r \sigma -3) \sin ^2(\theta ) r^4 \\
        & +a (3 \sigma ^2 r^4-18 \sigma  r^3+10 r^2-4 \sigma  r+1) \cos (\theta ) \sin ^2(\theta ) r^2 \\
        & +a (3 r^4+32 \sigma  r^3+8 (16 \sigma ^2-1) r^2-96 \sigma  r+16) \cos ^3(\theta )) \tanh ^3((r-\rho ) \sigma ) r \\
        & +32 \sigma  \cos ^2(\theta ) r + 4 a \cos (\theta ) \sin ^2(\theta ) r-432 \cos ^2(\theta )-8 \sin ^2(\theta ) \\
        & +(a (\sigma ^2 r^4-18 \sigma  r^3+30 r^2-8 \sigma  r+15) \cos (\theta ) \sin ^2(\theta ) r^3 \\
        & -2 (\sigma ^2 r^4-12 \sigma  r^3+8 r^2-12 \sigma  r+5) \sin ^2(\theta ) r^2\\
        & +a (9 r^4+64 \sigma  r^3 +16 (8 \sigma ^2-5) r^2-432 \sigma  r+88) \cos ^3(\theta ) r\\
        & -2 (3 r^4+32 \sigma  r^3+4 (32 \sigma ^2-3) r^2-112 \sigma  r +24) \cos ^2(\theta )) \tanh ^2((r-\rho ) \sigma )\\
        & +(a r (9 r^4+32 \sigma  r^3-136 r^2-352 \sigma  r+176) \cos ^3(\theta ) \\
        & -4 (3 r^4+16 \sigma  r^3-52 r^2-192 \sigma  r+72) \cos ^2(\theta )\\
        & -a r (6 \sigma  r^5-30 r^4+4 \sigma  r^3-27 r^2-4) \sin ^2(\theta ) \cos (\theta ) \\
        & +4 r^2 (3 \sigma  r^3-8 r^2+6 \sigma  r-17) \sin ^2(\theta )) \tanh ((r-\rho ) \sigma )) \text{sech}^4((r-\rho ) \sigma ) \\
        & -8 r \sigma  (-6 a^2 t^2 \cos ^3(\theta ) r^5-3 a^2 t^2 \cos (\theta ) \sin ^2(\theta ) r^5+12 a t^2 \cos ^2(\theta ) r^4+8 a \cos ^2(\theta ) r^4 \\
        & -2 a t^2 \sin ^2(\theta ) r^4+32 a^2 \cos ^3(\theta ) r^3+52 a^2 t^2 \cos ^3(\theta ) r^3-5 a^2 t^2 \cos (\theta ) \sin ^2(\theta ) r^3 \\
        & -\frac{1}{2} a^2 t^2 \tanh ^5((r-\rho ) \sigma ) r^2 \sigma  \cos (\theta ) \\
        & \times (-r^4+3 r^2+16 \sigma  r+(r^4+5 r^2+16 \sigma  r-8) \cos (2 \theta )-8) \\
        & -16 \cos (\theta ) r^3 -132 a t^2 \cos ^2(\theta ) r^2-192 a \cos ^2(\theta ) r^2+46 a t^2 \sin ^2(\theta ) r^2 \\
        & -\frac{1}{4} a t^2 (4 a \sigma  \cos (3 \theta ) r^5+4 \sigma  r^4 +3 a \cos (3 \theta ) r^4+36 a \sigma  \cos (3 \theta ) r^3-12 \sigma  r^2 \\
        & +48 a \sigma ^2 \cos (3 \theta ) r^2 -23 a \cos (3 \theta ) r^2-64 \sigma ^2 r -72 a \sigma  \cos (3 \theta ) r+32 \sigma \\
        & -a (4 \sigma  r^5-21 r^4-92 \sigma  r^3+(57-144 \sigma ^2) r^2+216 \sigma  r-24) \cos (\theta ) \\
        & -4 \sigma  (r^4+5 r^2+16 \sigma  r-8) \cos (2 \theta )+8 a \cos (3 \theta )) \tanh ^4((r-\rho ) \sigma ) r \\
        & -12 a^2 t^2 \cos ^3(\theta ) r-2 a^2 t^2 \cos (\theta ) \sin ^2(\theta ) r+256 \cos (\theta ) r \\
        & -2 a (2 a r (4 \sigma  r^3+t^2 (6 r^4+18 \sigma  r^3+4 (3 \sigma ^2-7) r^2-42 \sigma  r+9)) \cos ^3(\theta ) \\
        & -2 t^2 (3 r^4+14 \sigma  r^3+(16 \sigma ^2-13) r^2-36 \sigma  r+8) \cos ^2(\theta ) \\
        & -a r (r^2+1) t^2 (3 \sigma  r^3-6 r^2-1) \sin ^2(\theta ) \cos (\theta )\\
        & +r^2 t^2 (3 \sigma  r^3+r^2+11 \sigma  r-9) \sin ^2(\theta )) \tanh ^3((r-\rho ) \sigma )+80 a t^2 \cos ^2(\theta )+4 a t^2 \sin ^2(\theta ) \\
        & +2 a (-2 a r (9 t^2 r^4+8 (2 \sigma  t^2+\sigma ) r^3+(t^2 (4 \sigma ^2-54)-8) r^2-38 t^2 \sigma  r+15 t^2) \cos ^3(\theta ) \\
        & +2 ((9 t^2+2) r^4+2 (11 t^2+12) \sigma  r^3+(t^2 (8 \sigma ^2-59)-4) r^2-60 t^2 \sigma  r+36 t^2) \cos ^2(\theta ) \\
        & +a r (r^2+1) t^2 (2 \sigma  r^3-9 r^2-3) \sin ^2(\theta ) \cos (\theta ) \\
        & +t^2 (-3 \sigma  r^5-3 r^4-19 \sigma  r^3+41 r^2+2) \sin ^2(\theta )) \tanh ^2((r-\rho ) \sigma ) \\
        & +(-4 a^2 r (6 t^2 r^4+(5 t^2+4) \sigma  r^3-4 (11 t^2+4) r^2-12 t^2 \sigma  r+11 t^2) \cos ^3(\theta ) \\
        & +4 a ((9 t^2+4) r^4+2 (5 t^2+12) \sigma  r^3-(79 t^2+52) r^2-28 t^2 \sigma  r+48 t^2) \cos ^2(\theta ) \\
        & +r (a^2 (r^2+1) t^2 (\sigma  r^3-12 r^2-6) \sin ^2(\theta )-16 (r^2+8 \sigma  r-2)) \cos (\theta ) \\
        & -2 a t^2 (\sigma  r^5+3 r^4+9 \sigma  r^3-55 r^2-4) \sin ^2(\theta )) \tanh ((r-\rho ) \sigma )) \text{sech}^2((r-\rho ) \sigma ) \\
        & +(\tanh ((r-\rho ) \sigma )+1)^2 (2 a^2 r t^2 \cos (\theta ) (13 r^4-22 r^2 \\
        & +(11 r^4-26 r^2-5) \cos (2 \theta )-3) \tanh ^3((r-\rho ) \sigma ) \\
        & +a t^2 (33 a \cos (3 \theta ) r^5-20 r^4-78 a \cos (3 \theta ) r^3+88 r^2+3 a (37 r^4-70 r^2-11) \cos (\theta ) r \\
        & -15 a \cos (3 \theta ) r-4 (19 r^4-42 r^2+3) \cos (2 \theta )-20) \tanh ^2((r-\rho ) \sigma ) \\
        & +a (33 a t^2 \cos (3 \theta ) r^5-40 t^2 r^4-64 r^4-78 a t^2 \cos (3 \theta ) r^3+176 t^2 r^2+64 r^2 \\
        & +3 a (37 r^4-70 r^2-11) t^2 \cos (\theta ) r-15 a t^2 \cos (3 \theta ) r-40 t^2-8 ((19 t^2+8) r^4 \\
        & -2 (21 t^2+4) r^2+3 t^2) \cos (2 \theta )) \tanh ((r-\rho ) \sigma )+4 (4 a^2 r (3 r^4-6 r^2-1) t^2 \cos ^3(\theta ) \\
        & -8 a ((3 t^2+4) r^4-4 (2 t^2+1) r^2+t^2) \cos ^2(\theta )+r (a^2 (r^2+1)^2 t^2 \sin ^2(\theta ) \\
        & +64 (r^2-1)) \cos (\theta )+2 a (7 r^4-10 r^2-1) t^2 \sin ^2(\theta )))). \tag{C.1}  
    \end{align*}
    
    The remaining invariants may be found at
    \begin{enumerate}
        \item $r_1$ --- \\ \url{https://drive.google.com/file/d/1Lkww1oZizm-C7lifv-TfWIdJqQEHEn-e/view?usp=sharing}
        \item $r_2$ --- \\ \url{https://drive.google.com/file/d/1Rd7sXTKFE6F2v6p-uLUx-V3VFJ_A9r9o/view?usp=sharing}
        \item $w_2$ --- \\ \url{https://drive.google.com/file/d/17WnjSULu6PxcnHtk_FBlO1ZkFIkVcdMr/view?usp=sharing} 
    \end{enumerate}

%% file: References.tex
{}